\documentclass[11pt,twoside,letterpaper]{article} 
\newcommand{\rvec}{\mathrm {\mathbf {r}}} 
\newcommand{\pvec}{\mathrm {\mathbf {p}}}
\usepackage{times,fancyhdr}
\usepackage[dvips]{graphicx}
\usepackage{color}
\usepackage{amsfonts}
\usepackage{url}
\usepackage{subfigure}
\usepackage{subfigure}
\usepackage{xcolor}
\usepackage{amsmath}
\usepackage{enumerate}
\usepackage{tabularx}
\usepackage{array}
\usepackage[square,numbers]{natbib}


\makeatletter 
\def\cleardoublepage{\clearpage\if@twoside \ifodd\c@page\else%
    \hbox{}%
    \thispagestyle{empty}%
    \newpage%
    \if@twocolumn\hbox{}\newpage\fi\fi\fi} 
\makeatother

\def\figurename{Figure}
\makeatletter
\renewcommand{\fnum@figure}[1]{\figurename~\thefigure.}
\makeatother

\def\tablename{Table}
\makeatletter
\renewcommand{\fnum@table}[1]{\tablename~\thetable.}
\makeatother

\begin{document}
\title{
{\begin{flushleft}
\vskip 0.45in
{\normalsize\bfseries\textit{Chapter}}
\end{flushleft}
\vskip 0.45in
\bfseries\scshape Information analysis in free and confined harmonic oscillator}}

\author{\bfseries\itshape Neetik Mukherjee,  
\bfseries\itshape Amlan K. Roy\thanks{E-mail address: akroy@iiserkol.ac.in, akroy6k@gmail.com}\\
Department of chemical sciences,\\ 
Indian institute of science education and research Kolkata,\\
Mohanpur-741246, Nadia, India}
\date{} 
\maketitle
\thispagestyle{empty}
\setcounter{page}{1}
\thispagestyle{fancy}
\fancyhead{}
\fancyhead[L]{In: Harmonic Oscillators: Types, \\
Functions and Applications \\ 
Editor:Yilun Shang, pp. {\thepage-\pageref{lastpage-01}}} 
\fancyhead[R]{ISBN 978-1-53615-810-6 \\
\copyright~2019 Nova Science Publishers, Inc.}
\fancyfoot{}
\renewcommand{\headrulewidth}{0pt}

\vspace{2in}


\vspace{.08in} \noindent \textbf{Keywords:} R\'enyi entropy, Shannon entropy, Fisher information, Onicescu energy, Complexities, 
Confined isotropic harmonic oscillator, Particle in a symmetric box.

\clearpage 
\pagestyle{fancy}
\fancyhead{}
\fancyhead[EC]{\it  Neetik Mukherjee and Amlan K. Roy}
\fancyhead[EL,OR]{\thepage}
\fancyhead[OC]{\it  Free and confined harmonic oscillator $\cdots$}
\fancyfoot{}
\renewcommand\headrulewidth{0.0pt} 
\begin{abstract}
In this chapter we shall discuss the recent progresses of information theoretic tools in the context of free and confined 
harmonic oscillator (CHO). Confined quantum systems have provided appreciable interest in areas of physics, chemistry, biology, 
etc., since its inception. A particle under extreme pressure environment unfolds many fascinating, notable physical and 
chemical changes. The desired effect is achieved by reducing the spatial boundary from infinity to a finite region. Similarly, 
in the last decade, information measures were investigated extensively in diverse quantum problems, in both free and constrained 
situations. The most prominent amongst these are: Fisher information ($I$), Shannon entropy ($S$), R\'enyi entropy ($R$), Tsallis 
entropy ($T$), Onicescu energy ($E$) and several complexities. Arguably, these are the most effective measures of uncertainty, as 
they do not make any reference to some specific points of a respective Hilbert space. These have been invoked to explain several 
physico-chemical properties of a system under investigation. Kullback$–$Leibler divergence or relative entropy describes 
how a given probability distribution shifts from a reference distribution function. This characterizes a measure of discrimination 
between two states. In other words, it extracts the change of information in going from one state to another.

The one-dimensional confined harmonic oscillator (1DCHO), defined by $v(x)=\frac{1}{2}(x-d_{m})^{2}+v_{c}$ ($k$ is the force 
constant, $v_c=0,~x<x_c$ and $v_c=\infty,~x \ge x_c$), can be classified into two forms, (a) symmetrically confined harmonic 
oscillator (SCHO) (when $d_m=0$), (b) asymmetrically confined harmonic oscillator (ACHO) (corresponding to $d_m \ne 0$). Further, 
in latter case, confinement can be accomplished two different ways: (i) by changing the box boundary, keeping box length and 
$d_m$ fixed at zero; (ii) another route is to adjust $d_m$ by keeping box length and boundary fixed. SCHO can be treated as an 
intermediate between a particle-in-a-box (PIB) and a 1DQHO. Though the Schr\"odinger equation for SCHO 
can be solved \emph{exactly}, for ACHO it cannot be. We have employed two different methods for the latter, \emph{viz.}, (i) an 
imaginary time propagation (ITP), leading to minimization of an expectation value (ii) a variation-induced exact diagonalization
procedure that utilizes SCHO eigenfunctions as basis. It is found that, at very low $x_c$ region, $I,~S,~R,~T,~E$ remain 
invariant with change confinement length, $x_c$. At moderate $x_c$ region $S_{x},~R_{x},~T_{x}$ progress and $I_{x},~E_{x}$ 
decrease with rise in $x_c$. Additionally, special attention has been paid to study relative information in SCHO and ACHO.  

Analogously, a 3DCHO (radically confined within an impenetrable spherical well) can act as a bridge between particle in a spherical
box (PISB) and isotropic 3DQHO. The time-independent Schr\"odinger equation for D-dimensional harmonic oscillator in both free and 
confined condition can be solved exactly, within a Dirichlet boundary condition. Here a detailed exploration of information 
measures has been carried out in $r$ and $p$ 
spaces, for a 3DCHO. Some exact analytical expressions for these quantities have also been presented, wherever possible. Besides
we also consider some recent works on relative information and several complexity measures in composite $r$ and $p$ spaces. Also 
discussion is made of a recently proposed virial theorem in the context of confinement. In essence, this chapter provides a 
detailed in-depth investigation about the information theoretical analysis in 1DCHO and 3DCHO, as done in our laboratory. 
 
\end{abstract}
\tableofcontents
\section{Introduction}\label{sec:intro}
``A quantum particle inside an infinite impenetrable box" usually constitutes the first model problem taught in quantum 
mechanics curriculum. Later, the attention shifts towards the study of ``free" systems in some characteristic potential, 
such as a harmonic oscillator or a H atom in a Coulombic potential, which typify a quantum mechanical problem in a whole 
region of space. This is itself a recognition of the importance of sub-region $\Omega$ of the space, which in this 
particular scenario is achieved by modeling the boundary condition. Such a situation arises when one tries to explore systems 
in highly inhomogeneous media or in intense external field. The limit, when the box becomes very small, approaching quantum 
size is generally termed as ``quantum confinement". In this microscopic domain, confinement occurs on a scale comparable to 
atomic size (object and cavity sizes are commensurate), which is in sharp contrast to the ordinary confinement, where the 
cavity size is usually much larger than the atom. Classic examples of such macroscopic-size boxes are found in celebrated 
physics problems like kinetic theory of gas, black-body radiation, atom in a microwave enclosure, etc. 

The first seminal work in this direction is due to a pioneering article \cite{michels37} published in the fourth decade of 
twentieth century, where the effect of pressure on static polarizability of a H atom enclosed inside a spherical impenetrable 
hard enclosure was studied. It was realized that certain properties (such as energy/eigenvalue, orbital/eigenfunction, 
electron localization, shell-filling, polarizability, photonic ionization/absorption, etc., to name a few) of an atom could be 
significantly altered by ``squeezing" or ''shrinking" it by placing inside a hard sphere, which acts as an idealized 
classical piston. Depending on the geometrical form and dimension of cavity, such characteristic features of a caged-in atom 
exhibits numerous deviations from corresponding \emph{free} atom. Apparently there was some paucity in their immediate 
applications, but in the last few decades, these models have found wide-spread applications in various branches in physics and 
chemistry. Lately there has been a plethora of activities in harnessing their potential in various physico-chemical situations as 
evidenced by an almost exponential growth in the number of articles published in journals, and this continues to grow day by day. 
Excellent elegant reviews have been made available on the subject; we mention here a very selected set 
\cite{jaskolski96, dolmatov04,sabin09, sen14, leykoo18}. 

With the advancement of modern technology, confined quantum systems has emerged as an extremely important contemporary research 
area, from both theoretical and experimental perspectives. The discovery and development of modern experimental techniques have 
provided the desired impetus about these boxed-in systems. Constrained atoms, molecules inside various cavities has shed light on 
their electronic structure, chemical reactivity under high pressure. Apart from that, their relevance has been advocated in 
numerous other instances, such as impurity in semiconductor/nanostructure, atoms trapped in zeolite sieves or fullerene cage, 
solutes under solvent environment, artificial atom like quantum dot, quantum wire, quantum well, etc. Recently endohedral 
metallofullerene clusters where atoms/molecules are typically encapsulated in carbon (or fullerene-like inorganic and hybrid) 
cages of varying size, shape has found promising applications \cite{dolmatov09, charkin09, etindele17} in engineering. The subtle 
interaction between host cage and inserted guest molecule is manifested in their marked structural, energetic and spectroscopic 
properties of endocluster, in comparison to the isolated guest molecule or empty cavity. 

The confined harmonic oscillator (CHO) remains one of the oldest, heavily studied model of quantum confinement. In this case the 
quadratic potential acts through only a finite distance corresponding to the length of box; beyond this an infinite repulsive 
potential exists. This has been used in the context of various complicated physical situations ever since its inception. 
Some noteworthy applications include energy generation in dense stars from proton-deuteron transformation \cite{auluck41}, 
mass-radius relation in white dwarf theory \cite{auluck42}, rate of escape of a star from a galaxy or globular cumulus 
\cite{chandrasekhar43}, electric and magnetic properties of metals including small system of electrons within a cylinder 
\cite{dingle52, dingle52a}, vibronic spectra of point defects, impurities and luminescence in solids \cite{grinberg94}, specific 
heats of metal under high pressure, effect of high pressure on properties of materials, etc.  

Since the forty's decade, a vast amount of theoretical methods have been reported for accurate solution of a 1DCHO, placed inside 
a hard impenetrable symmetric box. The usual boundary condition of wave function vanishing at infinity of unenclosed or free 
harmonic oscillator (QHO) is replaced by the requirement that it vanishes at the walls of enclosure. Eigenfunctions in such systems 
could then be written in terms of Kummer's function \cite{auluck41, auluck42}, whereas the zeros of confluent hypergeometric 
function provided energy eigenvalues. Further, the authors employed various approaches to expand the latter in order to find an 
analytical expression in terms of box size. They found the correct qualitative behavior, i.e., energy levels of a 1DCHO increases 
rapidly as box length diminishes. An asymptotic expression \cite{auluck45} for energy valid in the region of small length of box 
was derived. An attempt \cite{baijal55} was made to locate the zeros of hypergeometric function numerically, albeit with 
limited success. It was further established that while in a QHO (wave function vanishing at $\infty$), transitions can occur
only between two adjacent levels, in a bounded oscillator, there occurs a non-zero transition probability between any two 
states of different parity. Simple analytical expressions \cite{hull56} of energy in certain special cases were put forth. 
The effect of finite boundary on energy levels has also been pursued graphically and from some series expansion \cite{dean66}. 
Later, within a semi-classical Wentzel-Kramers-Brillouin (WKB)-type approximation \cite{vawter68}, it was pointed out that 
eigenvalues of the constrained 
problem reduce to respective unenclosed oscillators, provided the classical turning points remain inside the potential 
enclosure and not near the walls. Moreover, when the separation of turning points remains large in comparison to the box size, 
they become plane-wave box eigenvalues. Subsequently, a series analytical solution \cite{vawter73} was proposed for 
eigenvalues by keeping the center of oscillation at the center of potential enclosure. Eigenvalues were also computed by a 
numerical procedure as the roots of a polynomial \cite{consortini76}. It was observed that when the box length remains below
a certain critical value (corresponding to the effective oscillation length for an unconfined oscillator of lowest energy), 
the individual energies dramatically increase from uncaged values. Closed-form energy expressions valid for boxes of 
any size were derived by constructing Pad\'e approximants \cite{navarro80} as interpolation between the perturbative and 
asymptotic solutions. Methods such as diagonal hyper-virial \cite{fernandez81, fernandez81a} as well as hyper-virial 
perturbation theory \cite{arteca83} were also employed for such enclosed systems for a variety of boundary conditions. 
A Rayleigh-Ritz variation method utilizing the trigonometric basis functions was suggested in \cite{taseli93}. 
A numerical scheme \cite{vargas96}, based on a set of theorems that guarantee that the solution of SE corresponding to a 
bounded system strongly converges in the norm of Hilbert space $L_2 (-\infty, \infty)$ to exact solution of respective 
unbounded problem. Spatially confined 1DQHO was treated by (WKB) along with modified airy function 
method \cite{sinha99}, as well as super-symmetric WKB approach \cite{sinha00}. Highly accurate eigenvalues were published in 
\cite{montgomery10} by numerically finding zeros of hypergeometric functions. Accurate eigenvalues were computed by power-series 
method \cite{campoy02}, perturbation theory \cite{amore10}, ITP method \cite{roy15} in conjunction 
with the minimization of an energy expectation value; in the former also the Einstein coefficients were considered. Very recently, 
this problem has been revisited by diagonalizing the Hamiltonian using a PIB basis \cite{aquino17}. However, all the above 
referenced works relate to the so-called \emph{symmetric confinement}; \emph{asymmetric confinement} studies are relatively less. 
For example, in an asymmetric CHO (ACHO), the energy spectrum have been reported by means of perturbation theory \cite{aquino01}, 
power-series method \cite{campoy02} and ITP \cite{roy15} method. Other properties such dipole moment \cite{aquino01} as well as 
Einstein coefficients \cite{aquino01, campoy02} were also considered. 
 
In parallel to the 1DCHO problem, much attention was also paid on 2D, 3D and $N$-dimensional counterparts, albeit with 
less vigour. A number of interesting unique phenomena occur in higher dimensions, especially in connection to \emph{simultaneous,
incidental and inter-dimensional degeneracy}, which are elaborated later in due course of time. In one of the earliest attempts, 
bounded multi-dimensional isotropic harmonic oscillators, including a 3DCHO were investigated using a hyper-virial treatment 
\cite{fernandez81b}. Energy eigenvalues of 2DCHO and 3DCHO in circular and spherical boxes respectively were reported through 
Rayleigh-Schr\"odinger perturbation expansion, considering the free PIB as corresponding unperturbed systems, as well as 
Pad\'e approximant solution \cite{navarro83}. A 3DCHO within an impenetrable spherical cavity was approached by a direct 
variational method \cite{marin91}, where the trial wave function was assumed as product of the ``free" solution of corresponding 
SE and a simple function obeying the respective boundary condition. For the 3DCHO, super-symmetric semi-classical approach 
\cite{dutt95} has 
been advocated. The radial SE corresponding to the 2DCHO and 3DCHO was solved quite efficiently by means of a variational procedure,
whereby the wave function was expanded into a Fourier-Bessel series \cite{taseli97a, taseli97b}, with matrix elements involving 
Bessel functions, that could be evaluated analytically. First detailed and systematic calculation of energies, eigenfunctions and 
spatial expectation values in isotropic 3DCHO was undertaken in \cite{aquino97}, employing a variational strategy. Like the 1DCHO 
case, a WKB approach was also put forth in 3DCHO by the same author \cite{sinha03}, where the centrifugal term is expanded 
perturbatively (in powers of $\hbar$), partitioning in two parts, \emph{viz.}, the usual centrifugal potential governed by 
classical laws and a quantum correction. A recipe was also provided for 3DCHO within the super-symmetric quantum mechanics 
\cite{filho03} in association with variational principle, as well as a generalized pseudo-spectral (GPS) method 
\cite{sen06, roy14mpla}. Eigenspectra of 2DCHO \cite{stevanovic08} and 3DCHO \cite{stevanovic08a} were investigated analytically 
in terms of annihilation and creation operators. A combination of semi-classical WKB method and a proper quantization rule 
\cite{serrano13} 
has also been suggested for the spherically confined harmonic oscillator.  
Some reports have also been made on higher dimensional CHO, although in much lesser intensity. Reasonably \cite{aljaber08} and 
very accurate \cite{montgomery07} eigenvalues have been reported for $D$-dimensional CHO from numerical calculations that find 
the roots of confluent hypergeometric functions satisfying the necessary degeneracy conditions. High-precision energies were 
presented \cite{montgomery10} by exploiting certain special properties of hypergeometric functions within MAPLE computer 
algebra system.

In the last few decades, quantum information theory has emerged as a subject of topical interest. This concept has been exploited 
to understand various phenomena in physics and chemistry, with potential applications in broad topics like thermodynamics 
\cite{poland2000,singer04,anton07}, quantum mechanics \cite{poland2000,singer04,anton07}, spectroscopy etc. Lately, it has been 
extensively used to understand quantum entanglement and quantum steering problems \cite{poland2000,singer04,anton07}. The 
information theoretic measures like Shannon entropy ($S$) \cite{birula75,shannon51}, R\'enyi entropy ($R^{\alpha}$) \cite{birula06},
Fisher information ($I$), Onicescu energy ($E$) and complexities ($C$) are invoked to get knowledge about diffusion of atomic 
orbitals, spread of electron density, periodic properties, correlation energy and so forth. Perhaps, these are the most effective 
measures of uncertainty \cite{birula75,birula06, duan2000, simon2000}, as 
they do not make any reference to some specific points of the respective Hilbert space. In principle, they can have any real 
value. However, $(-)$ve value in $R$ and $S$ indicates extreme localization, whereas, $E$ is always ($+$)ve. Likewise, changing 
the numerical values of $R^{\alpha},~S$ from $(-)$ve to ($+$)ve only interprets enhancement of de-localization. Another related 
concept is \emph{complexity}. A system has finite complexity when it is either in a state with less than some maximal order or not 
at a state of equilibrium. In a nutshell, it becomes \emph{zero} at two limiting cases, \emph{viz.,} when a system is (i) completely 
ordered (maximum distance from equilibrium) or (ii) at equilibrium (maximum disorder). Complexity has its contemporary interest
in chaotic systems, spatial patterns, language, multi-electronic systems, molecular or DNA analysis, social science, astrophysics 
and cosmology etc. 

In this chapter, we summarize the recent progress that has taken place in the information theoretical analysis in case of a 
quantum harmonic oscillator (QHO) contained in inside a hard, impenetrable cage. To this end, a thorough investigation is made for 
energy spectra and a host of information quantities, like $S, R, I, E, C$, in 1DCHO (symmetric and asymmetric) as well as a 
radially confined QHO. In the 1D case, we introduce two accurate methods for solution of relevant eigenvalue equations in 
presence of the Dirichlet boundary conditions, namely (i) ITP and (ii) variation-induced exact diagonalization--both developed in 
our laboratory. They provide accurate eigenvalues and eigenfunctions, which are used throughout. In the 3D counterpart, we employ
the very accurate GPS method. One important conclusion is that, a 1DCHO behaves as an intermediate between a PIB and 
an 1DQHO. Detailed results are compiled in tabular and graphical form for the afore-mentioned information-related quantities, 
always indicating the transition from bounded to free system. Additionally we also offer a brief account of the \emph{relative 
information} studies that has been studied by us. Finally, we examine the validity and feasibility of a recently proposed virial 
theorem in the context of such confined systems, which will consolidate the former's success further. Available literature results 
have been consulted as and when possible. A few concluding remarks are made at the end. 
       
\section{Theoretical Aspects}
This section is devoted to the theoretical methods that have been employed for calculation of relevant eigenvalues and 
eigenfunctions for the spatially confined CHO; both in 1D and 3D. Then we proceed for a discussion on the evaluation of 
momentum ($p$)-space wave functions from that in position space, as well as for the computation of information-related quantities.  
\subsection{Position-space wave function}
In 1DCHO problem can be categorized into two distinct forms, \emph{viz.,} (i) in the SCHO case, potential minimum is at the 
origin ($x=0$), leading to a \emph{symmetric} confinement ($R$ is length of box), 
($v_c(x)$ is the confinement potential), 
and (ii) ACHO, where the potential minimum is beyond the origin, as exemplified here through the following relations,
\[
-\frac{1}{2} \frac{d^2 \psi}{dx^2} + \frac{1}{2}(x-d)^2 \psi +V(x) \psi = E \psi,
\]
where $V(x)= +\infty$, for $|x| \geq R$ and $V(x)=0$, when $|x| < R$. It represents an infinite square well of width $2R$ with 
$d$ signifying position of minimum in the potential. The SCHO potential can be solved exactly, whereas in latter case 
one needs an appropriate numerical method to get the best possible wave function and energy. Like the SCHO case, 3DCHO is also 
exactly solvable and both wave functions are obtained in the form of \emph{Kummer confluent hypergeometric} function. It is 
crucial to point out that, in order to construct the exact wave functions of SCHO and 3DCHO for a specific state, however, one 
requires to provide energy eigenvalue of that state. In our calculation, these energies of SCHO and 3DCHO have been generated 
from ITP and GPS methods.   

\subsubsection{Exact Solution}
In this part we are going to discuss the exact solutions of SCHO and 3DCHO respectively.
\paragraph{SCHO:}
The time-independent non-relativistic SE is given by ($\alpha$ is force constant), 
\begin{equation} \label{eq:1}
-\frac{1}{2}\frac{d^{2}\psi_{n}(x)}{dx^{2}}+4\alpha^{2}x^{2}\psi_{n}(x)+v_{c} (x) \psi_{n} (x) = \mathcal{E}_{n}\psi_{n} (x),
\end{equation}
where the confining potential is defined as, $v_{c}(x)=0$ for $x<|x_{c}|$ and $v_{c}(x)=\infty$ for $x \ge |x_{c}|$, with $x_c$ 
signifying length of the impenetrable box. In case of an 1DQHO, $v_{c} (x) =0$. The exact analytical solution of Eq.~(\ref{eq:1}) 
for \emph{even} and \emph{odd} states are as follows, 
\begin{equation} \label{eq:2}
\begin{aligned}
\psi_{e}(x)= N_{e}\ _{1}F_{1}\left[\left(\frac{1}{4}-\frac{\mathcal{E}_{n}}{4\sqrt{2}\alpha}\right),
\frac{1}{2},2\sqrt{2}\alpha x^{2}\right]e^{-\sqrt{2}\alpha x^{2}}, \\
\psi_{o}(x)=N_{o} \ x \ _{1}F_{1}\left[\left(\frac{3}{4}-\frac{\mathcal{E}_{n}}{4\sqrt{2}\alpha}\right),
\frac{3}{2},2\sqrt{2}\alpha x^{2}\right]e^{-\sqrt{2}\alpha x^{2}}.
\end{aligned}
\end{equation} 
In this equation, $N_{e},N_{o}$ represent normalization constants for even, odd states respectively. It is imperative to mention 
that, in order to obtain the specific solution for a particular state at a definite $x_c$, one can adopt either of the following two 
procedures, \emph{viz.,} (i) use the Dirichlet boundary condition, $\psi_{n} (-x_{c})=\psi_{n}(x_{c})=0$ at a fixed $x_c$ 
to obtain $\mathcal{E}_{n}$ by solving the SE for certain $n$, and obtain wave function therefrom (ii) provide $\mathcal{E}_{n}$ as 
input to get the allowed $x_c$, which further leads to the wave function. Here, we have solved SE by using ITP procedure at a 
particular $x_c$ and calculated the respective energy spectrum. Then using this $\mathcal{E}_{n}$, the desired wave function is 
constructed to proceed for further calculation. In Eq.~(\ref{eq:2}), $_{1}F_{1}(a;b;x)$ symbolizes the \emph{Kummer confluent 
hypergeometric} or confluent hypergeometric function of \emph{1st kind} assuming following form,
\begin{equation} \label{eq:3}
_1F_1(a;b;y)=1+\frac{a}{b}y+\frac{a(a+1)}{b(b+1)}\frac{y^{2}}{2!}+ \cdots =\sum_{k=0}^{\infty}\frac{a_{k}}{b_{k}}\frac{y^{k}}{k!}.
\end{equation}
Here $a_{k},b_{k}$ denote the  Pochhammer symbols. It is noteworthy to indicate that, if one replaces $\mathcal{E}_{n}$ by the 1DQHO
energy, $(n+\frac{1}{2})2\sqrt{2}\omega$ in Eq.~(\ref{eq:3}), then $\psi_{e}(x), \psi_{o}(x)$ modify as,
\begin{equation} \label{eq:4}
\begin{aligned}
\psi_{e}(x)= N_{e}\ _{1}F_{1}\left[-\left(\frac{j}{2}\right),
\frac{1}{2},2\sqrt{2}\alpha x^{2}\right]e^{-\sqrt{2}\alpha x^{2}}=\\ N_{e}H_{j}(2\sqrt{2}\alpha x^{2})e^{-\sqrt{2}\alpha x^{2}}, \\
\psi_{o}(x)=N_{o} x \ _{1}F_{1}\left[-\left(\frac{m}{2}\right),
\frac{3}{2},2\sqrt{2}\alpha x^{2}\right]e^{-\sqrt{2}\alpha x^{2}}=\\ N_{o} x \ H_{m}(2\sqrt{2}\alpha x^{2})e^{-\sqrt{2}\alpha x^{2}}.
\end{aligned}
\end{equation} 
Here $H_{i}(2\sqrt{2}\alpha x^{2})$ signifies \emph{Hermite} polynomials and  $j, m$ correspond to \emph{even, odd} positive 
integers respectively. Equation~(\ref{eq:4}) clearly suggests that, in an 1DQHO the hypergeometric function reduces to 
\emph{Hermite} polynomial. 

\paragraph{Incidental degeneracy in SCHO:} In this occasion, quantization is outcome of the boundary condition, which is 
imposed by making the wave function vanish at $x_c$. Actually allowed energies are obtained by satisfying $_{1}F_{1}
[a;b;x]=0$ at $x_{c}$. The hypergeometric function takes the form given below, at $x=x_{c}$ (considering even-parity states),
\begin{equation} \label{eq:5}
\begin{aligned}
_{1}F_{1}\left[\left(\frac{1}{4}-\frac{\mathcal{E}_{n}}{4\sqrt{2}\alpha}\right),\frac{1}{2},2\sqrt{2}\alpha x_{c}^{2}\right]= 1+ 
\left(\frac{1}{4}-\frac{\mathcal{E}_{n}}{4\sqrt{2}\alpha}\right)4\sqrt{2}\alpha x_{c}^{2} + \\ \left(\frac{1}{4}-
\frac{\mathcal{E}_{n}}{4\sqrt{2}\alpha}\right)
\left(\frac{5}{4}-\frac{\mathcal{E}_{n}}{4\sqrt{2}\alpha}\right)\frac{32}{3}\alpha^{2} x_{c}^{4}+ \\ \left(\frac{1}{4}-
\frac{\mathcal{E}_{n}}{4\sqrt{2}\alpha}\right)
\left(\frac{5}{4}-\frac{\mathcal{E}_{n}}{4\sqrt{2}\alpha}\right)\left(\frac{9}{4}-\frac{\mathcal{E}_{n}}{4\sqrt{2}\alpha}\right)
\frac{128}{15}\alpha^{3} x_{c}^{6}+ \cdots
\end{aligned}
\end{equation}   
Let us consider only two terms in the right-hand series of Eq.~(\ref{eq:5}). Then we get, 
\begin{equation} {\label{eq:6}}
\left(\frac{5}{4}-\frac{\mathcal{E}_{n}}{4\sqrt{2}\alpha}\right)=0, \ \ \ \ \ \ \ \ \ \ \ \ \ \mathcal{E}_{n}=5\sqrt{2}\alpha.
\end{equation}
Now, putting the value of $\mathcal{E}_{n}=5\sqrt{2}\alpha$ and $x=x_{c}$ in the boundary condition we get,
\begin{equation} {\label{eq:7}}
\begin{aligned}
1+\left(\frac{1}{4}-\frac{E_{n}}{4\sqrt{2}\alpha}\right)4\sqrt{2}\alpha x_{c}^{2}=0,  \\
1+\left(\frac{1}{4}-\frac{5\sqrt{2}\alpha}{4\sqrt{2}\alpha}\right)4\sqrt{2}\alpha x_{c}^{2}=0, \\
(1-4\sqrt{2}\alpha x_{c}^{2})=0, \\
4\sqrt{2}\alpha x_{c}^{2}=1, \\
x_{c}=\pm \frac{1}{2\sqrt{\sqrt{2}\alpha}}.
\end{aligned}
\end{equation}
Since box length is positive, we choose $x_{c}=\frac{1}{2\sqrt{\sqrt{2}\alpha}}$. This shows that, at this $x_c$, 
$_{1}F_{1}\left[\left(\frac{1}{4}-\frac{\mathcal{E}_{n}}{4\sqrt{2}\alpha}\right),\frac{1}{2},2\sqrt{2}\alpha x^{2}\right]$ with 
two terms always represents a non-degenerate ground state having $\mathcal{E}_{0}=5\sqrt{2}\alpha$.
Next, let us proceed to the function having three terms. Therefore,
\begin{equation}
\frac{9}{4}-\frac{\mathcal{E}_{n}}{4\sqrt{2}\alpha}=0, \ \ \ \ \ \ \ \ \ \ \ \ \ \  \mathcal{E}_{n}=9\sqrt{2}\alpha.
\end{equation}  
Using the boundary condition and putting $\mathcal{E}_{n}=9\sqrt{2}\alpha$ and $x=x_{c}$ we obtain,
\begin{equation} {\label{eq:8}}
\begin{aligned}
1+2\left(\frac{1}{4}-\frac{E_{n}}{4\sqrt{2}\alpha}\right)2\sqrt{2} \alpha x_{c}^{2}+\\ \left(\frac{1}{4}-\frac{E_{n}}{4\sqrt{2}
\alpha}\right)
\left(\frac{1}{4}-\frac{E_{n}}{4\sqrt{2}\alpha}+1\right)\frac{32}{3}\alpha^{2} x_{c}^{4}=0, \\
64\alpha^{2} x_{c}^{4}-24\sqrt{2} \alpha x_{c}^{2}+3=0,\\
x_{c}^{2}=\frac{3+\sqrt{3}}{8\sqrt{2}\alpha} \ \ \  \rightarrow \ \ \  x_{c}=\pm \sqrt{\frac{3+\sqrt{3}}{8\sqrt{2}\alpha}}, \\
\ \ \ \ \ \ \ \ 
x_{c}^{2}=\frac{3-\sqrt{3}}{8\sqrt{2}\alpha} \ \ \  \rightarrow \ \ \  x_{c}=\pm \sqrt{\frac{3-\sqrt{3}}{8\sqrt{2}\alpha}}.
\end{aligned}
\end{equation}

Equation~(\ref{eq:8}) suggests that, in SCHO, $\mathcal{E}_{n}=9\sqrt{2}\alpha$ represents a pair of states; one at 
$x_c=\sqrt{\frac{3+\sqrt{3}}{8\sqrt{2}\alpha}}$ and other at $x_{c}= \sqrt{\frac{3-\sqrt{3}}{8\sqrt{2}\alpha}}$. The former 
represents a second excited state and latter symbolizes ground state. Similar analysis will reveal that, a function 
$_{1}F_{1}[a,b,y]$ with $(n+1)$ terms has $n$ number of such degenerate states at $n$ different $x_c$ values. Additionally, 
in such degenerate set, the smallest box length contains the ground state and longest box length holds the $n$th state. This 
procedure helps us to identify incidental degeneracy in SCHO at different box lengths.    
  
\paragraph{3DCHO:}
The time-independent, non-relativistic wave function for a 3DCHO system, in $r$ space may be written as ($n_r$ identifies 
radial quantum number), 
\begin{equation} \label{eq:9}
\Psi_{n_r,\ell,m} (\rvec) = \psi_{n_r, \ell}(r)  \ Y_{\ell,m} (\Omega), 
\end{equation} 
with $r$ and $\Omega$ illustrating radial distance and solid angle successively. Here $\psi_{n_{r},\ell}(r)$ represents the 
radial part and $Y_{\ell,m}(\Omega)$ identifies spherical harmonics of the wave function. The pertinent radial Schr\"odinger 
equation under the influence of confinement, with $v(r)=\frac{1}{2}\omega^{2}r^{2}$, is (atomic unit employed unless
otherwise mentioned), 
\begin{equation} \label{eq:10}
\left[-\frac{1}{2} \ \frac{d^2}{dr^2} + \frac{\ell (\ell+1)} {2r^2} + v(r) +v_c (r) \right] \psi_{n_r,\ell}(r)=
\mathcal{E}_{n_r,\ell}\ \psi_{n_r,\ell}(r).
\end{equation}
Our required confinement effect is introduced by invoking the following potential: $v_c(r) = +\infty$ for $r > r_c$, and $0$ 
for $r \leq r_c$, where $r_c$ signifies radius of confinement.

The \emph{exact} generalized radial wave function is mathematically expressed \cite{montgomery07} as, 
\begin{equation} \label{eq:11}
\psi_{n_{r}, \ell}(r)= N_{n_{r}, \ell} \ r^{\ell} \ _{1}F_{1}\left[\frac{1}{2}\left(\ell+\frac{3}{2}-
\frac{\mathcal{E}_{n_{r},\ell}}{\omega}\right), (\ell+\frac{3}{2}),\omega r^{2}\right] e^{-\frac{\omega}{2}r^{2}}.
\end{equation}
Here $N_{n_r, \ell}$ represents normalization constant, $\mathcal{E}_{n_r,\ell}$ corresponds to energy of a given state 
characterized by quantum numbers $n_r,\ell$, whereas $_1F_1\left[a,b,r\right]$ signifies confluent hypergeometric 
function. Allowed energies are computed by applying the boundary condition $\psi_{n_r,\ell} (0)= \psi_{n_r,\ell} \ (r_c)=0$. 
In this work, the GPS method has been used to calculate eigenvalues and eigenfunctions of these states. It has 
provided highly accurate results for various model and real systems including atoms, molecules, some of which could be found 
in the references \cite{roy04pla, roy04jpg, roy04jpbhollowb, roy05pramana, roy05ijqc, roy05jpbhollowb, sen06, roy07, roy08jmc,
roy08ijqc, roy08theochem, roy13ijqc, roy13rinp, roy14fbsys, roy14ijqc, roy14mpla_manning, roy14jmc, roy14mpla, roy15ijqc, 
roy16ijqc}.

\paragraph{Incidental degeneracy in 3DCHO:}
Just like the 1DCHO, quantization in a D-dimensional CHO is also an manifestation of the effect of making radial wave function vanishing
at $r_{c}$. In this occasion, allowed energies are obtained when,
\begin{equation} \label{eq:12}
_{1}F_{1}\left[\frac{1}{2}\left(\ell+\frac{D}{2}-\frac{\mathcal{E}_{n_{r},\ell}}{2\sqrt{2}\alpha}\right);\left(\ell+
\frac{D}{2}\right); 2\sqrt{2}\alpha r_{c}^{2}\right]=0,
\end{equation} 
where, at a fixed $\ell$, the successive roots are numbered $n_{r}=0,1,2,...$. Note that, the levels are designated by $n_{r} +1$ 
and $\ell$ values, such that $n_{r} =\ell = 0$, $n_r = \ell=2$ correspond to $1s$ and $3d$ states respectively. The radial quantum 
number $n_r$ relates to $n$ as $n = 2n_r + \ell$. Invoking the series expansion \cite{abramowitz64} in left-hand side, we get, 
\begin{equation} \label{eq:13}
\begin{aligned}
 & _{1}F_{1}\left[\frac{1}{2}\left(\ell+\frac{D}{2}-\frac{\mathcal{E}_{n_{r},\ell}}{2\sqrt{2}\alpha}\right);\left(\ell+ 
\frac{D}{2}\right); 2\sqrt{2}\alpha r_{c}^{2}\right]  =\\ 
&1+\frac{\left(\frac{\ell}{2}+\frac{D}{4}-\frac{\mathcal{E}_{n_{r},\ell}}
{4\sqrt{2}\alpha}\right)}{\left(\ell+ \frac{D}{2}\right)}2\sqrt{2}\alpha r_{c}^{2}+ \\
 &  \frac{\left(\frac{\ell}{2}+\frac{D}{4}-\frac{\mathcal{E}_{n_{r},\ell}}{4\sqrt{2}\alpha}\right) \left(\frac{\ell}{2}+\frac{D}{4}-
\frac{\mathcal{E}_{n_{r},\ell}}{4\sqrt{2}\alpha}+1\right)}{\left(\ell+\frac{D}{2}\right) \left(\ell+\frac{D}{2}+1\right)}\frac{8}
{2!}\alpha^{2} r_{c}^{4}+ \cdots
\end{aligned}
\end{equation}
In order to terminate the series after 2nd term we need,
\begin{equation} \label{eq:14}
\begin{aligned}
\left(\frac{\ell}{2}+\frac{D}{4}-\frac{\mathcal{E}_{n_{r},\ell}}{4\sqrt{2}\alpha}+1\right)=0 \\
\frac{\mathcal{E}_{n_{r},\ell}}{2\sqrt{2}\alpha} = 1+\frac{\ell}{2}+\frac{D}{4} \\
\mathcal{E}_{n_{r},\ell}=\sqrt{2} \alpha \left(D+2\ell+4\right).
\end{aligned}
\end{equation}
Now, it is important to determine the corresponding $r_{c}$ values at which Eq.~(\ref{eq:14}) is valid. According to Eq.~(\ref{eq:13}), 
this corresponds to the requirement that,
\begin{equation}
\begin{aligned}
1+\frac{\left(\frac{\ell}{2}+\frac{D}{4}-\frac{\mathcal{E}_{n_{r},\ell}}{4\sqrt{2}\alpha}\right)}{\left(\ell+\frac{D}{2}\right)}
2\sqrt{2}\alpha r_{c}^{2}=0 \\
1+\frac{\left(\frac{\ell}{2}+\frac{D}{4}-\frac{\sqrt{2}\alpha(D+2\ell+4)}{4\sqrt{2}\alpha}\right)}{\left(\ell+\frac{D}{2}\right)}
2\sqrt{2}\alpha r_{c}^{2}=0 \\ r_{c}=\sqrt{\frac{1}{2\sqrt{2}\alpha}\left(\ell+\frac{D}{2}\right)}. 
\end{aligned}
\end{equation}  
This shows that, degeneracy depends on $\ell$ and $D$ values. Let us consider some examples,
\begin{enumerate}
\item
$l=1,~D=2$, $\mathcal{E}_{n_{r},1}=8\sqrt{2}\alpha$, \ \ \
$l=0,~D=4$, $\mathcal{E}_{n_{r},0}=8\sqrt{2}\alpha$ \ \ at \ $r_{c}=\sqrt{\frac{1}{\sqrt{2}\alpha}}$,
\item
$l=1,~D=3$, $\mathcal{E}_{n_{r},1}=9\sqrt{2}\alpha$, \ \ \
$l=0,~D=5$, $\mathcal{E}_{n_{r},0}=9\sqrt{2}\alpha$ \ \ at \ $r_{c}=\sqrt{\frac{5}{4\sqrt{2}\alpha}}$,
\item
$l=1,~D=2$, $\mathcal{E}_{n_{r},2}=10\sqrt{2}\alpha$, \ \ \
$l=2,~D=4$, $\mathcal{E}_{n_{r},1}=10\sqrt{2}\alpha$ \ \ at \ $r_{c}=\sqrt{\frac{3}{2\sqrt{2}\alpha}}$,
\item
$l=1,~D=3$, $\mathcal{E}_{n_{r},2}=11\sqrt{2}\alpha$, \ \ \
$l=2,~D=5$, $\mathcal{E}_{n_{r},1}=11\sqrt{2}\alpha$ \ \ at \ $r_{c}=\sqrt{\frac{7}{4\sqrt{2}\alpha}}$,
\item
$l=2,~D=4$, $\mathcal{E}_{n_{r},2}=12\sqrt{2}\alpha$, \ \ \
$l=3,~D=2$, $\mathcal{E}_{n_{r},1}=12\sqrt{2}\alpha$ \ \ at \ $r_{c}=\sqrt{\frac{2}{\sqrt{2}\alpha}}$.
\end{enumerate}
From above instances it is clear that, degeneracy appears at a certain $r_{c}$. However, the selection rule for this 
degeneracy can be achieved from the following relation. Suppose, at a fixed $r_{c}$ we have a pair of states, characterized by 
quantum numbers $l_{1},l_2$, in dimensions $D_{1}, D_2$ respectively, having same energy. As a consequence of the above, we have, 
\begin{equation} \label{eq:15}
\sqrt{2} \alpha (D_{1}+2l_{1}+4)= \sqrt{2} \alpha (D_{2}+2l_{2}+4).
\end{equation}
\begin{equation} \label{eq:16}
\sqrt{\frac{1}{2\sqrt{2}\alpha}\left(l_{1}+\frac{D_{1}}{2}\right)}= \sqrt{\frac{1}{2\sqrt{2}\alpha}\left(l_{2}+\frac{D_{2}}{2}
\right)}
\end{equation}
Equations~(\ref{eq:15}) and (\ref{eq:16}) lead to the common relation,
\begin{equation} \label{eq:17}
(D_{2}-D_{1})=2(l_{1}-l_{2}).
\end{equation} 
It is noticeable from Eq.~(\ref{eq:17}) that, if $D_{2}>D_{1}$ then $l_{1}>l_{2}$ and vice-versa. Thus the incidental degeneracy 
selection rule is,
\begin{equation} \label{eq:18}
(n_{r},l,D,r_{c}) \ \ \rightarrow \ \ (n_{r},l \pm j,D \mp 2j, r_{c}),
\end{equation} 
where $j$ is an integer. It is noticed that, Eq.~(\ref{eq:18}) may further be written in following form,
\begin{equation}
\begin{aligned}
\mathcal{E}_{n_{r},(l+j)}^{D}(r_{c})= \mathcal{E}_{n_{r},l}^{(D+2j)}(r_{c}), \\
\mathcal{E}_{n_{r},l}^{(D-2j)}(r_{c})= \mathcal{E}_{n_{r},(l-j)}^{D}(r_{c}).
\end{aligned}
\end{equation} 
Thus, in a 3DCHO, incidental and inter-dimensional degeneracy occur simultaneously.

\paragraph{Simultaneous degeneracy in 3DCHO:}
It can be proved that, at $r_{c}=\sqrt{\frac{1}{2\sqrt{2}\alpha}\left(l+\frac{D}{2}\right)}$, the pair of energies 
$\mathcal{E}_{n_{r},(l+2)}^{D}$ and $\mathcal{E}_{(n_{r}+1),l}^{D}$ are related by the following relation,
\begin{equation}
\mathcal{E}_{(n_{r}+1),l}^{D}\left(\sqrt{\frac{1}{2\sqrt{2}\alpha}\left(l+\frac{D}{2}\right)}\right) =
\mathcal{E}_{n_{r},(l+2)}^{D}\left(\sqrt{\frac{1}{2\sqrt{2}\alpha}\left(l+\frac{D}{2}\right)}\right). 
\end{equation}
\paragraph{Inter-dimensional degeneracy in 3DCHO:}
Such a degeneracy occurs when, 
\begin{equation}
(n_{r},l,D) \rightarrow (n_{r},l \pm j, D \mp 2j).
\end{equation}
This can be demonstrated by the example as given below, 
\begin{equation}
\mathcal{E}_{0,4}^{2}(r_{c})= \mathcal{E}_{0,3}^{4}(r_{c}) = \mathcal{E}_{0,2}^{6}(r_{c}) =\mathcal{E}_{0,1}^{8}(r_{c})
\end{equation}
In this scenario, the terms inter-dimensional and incidental degeneracy are synonymous. 

\subsubsection{Imaginary-time propagation (ITP) method}
In this subsection, we discuss the general concepts of the imaginary-time evolution technique, as employed here for a particle under 
confinement. The scheme has been found to provide accurate bound-state solutions through a transformation of the appropriate 
time-dependent (TD) SE into a diffusion equation in imaginary time.  This resulting equation is numerically solved through a 
finite-difference procedure, in amalgamation with a minimization of expectation value to hit the ground state. After the original 
proposal that came several decades ago, a number of successful implementations \cite{anderson75, kosloff86, lehtovaara07, chin09, 
sudiarta09, strickland10, luukko13} have been reported in the literature since then. In this work we have adopted an implementation, 
which has been successfully applied to a number of physical systems, such as atoms, diatomic molecules within a quantum fluid 
dynamical density functional theory (DFT) \cite{roy99, roy02}, as well as some model (harmonic, anharmonic, self-interacting, 
double-well, 
spiked oscillators) potentials \cite{roy02a, gupta02, wadehra03, roy05, roy14jmc_itp}, in both 1D, 2D and 3D. Recently this was also 
extended to confinement (SCHO and ACHO) problems \cite{roy15} with very good success. 

In the following, we provide a short account of the essentials of the methodology; more details could be found in the references 
quoted above. Let us begin with the TDSE, which for a particle under the influence of a potential $v(x)$ in 1D, in atomic unit, is 
given by ($v(x)=4 \omega^2 x^2$ in an SCHO),  
\begin{equation} \label{eq:itp1}
i \frac{\partial}{\partial t} \psi (x, t) = H \psi(x, t) = 
\left[ -\frac{1}{2} \frac{d^2}{dx^2} + v(x) + v_c(x) \right] \psi(x, t). 
\end{equation}
The Hamiltonian operator consists of usual kinetic and potential energy operators. This method is, \emph{in principle, exact}. 
Here we have provided the equations for 1D SCHO; however this has been easily extended to higher dimensions (see, e.g., 
\cite{roy05,roy14jmc_itp}. The confinement condition can be achieved by reducing the boundary from \emph{infinity} to finite region, 
as expressed in the following equation (symmetric box of length $2R$),  
\begin{equation} v_c(x) = \begin{cases}
0,  \ \ \ \ \ \ \ -R < x < +R   \\
+\infty, \ \ \ \ |x| \geq R.  \\
 \end{cases} 
\end{equation}
Equation (\ref{eq:itp1}) can be expressed in imaginary time, $\tau= i t$ ($t$ is real time) to yield a non-linear diffusion-type 
equation similar to a diffusion quantum Monte Carlo equation \cite{hammond94},  
\begin{equation} \label{eq:itp2}
- \ \frac{d \psi(x, \tau) }{d \tau} =  H \psi(x, \tau).
\end{equation}
Its formal solution can be written as,  
\begin{equation}
\psi(x, \tau) = \sum_{k=0}^{\infty} c_k \psi_k(x) \exp{(-\epsilon_k \tau)}.
\end{equation}
The initial guessed wave function has the form, $\psi(x, \tau)$ at $\tau=0$ and if propagated for a sufficiently long time, it will 
finally converge towards the desired stationary ground-state wave function; 
$                       
\lim_{\tau \rightarrow \infty} \psi(x, \tau) \approx c_0 \psi_0 (x) e^{-\epsilon_0 \tau}.
$
Thus, provided $c_0 \neq 0$, apart from a normalization constant, this leads to the global minimum corresponding to an 
expectation value $\langle \psi(x, \tau) |H| \psi(x,\tau) \rangle$. 

The numerical solution of Eq.~(\ref{eq:itp2}) can be obtained by using a Taylor series expansion of 
$\psi(x,\tau+\Delta \tau)$ around time $\tau$ as follows,
\begin{equation} \label{eq:itp3}
\psi(x, \tau+\Delta t)= e^{-\Delta \tau H} \psi(x, \tau).
\end{equation}
Here, the exponential in right side represents a time-evolution operator, which propagates diffusion function $\psi(x, \tau)$ 
at an initial time $\tau$ to an advanced time step to $\psi(x, \tau+\Delta \tau)$. Since this is a non-unitary operator, 
normalization of the function at a given time $\tau$ does not necessarily preserve the same at a future time $\tau+\Delta \tau$. 
Transformation of Eq.~(\ref{eq:itp3}) into an equivalent, symmetrical form leads to ($j,n$ signify space, time indices 
respectively), 
\begin{equation}
e^{(\Delta \tau/2) H_j} \ \psi_j^{'(n+1)}= e^{-(\Delta \tau/2)H_j} \ \psi_j^n.
\end{equation}
A prime is introduced in above equation to indicate the \emph{unnormalized} diffusion function. Taking the full form of 
Hamiltonian from Eq.(\ref{eq:itp1}), one can further write, 
\begin{equation}
e^{(\Delta \tau/2) [-\frac{1}{2}D_x^2+v(x_j)]} \ \psi_j^{'(n+1)} = e^{-(\Delta \tau/2) [-\frac{1}{2}D_x^2+v(x_j)]} 
\ \psi_j^n,
\end{equation}
where the spatial second derivative has been defined as $D_x^2= \frac{d^2}{dx^2}$. Now, expanding the exponentials on both sides, 
followed by truncation after second term and approximation of second derivative by a five-point difference formula 
\cite{abramowitz64} ($\Delta x= h$),   
\begin{equation}
D_x^2 \ \psi_j^n  \approx  \frac{-\psi_{j-2}^n +16\psi_{j-1}^n -30\psi_j^n+ 16\psi_{j+1}^n-\psi_{j+2}^n}{12 h^2}, 
\end{equation}
yields a set of $N$ simultaneous equations, as follows, 
\begin{equation}
\alpha_j \psi_{j-2}^{'(n+1)} + \beta_j \psi_{j-1}^{'(n+1)} + \gamma_j \psi_{j}^{'(n+1)} 
\delta_j \psi_{j+1}^{'(n+1)} + \zeta_j \psi_{j+2}^{'(n+1)} = \xi_j^n. 
\end{equation}
After some straightforward algebra, the quantities $\alpha_j, \beta_j, \gamma_j, \delta_j, \zeta_j, \xi_j^n$ are identified as, 
\begin{eqnarray}
\alpha_j & = & \zeta_j = \frac{\Delta \tau}{48 h^2}, \ \  
\beta_j = \delta_j = -\frac{\Delta \tau}{3 h^2}, \ \ \  
\gamma_j =  1+\frac{ 5\Delta \tau}{8 h^2} + \frac{\Delta \tau }{2} \ v(x_j), \\ \nonumber 
\xi_j^n & = & \left[ -\frac{\Delta \tau}{48 h^2} \right] \psi_{j-2}^n + \left[ \frac{\Delta \tau}{3 h^2} \right] \psi_{j-1}^n +
    \left[ 1- \frac{5 \Delta \tau}{8 h^2} - \frac{\Delta \tau}{2} \ v(x) \right] \psi_j^n     \\
& & + \left[ \frac{\Delta \tau}{3 h^2} \right] \psi_{j+1}^n + \left[ -\frac{\Delta \tau}{48 h^2} \right] \psi_{j+2}^n. \nonumber 
\end{eqnarray}

Since discretization and truncation occur on both sides, there may be some cancellation of errors. Here, $\psi^{'(n+1)}$ signifies 
the unnormalized diffusion function at some future time $\tau_{n+1}$. The quantities like $\alpha_j, \beta_j, \gamma_j, \delta_j, 
\zeta_j$, after some algebraic manipulation, can be easily expressed in terms of space and time spacings as well as the potential 
(appears only in $\gamma_j$ and $\xi_j^n$).  The latter also requires knowledge of normalized diffusion functions $\psi_{j-2}^n, 
\psi_{j-1}^n, \psi_j^n, \psi_{j+1}^n, \psi_{j+2}^n$ at spatial grid points $x_{j-2}, x_{j-1}, x_j, x_{j+1}, x_{j+2}$ at an earlier time 
$\tau_n$. It is convenient to recast this in a pentadiagonal matrix form, as below, 
\begin{equation}
\left[ \begin{array}{ccccccc}
\gamma_1  &  \delta_1  & \zeta_1    &               &                &                 &      (0)       \\
\beta_2   &  \gamma_2  & \delta_2   &  \zeta_2      &                &                 &                \\
\alpha_3  &  \beta_3   & \gamma_3   & \delta_3      &  \zeta_3       &                 &                \\
          &  \ddots    & \ddots     & \ddots        &  \ddots        &  \ddots         &                \\
          &            & \ddots     & \ddots        & \ddots         &  \ddots         & \zeta_{N-2}    \\
          &            &            & \alpha_{N-1}  & \beta_{N-1}    &  \gamma_{N-1}   & \delta_{N-1}   \\
    (0)   &            &            &               & \alpha_N       & \beta_N         & \gamma_N       \\
\end{array} \right]  
\left[ \begin{array}{c}
\psi_1^{'(n+1)} \\ 
\psi_2^{'(n+1)} \\ 
\psi_3^{'(n+1)} \\ 
\vdots \\
\psi_{N-2}^{'(n+1)} \\ 
\psi_{N-1}^{'(n+1)} \\ 
\psi_N^{'(n+1)} 
\end{array} \right]
=
\left[ \begin{array}{c}
\xi_1^n \\ 
\xi_2^n \\ 
\xi_3^n \\ 
\vdots \\
\xi_{N-2}^n \\ 
\xi_{N-1}^n \\ 
\xi_N^n 
\end{array} \right].
\end{equation}
The above matrix equation can be readily solved for $\{\psi^{'(n+1)}\}$ using standard routine, satisfying the required boundary condition 
$\psi_1^n=\psi_N^n=0$, at all time. Hence, from an initial trial function $\psi_j^0$ at $n=0$ time step, the diffusion function can 
be propagated according to Eq.~(29) following the sequence of steps as delineated above. Then at a given time level $(n+1)$, 
the following series of instructions are carried out, \emph{viz.}, (a) $\psi_j^{'(n+1)}$ is normalized to $\psi_j^{(n+1)}$ (b) if one 
is interested in an excited-state calculation, then $\psi_j^{'(n+1)}$ is required to be made orthogonalized with respect to all lower 
states (we use the standard Gram-Schmidt method) (c) desired expectation values are computed as $\epsilon_0 = \langle \psi^{(n+1)} | H | 
\psi^{(n+1)} \rangle$ (d) difference in the observable expectation values between two successive time steps, $\Delta \epsilon = \langle H 
\rangle^{(n+1)} - \langle H \rangle^n$, is monitored, and (e) until the above discrepancy $\Delta \epsilon$, goes below a certain 
threshold tolerance limit, one proceeds with the calculation of $\psi_j^{(n+2)}$ at the next time level iteratively. The guessed functions 
for even and odd states were chosen as simple Gaussian-type functions such as $e^{-x^2}$ and $xe^{-x^2}$ respectively. All the integrals 
were evaluated with the help of standard Newton-Cotes quadratures \cite{abramowitz64}. 

\subsubsection{Generalized pseudospectral (GPS) method}
All the 3DCHO calculations in this work, were done by using the the GPS formalism. It provides accurate eigenvalues and eigenfunctions 
easily various $r_c$. By means of an optimal, non-uniform spatial grid, it leads to a symmetric eigenvalue problem, which can be 
easily solved by standard diagonalization routines available. It has been applied to a series of model and real systems, in both 
\emph{free and confined} cases, \emph{viz.}, spiked harmonic oscillators \cite{roy04pla, roy08theochem}, power-law and logarithmic
\cite{roy04jpg}, H\'ulthen and Yukawa \cite{roy05pramana}, Hellmann \cite{roy08jmc}, rational \cite{roy08ijqc}, exponential-screened 
Coulomb \cite{roy13ijqc}, Morse \cite{roy13rinp}, hyperbolic \cite{roy14fbsys}, Deng-Fan \cite{roy14ijqc}, Manning-Rosen 
\cite{roy14mpla_manning}, Tietz-Hua \cite{roy14jmc}, other singular \cite{roy05ijqc} potentials, as well as many-electron systems 
within the broad domain of DFT \cite{roy02exc, roy02qfd, roy02, roy04jpbhollowb,roy05jpbhollowb, roy07}. Of late
this has also produced excellent quality results in various radial confinement \cite{sen06,roy14mpla, roy15ijqc, roy16ijqc} studies 
in several Coulombic systems, as well as in 3DCHO. 

The key characteristic of this method is to approximate an \emph{exact} wave function $f(x)$ defined in the period $[-1,1]$
by the $N$th-order polynomial $f_{N}(x)$,
\begin{equation} \label{eq:23}
f(x) \cong f_{N}(x)=\sum_{j=0}^{N} f(x_{j})g_{j}(x),
\end{equation}     
and confirm the estimation to be \emph{exact} at the \emph{collocation points} $x_{j}$,
\begin{equation}
f_{N}(x_{j})=f(x_{j}). 
\end{equation}
In this work, the authors have employed the Legendre pseudo-spectral method where $x_{0}=-1$,$x_{N}=1$, and $x_{j}(j=1,....,N-1)$ are 
defined by the roots of first derivative of Legendre polynomial $P_{N}(x)$, with respect to $x$, namely,
\begin{equation}
P_{N}^{'}(x_{j})=0.
\end{equation}
In Eq.~(\ref{eq:23}), $g_{j}(x)$ are termed \emph{cardinal functions}, and as such, are given by,
\begin{equation}
g_{j}(x)=-\frac{1}{N(N+1)P_{N}(x_{j})}\frac{(1-x^{2})P_{N}^{'}(x)}{(x-x_{j})},
\end{equation} 
fulfilling the unique property that, $g_{j}(x_{j^{'}})=\delta_{j^{'},j}$. 

\begingroup           
\begin{table}
\caption{The co-efficients $a_k, b_j$ for even-$l$ $p$-space wave functions in central potential \cite{mukherjee18c}.} 
\centering
\begin{tabular}{>{\small}l|c>{\small}c>{\small}c>{\small}c>{\small}c |>{\small}c>{\small}c>{\small}c>{\small}c}
\hline
$l$ & $b_{0}$ & $b_{2}$ & $b_{4}$ & $b_{6}$ & $b_{8}$ & $a_{1}$ & $a_{3}$ & $a_{5}$ & $a_{7}$   \\
\hline
$0$ & $\frac{1}{\sqrt{\pi}}$    & --         & --        & --     & --     & --     & --      & --       & --              \\
$2$ & $\frac{1}{\sqrt{\pi}}$    & $-$$\frac{3}{\sqrt{\pi}}$      & --        & --         & --        & $\frac{3}{\sqrt{\pi}}$     
    & --         & --       & --              \\
$4$ & $\frac{1}{\sqrt{\pi}}$  & $-$$\frac{105}{\sqrt{\pi}}$    & $\frac{315}{\sqrt{\pi}}$      & --         & --        
    & $\frac{30}{\sqrt{\pi}}$    & $-$$\frac{315}{\sqrt{\pi}}$    & --       & --              \\
$6$ & $\frac{1}{\sqrt{\pi}}$   & $-$$\frac{210}{\sqrt{\pi}}$    & $\frac{4725}{\sqrt{\pi}}$    & $-$$\frac{10395}{\sqrt{\pi}}$  
    & --  & $\frac{21}{\sqrt{\pi}}$    & $\frac{-1260}{\sqrt{\pi}}$   & $\frac{10395}{\sqrt{\pi}}$   & --         \\
$8$ & $\frac{1}{\sqrt{\pi}}$  & $-$$\frac{630}{\sqrt{\pi}}$  & $\frac{51975}{\sqrt{\pi}}$  & $-$$\frac{945945}{\sqrt{\pi}}$ 
    & $\frac{2027025}{\sqrt{\pi}}$ & $\frac{36}{\sqrt{\pi}}$   & $-$$\frac{6930}{\sqrt{\pi}}$ & $\frac{270270}{\sqrt{\pi}}$ 
    & $-$$\frac{2027025}{\sqrt{\pi}}$    \\
\hline
\end{tabular}
\end{table}
\endgroup

In order to solve the radial SE for a central potential using finite-difference methods, it requires a significantly large number 
of points in an equal-spacing grid arrangement. However, in GPS method, this is alleviated by (i) mapping semi-infinite domain 
$r \in [0,\infty]$ onto a finite domain $x \in [-1,1]$ via a transformation $r=r(x)$, and then (ii) employing a Legendre 
pseudo-spectral discretization technique. Utilizing an algebraic non-linear mapping, 
\begin{equation} \label{eq:25}
r=r(x)=L\frac{(1+x)}{(1-x+\alpha)}, 
\end{equation}   
where $L, \alpha=\frac{2L}{r_{max}}$ denote two mapping parameters, plus a symmetrization procedure, 
\begin{equation}
\psi(r(x))=\sqrt{r^{'}(x)}f(x),
\end{equation}
leads to the transformed Hamiltonian, as given below, 
\begin{equation}
\hat{H}(x)=-\frac{1}{2}\frac{1}{r^{'}(x)}\frac{d^{2}}{dx^{2}}\frac{1}{r^{'}(x)}+V(r(x))+V_{m}(x),
\end{equation}
where,
\begin{equation}
V_{m}(x)=\frac{3(r^{''})^{2}-2r^{'''}r^{'}}{8(r^{'})^{4}}.
\end{equation}
This turns out to be a symmetric matrix eigenvalue problem. The mapping used in Eq.~(\ref{eq:25}) is such that $V_{m}(x)=0$. 
Eventually, we obtain following discrete set of coupled equations,
\begin{equation} \label{eq:26}
\sum_{j=0}^{N}\left[-\frac{1}{2}D^{(2)}_{j^{'}j}+\delta_{j^{'}j}V(r(x_{j}))+\delta_{j^{'}j}V_{m}(r(x_{j}))\right]A_{j}=
\mathrm{E}A_{j'}, \  J=1,...,N-1,
\end{equation}
\begin{equation}
A_{j}=R^{'}(x_{j})f(x_{j})[P_{N}(x_{j})]^{-1}= [r^{'}(x_{j})]^{\frac{1}{2}}\psi(r(x_{j}))[P_{N}(x_{j})]^{-1}.
\end{equation} 
Here, $D^{(2)}_{j^{'}j}$ signifies symmetrized second derivative of cardinal function with respect to $r$,
\begin{equation}
D^{(2)}_{j^{'}j}=[r^{'}(x_{j^{'}})]^{-1}d^{(2)}_{j^{'}j}[r^{'}(x_{j})]^{-1},
\end{equation}
\begin{equation} d^{(2)}_{j^{'}j} = \begin{cases}
\frac{(N+1)(N+2)}{6(1-x^{2}_{j})},  \ \ \ \  j=j^{'},   \\
\frac{1}{(x_{j}-x_{j^{'}})^{2}}, \ \ \ \ j \neq j^{'}.  \\
 \end{cases} 
\end{equation}
The symmetric eigenvalue problem can be easily and efficiently be solved by standard library routines, to obtain accurate 
eigenvalues and eigenfunctions. 

\begingroup           
\begin{table}
\small
\caption{The co-efficients $a_k,b_j$ for odd-$l$ $p$-space wave functions in central potential \cite{mukherjee18c}.} 
\centering
\begin{tabular}{>{\footnotesize}l
|>{\footnotesize}c>{\footnotesize}c>{\footnotesize}c>{\footnotesize}c>{\footnotesize}c<{\footnotesize}
  }
\hline
$l$ & $a_{0}$ & $a_{2}$ & $a_{4}$ & $a_{6}$ & $a_{8}$    \\
\hline 
$1$ & $\frac{1}{\sqrt{\pi}}$   & --     & --        & --            & --      \\
$3$ & $\frac{1}{\sqrt{\pi}}$  & $-$$\frac{15}{\sqrt{\pi}}$    & --        & --            &  --                    \\
$5$ & $\frac{1}{\sqrt{\pi}}$  & $-$$\frac{105}{\sqrt{\pi}}$    & $\frac{945}{\sqrt{\pi}}$   & --            & --    \\
$7$ & $\frac{1}{\sqrt{\pi}}$ & $-$$\frac{378}{\sqrt{\pi}}$   & $\frac{17325}{\sqrt{\pi}}$ & $-$$\frac{135135}{\sqrt{\pi}}$      
    & --   \\
$9$ & $\frac{1}{\sqrt{\pi}}$   & $-$$\frac{990}{\sqrt{\pi}}$ & $\frac{135135}{\sqrt{\pi}}$  & $-$$\frac{4729725}{\sqrt{\pi}}$ 
    & $\frac{34459425}{\sqrt{\pi}}$     \\
\hline
\end{tabular}

\begin{tabular}{>{\footnotesize}l
|>{\footnotesize}c>{\footnotesize}c>{\footnotesize}c>{\footnotesize}c>{\footnotesize}c<{\footnotesize}}
\hline
$l$   & $b_{1}$ & $b_{3}$ & $b_{5}$ & $b_{7}$ & $b_{9}$   \\
\hline 
$1$ & $-$$\frac{1}{\sqrt{\pi}}$     & --      & --           & --         & --              \\
$3$ & $-$$\frac{3!}{\sqrt{\pi}}$    & $\frac{15}{\sqrt{\pi}}$    & --           & --         & --              \\
$5$ & $-$$\frac{15}{\sqrt{\pi}}$    & $\frac{420}{\sqrt{\pi}}$  & $-$$\frac{945}{\sqrt{\pi}}$      & --     & --          \\
$7$ & $-$$\frac{28}{\sqrt{\pi}}$  & $\frac{3150}{\sqrt{\pi}}$ & $-$$\frac{2370}{\sqrt{\pi}}$  
    & $\frac{135135}{\sqrt{\pi}}$ & --          \\
$9$ & $-$$\frac{45}{\sqrt{\pi}}$  & $\frac{13860}{\sqrt{\pi}}$  & $-$$\frac{945945}{\sqrt{\pi}}$ 
    & $\frac{16216200}{\sqrt{\pi}}$ &  $-$$\frac{34459425}{\sqrt{\pi}}$  \\
\hline
\end{tabular}
\end{table}
\endgroup

\subsection{Momentum-space wave function}
The $p$-space wave function ($\pvec = \{ p, \Omega \}$) for a particle in a central potential is obtained from respective 
Fourier transform of its $r$-space counterpart \cite{mukherjee18c}, and as such, is given below,
\begin{equation} \label{eq:psi_p}
\begin{aligned}
\psi_{n,l}(p) & = & \frac{1}{(2\pi)^{\frac{3}{2}}} \  \int_0^\infty \int_0^\pi \int_0^{2\pi} \psi_{n,l}(r) \ \Theta(\theta) 
 \Phi(\phi) \ e^{ipr \cos \theta}  r^2 \sin \theta \ \mathrm{d}r \mathrm{d} \theta \mathrm{d} \phi,  \\
      & = & \frac{1}{2\pi} \sqrt{\frac{2l+1}{2}} \int_0^\infty \int_0^\pi \psi_{n,l} (r) \  P_{l}^{0}(\cos \theta) \ 
e^{ipr \cos \theta} \ r^2 \sin \theta  \ \mathrm{d}r \mathrm{d} \theta.  
\end{aligned}
\end{equation}
Note that $\psi(p)$ is not normalized; thus needs to be normalized. Integrating over $\theta$ and $\phi$ variables, 
Eq.~(\ref{eq:psi_p}) can be further reduced to, 
\begin{equation}
\psi_{n,l}(p)=(-i)^{l} \int_0^\infty \  \frac{\psi_{n,l}(r)}{p} \ f(r,p)\mathrm{d}r.    
\end{equation}
Depending on $l$, this can be rewritten in following simplified form ($m'$ starts with 0),  
\begin{equation}
\begin{aligned}
f(r,p) & = & \sum_{k=2m^{\prime}+1}^{m^{\prime}<\frac{l}{2}} a_{k} \ \frac{\cos pr}{p^{k}r^{k-1}} +  
  \sum_{j=2m^{\prime}}^{m^{\prime}=\frac{l}{2}} b_{j} \ \frac{\sin pr}{p^{j}r^{j-1}}, \ \ \ \ \mathrm{for} \ 
  \mathrm{even} \ l,   \\
f(r,p) & = & \sum_{k=2m^{\prime}}^{m^{\prime}=\frac{l-1}{2}} a_{k} \ \frac{\cos pr}{p^{k}r^{k-1}} +  
\sum_{j=2m^{\prime}+1}^{m^{\prime}=\frac{l-1}{2}} b_{j} \ \frac{\sin pr}{p^{j}r^{j-1}}, \ \ \ \ \mathrm{for} \ \mathrm{odd} \ l.
\end{aligned} 
\end{equation}
The co-efficients $a_{k}$, $b_{j}$ of even- and odd-$l$ states are collected in Tables~1 and 2 respectively. 

\section{Formulation of Information-theoretical quantities}
In this section we shall briefly discuss the various information-theoretic quantities along with their mathematical form.s
This will provide the context where these quantities are defined and the relations between them.

\subsection{Shannon entropy ($S$)}
Information is carried over from one place to another. We seek information when there are more than one alternatives, and we are 
not certain about the outcome of the event. If an event occurs in just one way, there is no uncertainty and no information is called 
for. In summary, we get some information due to the occurrence of an event if there was some uncertainty prevailing before the 
occurrence \cite{birula75,cover06,nielsen10}.
\begin{equation}
\begin{aligned}
\mathrm{Information~received~by~occurrence~of~an~event~} = \mathrm{Amount~of~uncertainty~} \\
\mathrm{prevailed~before~its~occurrence}
\end{aligned}
\end{equation}   
Let us consider a discrete probability distribution ($p_{1},p_{2},p_{3},...,p_{n}$) consisting of $n$ different events. To 
quantify the uncertainty in results, Shannon in 1948 defined the information entropy as \cite{shannon51},
\begin{equation}
S=-k\sum{p_{i}~\mathrm{ln}~p_{i}}, 
\end{equation}
where $k$ is a positive constant depending on the choice of unit. This definition can be explained by choosing two limiting cases: 
(i) at first, when any of the $p_{i}=1$ and others are \emph{zero}, then for this certain event $S=0$, which is minimum (ii) 
in case of equiprobability, where all $p_{i}=\frac{1}{n}$, and the uncertainty of the outcome is maximum, then $S$ is also maximum 
($S=k\ln n$) \cite{cover06}. In essence, it can be said that, for a given distribution, lesser the probability of occurring an 
event, higher will be the uncertainty associated with it. Hence, after occurrence of that event, more information will come out. 
Arguably, $S$ is the best measure of information \cite{birula75}.

In wave mechanics, the concept of $S$ has been used to explain and interpret various phenomena. Illustrative examples involve in 
illuminating Colin conjecture \cite{ramirez97,delle15}, atomic avoided crossing \cite{ghiringhelli10,ghiringhelli10a}, orbital-free 
DFT \cite{nagy14,alipour15}, electron correlation \cite{delle09,delle09a,mohajeri09,grassi11,gallegos16,
alipour18}, configuration-interaction, entanglement in artificial atoms \cite{barghathi18}, aromaticity \cite{noorizadeh10} in 
many-electron systems, etc. Moreover, it can be proved that a stronger version (compared to the traditional position-momentum 
uncertainty, due to Heisenberg), has been derived by adopting the idea of $S$ \cite{birula75},
\begin{equation} \label{eq:20} 
S_{\rvec}+S_{\pvec} \ge D(1+\mathrm{ln}~\pi), 
\end{equation}  
where $D$ refers to dimensionality of the system. Here, $S_{\rvec}$, $S_{\pvec}$ have the form,
\begin{equation}
\begin{aligned} 
S_{\rvec}  & =  -\int_{{\mathcal{R}}^3} \rho(\rvec) \ \ln [\rho(\rvec)] \ \mathrm{d} \rvec   = 
2\pi \left(S_{r}+S_{(\theta,\phi)}\right), \\   
S_{\pvec}  & =  -\int_{{\mathcal{R}}^3} \Pi(\pvec) \ \ln [\Pi(\pvec)] \ \mathrm{d} \pvec   = 
2\pi \left(S_{p}+S_{(\theta, \phi)}\right), \\ 
S_{r} & =  -\int_0^\infty \rho(r) \ \ln [\rho(r)] r^2 \mathrm{d}r, \ \ \ \ \ \ \ \ \rho(r) = |\psi_{n,l}(r)|^{2}, \\
S_{p} & =  -\int_{0}^\infty \Pi(p) \ln [\Pi(p)]  \ p^2 \mathrm{d}p, \ \ \ \ \ \ \ \ \Pi(p) = |\psi_{n,l}(p)|^{2}, \\
S_{(\theta, \phi)}  & =   -\int_0^\pi \chi(\theta) \ \ln [\chi(\theta)] \sin \theta \mathrm{d} \theta, \ \ \ \ \ \
\chi(\theta)   =  |\Theta(\theta)|^2.  \\  
\end{aligned} 
\end{equation}

\subsection{R\'enyi entropy ($R$)}
R\'enyi entropy is a one-parameter extension of $S$. R\'enyi in 1961 defined this quantity as, ``the measure of information of order 
$\alpha$ associated with the probability distribution $P=(p_{1},p_{2},....,p_{n})$" \cite{renyi61}. The mathematical form of $R$ is,
\begin{equation}
R_{\alpha}=\frac{1}{1-\alpha} \ln \left(\sum_{k} p_{k}^{\alpha}\right).
\end{equation} 
It is important to note that, $R_{\alpha}$ is considered as another measure of information because (i) $R_{\alpha}$ is the 
exponential mean of information entropy whereas $S$ provides the arithmetic mean of it, and (ii) at $\alpha \rightarrow \infty$, 
$R_{\alpha}$ reduces to $S$ \cite{birula75}. 

It is well known that, $R^{\alpha}$, the so-called information generating functional, is closely related to $\alpha$ order
entropic moments, and completely characterize density $\rho(\rvec)$. In case of continuous density distribution, it is expressed in 
terms of expectation values of density, in the following conventional form \cite{renyi61,renyi70},
\begin{equation}
R^{\alpha}[\rho (\rvec)]  =  \frac{1}{(1-\alpha)} \ \mathrm{ln} \ \langle \rho(\rvec)^{(\alpha-1)}\rangle,  \ \ \ \ 0 < \alpha 
< \infty, \ \ \alpha \neq 1. 
\end{equation}
Quantum mechanical uncertainty principle such as, $\Delta x \Delta p \ge \hbar$, does not comment about the accuracy of our 
measuring instruments. On the contrary, entropic uncertainty relations depend on the accuracy of a given measurement. Because 
they manifest the area of phase-space ($\delta \rvec \delta \pvec$) obtained by the resolution of measuring instruments. It 
suggests that, with an increase of localization of a particle in phase space, the sum of uncertainties in position and momentum 
space enhances. The quantum mechanical uncertainty relation containing the phase space is as follows,
\begin{equation} \label{eq:21}
R_{\rvec}^{\alpha}+R_{\pvec}^{\alpha} \ge -\frac{1}{2}\left(\frac{\ln~\alpha}{1-\alpha}+\frac{\ln~\beta}{1-\beta}\right) - 
\ln~\frac{\delta \rvec \delta \pvec}{\pi \hbar};  \ \ \ \ \ \left(\frac{1}{\alpha}+\frac{1}{\beta}\right)=2.
\end{equation} 
In the limit, when $\alpha \rightarrow 1$ and $\beta \rightarrow 1$, this uncertainty relation reduces to the relation given in 
Eq.~(\ref{eq:20}). The relation given in Eq.~(\ref{eq:21}) provides better idea about uncertainty as it contains all-order entropic 
moments \cite{birula06}. But its improvement is still necessary, which remains an open challenging problem. Interestingly, 
Eq.~(\ref{eq:21}) becomes sharper and sharper when the relative size of phase-space area $\left(\frac{\delta \rvec \delta 
\pvec}{\pi \hbar}\right)$ defined by the experimental resolution decreases; as it is when we enter deeper and deeper into the 
quantum regime.        

In quantum mechanics, $R^{\alpha}$ has been successfully employed to investigate and predict various quantum properties and 
phenomena like entanglement, communication protocol, correlation de-coherence, measurement, localization properties of Rydberg 
states, molecular reactivity, multi-fractal thermodynamics, production of multi-particle in high-energy collision, disordered 
systems, spin system, quantum-classical correspondence, localization in phase space \citep{varga03,renner05, levay05,verstraete06,
bialas06,salcedo09,nagy09,mcminis13,liu15,kim18}, etc. 

R{\'e}nyi entropies of order $\lambda (\neq 1)$ ($\lambda$ is either $\alpha$ or $\beta$) are obtained by taking logarithm of 
$\lambda$-order entropic moment \cite{renyi70,birula06}. In spherical polar coordinate these can be written in following simplified form by some 
straightforward mathematical manipulation \cite{mukherjee18c}, 
\begin{equation}
\begin{aligned} 
R_{\rvec}^{\lambda}  = & \frac{1}{1-\lambda} \ln \left(\int_{{\mathcal{R}}^3} \rho^{\lambda}(\rvec)\mathrm{d} \rvec \right) \\ = &
\frac{1}{(1-\lambda)} \ln \left(2\pi\int_0^\infty [\rho(r)]^{\lambda} r^2 \mathrm{d}r \int_0^\pi [\chi(\theta)]^{\lambda} \sin 
\theta \mathrm{d}\theta \right) \\ 
 = & \frac{1}{(1-\lambda)}\left( \ln 2\pi + \ln [\omega^{\lambda}_r] + \ln [\omega^{\lambda}_{(\theta, \phi)}] \right),  \\
R_{\pvec}^{\lambda}  = & \frac{1}{1-\lambda} \ln \left[\int_{{\mathcal{R}}^3} \Pi^{\lambda}(\pvec)\mathrm{d} \pvec \right] \\ = &
\frac{1}{(1-\lambda)} \ln \left(2\pi \int_{0}^\infty [\Pi(p)]^{\lambda} p^2 \mathrm{d}p \int_0^\pi [\chi(\theta)]^{\lambda} \sin 
\theta \mathrm{d}\theta \right) \\
 = & \frac{1}{(1-\lambda)}\left( \ln 2\pi + \ln [\omega^{\lambda}_p] + \ln [\omega^{\lambda}_{(\theta, \phi)}] \right).
\end{aligned} 
\end{equation}          
Here $\omega^{\lambda}_{\tau}$s are entropic moments in $\tau$ ($r$ or $p$ or $\theta$) space with order $\lambda$, having forms,
\begin{equation}
\omega^{\lambda}_r= \int_0^\infty [\rho(r)]^{\lambda} r^2 \mathrm{d}r, \ \ \  
\omega^{\lambda}_p= \int_{0}^\infty [\Pi(p)]^{\lambda} p^2 \mathrm{d}p, \ \ \  
\omega^{\lambda}_{(\theta, \phi)}= \int_0^\pi [\chi(\theta)]^{\lambda} \sin \theta \mathrm{d}\theta. 
\end{equation}          

\begin{figure}                         
\begin{minipage}[c]{0.6\textwidth}\centering
\includegraphics[scale=0.6]{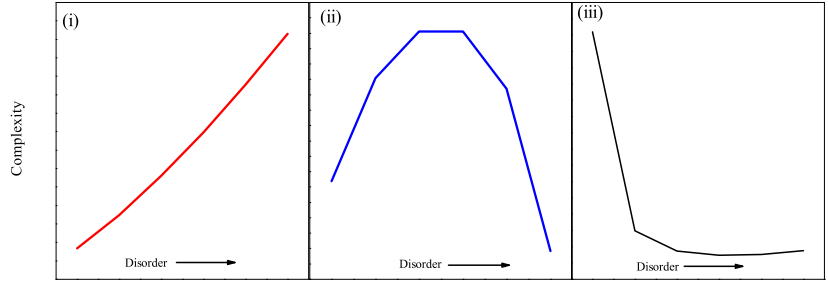}
\end{minipage}%
\caption{Three different categories of Complexities, as functions of disorder.}
\end{figure} 

\subsection{Fisher information ($I$)}
The idea of entropy can adequately explain the degree of disorder of a given phenomenon. Apart from that, however, it is necessary 
to find out a suitable measure of disorder whose variation derives the event. The concept of entropy is not able to do this. 
However, $I$ can serve as a good candidate in this context, having the ability to estimate a parameter. Hence it becomes a 
cornerstone of the statistical field of study, called parameter estimation \cite{frieden04}. $I$ measures the expected error 
in a smart measurement. Let $e^{2}$ be the mean-square error in an estimation of $\hat{\theta}$ then it obeys the relation,
\begin{equation} \label{eq:22}
\begin{aligned}
e^{2}I \ge 1; \ \ \ \
I \ge \frac{1}{e^{2}}.
\end{aligned}
\end{equation}
Equation~(\ref{eq:22}) suggests that, $I$ is always greater than the reciprocal of $e^{2}$; only in case of \emph{Gaussian} 
distribution it becomes equal to inverse of $e^{2}$. The general form of $I$ is,
\begin{equation}
I=\int \frac{|\nabla \rho(\tau)|^{2}}{\rho(\tau)} \mathrm{d}\tau, 
\end{equation}     
which is a gradient functional of density, measuring the local density fluctuation in a given space. In case of a sharp 
distribution $I$ is higher; whereas, for a flat distribution it is lower. Thus, it can be concluded 
that, with a rise in uncertainty, $I$ decreases. It resembles the Weizs\"acker kinetic energy functional, $T_{\omega}[\rho]$ often 
used in DFT \cite{sen11,nagy14}. 

In case of a central potential $I_{\rvec}$ and $I_{\pvec}$, the net Fisher information, in $\rvec$ and $\pvec$ spaces respectively, 
are expressed as \cite{romera05},
\begin{equation}
\begin{aligned} 
I_{\rvec}  =  \int_{{\mathcal{R}}^3} \left[\frac{|\nabla\rho(\rvec)|^2}{\rho(\rvec)}\right] \mathrm{d}\rvec  =  
4\langle p^2\rangle - 2(2l+1)|m|\langle r^{-2}\rangle \\ 
I_{\pvec} =  \int_{{\mathcal{R}}^3} \left[\frac{|\nabla\Pi(\pvec)|^2}{\Pi(\pvec)}\right] \mathrm{d} \pvec  = 
4\langle r^2\rangle - 2(2l+1)|m|\langle p^{-2}\rangle, \\
\langle r^2\rangle=\int_{0}^{r_c} \psi_{n_{r}, l}^{*}(r) r^{4} \ \psi_{n_{r}, l}(r) \mathrm{d}r, \  \  \ \ \
\langle p^2\rangle=\int_{0}^{r_c} \psi_{n_{r}, l}^{*}(r) [-\nabla^{2}\psi_{n_{r}, l}(r)] r^2 \mathrm{d}r  \\
\left\langle \frac{1}{r^2}\right\rangle=\int_{0}^{r_c} \psi_{n_{r}, l}^{*}(r) \psi_{n_{r}, l}(r) \mathrm{d}r, \  \  \ \ \
\left\langle \frac{1}{p^2}\right\rangle=\int_{0}^{r_c} \psi_{n_{r}, l}^{*}(p) \psi_{n_{r}, l}(p) \mathrm{d}p. 
\end{aligned}
\end{equation}
The above equations can be further recast in following equivalent forms \cite{mukherjee18d, mukherjee18},
\begin{equation}
\begin{aligned}
I_{\rvec} & = 8\mathcal{E}_{n,l}-8\langle v(r)\rangle-2(2l+1)|m|\langle r^{-2}\rangle \\
I_{\pvec} & = 8\mathcal{E}_{n,l}-8\langle T\rangle-2(2l+1)|m|\langle p^{-2}\rangle \\
          & = \frac{8}{\omega^{2}}\langle v(r)\rangle-2(2l+1)|m|\langle p^{-2}\rangle,
\end{aligned}
\end{equation}
where $v(p)$ is the $p$-space counterpart of $v(r)$. 

In case of a 3DCHO, $I$'s in $r$ and $p$ space can be expressed analytically as \cite{patil07}, 
\begin{equation}
\begin{aligned} 
I_{\rvec}(\omega)  =\frac{\omega}{\sqrt{2}} I_{\rvec}(\omega=1), \hspace{3mm} \ \ \ \  I_{\pvec}(\omega) = 
\frac{\sqrt{2}}{\omega}I_{\pvec}(\omega=1).
\end{aligned}
\end{equation}
Thus, an increase in $\omega$ leads to rise in $I_{\rvec}(\omega)$ and fall in $I_{\pvec}(\omega)$. However, it is obvious that 
$I_{t} \ (=I_{\rvec} I_{\pvec})$ remains invariant with $\omega$. Throughout the article, for brevity, $I_{\rvec}(\omega=1)$ and 
$I_{\pvec}(\omega=1)$ will be symbolized as $I_{\rvec}$, $I_{\pvec}$ respectively.

When $m=0$, $I_{\rvec}$ and $I_{\pvec}$ in Eq.~(62) reduce to further simplified forms as below,  
\begin{equation}
\begin{aligned} 
I_{\rvec}  =  4\langle p^2\rangle, \hspace{3mm} \ \ \ \  I_{\pvec} =4\langle r^2\rangle.
\end{aligned}
\end{equation}
It is seen that, at fixed $n_{r}, l$, both $I_{\rvec}, I_{\pvec}$ provide maximum values when $m=0$, and both of them 
decrease with rise in $m$. Hence one obtains the following upper bound for $I_t$, 
\begin{equation}
 I_{\rvec} I_{\pvec} \ (=I_t) \leq 16 \langle r^2\rangle \langle p^2\rangle .
\end{equation}
Further adjustment using Eq.~(65) leads to following uncertainty relations \cite{romera05}, 
\begin{equation}
\frac{81}{\langle r^2\rangle \langle p^2\rangle} \leq I_{\rvec} I_{\pvec} \leq 16 \langle r^2\rangle \langle p^2\rangle.  
\end{equation}
Therefore, in a central potential, $I$-based uncertainty product is bounded by both upper as well as lower limits. They are 
state-dependent, varying with $n_{r}$ and $l$ quantum numbers.
 
\begin{figure}                         
\begin{minipage}[c]{0.5\textwidth}\centering
\includegraphics[scale=0.65]{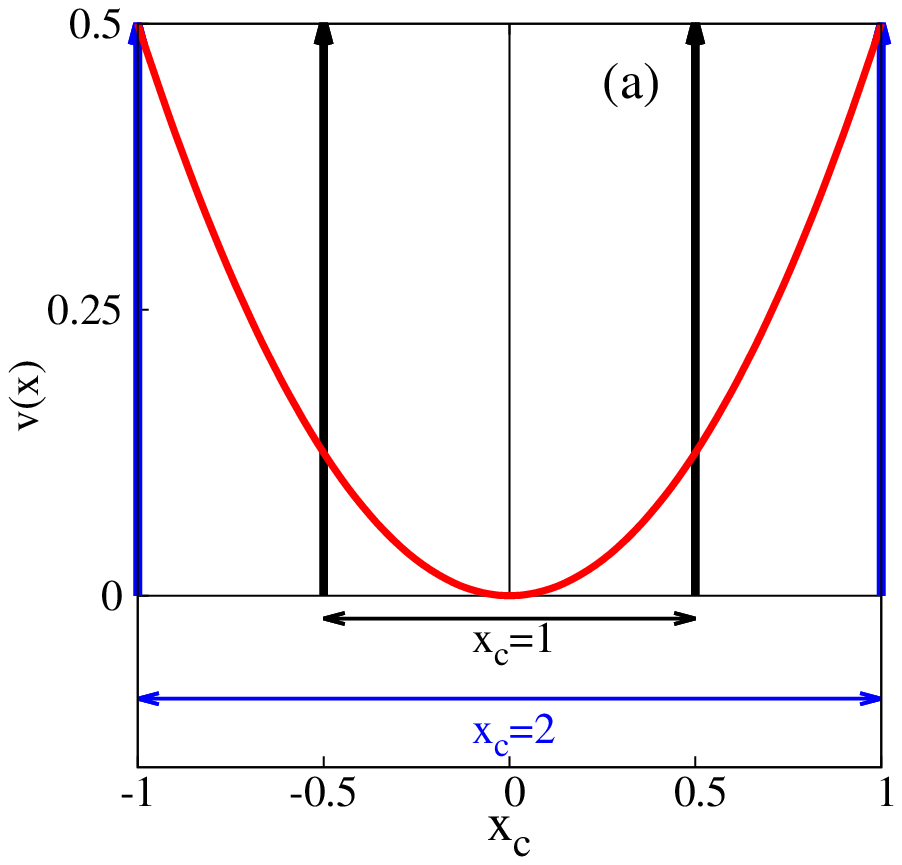}
\end{minipage}%
\hspace{0.1in}
\begin{minipage}[c]{0.5\textwidth}\centering
\includegraphics[scale=0.65]{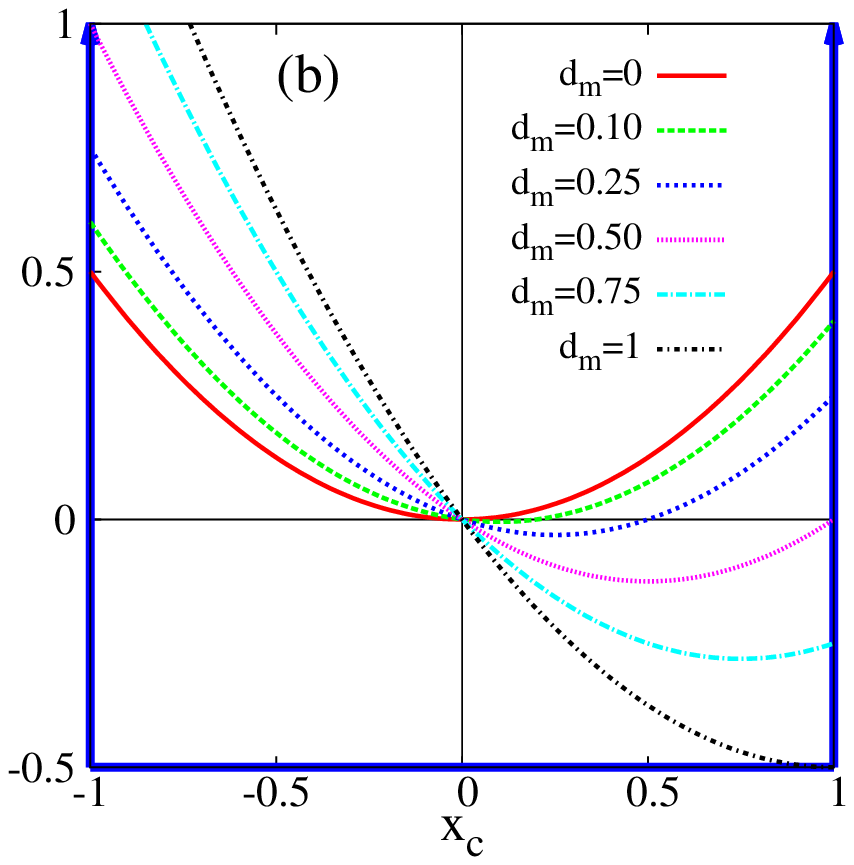}
\end{minipage}%
\caption{Schematic representation of a confined 1DQHO potential: (a) SCHO at two separate box lengths
(b) ACHO at six chosen values of $d_m$ \cite{ghosal16}.}
\end{figure} 

\subsection{Onicescu energy ($E$)}
Information energy is the counterpart of information entropy. In 1966, Onicescu defined this quantity as \cite{onicescu66},
\begin{equation}
E=\sum_{i} p_{i}^{2}.
\end{equation} 
It is interesting to note that, in case of a certain event $E$ is maximum. On the contrary, for an equiprobable distribution it is 
minimum \cite{pardo86,sen11,bhatia97}. Qualitatively it is inverse of $S$. When $S$ increases, $E$ decreases and vise-versa. In case 
of continuous probability distribution, $E$ is an expectation value of probability density. Its mathematical form is,
\begin{equation}
E=\int \rho^{2}(\tau) \ \mathrm{d}\tau.
\end{equation}
In addition, \emph{Disequilibrium} ($D_e$) for a continuous probability distribution is defined as,
\begin{equation}
De=\int_{-\infty}^{+\infty} \rho^{2}(\tau) \mathrm{d}\tau. 
\end{equation} 
Therefore, in this occasion, disequilibrium and information energy have same form. Like the previous measures, it is also 
utilized in orbital-free DFT \cite{alipour15}, testing normality \cite{noughabi15}, electron correlation 
\cite{gallegos16}, Colin conjecture \cite{ramirez97,delle15}, configuration interaction \cite{alcoba16} etc.  

By definition, $E$ refers to the 2nd-order entropic moment \cite{sen11}; for central potential it assumes the form ($E_{t}$ is 
the Onicescu energy product),
\begin{equation}
\begin{aligned}
E_{r} & = \int_0^\infty [\rho(r)]^{2} r^2 \mathrm{d}r, \ \   
E_{p}  = \int_{0}^\infty [\Pi(p)]^{2} p^2 \mathrm{d}p, \\
E_{\theta, \phi} & = \int_0^\pi [\chi(\theta)]^{2} \sin \theta \mathrm{d}\theta.  \ \
E_{t} = E_{r}E_{p}E_{\theta, \phi}^{2}.
\end{aligned} 
\end{equation}
Uncertainty product for such measures are studied in \cite{zozor07}.

\subsection{Complexities}
In a system, statistical complexity arises due to breakdown of symmetry. It illustrates a competing effect of two complementary 
quantities, offering a qualitative idea of the organization, structure and correlation. According to \cite{shiner99} 
it can be broadly characterized into three categories: (i) advances monotonically with disorder (ii) reaches its minimal 
value for both completely ordered and disordered systems, and a maximum at some intermediate level (iii) increases with order. It 
has finite value in a state lying between two limiting cases of complete order (maximum distance from equilibrium) and maximum 
disorder (at equilibrium). Figure~1 pictorially demonstrates above three different kind of complexities as a function 
of disorder.  

The statistical measure of complexity ($C_{LMC}$) is nothing but the product of the information content ($H$) (such as $S$, $R$ 
etc.) and concentration of spatial distribution ($D$), and can be written as $C_{LMC} = H. D$. This was later criticized 
\cite{catalan02} and modified \cite{sanchez05} to the form of $C_{LMC} = D. e^{S}$, in order to satisfy few conditions such as 
reaching minimal values for both extremely ordered and disordered limits, invariance under scaling, translation and replication. 
Various definitions were put forth in literature. Some notable ones include Shiner, Davidson and Landsberg (SDL) \cite{landsberg98,
shiner99}, Fisher-Shannon $(C_{IS})$ \cite{romera04,sen07,angulo08}, Cram\'er-Rao \cite{angulo08,antolin09}, generalized 
R\'enyi-like \cite{calbet01,martin06,romera08} complexity, etc. Amongst these, $C_{IS}$ corresponds to a measure which probes a 
system in terms of complementary global and local factors, and also satisfies certain desirable properties in complexity \cite{sen11}, 
like invariance under translations and re-scaling transformations, invariance under replication, near-continuity, etc. This has 
remarkable applications in the study of atomic shell structure, ionization processes \cite{sen07,angulo08,angulo08a}, as well as 
in molecular properties like energy, ionization potential, hardness, dipole moment in the localization-delocalization plane showing 
chemically significant pattern \cite{esquivel10}, molecular reactivity studies \cite{welearegay14}. Some elementary chemical 
reactions such as hydrogenic-abstraction reaction \cite{esquivel11}, identity $SN^{2}$ exchange reaction \cite{molina12}, and also 
concurrent phenomena occurring at the transition region \cite{esquivel12} of these reactions have been investigated through 
composite information-theoretic measures in conjugate spaces.

Without any loss of generality, let us define complexity in following general form $C = Ae^{b.B}$. The order ($A$) and disorder 
parameters ($B$) may include ($E, I$) and ($R, S$) respectively. With this in mind, we are interested in the following four
quantities, 
\begin{equation}
\begin{aligned}
C_{ER} & = E e^{bR}, \ \ \ \ \ \ \ \ C_{IR} = Ie^{bR}, \ \ \ \ \ \ \ C_{ES} & = E e^{bS}, \ \ \ \ \ \ \ \ C_{IS} = Ie^{bS}. 
\end{aligned} 
\end{equation}

\begin{figure}                         
\small
\begin{minipage}[c]{0.47\textwidth}\centering
\includegraphics[scale=0.60]{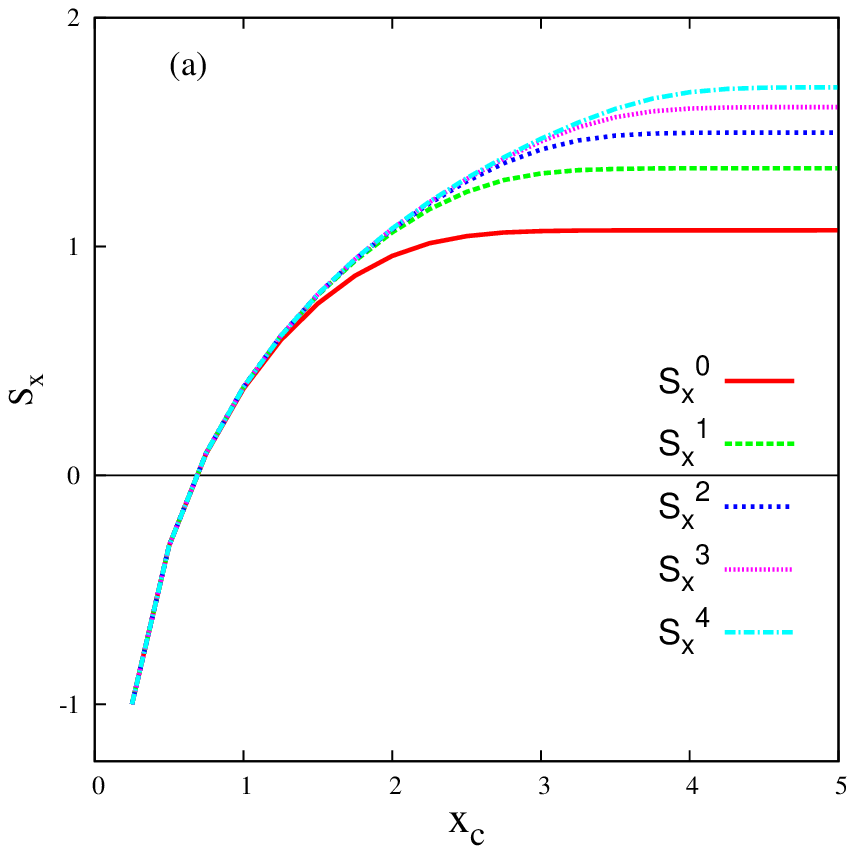}
\end{minipage}%
\hspace{0.1in}
\begin{minipage}[c]{0.47\textwidth}\centering
\includegraphics[scale=0.60]{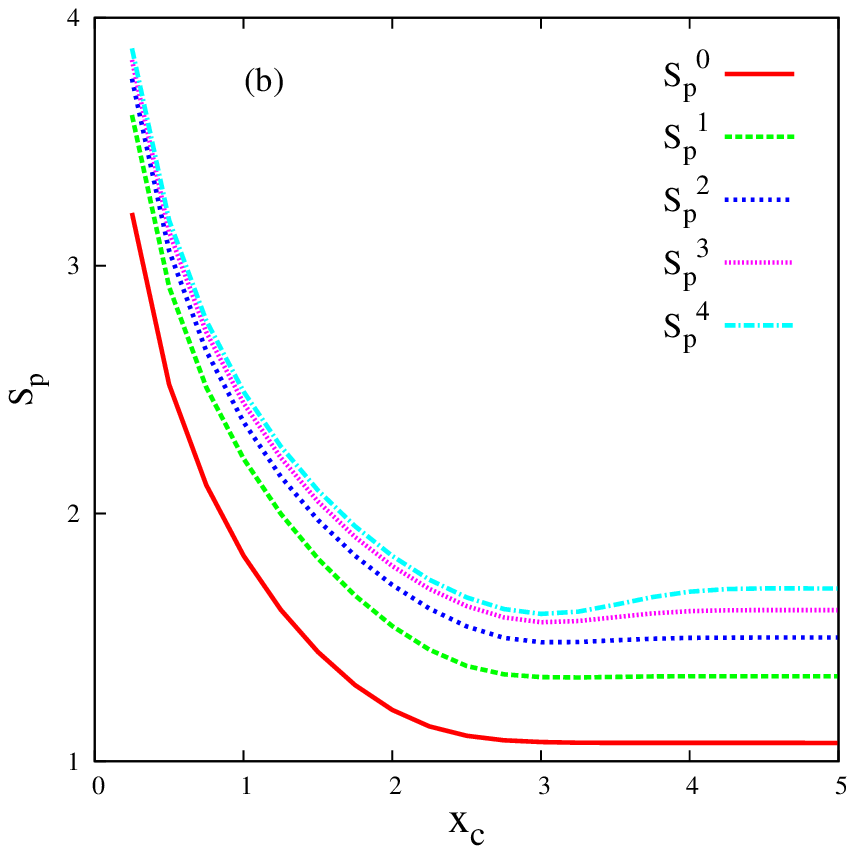}
\end{minipage}%
\caption{Plot of (a) $S_x$ (b) $S_p$, of SCHO potential, as function of $x_c$, for first five states \cite{ghosal16}.}
\end{figure} 

\subsection{Relative information}
Kullback-Leibler divergence or relative entropy is a descriptor or quantifier of how a measured probability distribution function 
deviates from a given reference distribution \cite{kullback51,kullback78}. In quantum mechanics, this characterizes a measure of 
distinguishability between two states. It actually quantifies the change of information from one state to other \cite{frieden04}. 
Relative $R$ and $S$ was explored for various atomic systems using H atom ground state as reference \cite{sagar08}. A detailed 
investigation reveals  that, they are directly related with atomic radii and quantum capacitance \cite{nagy09a,sagar08}. Another 
interesting measure is the relative Fisher information (IR) \cite{villani00}. It has importance in different topics of physics 
and chemistry, such as to calculate phase-space gradient of dissipated work and information \cite{yamano13}, deriving Jensen 
divergence \cite{sanchez12}, relation with score function \cite{toscani17}, in the context of study of probability current 
\cite{yamano13a}, in thermodynamics \cite{frieden99}, etc. Of late it has been successfully used in formulating atomic densities 
\cite{antolin09a} and deriving density functionals under local-density and generalized-gradient approximations \cite{levamaki17}. 
Further, IR along with Hellmann-Feynman and virial theorem has been used to develop a Legendre transform structure related to 
SE \cite{flego11}. It has been designed self-consistently on the basis of estimation theory \cite{frieden10}. In quantum chemistry 
perspective, it has been successfully formulated using above two theorems and entropy maximization principle \cite{flego11a,
venkatesan14,venkatesan15}. Very recently, IR for some exactly solvable potentials including 1D and 3D QHO in both position and 
momentum spaces has been estimated analytically using ground state of definite symmetry (for example, $l=0$ for $s$ orbitals, $l=1$ 
for $p$ orbital, and so on) as reference \cite{mukherjee18b}. Before that, a numerical estimation of IR in position space is done 
for H-atom using $1s$ orbital as basis \cite{yamano18}.   

For two normalized probability densities $\rho_{n,l,m}(\tau), \rho_{n_{1},l_{1},m_{1}}(\tau)$, IR is expressed as,
\begin{equation}
\mathrm{IR} \ [\rho_{n,l,m}(\tau)|\rho_{n_{1},l_{1},m_{1}}(\tau)]=\int_{{\mathcal{R}}^3}\rho_{n,l,m}(\tau)\left|\nabla \ 
\mathrm{ln} \left\{ \frac{\rho_{n,l,m}(\tau)}{\rho_{n_{1},l_{1},m_{1}} (\tau)}\right \} \right|^{2}  \mathrm{d}\tau.
\end{equation}
Here $n,l,m$ and $n_{1},l_{1},m_{1}$ are the descriptors of target and reference states respectively, while $\tau$ is a 
generalized variable. In case of central potential, these probability densities $\rho_{n,l,m}(\tau)$, $\rho_{n_{1},l_{1},m_{1}}
(\tau)$ can be expressed in following forms, without any loss of generality, 
\begin{equation}
\begin{aligned}
\rho_{n,l,m}(\tau) & = R_{n,l}^{2}(s) \Theta_{l,m}^{2}(\theta);  \      \hspace{3mm} 
\rho_{n_{1},l_{1},m_{1}}(\tau) = R_{n_{1},l_{1}}^{2}(s) \Theta_{l_{1},m_{1}}^{2}(\theta). 
\end{aligned}
\end{equation}
In the above equation, $R_{n,l}(s), R_{n_{1},l_{1}}(s)$ signify radial parts, $\Theta_{n,l}(\theta), \Theta_{n_{1},l_{1}}
(\theta)$ represent angular contributions of two wave functions, whereas $``s"$ implies either $r$ or $p$ variable in respective 
radial functions. Thus Eq.~(73) may be rewritten as,
{\footnotesize
\begin{multline}
\mathrm{IR} \ [\rho_{n,l,m}(\tau)|\rho_{n_{1},l_{1},m_{1}}(\tau)]= \mathrm{IR} \ [\rho_{n,l}(s)|\rho_{n_{1},l_{1}}(s)]+
\left\langle \frac{1}{s^{2}} \right\rangle \mathrm{IR} \ [\Theta_{l,m}^{2}(\theta)|\Theta_{l_{1},m_{1}}^{2}(\theta)]+\\
2\int_{0}^{\infty}s~R_{n,l}^{2}(s)\left[\frac{d}{ds}\mathrm{\ln}\left\{ \frac{R^{2}_{n,l}(s)}{R^{2}_{n_{1},l_{1}}(s)}\right\}
\right]\mathrm{d}s
\int_{0}^{\pi}\Theta_{l,m}^{2}(\theta)\left[\frac{d}{d\theta}\mathrm{ln}\left\{ \frac{\Theta_{l,m}^{2}(\theta)}
{\Theta_{l_{1},m_{1}}^{2}(\theta)}\right\} \right] \sin\theta \mathrm{d}\theta
\end{multline}
}
where the following quantities have been defined, 
\begin{equation}
\begin{aligned}
\mathrm{IR} \ [\rho_{n,l}(s)|\rho_{n_{1},l_{1}}(s)]& =\int_{0}^{\infty}s^{2}~\rho_{n,l}(s)\left|\frac{d}{ds}\mathrm{ln}
\left\{ \frac{\rho_{n,l}(s)}{\rho_{n_{1},l_{1}}(s)}\right\} \right|^{2}\mathrm{d}s \\ 
\mathrm{IR} \ [\Theta_{l,m}^{2}(\theta)|\Theta_{l_{1},m_{1}}^{2}(\theta)]&=\int_{0}^{\pi}\Theta_{l,m}^{2}(\theta)\left|
\frac{d}{d\theta}\mathrm{ln}\left\{ \frac{\Theta_{l,m}^{2}(\theta)}{\Theta_{l_{1},m_{1}}^{2}(\theta)}\right\} \right|^{2} 
\sin\theta~\mathrm{d}\theta      
\end{aligned}
\end{equation}

\begin{figure}[h]             
\small
\centering
\begin{minipage}[c]{0.47\textwidth}\centering
\includegraphics[scale=0.52]{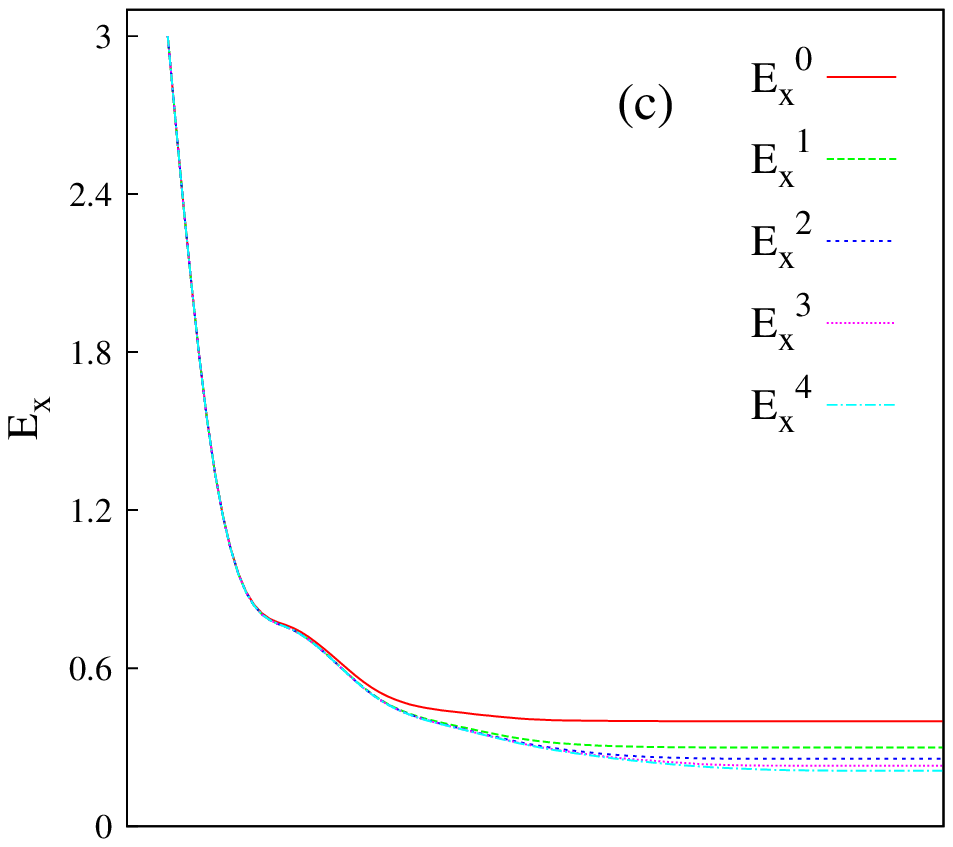}
\end{minipage}
\hspace{5pt}
\begin{minipage}[c]{0.47\textwidth}\centering
\includegraphics[scale=0.52]{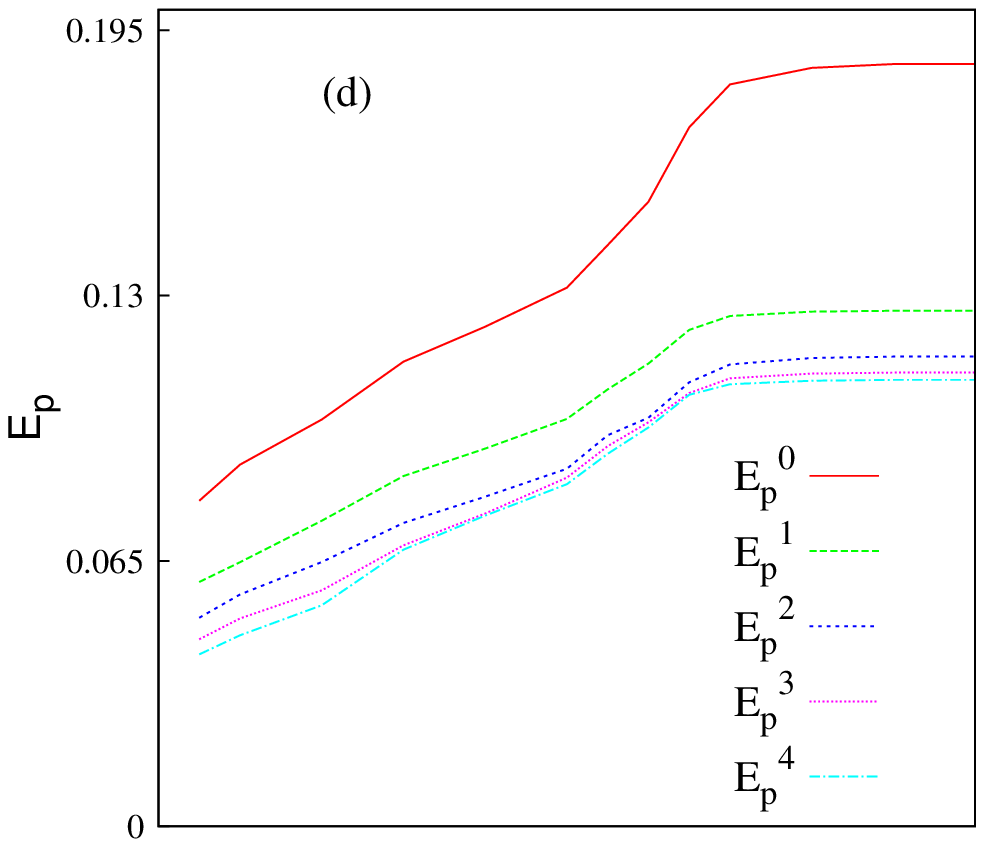}
\end{minipage}
\\[5pt]
\begin{minipage}[c]{0.47\textwidth}\centering
\includegraphics[scale=0.580]{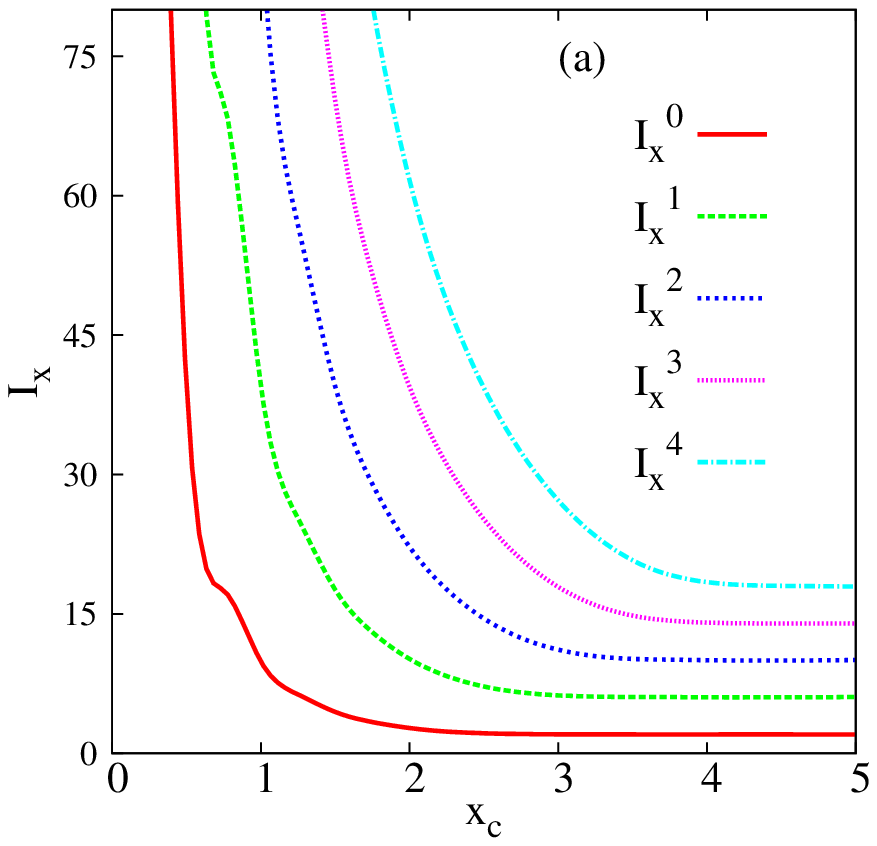}
\end{minipage}
\hspace{10pt}
\begin{minipage}[c]{0.47\textwidth}\centering
\includegraphics[scale=0.580]{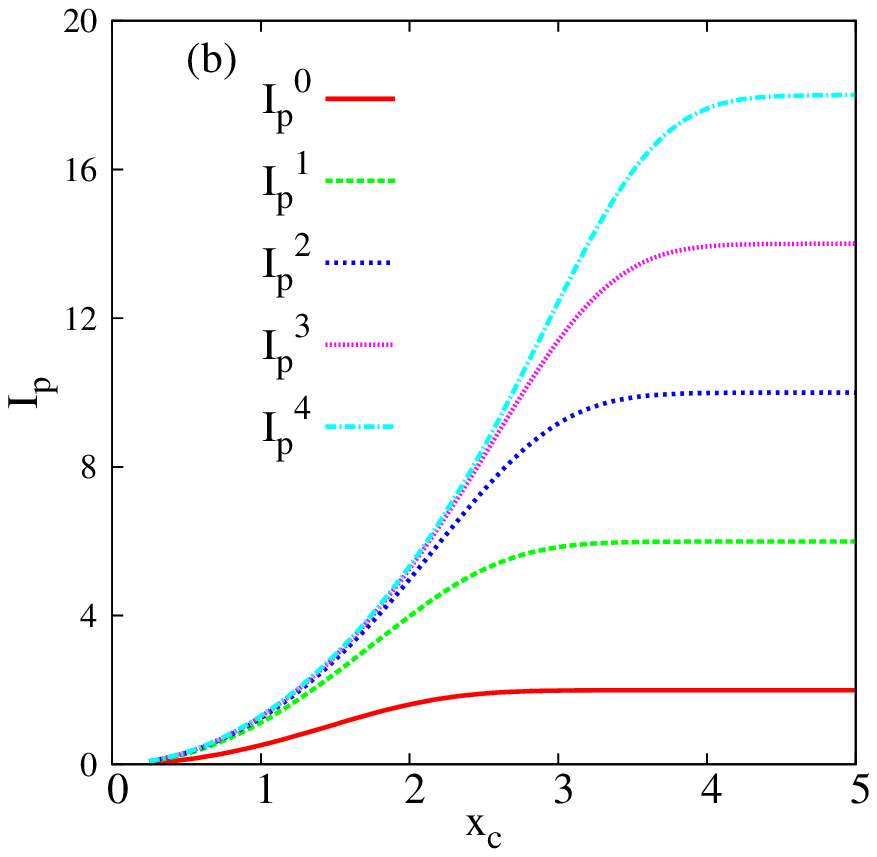}
\end{minipage}
\caption{Plot of $I_x$,~$I_p$, $E_x$,~$E_p$ for first five states of SCHO potential, as functions of $x_c$, in panels 
(a), (b), (c) and (d). See text for details \cite{ghosal16}.}
\end{figure}
 
\section{Result and Discussion}
\subsection{ 1D Confined harmonic oscillator (1DCHO)}
Panel (a) of Fig.~2 represents an SCHO scenario for two box lengths $1$ and $2$, respectively. Next, the asymmetric confinement 
in a 1DQHO can be invoked in two different ways: (i) by varying the box boundary, keeping box length and $d_m$ fixed at zero (ii) 
the other way is to change $d_m$ by keeping box length and boundary constant. We have chosen the second option; where a rise in 
$d_m$ transfers the minimum towards the right of origin, keeping box length, $b_2-b_1 = 2$ fixed, while left and right 
boundaries are residing at $b_1 = -1$ and $b_2 = 1$. Moreover, since $\frac{1}{2} k(x \pm d_{m})^{2}$ illustrates a mirror-image pair 
potential trapped within $b_1$ and $b_2$, their eigenvalues and expectation values are same for all states. As the respective 
wave functions are mirror images of each other, it suffices to consider the behavior of either one of them. The other one 
automatically follows from it. The right-hand panel (b) of Fig.~1 shows a schematic representation of an ACHO potential, at 
five different $d_m$ values.

\begingroup      
\begin{table}
\centering
\caption{Comparison of energies of first six states of ACHO potential at four distinct $d_m$. Length of the box is 
fixed at 2. PR implies Present Result \cite{ghosal16}. See text for details.} 
{\scriptsize
\begin{tabular}{>{\scriptsize}l|>{\scriptsize}l>{\scriptsize}l|>{\scriptsize}l>{\scriptsize}l}
\hline 
$d_m$ & $\epsilon_0$ (PR) &  $\epsilon_0$ (Ref.) & $\epsilon_1$ (PR) &  $\epsilon_1$(Ref.) \\
\hline
0.36 & 2.7177633960054 &  2.7177633960054$^\ddag$         
     & 10.283146010610 &  10.283146010610$^\ddag$         \\
1.92 & 6.0383021056781 &  6.0383021056781$^\ddag$             
     & 13.901445986629 &  13.901445986629$^\ddag$         \\
5.00 & 26.065225076406 & 26.065225076406$^\ddag$ & 35.462261039378 & 35.462261039378$^\ddag$ \\
10.0 & 97.474035270680 & 97.474035270680$^\ddag$ & 110.51944554927 & 110.51944554927$^\ddag$ \\
   \hline
    & $\epsilon_2$ (PR) &  $\epsilon_2$ (Ref.) & $\epsilon_3$ (PR) &  $\epsilon_3$(Ref.) \\
    \hline
0.36 & 22.648848755052 &  22.648848755052$^\ddag$           
     & 39.929984298830 &  39.929984298830$^\ddag$         \\
1.92 & 26.249310409373 &  26.249310409373$^\ddag$          
     & 43.513981920357 &  43.513981920357$^\ddag$          \\
5.00 & 47.817024422796 & 47.817024422796$^\ddag$  & 64.900200447511 & 64.900200447511$^\ddag$  \\
10.0 & 123.593144939095& 123.593144939095$^\ddag$ & 140.555432078323  & 140.555432078323$^\ddag$  \\
    \hline
    & $\epsilon_4$ (PR) &  $\epsilon_4$ (Ref.) & $\epsilon_5$ (PR) &  $\epsilon_5$(Ref.) \\
    \hline
0.36 & 62.140768627508 & 62.140768627508$^\ddag$ & 89.284409553063  & 89.284409553063$^\ddag$ \\
1.92 & 65.715672311936 & 65.715672311936$^\ddag$ & 92.854029622882  & 92.854029622882$^\ddag$ \\
5.00 & 87.137790461503 & 87.137790461503$^\ddag$ & 114.244486402564 & 114.244486402564$^\ddag$  \\
10.0 & 162.519960161732 & 162.519960161732$^\ddag$ & 189.515389275133 & 189.515389275133$^\ddag$  \\
\hline
\end{tabular}
}
\begin{tabbing}
$^\ddag$ITP result \cite{roy15}. \hspace{35pt} \=
\end{tabbing}
\end{table} 
\endgroup  

\subsubsection{Symmetrically confined harmonic oscillator (SCHO)}
In this case, all calculations are performed involving the wave function given in Eq.~(2). Figure~3 shows 
plots of $S_{x}$, $S_{p}$ versus $x_c$, for first five states of SCHO. In panel (a), $S_{x}$ progresses with box 
length and converges to a constant value ($S_{x}$ of 1DQHO) at sufficiently large box length. At small $x_c$ region, $S_x$ changes 
rather insignificantly with $n$, corresponding to the behavior of a PIB problem (where $S_x$ remains unchanged with $n$). 
Next, panel (b), imprints that $S_{p}$ abates with increase in box length and finally merges to corresponding 1DQHO value. It is 
important to note that, there appears a minimum in $S_p$ for all excited states. Appearance of such minimum may be ascribed to 
the competing effect in $p$ space. In fact, three possibilities could be contemplated ($l_{x}$, $l_{p}$ are box lengths in $x$, 
$p$ space):
\begin{enumerate}[(a)]
\item
When $l_{x} \rightarrow 0,$ then $l_{p} \rightarrow \infty$.
\item
When $l_{x}$ is finite, then $l_{p}$ is also finite. But an increase in $l_{x}$ leads to a decrease in $l_{p}$.
\item
When $l_{x} \rightarrow \infty,$ then also $l_{p} \rightarrow \infty$. 
\end{enumerate}
These plots suggest that, at the beginning, with increase in $l_{x}$, particle gets localized in $p$ space ($S_{p}$ decreases), 
but when potential behaves like 1DQHO, de-localization predominates. Thus, existence of minimum in $S_{p}$ is due to presence 
of two conjugate forces. These variations of $S_x$, $S_p$ with $x_c$ are in complete agreement with the findings of 
\cite{laguna14}. 

The behavior of $I_{x}$, $I_{p}$ with respect to $x_{c}$ of first five stationary states are exhibited in two bottom panels (a), (b) 
of Fig.~(4). One can see that initially $I_{x}$ decreases sharply, extent of which is maximum for the lowest state and successively 
decreases with $n$. Then it attains a constant value after a certain $x_c$. On the contrary, in panel (b), $I_{p}$ strongly grows 
with $x_{c}$ at first; the extent increases with $n$, lowest state producing lowest. Finally, for all states, it eventually  
reaches a state-dependent constant value as in $I_x$. It is pertinent to mention that, behavior of $I_x$, $I_p$ under confinement 
agrees with those of $\Delta p$, $\Delta x$ \cite{laguna14}. Next, panel (c) reveals that, $E_x$ for all these states remain 
very close to each other at smaller $x_c$; then as $x_c$ progresses, $E_x$ for individual states branch out and decreases 
indicating delocalization. In the end, it reaches some constant value for all $n$. Lastly, in panel (d), $E_p$s tend to 
advance in the beginning for all states at low $x_c$ region, eventually becoming smooth after some threshold $x_c$. These 
calculation have further consolidated the conclusions obtained from the study of $S_x$,~$S_p$, $S$ demonstrated earlier.

\begin{figure}                         
\small
\begin{minipage}[c]{0.3\textwidth}\centering
\includegraphics[scale=0.4]{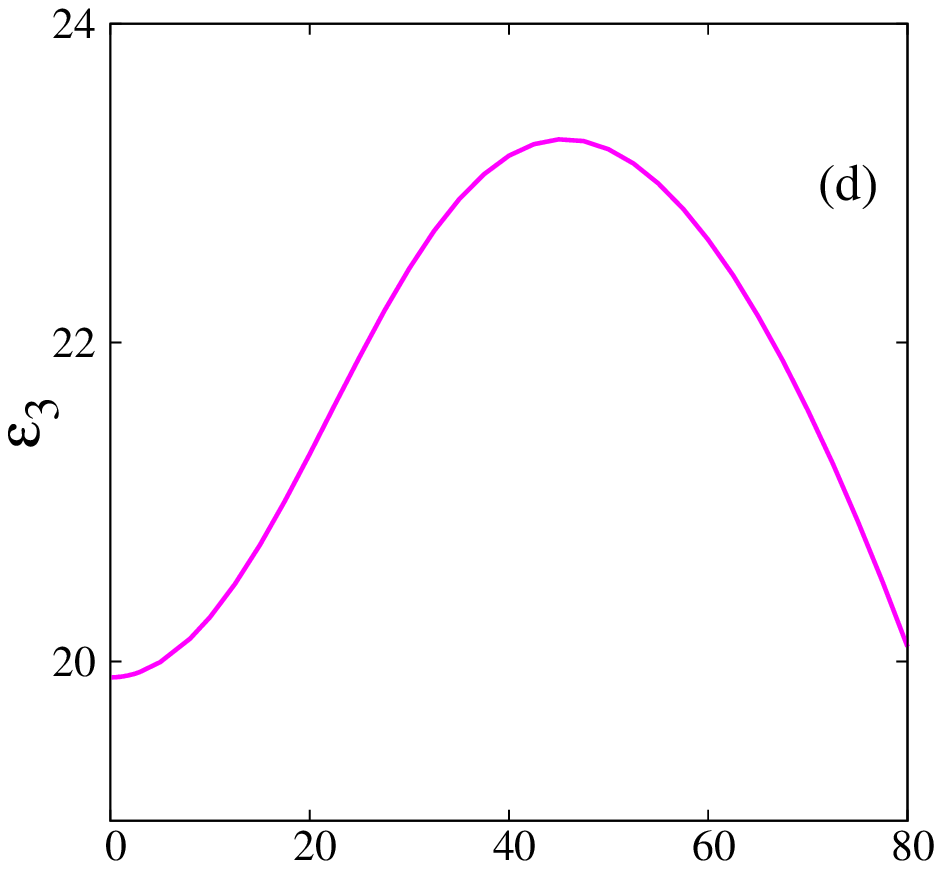}
\end{minipage}%
\hspace{0.15in}
\begin{minipage}[c]{0.3\textwidth}\centering
\includegraphics[scale=0.4]{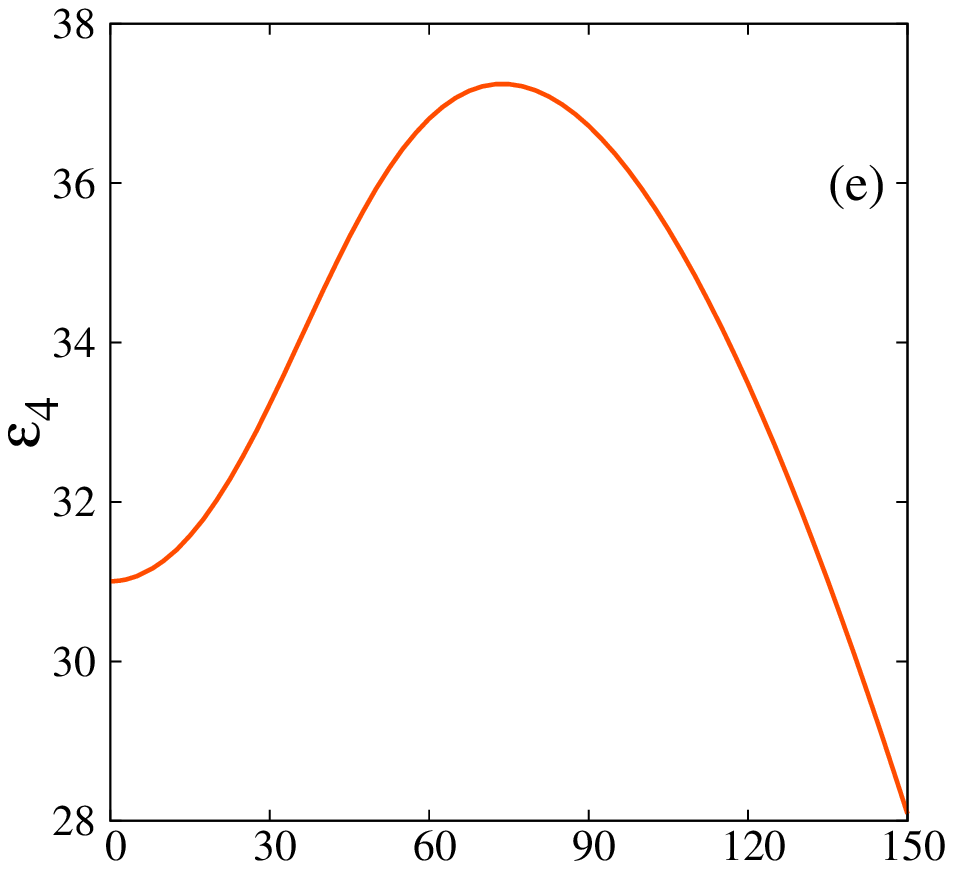}
\end{minipage}%
\hspace{0.15in}
\begin{minipage}[c]{0.3\textwidth}\centering
\includegraphics[scale=0.4]{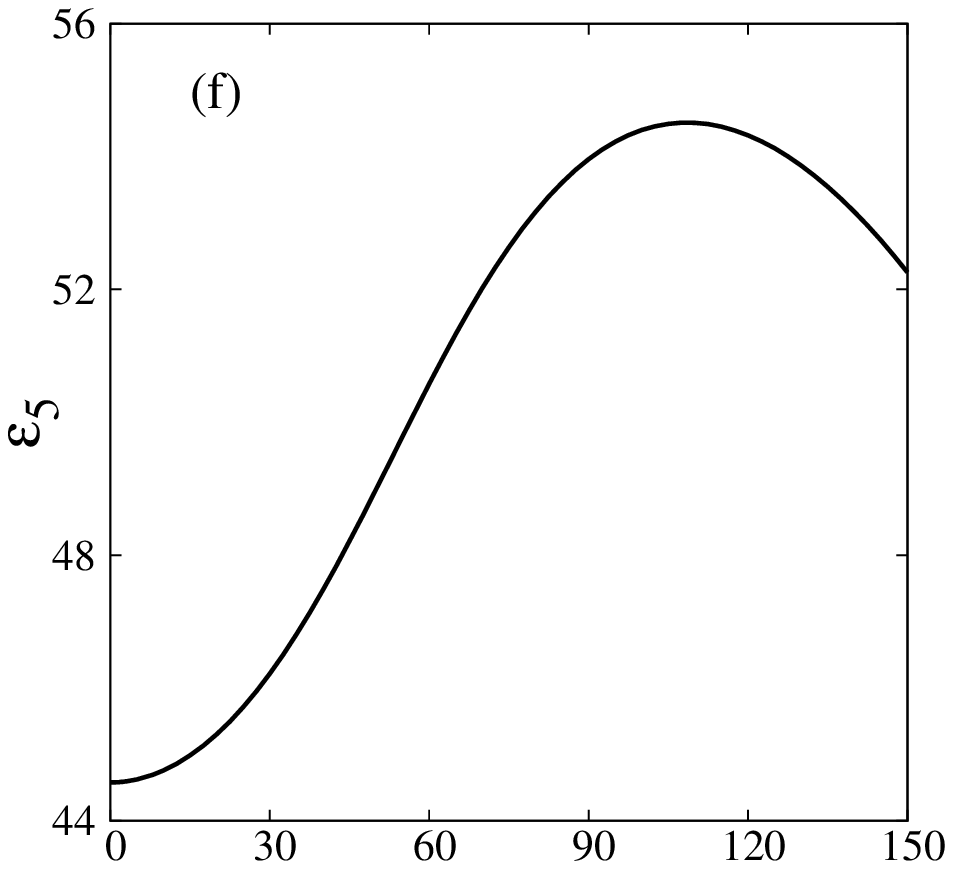}
\end{minipage}%
\hspace{0.01in}
\begin{minipage}[c]{0.30\textwidth}\centering
\includegraphics[scale=0.4]{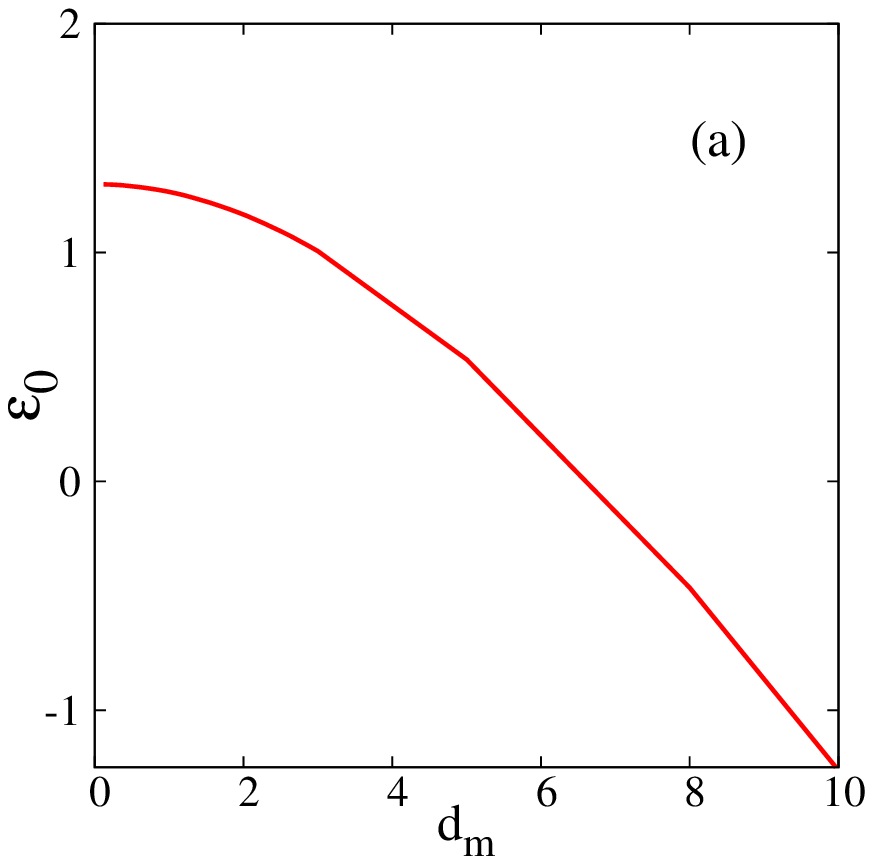}
\end{minipage}
\hspace{0.2in}
\begin{minipage}[c]{0.30\textwidth}\centering
\includegraphics[scale=0.4]{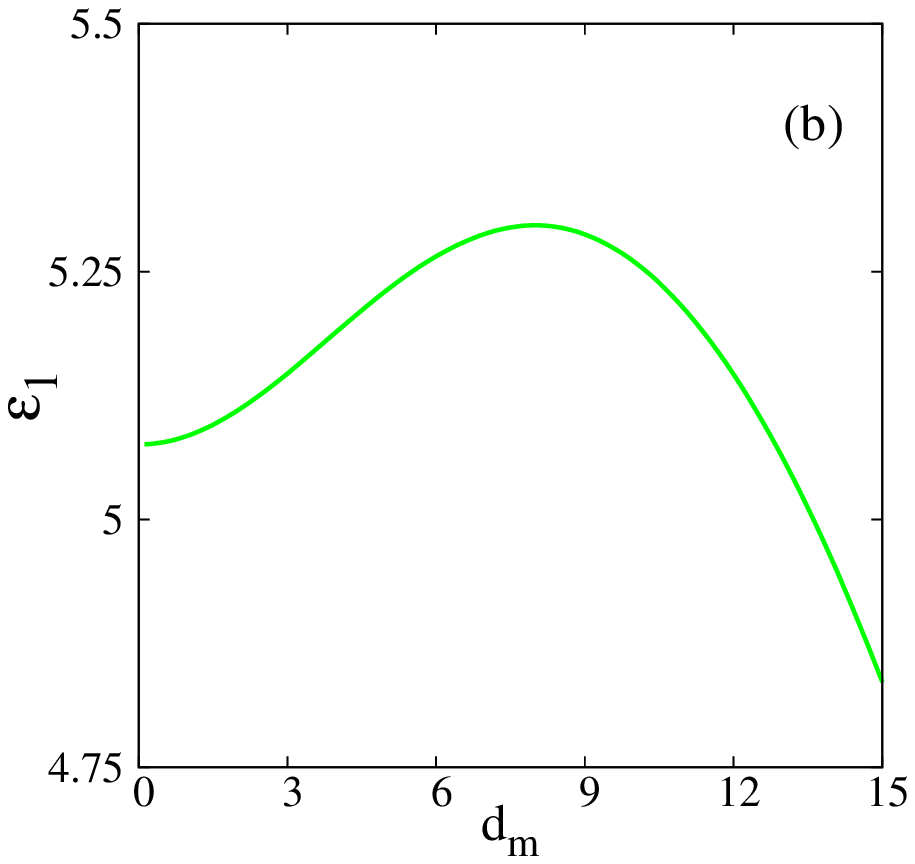}
\end{minipage}%
\hspace{0.21in}
\begin{minipage}[c]{0.30\textwidth}\centering
\includegraphics[scale=0.4]{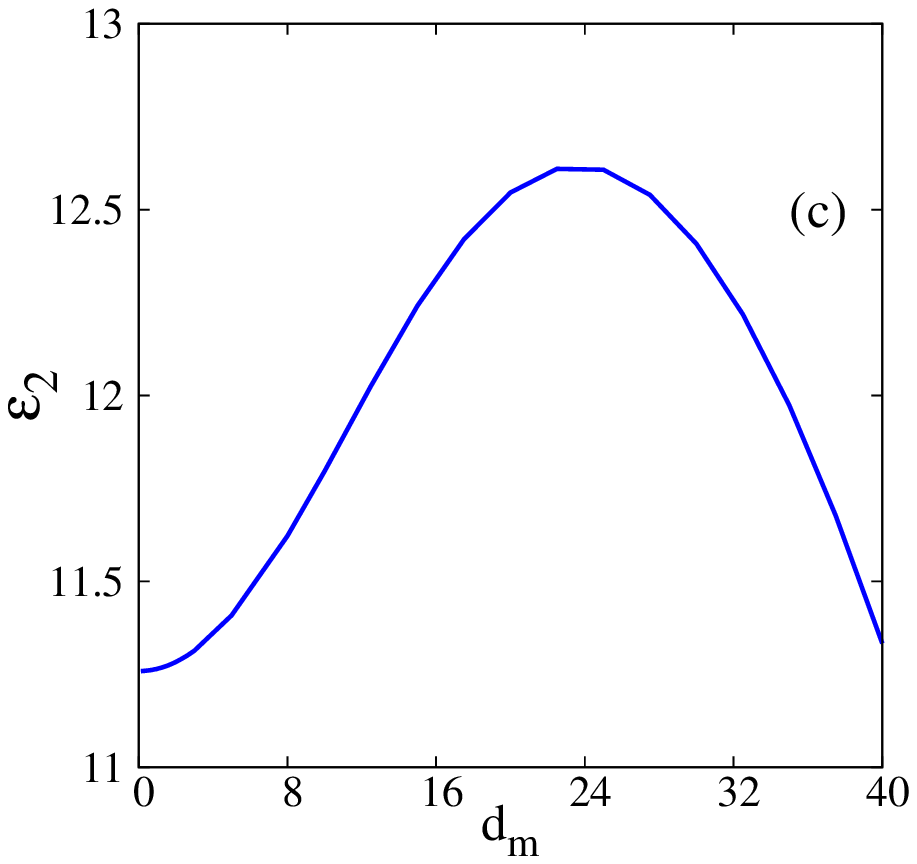}
\end{minipage}%
\caption{Variation of $\epsilon_n$ ($n \! = \! 0-5$) of ACHO potential as function of $d_m$ \cite{ghosal16}.}
\end{figure}

\subsubsection{Asymmetrically confined harmonic oscillator (ACHO)}
Now, the focus is on the ACHO case. This presentation is based on results of $\epsilon_n$; $S_x, S_p$; $I_x, I_p$; $E_x, E_p$ 
as functions of $d_m$, for some low-lying states. Throughout the whole presentation, the box length has been kept fixed at 2 and 
boundaries are placed at $-$1 and 1. At the outset it is important to note that, an increase in $d_m$ leads to localization in real 
space. Thus it can be expected that, there will be delocalization in $p$ space. 

\begin{figure}                         
\small
\begin{minipage}[c]{0.28\textwidth}\centering
\includegraphics[scale=0.364]{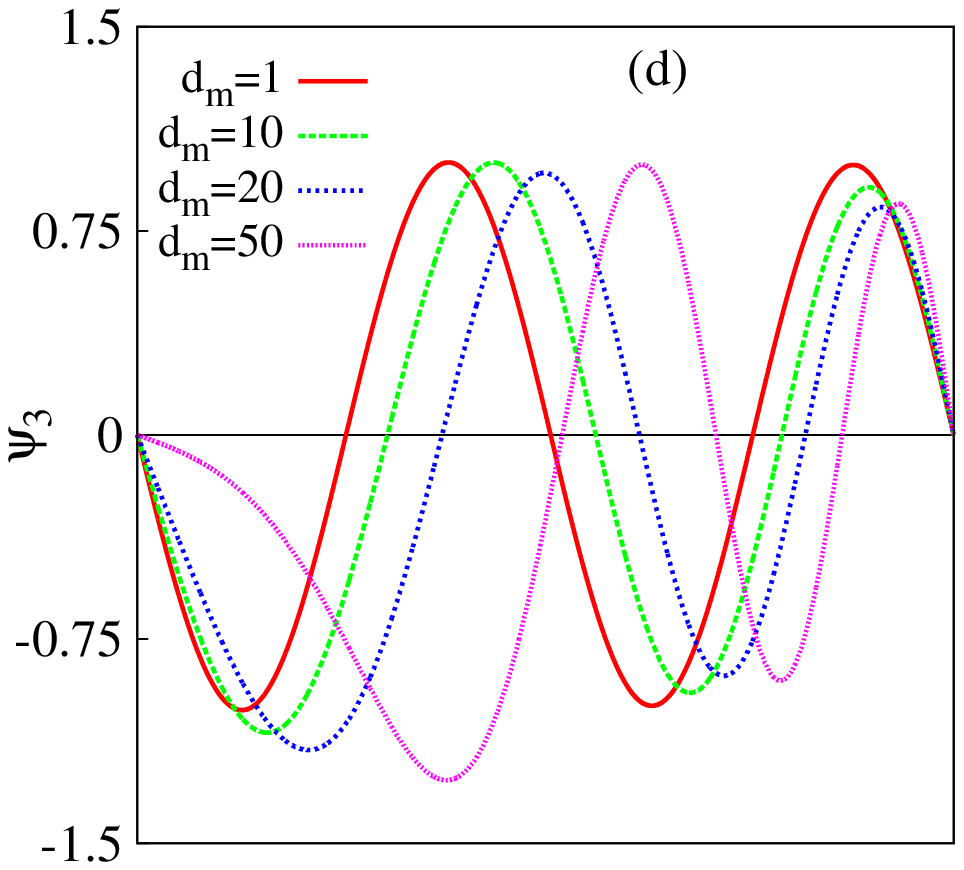}
\end{minipage}%
\hspace{-0.21in}
\begin{minipage}[c]{0.2\textwidth}\centering
\includegraphics[scale=0.364]{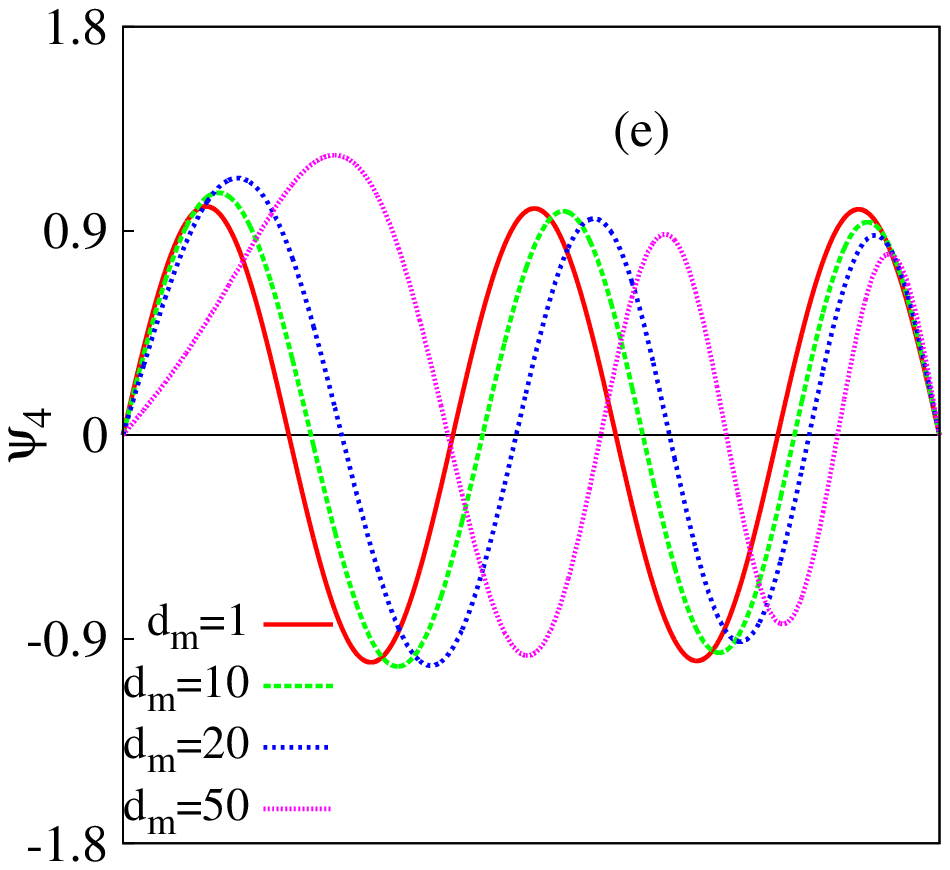}
\end{minipage}%
\hspace{-0.1in}
\begin{minipage}[c]{0.4\textwidth}\centering
\includegraphics[scale=0.364]{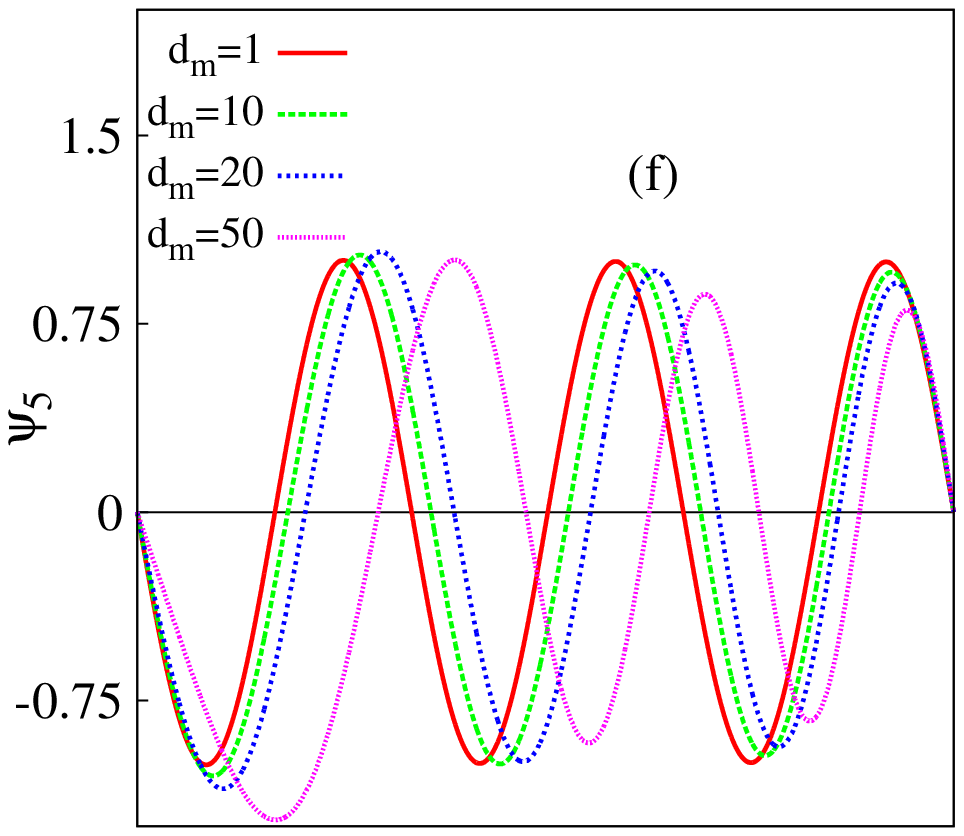}
\end{minipage}%
\hspace{0.2in}
\begin{minipage}[c]{0.273\textwidth}\centering
\includegraphics[scale=0.4]{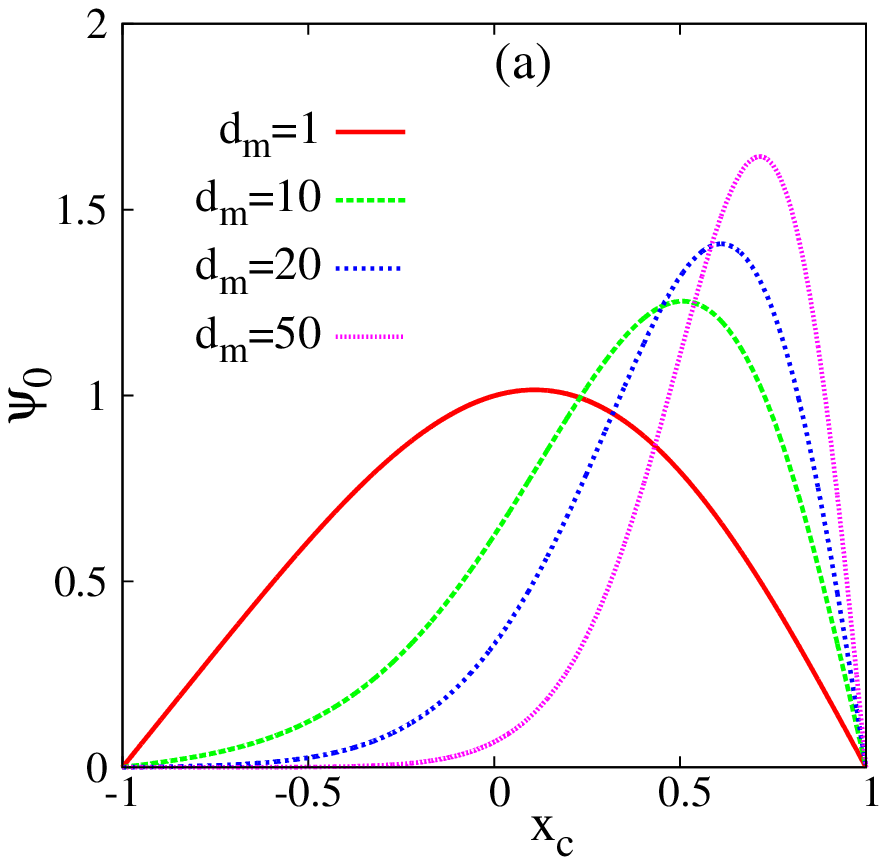}
\end{minipage}%
\hspace{0.2in}
\begin{minipage}[c]{0.273\textwidth}\centering
\includegraphics[scale=0.4]{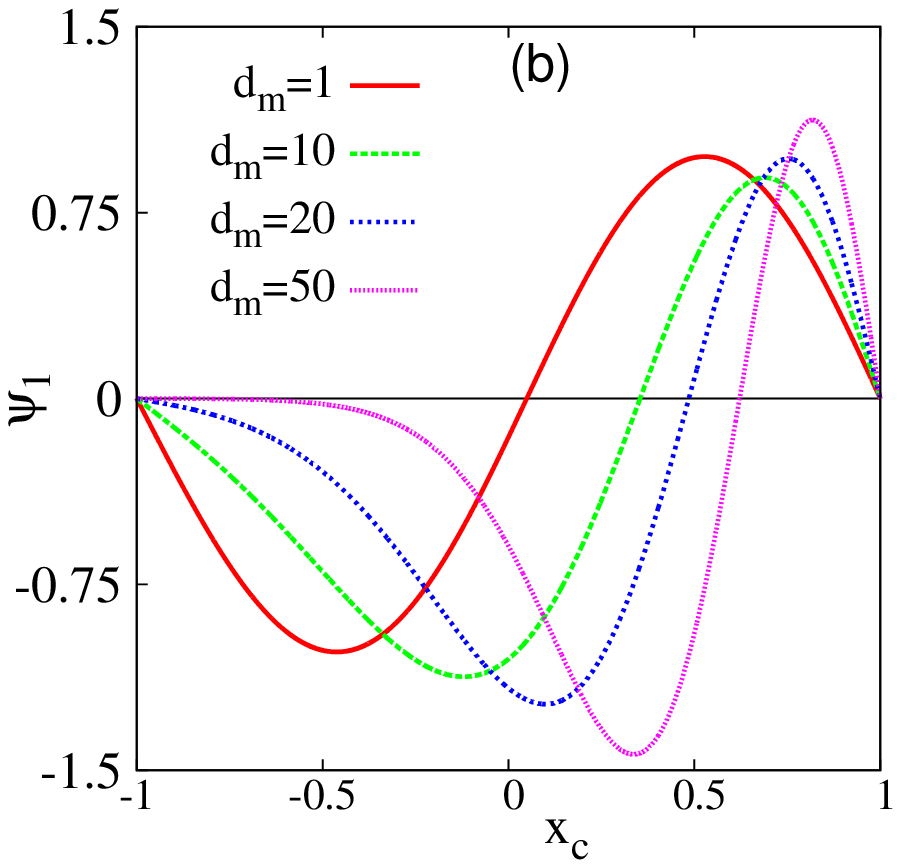}
\end{minipage}%
\hspace{0.25in}
\begin{minipage}[c]{0.273\textwidth}\centering
\includegraphics[scale=0.4]{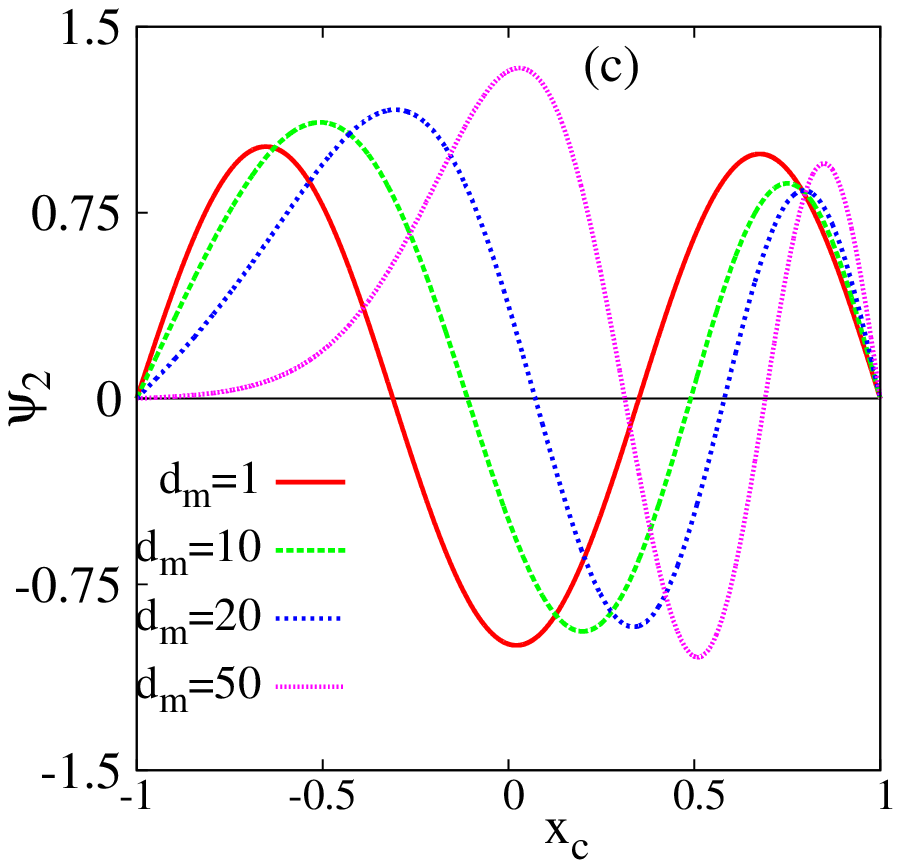}
\end{minipage}%
\caption{Wave functions of lowest six states of ACHO potential at four specific values of $d_m$, namely 1,
10, 20, 50. Panels (a)--(f) represent $n \! = \! 0-5$ states \cite{ghosal16}. See text for details.}
\end{figure}  

Before looking into the behavior of IE in ACHO, it is pertinent to make some comments about its eigenvalues and eigenfunctions. 
ACHO has been studied using power-series solution \cite{campoy02} and ITP method \cite{roy15}. In this section, we shall, however,
present another simple method, which produces quite accurate results. In this \emph{variation induced 
exact diagonalization} procedure \cite{griffiths04}, an energy functional is minimized using an SCHO basis set. Recently this 
method has been successfully employed in studying symmetric and asymmetric double-well potentials \cite{mukherjee15, mukherjee16}, where 
a 1DQHO basis was utilized. 

\begin{figure}                      
\begin{minipage}[c]{0.4\textwidth}\centering
\includegraphics[scale=0.56]{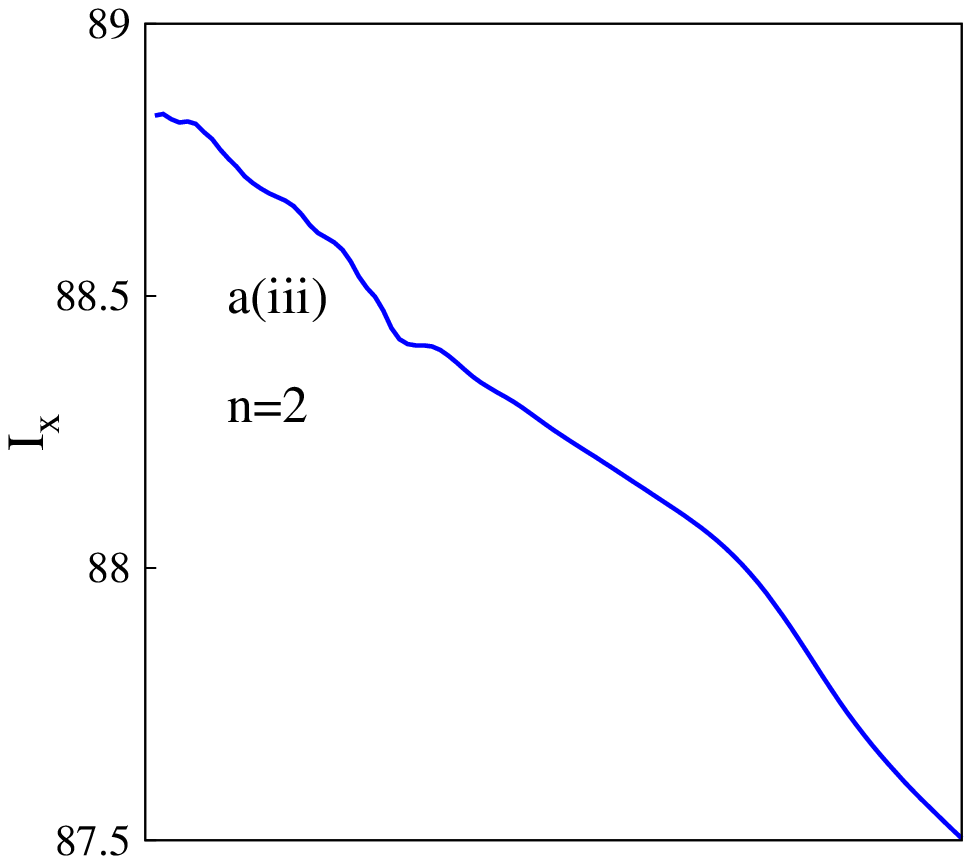}
\end{minipage}
\hspace{0.5in}
\begin{minipage}[c]{0.4\textwidth}\centering
\includegraphics[scale=0.56]{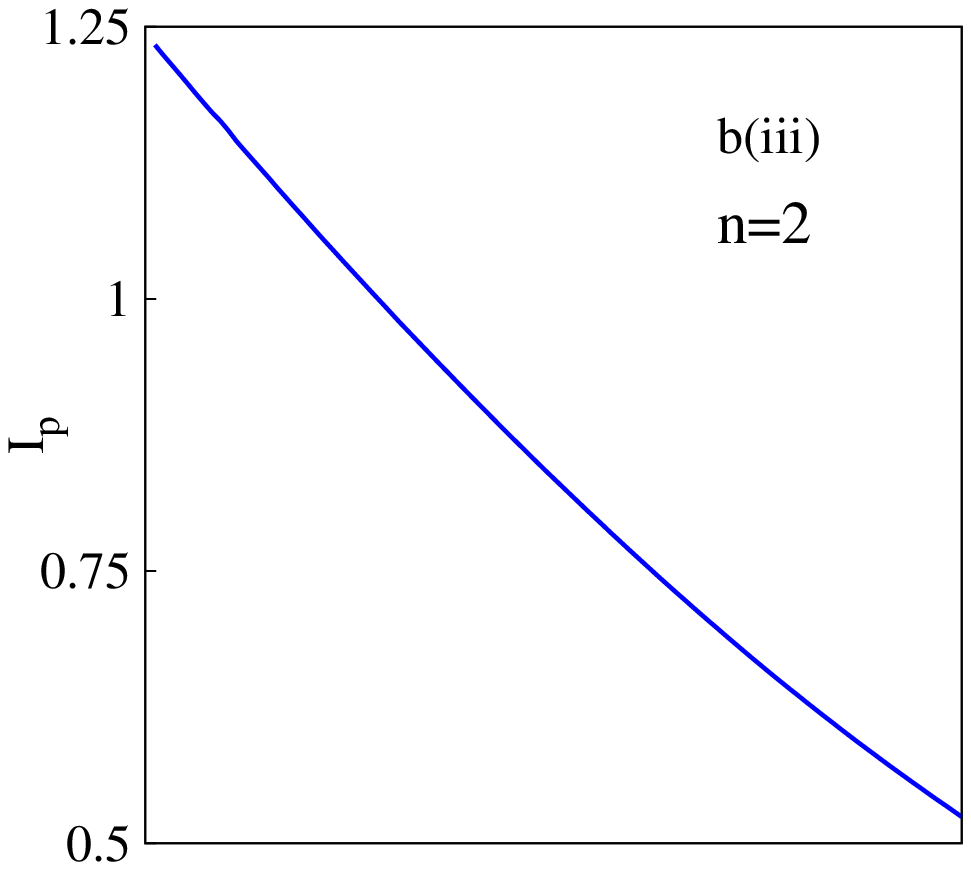}
\end{minipage}
\\[10pt]
\begin{minipage}[c]{0.4\textwidth}\centering
\includegraphics[scale=0.56]{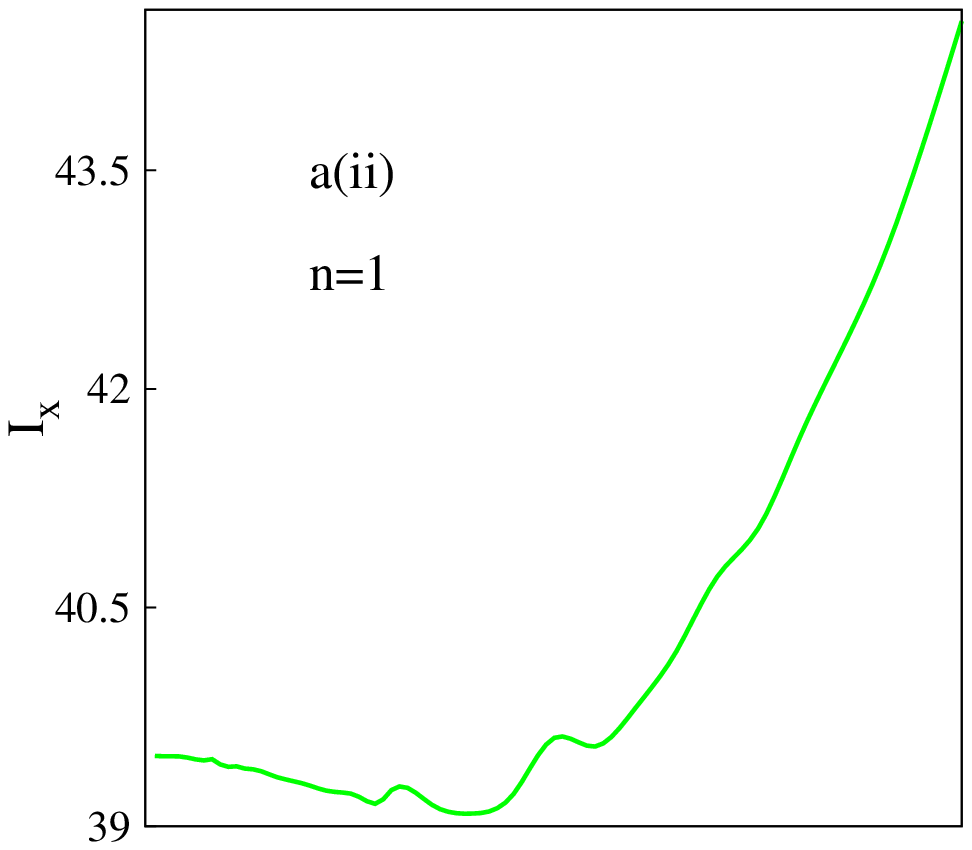}
\end{minipage}
\hspace{0.5in}
\begin{minipage}[c]{0.4\textwidth}\centering
\includegraphics[scale=0.56]{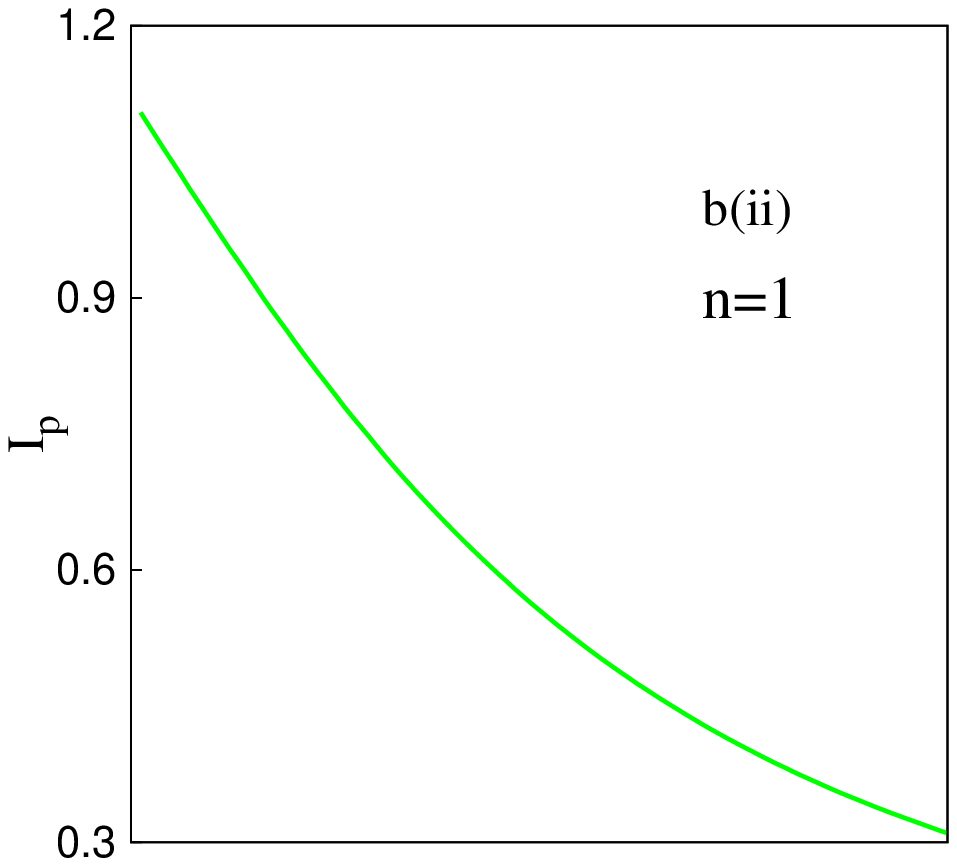}
\end{minipage}
\\[10pt]
\begin{minipage}[c]{0.42\textwidth}\centering
\includegraphics[scale=0.6]{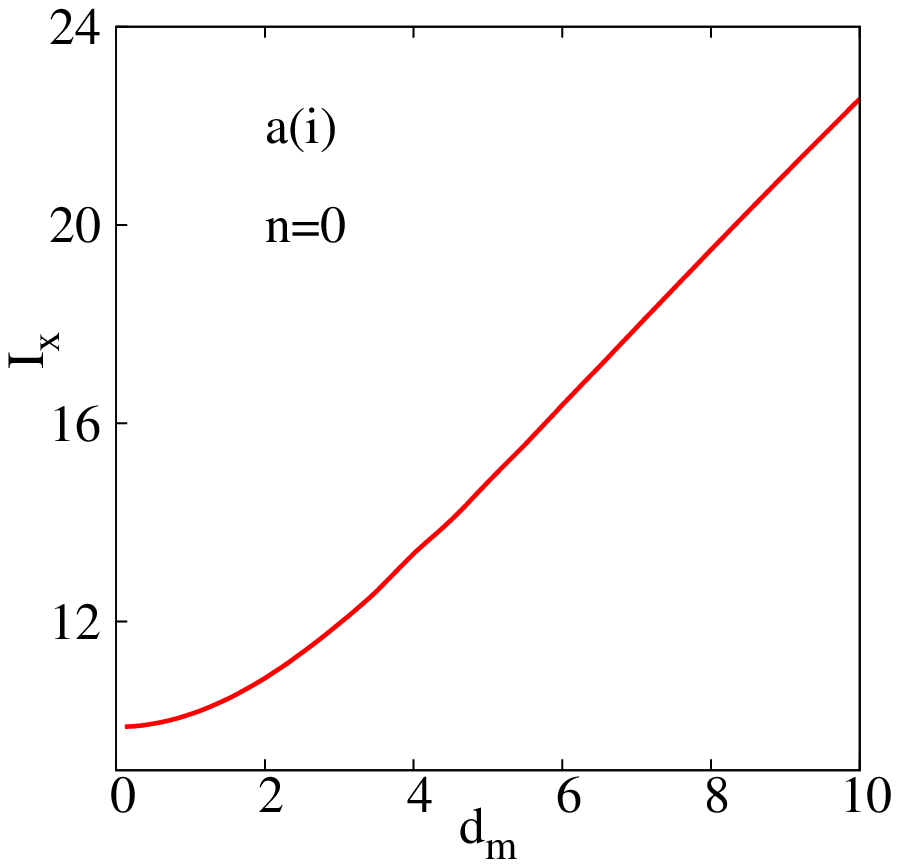}
\end{minipage}
\hspace{0.5in}
\begin{minipage}[c]{0.42\textwidth}\centering
\includegraphics[scale=0.6]{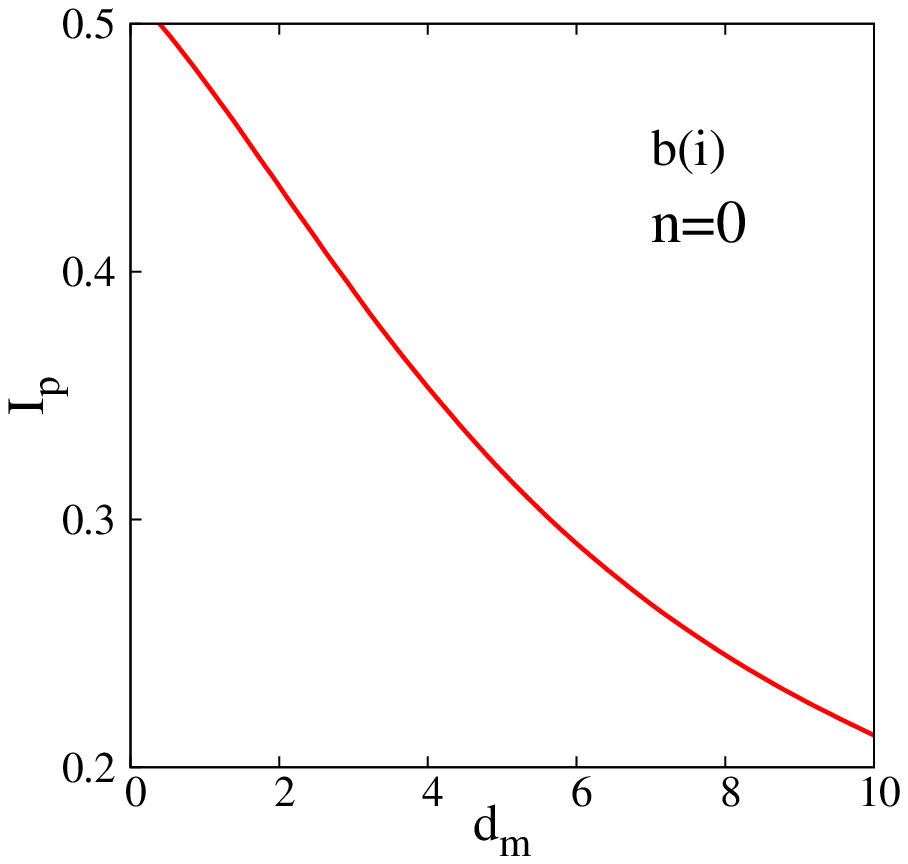}
\end{minipage}
\caption{Plot of $I_x$, $I_p$, $I$ of first five states of ACHO potential, as function of $d_m$, in left (a), middle (b), 
right (c) columns; (i)--(iii) represent $n=0-2$ states \cite{ghosal16}. See text for details.}
\end{figure}

The Hamiltonian matrix elements are evaluated by using functions given in Eq.~(2). Presence of a single non-linear parameter $\alpha$ 
allows us to adopt 
a coupled variation procedure. Further it is easy to confirm the convergence of results with respect to basis dimension $N$. Thus 
it provides a secular equation at each $\alpha$. The kinetic energy part blows up to $\infty$ when $\alpha \rightarrow \infty$, 
whereas at $\alpha \rightarrow 0$, potential energy part behaves in a similar fashion. This qualitative analysis through 
uncertainty principle, confirms the existence of such basis. Diagonalization of $H_{mn} = \langle m | \hat{H}| n \rangle$ leads 
to accurate eigenvalues and eigenfunctions, which is attained through MATHEMATICA package. In principle, to achieve exact solution, 
one needs to employ the \emph{complete} basis; however for practical purposes a truncated basis of finite dimension is envisaged. 
Here, $N=50$ appears adequate; with further increase in basis, result improves. A cross-section of energies obtained from above scheme 
is produced in Table~3, for lowest six states of ACHO at four selected $d_m$. Some literature results are provided for comparison, 
wherever available. For all four $d_m$, present energies with SCHO basis, practically coincide with ITP results \cite{roy15}, 
for all digits reported. It is expected that this SCHO basis may be useful in future for other confined and free 1D potentials.

Our calculated ACHO energies of Table~3 are presented in Fig.~5, for first six states with respect to $d_m$, in panels (a)-(f). 
Ranges of $\epsilon_n$ and $d_m$ are different for each $n$. Variation in $\epsilon_0$ is unique amongst all $\epsilon_n$, where, from 
an initial higher value at small $d_m$, ground-state energy continuously falls. In all excited states, however, it passes through 
a maximum; with increase in $n$, this maximum shifts to higher values of $d_m$. Next Fig.~6 portrays our computed wave functions for 
$n \! = \! 0-5$ states of ACHO at four selected $d_m$ values. Clearly, the maximum, minimum and nodal positions of all $\psi_n$'s 
switch towards right with advancement in $d_m$. These plots suggest that particle gets localized in $x$ space as $d_m$ progresses. 

\begin{figure}                      
\begin{minipage}[c]{0.42\textwidth}\centering
\includegraphics[scale=0.565]{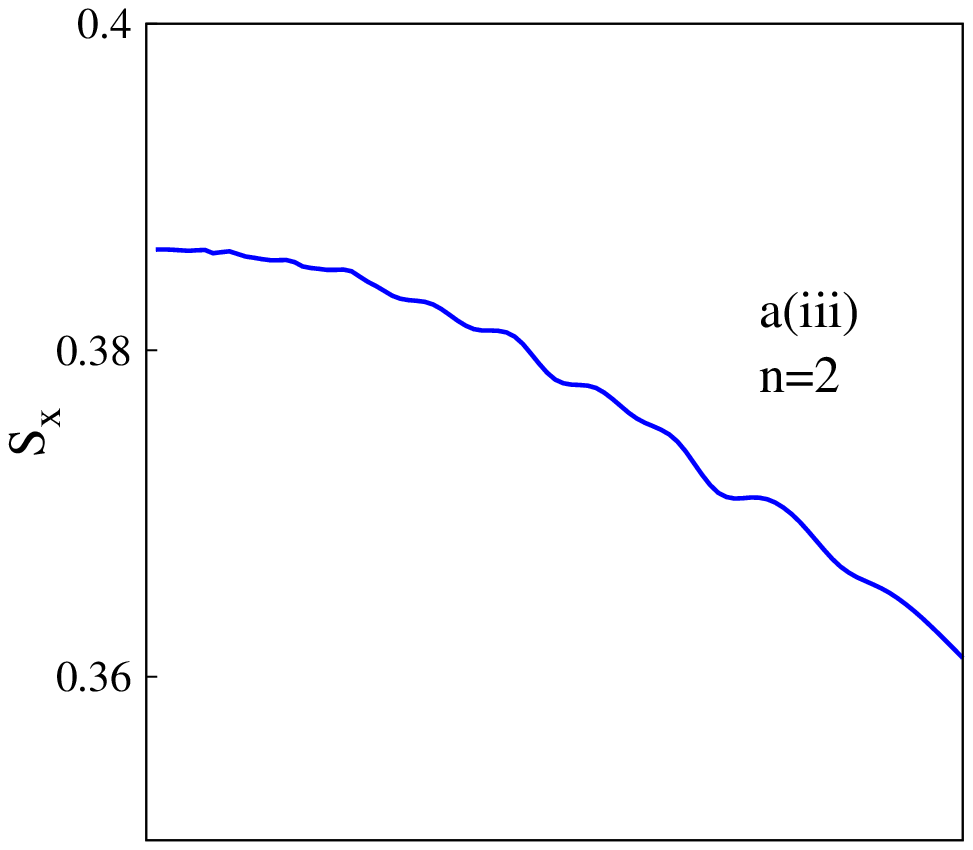}
\end{minipage}%
\hspace{0.4in}
\begin{minipage}[c]{0.42\textwidth}\centering
\includegraphics[scale=0.565]{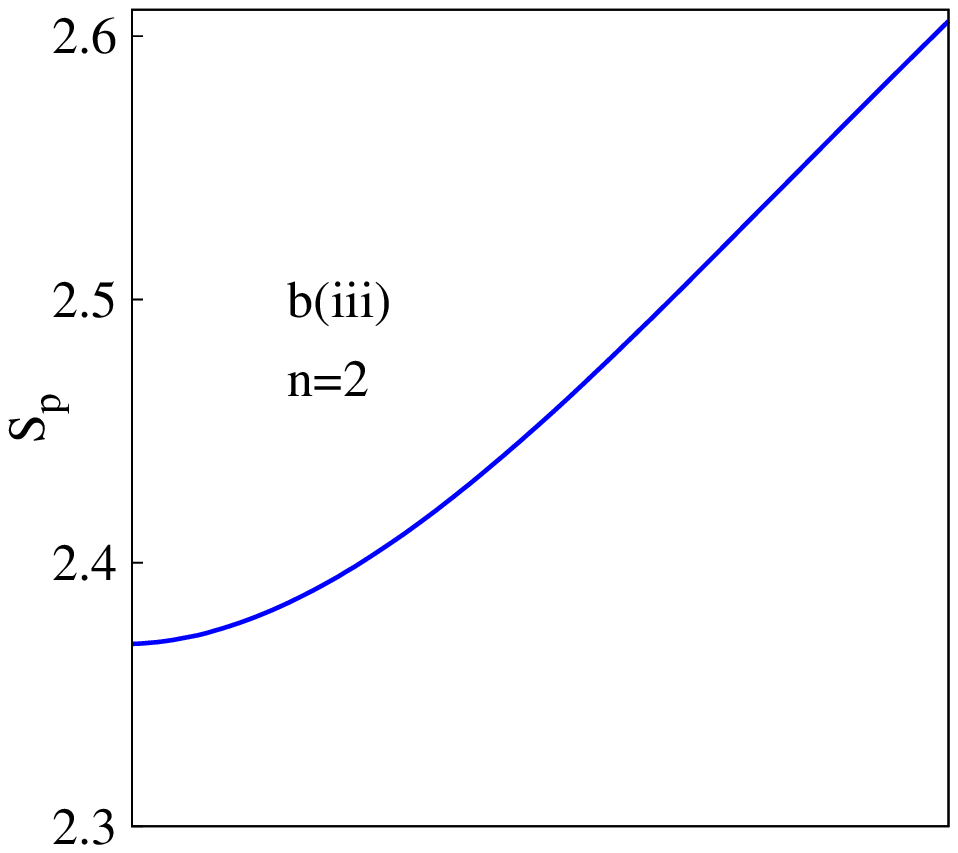}
\end{minipage}%
\\[10pt]
\begin{minipage}[c]{0.42\textwidth}\centering
\includegraphics[scale=0.565]{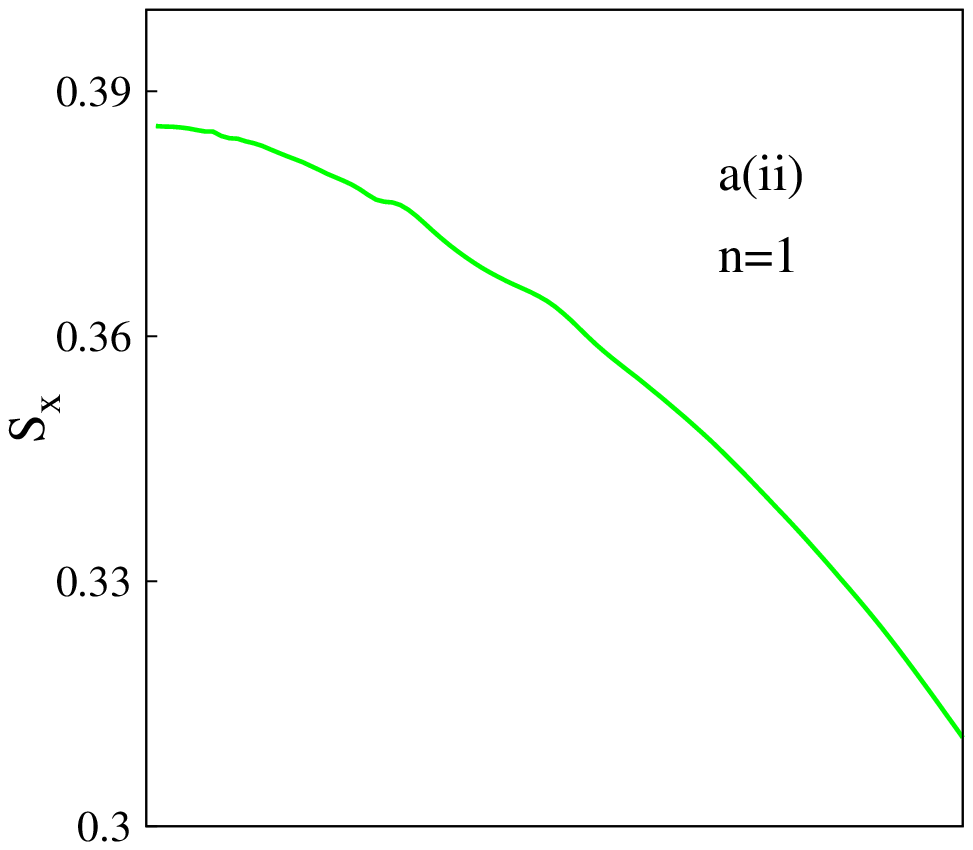}
\end{minipage}%
\hspace{0.4in}
\begin{minipage}[c]{0.42\textwidth}\centering
\includegraphics[scale=0.565]{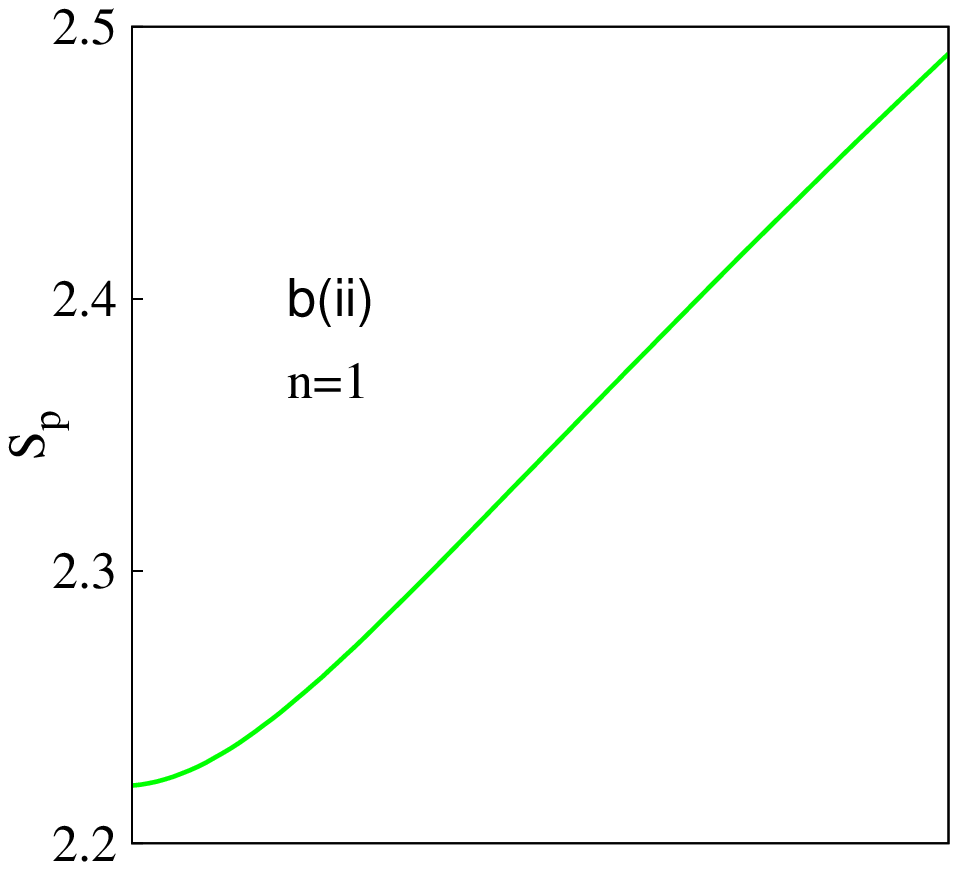}
\end{minipage}%
\\[10pt]
\begin{minipage}[c]{0.42\textwidth}\centering
\includegraphics[scale=0.6]{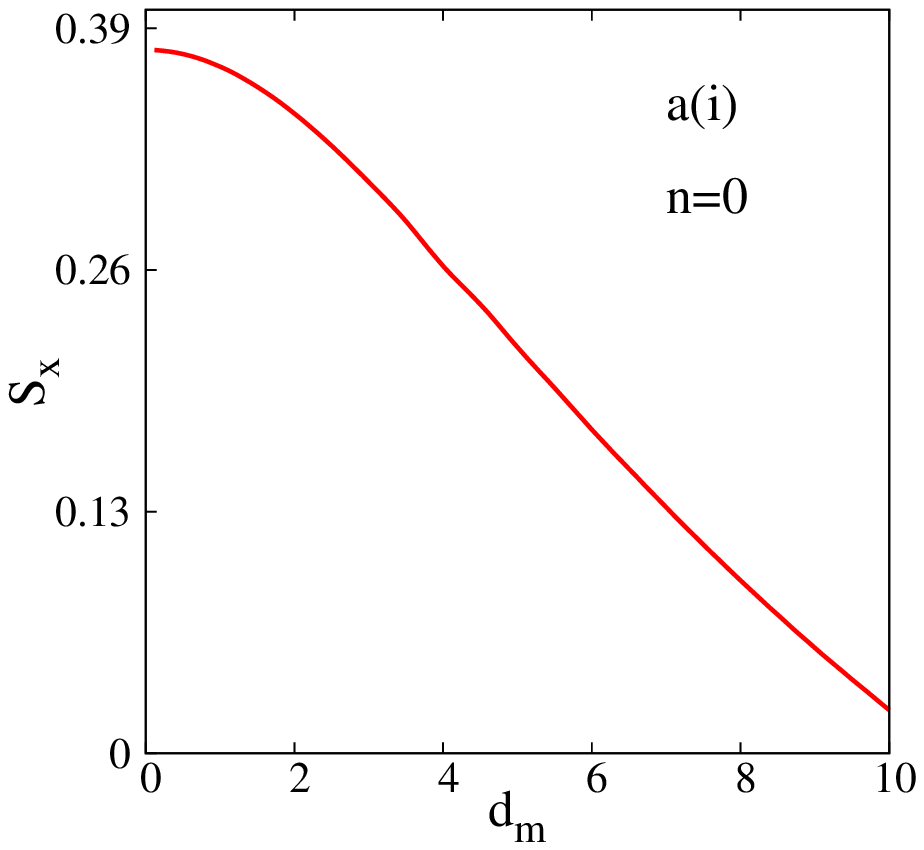}
\end{minipage}%
\hspace{0.4in}
\begin{minipage}[c]{0.42\textwidth}\centering
\includegraphics[scale=0.6]{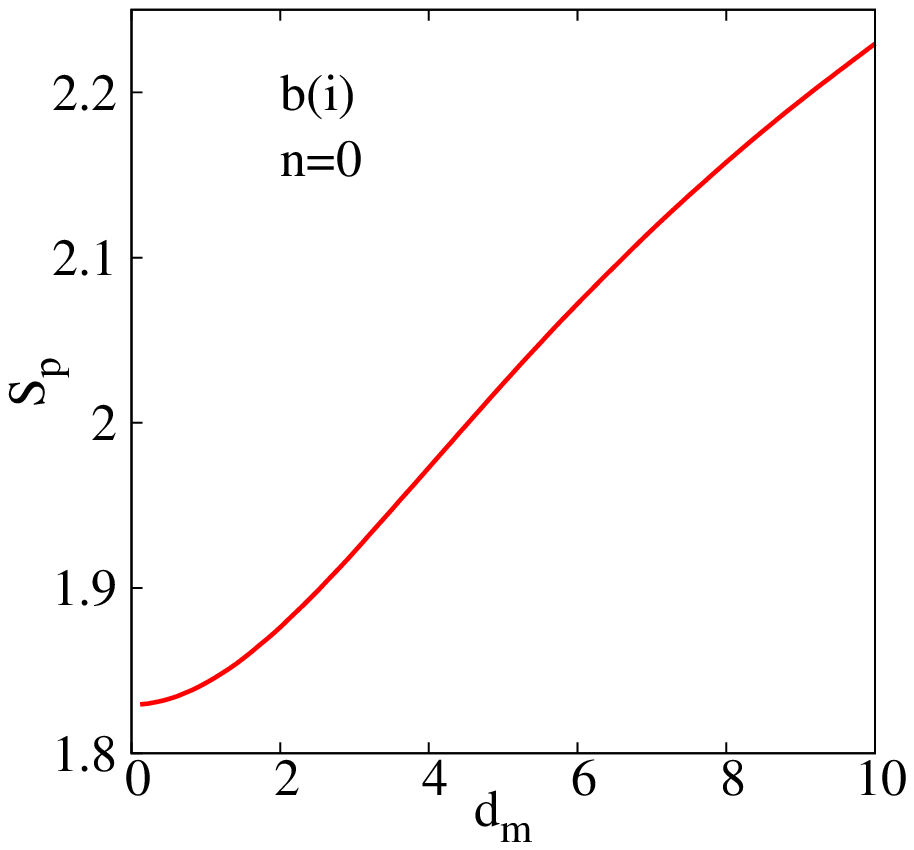}
\end{minipage}%
\caption{Plot of $S_x$, $S_p$, $S$ of first three states of ACHO potential, as function of $d_m$, in left (a), middle (b), 
right (c) columns; (i)--(iii) represent $n=0-2$ states \cite{ghosal16}. See text for details.}
\end{figure}

Now we move on to $I_x,I_p$ variations with change of $d_m$. These are shown pictorially in Fig.~7 left (a), and right (b) panels 
display these in  conjugate $x$, $p$ spaces respectively. For ground and first excited state $I_x$ tends to increase with $d_m$, 
on the whole, whereas for ($n = 2$) state, the same decreases. Thus, for first two states, it indicates localization in 
$x$ space and delocalization for 2nd excited state. A careful investigation of panels b(i)--b(iii) of Fig.~7 reveals that, $I_{p}$, 
for all $n$ under consideration, consistently decreases with increase of $d_m$, signifying a delocalization of particle in $p$ 
space. 

Next, it is imperative to explore changes in behavior of $S_x, S_p$ with respect to $d_m$. These are offered in Fig.~8. It is clear 
from left panels a(i)--a(iii) of Fig.~8 that, $S_x$ for all these three states generally decreases monotonically with growth in $d_m$, 
signifying localization of particle in position space. Likewise, a scrutiny of right plots b(i)--b(iii) of Fig.~8 summarizes an opposite 
trend in $S_p$ with increase in $d_m$. A sample of $S_x, S_p$ are reported in 
Table~4 for $n \! = \! 0,1,2$ states at five selected $d_m$ values, namely, 0.12, 2.04, 5, 7 and 10. Evidently, these entries 
corroborate the outcomes of Fig.~8. 

Now, we move on to an analysis of $E$, using Fig.~9, for $n=0-2$. Panels a(i)--a(iii) suggest that, for all three states there 
is an overall increase of $E_{x}$ with rise in $d_m$. This indicates localization in $x$ space. Panels b(i)--b(iii) portray that, 
general trend of $E_{p}$ is a gradual decrease with rise in $d_m$.  

On the basis of above discussion, it is clear that, only $S_{x}, S_{p}$ can adequately explain localization-delocalization 
phenomena in an ACHO. Study of $E_{x}, E_{p}$ can also offer valuable knowledge about the dual nature of $E$ in composite $x$, 
$p$ space. But, $I_{x}, I_{p}$ appear to be inadequate in explaining the contrasting phenomena in ACHO.
   
\begin{figure}                         
\begin{minipage}[c]{0.35\textwidth}\centering
\includegraphics[scale=0.57]{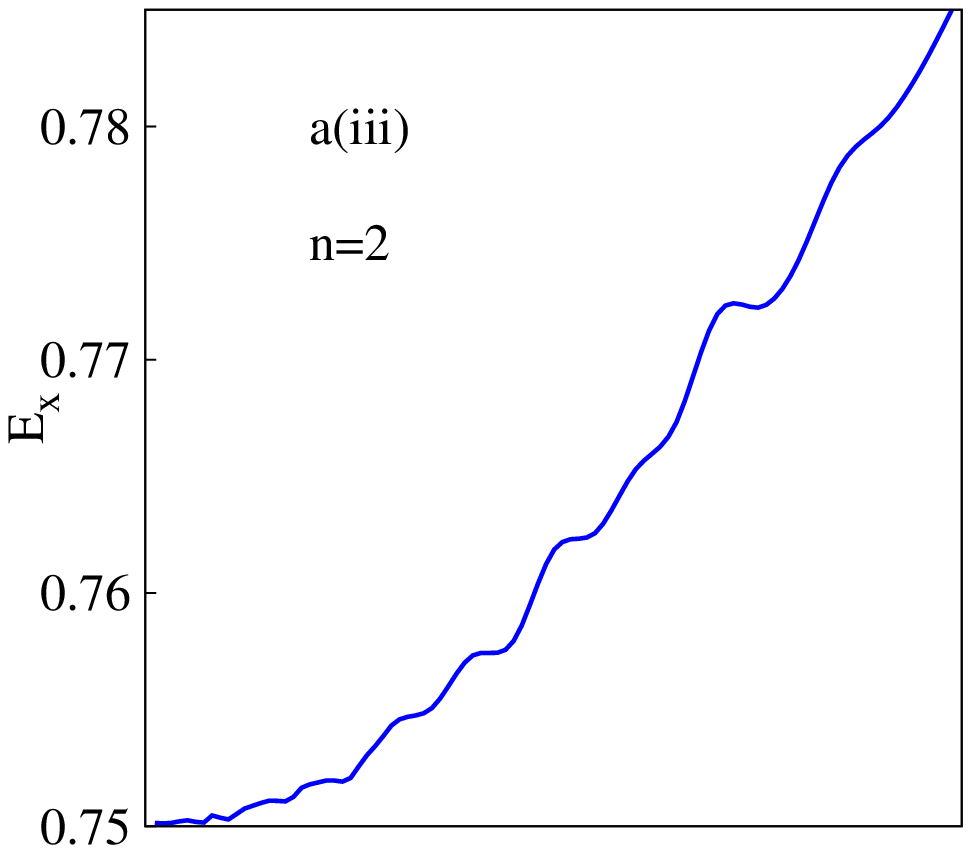}
\end{minipage}%
\hspace{0.75in}
\begin{minipage}[c]{0.35\textwidth}\centering
\includegraphics[scale=0.57]{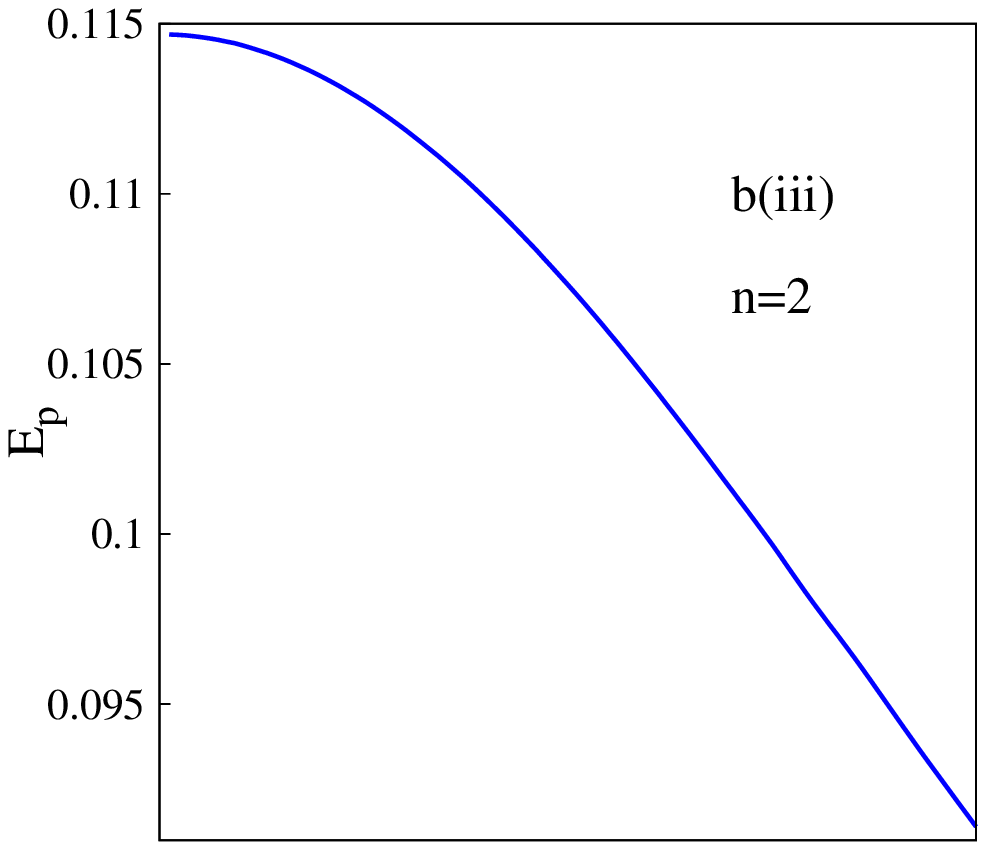}
\end{minipage}%
\\[10pt]
\begin{minipage}[c]{0.4\textwidth}\centering
\includegraphics[scale=0.57]{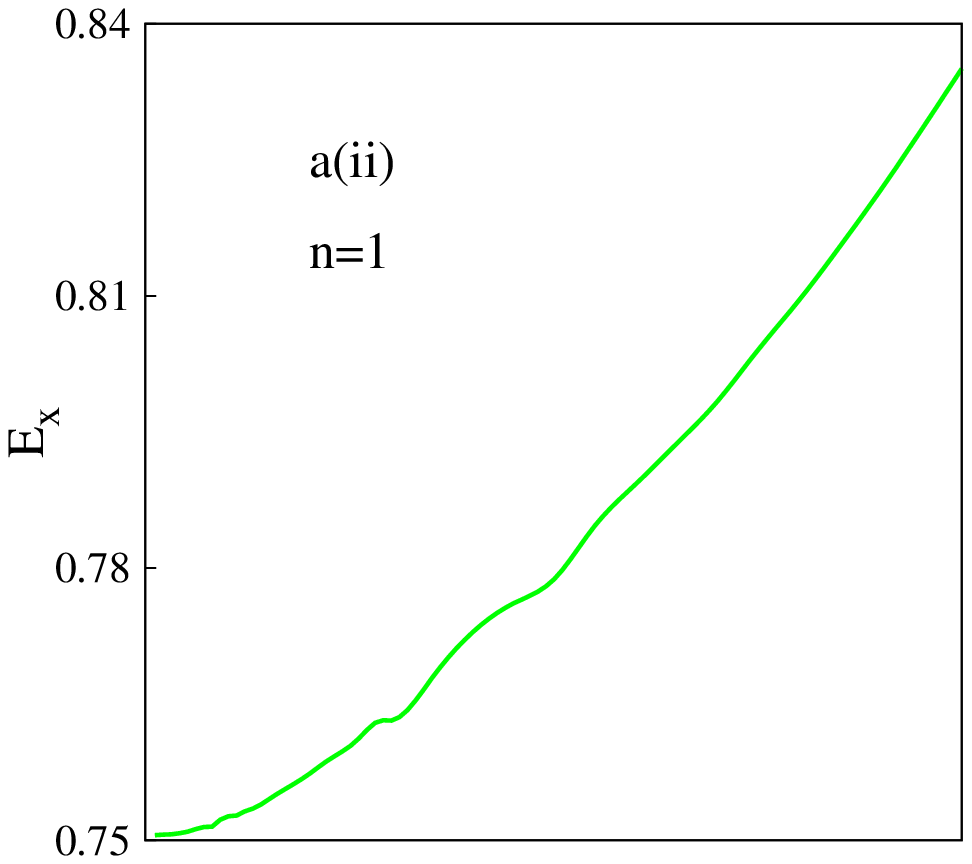}
\end{minipage}%
\hspace{0.45in}
\begin{minipage}[c]{0.4\textwidth}\centering
\includegraphics[scale=0.57]{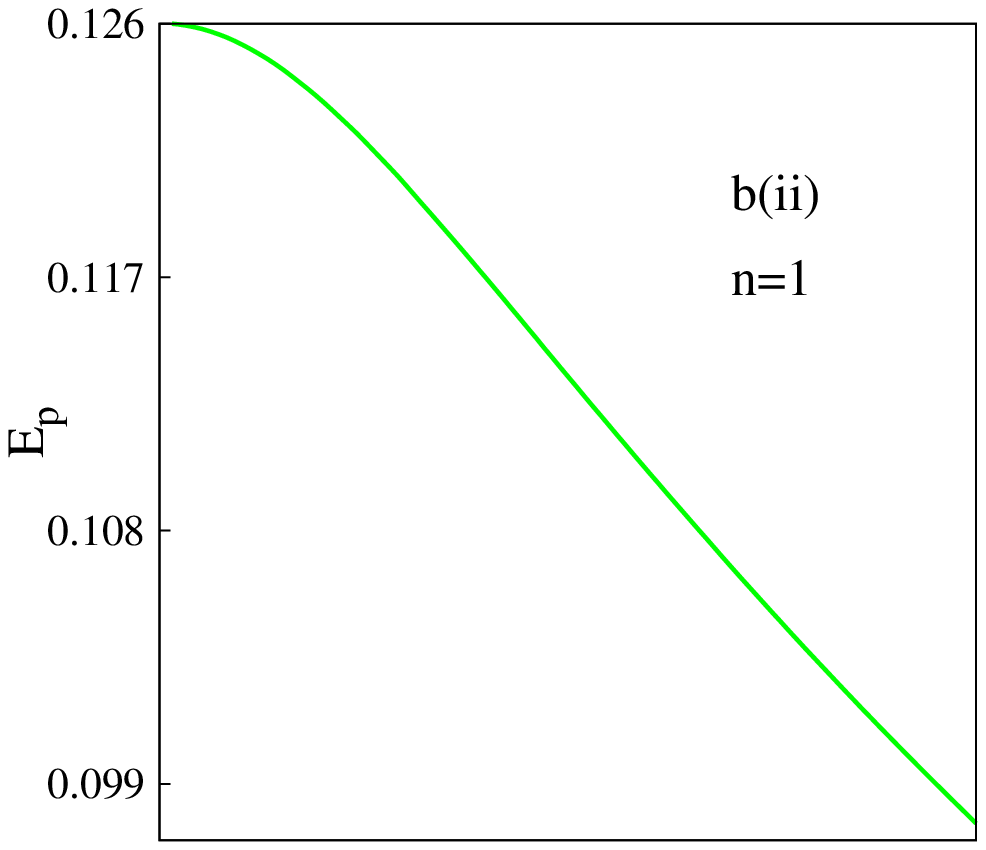}
\end{minipage}%
\hspace{0.45in}
\\[10pt]
\begin{minipage}[c]{0.35\textwidth}\centering
\includegraphics[scale=0.65]{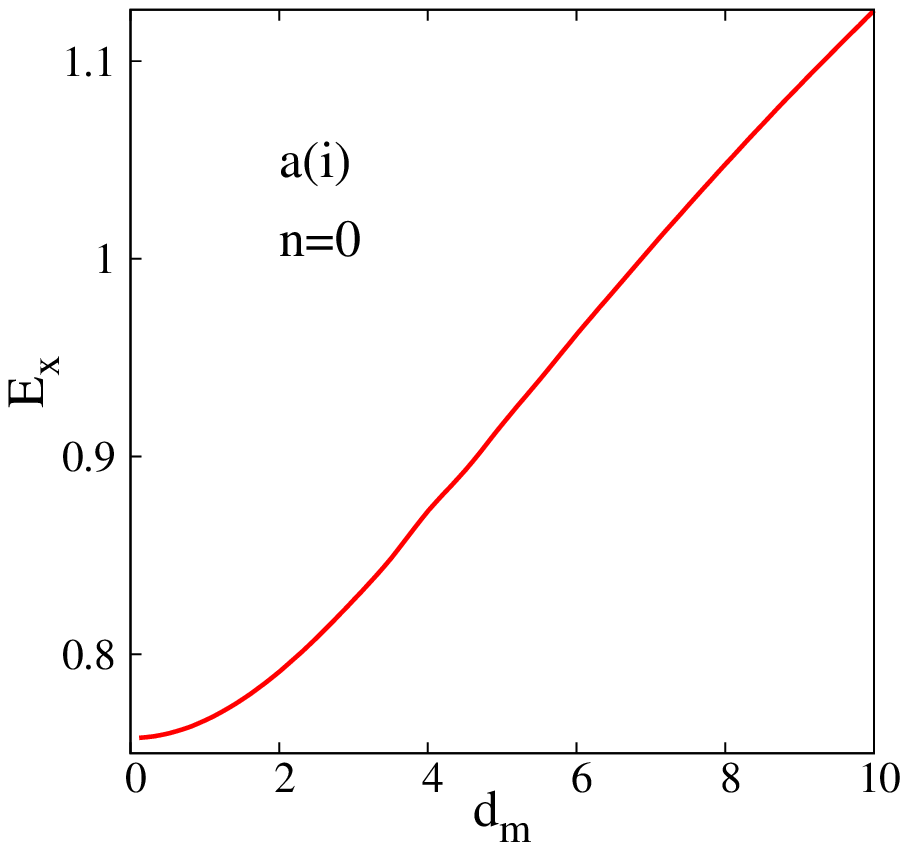}
\end{minipage}%
\hspace{0.7in}
\begin{minipage}[c]{0.35\textwidth}\centering
\includegraphics[scale=0.65]{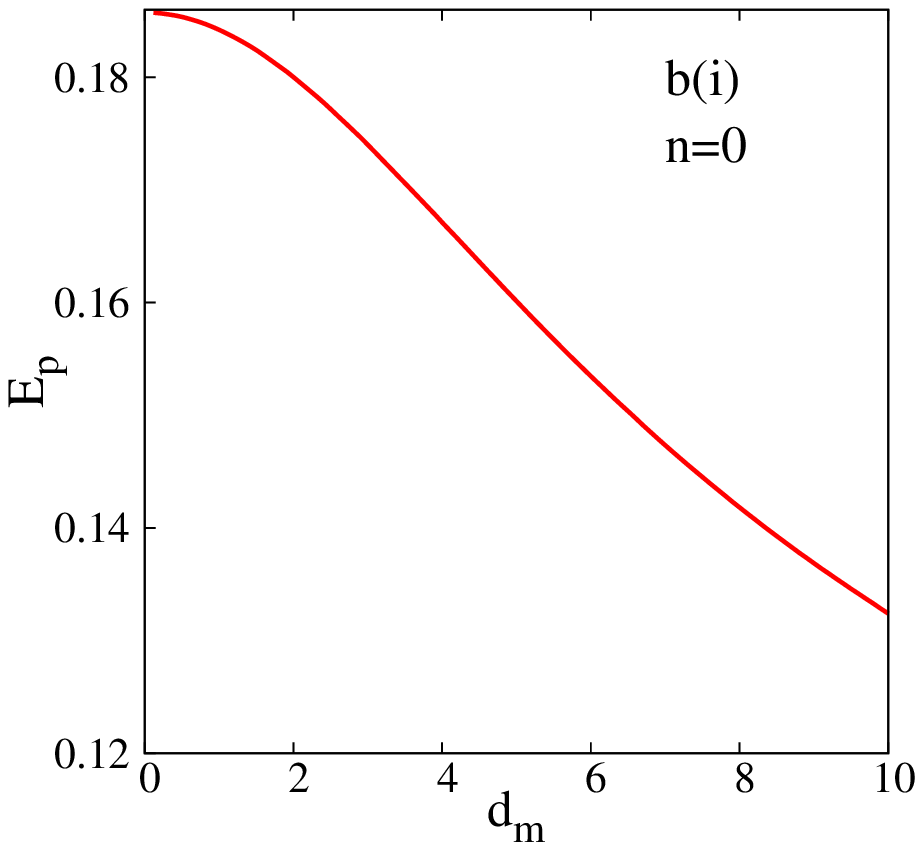}
\end{minipage}%
\caption{Plot of $E_x$, $E_p$, $E$ of first three states of ACHO potential, as function of $d_m$, in left (a), middle (b), 
right (c) columns; (i)--(iii) represent $n=0-2$ states \cite{ghosal16}. See text for details.}
\end{figure}     

\subsection{3D Confined harmonic oscillator (3DCHO)}
At the onset, it is convenient to point out that, \emph{net} information measures in conjugate $r$ and 
$p$ space may be segmented into radial and angular parts. In a given space, the results provided correspond to \emph{net} measures 
including the \emph{angular} contributions. One can transform the 3DQHO into a 3DCHO by squeezing the radial boundary of former from 
infinity to a finite region. This alteration in radial environment does not affect the \emph{angular} boundary conditions. 
Therefore, angular part of the information measures in free and confined systems remain unchanged in both spaces. Further 
as we are solely focused in \emph{radial} confinement, this will also not influence the characteristics of a given measure as one 
changes $r_c$. Throughout this investigation, magnetic quantum number $m$ remains fixed to $0$ for $R, S, E$ calculation. However, $I$ 
has been studied for \emph{non-zero} $m$ states. The radial wave function in $r$ and $p$ spaces depend only on $n_{r}, l$ quantum 
numbers. Hence, in both space, radial wave function can be obtained by considering $m=0$. Further, a change in $m$ from zero to 
non-zero value will not influence the expression of radial wave function in $p$ space. Also note that, we have followed the 
spectroscopic notation, i.e., the levels are denoted by $n_r+1$ and $l$ values (see, e.g., \cite{roy08ijqc}). Therefore, $n_r=0$ and 
$l=4$ signifies $1g$ state. The radial quantum number $n_r$ relates to $n$ as $n=2 n_{r}+l$.

\begingroup      
\tiny             
\begin{table}
\centering
\caption{$S_{x}, S_{p}, S$ for $n \! = \! 0-2$ states of ACHO at five specific $d_m$ \cite{ghosal16}. See text for details.} 
\centering
\begin{tabular} {>{\scriptsize}c|>{\scriptsize}c>{\scriptsize}c>{\scriptsize}c<{\scriptsize}|>{\scriptsize}c>{\scriptsize}c>{\scriptsize}c<{\scriptsize}|>{\scriptsize}
c>{\scriptsize}c>{\scriptsize}c<{\scriptsize}}
\hline
$d_m$ & $S_{x}^{0}$ & $S_{p}^{0}$ & $S^{0}$ & $S_{x}^{1}$ & $S_{p}^{1}$ & $S^{1}$ & $S_{x}^{2}$ & $S_{p}^{2}$ & $S^{2}$ \\
\hline
0.12 & 0.3783 & 1.8296 & 2.2079 & 0.3857 & 2.2212 & 2.6069 & 0.3862 & 2.3692 & 2.7554 \\
2.04 & 0.3428 &	1.8779 & 2.2208 & 0.3807 & 2.2512 & 2.6319 & 0.3850 & 2.3852 & 2.7703 \\
5.0 & 0.2184 & 2.0238 & 2.2422 & 0.3636 & 2.3390	& 2.7027 & 0.3782 &	2.4512 & 2.8295  \\
8.0 & 0.093 & 2.1576 & 2.2506 & 0.3360 & 2.4313 & 2.7674 & 0.3695 & 2.5429 &	2.9124  \\
10.0 & 0.0233 & 2.2294 & 2.2527 & 0.3108 & 2.4900 & 2.8009 & 0.3611 & 2.6057 & 2.9668 \\
\hline
\end{tabular}
\end{table} 
\endgroup 

$I_{\rvec}, I_{\pvec}$ values are obtained from Eq.~(62). In all occasions, it has been verified that, in both spaces, 
as $r_c \rightarrow \infty$, $I_{\rvec}$ and $I_{\pvec}$ merge to respective 3DQHO limit. The net $I$ in $r$ and $p$ spaces are 
divided into radial and angular part. But in both $I_{\rvec}$ and $I_{\pvec}$ expressions, angular contribution is normalized 
to unity. Hence, evaluation of all these targeted quantities using only radial part will serve our purpose. Pilot calculations 
are done for $1s$-$1g$ and $2s$-$2m$ states, with $r_c$ varying from 0.1 to 7 a.u. The former set is chosen as they represent 
node-less ground states corresponding to various $l$, whereas, $2s$-$2m$ states are considered to perceive the effect of nodes on 
$I$ at non-zero $m$.

It is worthwhile noting that, $I$'s can be accurately calculated from a knowledge of $\langle p^2\rangle, \left\langle 
\frac{1}{r^2}\right\rangle, \langle r^2\rangle, \left\langle \frac{1}{p^2}\right\rangle$. Two possibilities may be invoked: (i) 
first three quantities evaluated in $r$ space, while $\langle p^{-2}\rangle$ in $p$ space (ii) $\langle r^2\rangle, 
\langle r^{-2}\rangle$ in $r$ space, while $\langle p^2\rangle, \langle p^{-2}\rangle$ in $p$ space. Here we have opted for 
first route removing the necessity to do numerical differentiation in either spaces. This has been discussed in 
detail in a recent article \cite{mukherjee18c}.

Exact analytical form of $I_{\rvec}$ and $I_{\pvec}$ in isotropic 3DQHO was given in \cite{romera05},
\begin{equation}
I_{\rvec}(\omega)= 4\omega \left(2n_{r}+l-|m|+\frac{3}{2}\right), \ \ \ 
I_{\pvec}(\omega)=\frac{4}{\omega}\left(2n_{r}+l-|m|+\frac{3}{2} \right). 
\end{equation}
Thus, at a certain $m$, both $I_{\rvec}(\omega), I_{\pvec}(\omega)$ increase as $n_{r}$ and $l$ approach higher values. 
Similarly, for specific $n_{r}$ and $l$, both $I_{\rvec}(\omega), I_{\pvec}(\omega)$ regress with growth in $|m|$. Effect of 
$\omega$ on $I_{\rvec}(\omega), I_{\pvec}(\omega)$ is quite straightforward. $I_{\rvec}(\omega)$ advances and $I_{\pvec}
(\omega)$ decreases with rise of $\omega$. By putting $\omega=1$ in Eq.~(77) one easily recovers expressions for $I_{\rvec}, I_{\pvec}$ 
in a 3DQHO. 

\begingroup            
\begin{table}
\small
\caption{$\mathcal{E}_{n_r,l} (n_r =0)$ of lowest five circular states of 3DCHO, PISB at five $r_c$, from \cite{roy15, mukherjee18}.}
\centering
\begin{tabular}{>{\scriptsize}l|>{\scriptsize}l>{\scriptsize}l>{\scriptsize}l>{\scriptsize}l>{\scriptsize}l}
\hline
   $l$ & $r_c=0.01$  &  $r_c=0.05$   &  $r_c=0.1$ &   $r_c=0.2$   &   $r_c=0.5$    \\
\hline
\multicolumn{6}{c}{3DCHO}  \\
\hline
 0   &   49348.02202373  &  1973.92123372 &    493.48163345 &   123.37570844  &   19.77453418  \\
 1   &  100953.64280465  &  4038.14617967 &   1009.53830088 &   252.39159906  &   40.42827649  \\
 2   &  166087.30959293  &  6643.49293120 &   1660.87528919 &   415.22704789  &   66.48975653  \\
 3   &  244155.96823805  &  9753.60193720 &   2441.56211674 &   610.39965899  &   97.72324914  \\
 4   &  334771.55964446  & 13390.86304096 &   3347.71822121 &   836.93939916  &  133.97424683 \\
\hline
\multicolumn{6}{c}{PISB}  \\
\hline
 0   &   49348.022005446   &  1973.92088021   & 493.48022005    &   123.37005501    &     19.73920880  \\
 1   &   100953.64278213   &  4038.14571128   & 1009.53642782   &   252.38410695    &     40.38145711  \\
 2   &   166087.30957134   &  6643.49238285   & 1660.87309571   &   415.21827392    &     66.43492382 \\
 3   &   244155.96821809   &  9753.60153136   & 2441.55968218   &   610.38992054    &     97.66238728  \\
 4   &   334771.55962552   &  13390.86238502  & 3347.71559625   &   836.92889906    &    133.90862385 \\
\hline
\end{tabular}
\end{table}
\endgroup

At first we illustrate the behavior of 3DCHO at $r_c \rightarrow 0$. A diligent study reveals that, at small $r_c$ region, 3DCHO has an 
energy spectrum comparable to that of a particle in a spherical box (PISB). This leaning generally holds good for all other 
states as well. A cross-section of eigenvalues ($1s$-$1g$) of lowest five circular states corresponding to $l=0$-4, of 3DCHO and 
PISB, given in Table~5, at five selected $r_c$, \emph{viz.}, $0.01, 0.05, 0.1, 0.2, 0.5$, supports this fact. However this 
observation should not be misinterpreted to conclude that at $r_c \rightarrow 0$, 3DCHO leads to PISB. Because that can happen only 
when both systems have nearly equal kinetic energy as well as potential energy components. This is not apparent from this table. 
At this point, it is noteworthy to mention that, like 3DCHO, PISB is also exactly solvable; eigenfunctions are explicable directly 
in terms of first-order Bessel function, and given as, 
\begin{equation}
J_{l}(Z)=(-1)^{l}Z^{l}\left(\frac{1}{Z}\frac{d}{dZ}\right)^{l}\left(\frac{\sin{Z}}{Z}\right), \\
\end{equation} 
where $Z=\sqrt{\mathcal{E}_{n_{r},l}}\ r$. At the boundary when $r = r_c$, $Z=Z_{n_r,l}$ and $J_{l}(Z_{n_r,l})=0$. Moreover, 
at $r = r_c$, the energy of a $(n_r,l)$ state is written as $\mathcal{E}_{n_{r},l}=\frac{Z_{n_r,l}^{2}}{r_{c}^{2}}$ 
\cite{cohen78}.  This $J_{l}(Z_{n_r,l})=0$ is a transcendental equation and at a constant $n_{r},l$, this $Z_{n_{r},l}$ is evaluated 
by the help of MATHEMATICA program package. Thus all the PISB energies in lower segment of this table have been calculated 
following the above procedure.  

\begingroup            
\begin{table}[h]
\caption{$I_{\rvec}, I_{\pvec}$ of lowest five circular states of 3DCHO and PISB at five $r_c$ at fixed $m(0)$ \cite{mukherjee18}.}
\centering
\begin{tabular}{>{\tiny}l|>{\tiny}l|>{\tiny}l>{\tiny}l>{\tiny}l>{\tiny}l
>{\tiny}l}
\hline
 System &  $l$ & $r_c=0.01$  &  $r_c=0.05$   &  $r_c=0.1$ &   $r_c=0.2$   &   $r_c=0.5$    \\
\hline
\multicolumn{7}{c}{$I_{\rvec}$}  \\
\hline
     & 0   &   394784.1761898  & 15791.36986976 &   3947.84176    &    986.960440    &   157.913740    \\
     & 1   &   807629.1424372  & 32305.16943736 &   8076.29142    &   2019.072855   &    323.05170823  \\
 3DCHO & 2   &  1328698.4767434  & 53147.9434496  &   13286.984765  &   3321.7461915  &    531.4794285   \\
     & 4   &  2678172.4771556  & 107126.9043276 &   26781.72476   &   6695.431192   &   1071.269013    \\
\hline
     & 0   &    394784.1760435  &  15791.3670417  &  3947.8417604  &  986.9604401  & 157.9136704   \\
     & 1   &    807629.1422570  &  32305.1656902  &  8076.2914225  & 2019.0728556 &  323.0516569   \\
PISB & 2   &   1328698.4765707  &  53147.9390628  & 13286.9847657  & 3321.7461914 &  531.4793906  \\
     & 4   &   2678172.4770042  & 107126.8990801  & 26781.7247700  & 6695.4311925 & 1071.2689908 \\
\hline
\multicolumn{7}{c}{$I_{\pvec}$}  \\
\hline
     & 0   &   0.0001130690  &  0.0028267272 &   0.011306900 &   0.0452270673  &  0.282533301  \\
     & 1   &   0.0001498425  &  0.0037460639 &   0.014984249 &   0.0599366038  &  0.374503742 \\
3DCHO & 2   &    0.0001754798  &  0.0043869959 &   0.017547979 &   0.0701916254  &  0.438623720 \\
     & 4   &   0.0002100025  &  0.0052500632 &   0.021000250 &   0.0840008284  &  0.524961467 \\
\hline
     & 0   &   0.0001130690     &  0.002826727 & 0.011306909 &  0.045227638  & 0.282672741 \\
     & 1   &   0.0001498425     &  0.003746064 & 0.014984256 &  0.059937024  & 0.374606402 \\
PISB & 2   &   0.0001754798     &  0.004386996 & 0.017547984 &  0.070191936  & 0.438699601 \\
     & 4   &   0.0002100025     &  0.005250063 & 0.021000253 &  0.084001012  & 0.525006325 \\
\hline
\end{tabular}
\end{table}
\endgroup

The upper portion of Table~6 portrays $I_{\rvec}$ of 3DCHO and PISB at five selected $r_c$ values introduced before; the 
respective $I_{\pvec}$ of two systems are reported in lower portion. It is interesting to point out that, for 3DQHO and 3DCHO 
in a state having $m=0$, $I_{\rvec}$ and $I_{\pvec}$ are directly connected to expectation values of kinetic and potential energy as;
\begin{equation}
I_{\rvec}(\omega)= 8\langle T \rangle, \ \ \ I_{\pvec}(\omega)=\frac{8}{\omega^{2}}\langle v(r)\rangle. 
\end{equation}
But for a PISB, total energy is exclusively kinetic energy as the potential energy is zero. However, one can evaluate $I_{\rvec}, 
I_{\pvec}$ for PISB and collate the results with 3DCHO. Because, a pair of systems possessing same $I_{\rvec}, I_{\pvec}$ for all states 
indicate identical physical and chemical environment. Hence, here we have exploited $I$ to investigate the characteristics of PISB and 
3DCHO at $r_c \rightarrow 0$. The table clearly manifests that at $r_c \rightarrow 0$, a 3DCHO has comparable $I_{\rvec}, 
I_{\pvec}$ values with that of PISB, thus confirming our presumption that, at $r_c \rightarrow 0$, 3DCHO behaves like a PISB. One 
also marks that, with reduction in $r_{c}$, $I_{\pvec}$ (and also potential energy) approaches zero. On the other hand, as 
$r_c$ grows, the separation between $I_{\rvec}$ (also $I_{\pvec}$) values of 3DCHO and PISB tends to increase significantly. Moreover, 
as expected, at $r_c \rightarrow \infty$, 3DCHO reduces to 3DQHO. In previous section it was pointed out that, a 
1DCHO may be treated as a two-mode system; at smaller (approaching zero) and larger (tending infinity) confinement lengths it 
behaving like a particle in a box and an 1DQHO respectively \cite{Gueorguiev06,laguna14}. Here, also we observe resembling behavior. At $r_c \rightarrow 0$ and 
$\infty$, 3DCHO leads to PISB and a 3D isotropic QHO respectively. We also note that, larger the value of $\omega$ higher will be 
$\langle v(r)\rangle$; as a matter of fact 3DCHO is more prone to 3DQHO in such a case. Conversely, lesser the $\omega$ value, 
$\langle v(r)\rangle$ is smaller and 3DCHO, in that occasion, is inclined towards a PISB. Therefore at a fixed $r_c$, by controlling 
$\omega$ values one can inquire the properties of all three systems starting from PISB to 3DQHO through 3DCHO.  

\begingroup            
\begin{table}
\caption{$I_{\rvec}, I_{\pvec}$ for lowest five $n_r$ ($l=0$) values at six different $r_c$ \cite{mukherjee18}. See text for details.}
\centering
\begin{tabular}{>{\tiny}l<{\tiny}|>{\tiny}l>{\tiny}l>{\tiny}l>{\tiny}l
>{\tiny}l>{\tiny}l<{\tiny}}
\hline
   $n_{r}$ & $r_c=0.1$  &  $r_c=0.5$   &  $r_c=1$ &   $r_c=2$   &   $r_c=7$  & $r_c=\infty$    \\
\hline
\multicolumn{7}{c}{$I_{\rvec}$}  \\
\hline
 0   &   3947.84176  &  157.9137401 &    39.48285935 &  10.130828577  &  6.00000000  & 6  \\
 1   &  15791.3670  &  631.654662 &   157.91245186   &  39.42241043   & 14.00000000  & 14 \\
 2   &  35530.57584  &  1421.2230213 &   355.3049651 &  88.77709457   & 22.00000000  & 22 \\
 3   &  63165.46816  &  2526.6187189 &   631.6541846 &  157.88183967  & 30.00000000  & 30  \\
 4   &  98696.0440  & 3947.8417551 &   986.960106    &  246.718681361 & 38.00000000  & 38 \\
\hline
\multicolumn{7}{c}{$I_{\pvec}$}  \\
\hline
 0   &  0.0113069007    &  0.2825333012  & 1.1217967676   & 3.9877029335    & 6.0000000      & 6  \\
 1   &  0.01282672989   & 0.32070687721 & 1.28512015257   & 5.25470219890   & 14.0000000     & 14 \\
 2   &  0.0131081767    & 0.3277292039  &  1.3124046596   &  5.34276239849  & 22.00000000    & 22 \\
 3   &  0.0132066828    & 0.3301825760   & 1.3216621568   & 5.3463257844    & 30.00000000    & 30 \\
 4   &  0.0132522770    & 0.33131731951   & 1.3258940221   & 5.3437134181   & 38.00000000    & 38 \\
\hline
\end{tabular}
\end{table}
\endgroup

So far, we have explored the limiting trend of 3DCHO. Now we look into its behavior at intermediate $r_c$ region. For that, at 
first, the dependence of $I_{\rvec}, I_{\pvec}$ on quantum number $n_r$ is recorded in Table~7. It tabulates these quantities for 
lowest five $n_r$ (0-4) at six representative $r_c$ values. This clearly implies that, at fixed $m,l$ and $r_c$, both $I_{\rvec}, 
I_{\pvec}$ in 3DCHO get incremented as $n_{r}$ attains higher values. Henceforth, the role of $n_{r}$ on these measures is not 
discussed any further.   

\begingroup            
\begin{table}
\caption{$I_{\rvec}, I_{\pvec}$ of $1p, 1d,1f$ states of 3DCHO at six $r_c$, with varying $m$. Last column correspond 
to respective 3DQHO values, computed from Eq.~(77) \cite{mukherjee18}.}
\centering
\begin{tabular}{>{\scriptsize}l<{\scriptsize}|>{\scriptsize}l>{\scriptsize}l>{\scriptsize}l>{\scriptsize}l
>{\scriptsize}l>{\scriptsize}l<{\scriptsize}}
\hline
 $|m|$ &  $r_c=0.1$ & $r_c=0.5$  & $r_c=1$   &   $r_c=2$ &   $r_c=7$  &   $r_c=\infty$    \\
\hline
\multicolumn{6}{c}{$I_{\rvec}(1p)$}  \\
\hline
 0   &   8076.29142  & 323.0517082 &  80.76619765  & 20.39764116 & 10.00000000  &  10 \\
 1   &   5736.542528 & 229.4242339 &  57.21792995  & 13.91660078 & 6.00000000  &   6 \\
\hline
\multicolumn{6}{c}{$I_{\pvec}(1p)$}  \\
\hline
 0   &   0.014984249  & 0.374503742 & 1.491857857  & 5.577935621 & 10.00000000  & 10 \\
 1   &   0.01003      & 0.2507      & 0.9980       & 3.975       & 6.0000      &  6 \\
\hline
\multicolumn{6}{c}{$I_{\rvec}(1d)$}  \\
\hline
 0   &   13286.984765  &   531.4794285  & 132.8722779   & 33.37450170  & 14.00000000  & 14  \\
 1   &   10419.849672  &   416.7666252  & 104.0908347   & 25.74558754  & 10.00000000  & 10  \\
 2   &   7552.714578   &   302.0538220  & 75.3093915    & 18.11667339  &  6.00000000   &  6  \\
\hline
\multicolumn{6}{c}{$I_{\pvec}(1d)$}  \\
\hline
 0   &   0.017547979  &  0.43862372 & 1.74993894  & 6.70582522 & 14.0000000  &  14 \\
 1   &   0.0133333       &  0.33326        & 1.32905         & 5.0595         & 10.00000      &  10 \\
 2   &   0.0091187       &  0.2279         & 0.9081          & 3.4132         &  6.00000      &   6 \\
\hline
\multicolumn{6}{c}{$I_{\rvec}(1f)$}  \\
\hline
 0   &   19532.47745   &    781.2991270 &  195.3266196   & 48.95155358 &  18.000000000 &  18 \\
 1   &   16156.649498  &    646.2448005  & 161.48315319  & 40.15669405  & 14.00000000 &  14 \\
 2   &   12780.8215    &    511.1904739 &  127.63968669  & 31.36183453  & 10.00000000 &  10 \\
 3   &   9404.993581   &    376.1361474 &  93.79622020   & 22.566975007 & 6.000000000  &   6 \\
\hline
\multicolumn{6}{c}{$I_{\pvec}(1f)$}  \\
\hline
 0   &   0.0194769435    & 0.486866098     & 1.944005783    & 7.551236509    & 18.0000000  &  18 \\
 1   &   0.0157907758    & 0.39471708     & 1.57572258      & 6.0995039      & 14.00000   &  14 \\
 2   &   0.012104608     & 0.30256807     & 1.20743938      & 4.6477714      & 10.00000   &   10 \\
 3   &   0.008418440     & 0.2104190      & 0.8391561       & 3.196038       &  6.00000   &   6 \\
\hline
\end{tabular}
\end{table}
\endgroup  

Now, in order to get a clear picture of the effect of magnetic quantum number, Table~8 gives $I_{\rvec}, I_{\pvec}$ of 3DCHO 
for lowest three nodeless states having $l \neq 0$, i.e., $1p$-$1f$ respectively, for all allowed $m$ at 6 carefully selected 
$r_c$ ($0.1, 0.5, 1, 2, 7, \infty$). Likewise, Fig.~10 portrays these for $1g$ state in panels (a) and (b), for all possible $|m|$. 
It is to be noted that, the quantities in last column at $r_c = \infty$ are given from Eq.~(77) considering $\omega=1$. 
For this special $\omega$, Eq.~(77) dictates that, $I_{\rvec}=I_{\pvec}$ for 3DQHO. One notices that, behavior of $I_{\rvec}$ and 
$I_{\pvec}$ in 3DCHO is always in agreement with 3DQHO; an explicit analysis suggests identical patterns. In general, $I_{\rvec}$ 
decays monotonically while $I_{\pvec}$ progresses with rise in $r_c$; this is found to hold true for all $|m|$. This is 
to be expected, as a growth in $r_c$ promotes delocalization in $r$ space and localization in $p$ space. At a fixed $r_c$ and 
fixed quantum numbers $n_{r}$ and $m$, $I_{\rvec}$ tends to rise with $l$; this is consistent to what is observed from Eq.~(77) in 3DQHO. 
Because, in both 3DCHO and 3DQHO, kinetic energy grows with $n_{r},l$. Similarly, at fixed $n_{r}, l$, in both 3DCHO and 3DQHO, 
$I_{\rvec}$ falls down as one descends down the table (increment in $|m|$). One striking fact is that, for all these three states 
considered, both 3DCHO and 3DQHO portray comparable pattern with respect to $m$. Next, $I_{t}$ for $1p$-$1f$ states of 3DCHO are offered
in Table~9. It seems to show an inclination towards $I_{\rvec}$ in the behavioral pattern. In all instances, the lower and
upper bounds stated by Eq.~(67) is satisfied. For an arbitrary state distinguished by quantum numbers $n_r,l$, change in 
$I_{\rvec}, I_{\pvec}$ with $r_c$ in a 3DCHO preserves same qualitative orderings for various $m$ (general nature of the plots
remain unchanged) as depicted in these figures. This has been accomplished in a number of occasions, which are not reported here to save 
space. As usual, in all cases, they all eventually reach their respective 3DCHO-limit at some sufficiently large $r_c$, which 
alters from state to state.         

Now to understand the dependence of $I_{\rvec}, I_{\pvec}$ on $l$, we present Table~10, where these are given for $l=1-9$ states having 
$|m|=1$ and $n_{r}$ corresponding to 1, at same six chosen $r_c$ values of previous table. The last column again has same significance 
as Table~8. In accordance with Eq.~(77), here also for $n_rl \ (n_r=2)$ states, 
$I_{\rvec}=I_{\pvec}$ in 3DQHO. Dependence of $I_{\rvec}, I_{\pvec}$ of 3DCHO on $l$ compliments our observation in 3DQHO. At a fixed 
$n_{r}, m$ and $r_c$, they both enhance with rise of $l$ in 3DCHO and 3DQHO. This may occur probably because that, as $l$ 
advances, the density gets increasingly concentrated. Therefore, at a certain $n_{r}, m$, a state with higher $l$ undergoes 
greater fluctuation. Thus all our foregoing discussion leads to a general fact that,in a 3DCHO, the qualitative variations of 
$I_{\rvec}$ with all three quantum numbers remain quite analogous  to that of $I_{\pvec}$; also the patterns in 3DCHO and 3DQHO are 
similar. It may be appropriate to mention a parallel work \cite{mukherjee18d} along this direction for free and confined H atom 
inside an impenetrable spherical environment. One finds various significant deviations in the variation pattern between two systems 
there. Here we mention two of the most interesting facts, which are in complete contrast with a 3DCHO, e.g., $I_{\rvec}$ remains 
unchanged with respect to changes in $l$, whereas under confinement, it reduces with $l$ (at fixed $n,m$). Besides, for a given 
state having fixed $n,m$, $I_{\pvec}$ enhances when the atom is compressed, whereas declines in a free H atom. 

\begin{figure}                         
\begin{minipage}[c]{0.5\textwidth}\centering
\includegraphics[scale=0.65]{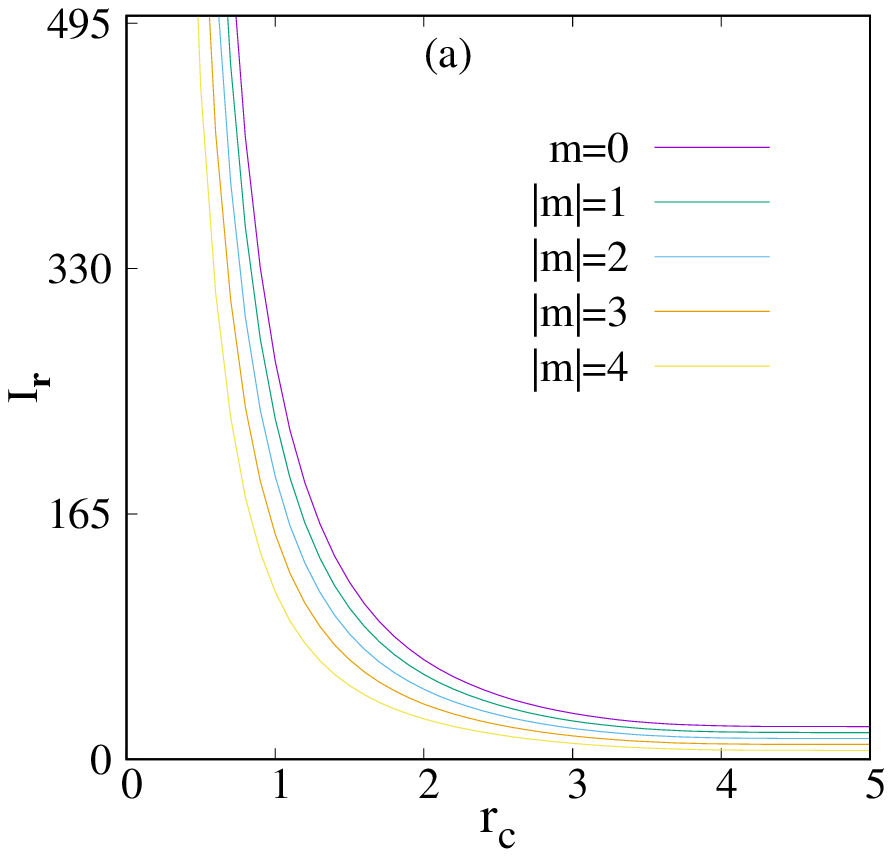}
\end{minipage}%
\begin{minipage}[c]{0.5\textwidth}\centering
\includegraphics[scale=0.65]{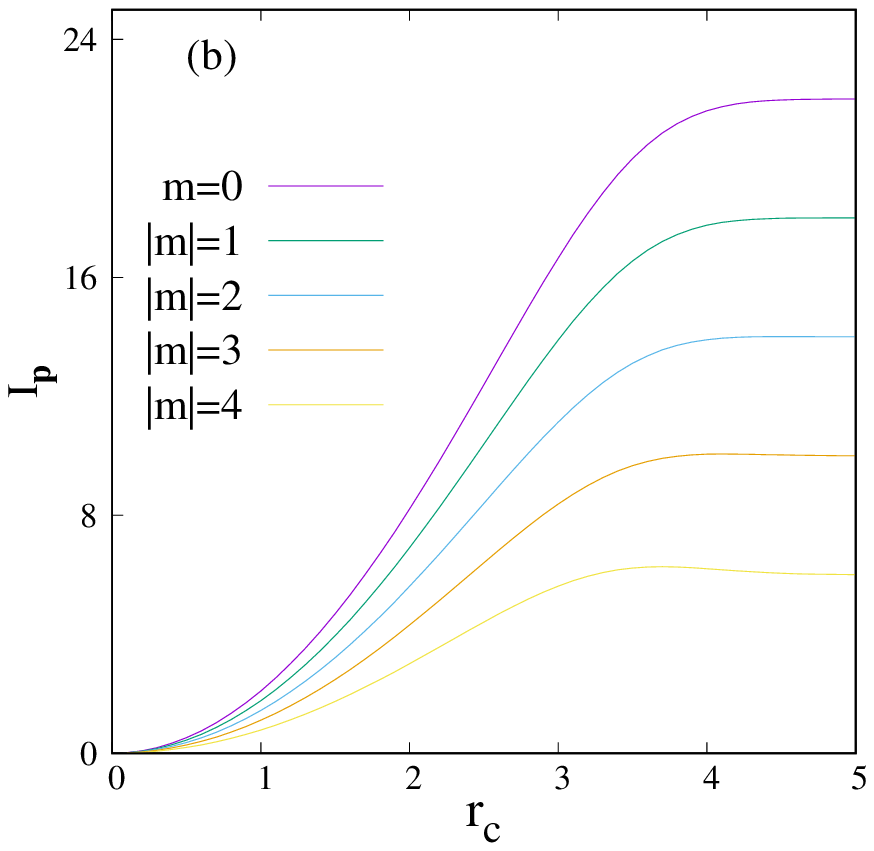}
\end{minipage}%
\caption{Variation of $I_{\rvec}$ and $I_{\pvec}$ in 3DCHO, with $r_c$, for all allowed $|m|$ values of $1g$ state, in panels 
(a) and (b) respectively \cite{mukherjee18}. See text for details.}
\end{figure}

At this juncture, a few words may be devoted to the influence of $\omega$ on $I_{\rvec}, I_{\pvec}$. In order to establish this, 
Fig.~11 depicts plots of $I_{\rvec}$ and $I_{\pvec}$ against $r_c$, for 5 selected $\omega^{2}$ ($1, 2, 4, 8, 32$), in bottom 
((a), (b)) and top ((c), (d)) panels. These are provided for $1p$ state; left and right panels characterize $|m|=0$ and 1 
respectively. Evidently at a fixed $r_c$, $I_{\rvec}$ grows and $I_{\pvec}$ decays with increment in $\omega$. At a certain 
$\omega$, dependence of these measures on $n_{r},l,m$ is similar to that in 3DQHO. As $\omega$ goes up, there is more localization, 
hence compactness in single-particle density increases with oscillation strength.

An in-depth analysis of $I$ reveals that, a gradual increase in $r_c$ should lead to a delocalization of the system in such a 
way that, at $r_c \rightarrow \infty$ it should approach towards an 3DQHO. On the contrary, when $r_c \rightarrow 0$ the impact of 
confinement is maximum. It suffices for us to vary $r_c$ from $0.1$ to $8$. This parametric increase in $r_c$ elicits the system 
from an extremely confined environment to a free 3DQHO scenario.

\begingroup            
\begin{table}
\caption{ $I_{t}$ for $1p$-$1f$ $(n_{r}=0)$ orbitals at six different $r_c$ \cite{mukherjee18}. See text for details.}
\centering
\begin{tabular}{>{\footnotesize}l<{\footnotesize}|>{\footnotesize}l>{\footnotesize}l>{\footnotesize}l>{\footnotesize}
l>{\footnotesize}l>{\footnotesize}l<{\footnotesize}}
\hline
 $|m|$ &  $r_c=0.1$ & $r_c=0.5$  & $r_c=1$   &   $r_c=2$ &   $r_c=7$  &   $r_c=\infty$    \\
\hline
\multicolumn{7}{c}{$I_{t}(1p)^{a}$}  \\
\hline
 $0^{\dag}$   & 121.01515 & 120.98286 &  120.491051  & 113.776614 & 100.00000000  & 100 \\
 1            & 57.53752  & 57.5166   &  57.1034     & 55.318     & 36.0000       &  36 \\
\hline
\multicolumn{7}{c}{$I_{t}(1d)^{b}$}  \\
\hline
$0^{\dag}$ & 233.15973  & 233.117506 & 232.518248 & 223.80357 & 196.000000000  & 196  \\
 1         & 138.93098  & 138.89164  & 138.3419   & 130.2598  & 100.00000      & 100  \\
 2         & 68.870938  &  68.8380   & 68.3884    &  61.8358  & 36.00000       &  36  \\
\hline
\multicolumn{7}{c}{$I_{t}(1f)^{c}$}  \\
\hline
$0^{\dag}$ & 380.432960  &  380.388051  & 379.716077 & 369.644758 & 324.0000000 &  324 \\
1          & 255.126029  &  255.083860  & 254.452650 & 244.935911 & 196.00000   &  196 \\
2          & 154.706834  &  154.669915  & 154.117184 & 145.762637 & 100.00000   &  100 \\
3          & 79.1753741  &  79.1461919  & 78.7096703 & 72.1249096 &  36.00000   &   36 \\
\hline
\end{tabular}
\begin{tabbing}
{$^\dag$ These also correspond to upper bounds, given in Eq.~(66).} \\
{$^a$ Lower bounds Eq.~(11), for $1p$ at 6 $r_c$ are: 10.70940, 10.71226, 10.75598,} \\ { 11.39074,12.96, 12.96.} \\
{$^b$ Lower bounds Eq.~(11), for $1d$ at 6 $r_c$ are: 5.55842, 5.559428, 5.573756, } \\ {5.79079,6.61224, 6.61224.} \\
{$^c$ Lower bounds Eq.~(11), for $1f$ at 6 $r_c$ are: 3.40664, 3.40704, 3.41307,} \\ { 3.506068, 4, 4.}   
\end{tabbing}
\end{table}
\endgroup

Next, Table~11 shows evaluated $R_{\rvec}^{\alpha}, R_{\pvec}^{\beta}, R_{t}^{(\alpha, \beta)}$ for first two ($n_r=1,2$) 
$s,p$ and $d$ states of 3DCHO at a selected set of $r_c$ values. $R_{\rvec}^{\alpha}$, starting from a $(-)$ve value at very low 
$r_c$, continuously increases, before finally reaching to the respective 3DQHO limit. This pattern in $R_{\rvec}^{\alpha}$ distinctly 
expresses delocalization of the system with growth in $r_c$. In contrast, $R_{\pvec}^{\beta}$, for all states generally tend 
to diminish monotonically with $r_c$, again converging to 3DQHO in the end. The analysis of $R_{t}^{(\alpha, \beta)}$ shows  
that, for $l=0$ states it lowers with $r_c$ to reach the unconfined result. However, for $1p$ state, it advances with $r_c$ to 
attain the 3DQHO limit. In case of $2p,1d$ and $2d$ states it initially increases with $r_c$, then passes through a maximum, and 
finally convenes to limiting 3DQHO result. At small $r_c$ (up to $ \approx 2$), $n_r=1$ states have higher $R_{\rvec}^{\alpha}$ 
compared to their $n_r=2$ counterparts. But at moderate $r_c$ ($ > 2$) region the situation alters; the $n_r=2$ states now 
possess larger $R_{\rvec}^{\alpha}$. Moreover this crossover region appears at larger $r_c$ with rise in $l$. This observation 
concludes that, at smaller $r_c$ region the effect of confinement is more pronounced for higher $n_r$ states (as $n_r=1$ states 
possess higher $R_{\rvec}^{\alpha}$ values than that of respective $n_r=2$ states). There is no such crossover in 
$R_{\pvec}^{\beta}$ and $R_{t}^{(\alpha, \beta)}$ in any of these reported states. The above observations are graphically demonstrated in Fig.~12, 
where in segments (a)-(c), 
$R_{\rvec}^{\alpha}, R_{\pvec}^{\beta}, R_{t}^{(\alpha, \beta)}$ of first five circular states are presented as function of $r_c$. 
Panel (a) leads to the fact that, for all of them, $R_{\rvec}^{\alpha}$'s quite steadily progress with $r_c$ and finally converge 
to 3DQHO. Similarly, from panel (b) it is clear that, $R_{\pvec}^{\beta}$ shows reverse pattern with $r_c$, before reaching 
3DQHO-limit. From the last panel, it appears that, for $l=0$ state, $R_{t}^{(\alpha, \beta)}$ abates with $r_c$, whereas an 
opposite trend occurs otherwise. Also one finds that major changes in $R_{t}^{(\alpha, \beta)}$ occurs in the range of $r_c$ 
varying from 2.65-5.30, with the higher state showing most predominant effect (see Figs.~15 and 17 later for $S_t$ and $E_t$, for 
similar effects). Some of these may not be evident from the figure as such, but becomes clear upon magnification of the plot, in 
consultation with the data given in Table~11.

\begingroup            
\begin{table}[h]
\caption{$I_{\rvec}, I_{\pvec}$ for $2p$-$2m$ ($|m|=1$) states in 3DCHO, at six selected $r_c$ \cite{mukherjee18}. }
\centering
\begin{tabular}{>{\tiny}l<{\tiny}|>{\tiny}l>{\tiny}l>{\tiny}l>{\tiny}l
>{\tiny}l>{\tiny}l<{\tiny}}
\hline
 $l$ &  $r_c=0.1$  & $r_c=0.5$  &   $r_c=1$ &   $r_c=2$   &   $r_c=7$ &   $r_c=\infty$   \\
\hline
\multicolumn{7}{c}{$I_{\rvec}$}  \\
\hline
 1   & 19544.75146    & 781.7758499    & 195.3902144   & 48.61008407   & 14.000000000    & 14  \\
 2   & 28201.56829    & 1128.0488516   & 281.9599554   & 70.272141773  & 18.000000000   & 18   \\
 3   & 37976.00550    & 1519.027385    & 379.70865634  & 94.73419843   & 22.0000000000    & 22  \\
 4   & 48838.86407    & 1953.5428786   & 488.3419262   & 121.91517166  & 26.0000000000    & 26  \\
 5   & 60766.49285    & 2430.649110    & 607.62258270  & 151.75456390  & 30.0000000001   & 30  \\ 
 6   & 73739.4732     & 2949.5692932   & 737.3562778   & 184.204055587 & 34.0000000743   & 34  \\
 7   & 87741.58879    & 3509.6547640   & 877.38084537  & 219.22364350  & 38.00000037     & 38  \\
 8   & 102759.07974   & 4110.3551463   & 1027.5587357  & 256.77944910  & 42.0000017797    & 42  \\
 9   & 118780.106607  & 4751.196872    & 1187.77161130 & 296.84230573  & 46.0000074341   & 46  \\
\hline
\multicolumn{7}{c}{$I_{\pvec}$}  \\
\hline
 1   & 0.01221      & 0.3054        & 1.222        & 4.917       & 14.000      & 14  \\
 2   & 0.0133333    & 0.333337      & 1.33356      & 5.34750     & 18.00000    & 18  \\
 3   & 0.014439157  & 0.36097833    & 1.4438781    & 5.7743588   & 21.999997   & 22 \\
 4   & 0.0154743767 & 0.38685628    & 1.54723822   & 6.1782594   & 26.0000000  & 26  \\
 5   & 0.016426161  & 0.4106495     & 1.6423270    & 6.55303527  & 30.000000   & 30  \\ 
 6   & 0.017296672  & 0.4324115849  & 1.7293332625 & 6.898114452 & 34.00000    & 34  \\
 7   & 0.0180927445 & 0.4523131126  & 1.808922468  & 7.21517631  & 38.00000    & 38  \\
 8   & 0.0188222240 & 0.470550071   & 1.88186836   & 7.50667165  & 41.9999991  & 42  \\
 9   & 0.019492648  & 0.487310795   & 1.94891804   & 7.775182    & 45.99998    & 46  \\
\hline
\end{tabular}
\end{table}
\endgroup

To gain further insight, Fig.~13 delineates $R^{\alpha}_{\rvec}, R^{\beta}_{\pvec},  R_{t}^{(\alpha, \beta)}$, in left (a),
middle (b), right (c) panels, for lowest five circular states with changes in $n_r$ (maximum being 9). Four different $r_c$'s are 
chosen, namely, $0.1, 3, 5, \infty$ in segments (A)-(D) from bottom to top. At the lowest $r_c$, for all $l$, 
$R^{\alpha}_{\rvec}$'s gradually falls off with advancement in $n_r$, providing highest values for $l=0$. But for \emph{non-zero} 
$l$, it increases with $l$ for smaller $n_r$. However, at $n_r>6$, there seems to be a crossover between $l=1$ and $l=2$ states. 
Hence it can be concluded that, effect of confinement is maximum, minimum for $l=1$, $l=0$ states respectively. At a fixed $l$ and 
$r_c$ (lower region), higher $n_r$-states encounter the confinement to a greater extent. On the other hand, both 
$R^{\beta}_{\pvec}$ (b) and $R_{t}^{(\alpha, \beta)}$ (c) show inverse trend. For smaller $n_r$, both these quantities follow the 
sequence $1g>1f>1d>1p>1s$, which however, gets reversed ($1s>1p>1d>1f>1g$) at higher $n_r$. These results infers that, at lower 
$r_c$ region quantum effect gets amplified as information content decreases, whereas total information increases with $n_r$. The 
first column (a) interestingly shows the appearance of maximum in $R^{\alpha}_{\rvec}$ at some moderate $r_c$, whose positions 
get right shifted as $r_c$ enhances. Apparently there exists an interplay between two opposing effects: (i) radial confinement 
(localization) (ii) accumulation of nodes with $n_r$ (delocalization). As $r_c$ progresses, delocalization prevails for lower 
$n_r$. Hence with gradual relaxation in confinement, states with larger $n_r$ get delocalized. At $r_c \rightarrow \infty$, 
the second effect predominates; so the system behaves like 3DQHO. In second and third columns, one sees that, both 
$R^{\beta}_{\pvec}, R_{t}^{(\alpha, \beta)}$ enhance with $n_r$. At $r_c \rightarrow \infty$, these two quantities approach the 
respective 3DQHO-limit, as predicted.

\begin{figure}                   
\begin{minipage}[c]{0.45\textwidth}\centering
\includegraphics[scale=0.65]{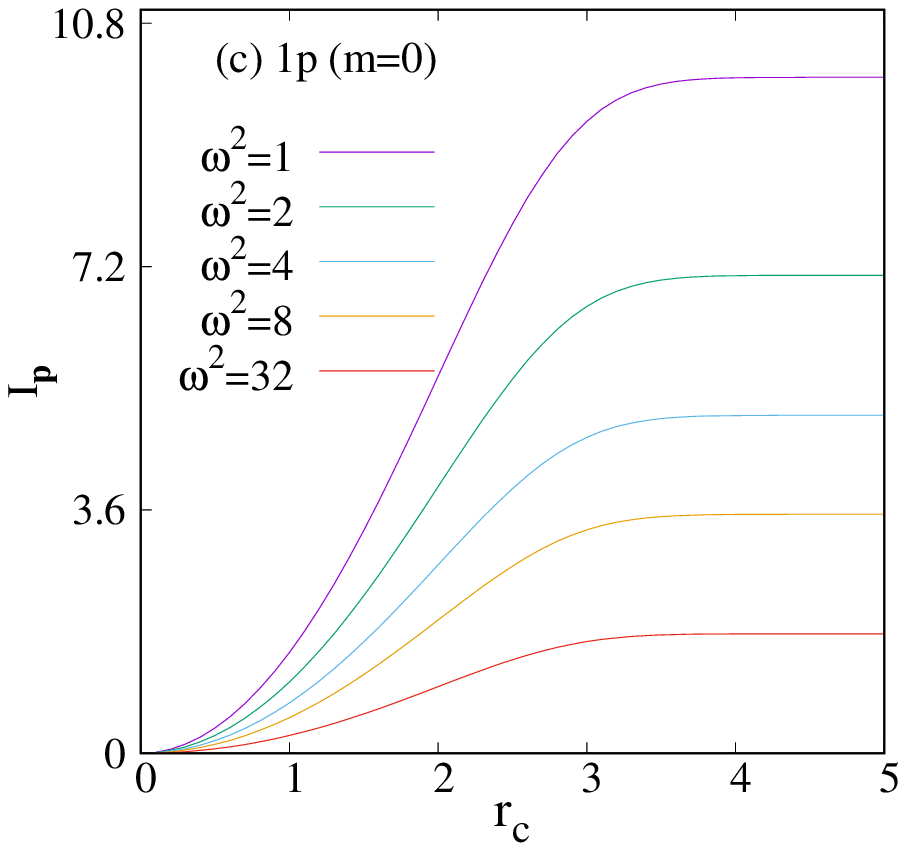}
\end{minipage}%
\hspace{0.2in}
\begin{minipage}[c]{0.45\textwidth}\centering
\includegraphics[scale=0.65]{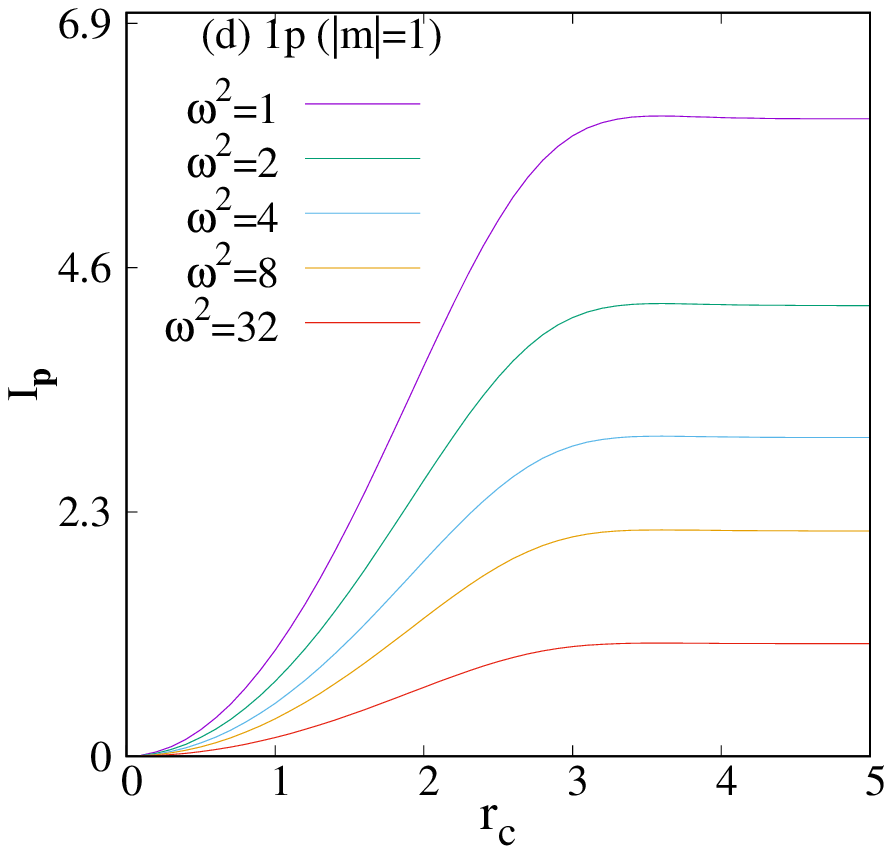}
\end{minipage}%
\\[15pt]
\begin{minipage}[c]{0.45\textwidth}\centering
\includegraphics[scale=0.65]{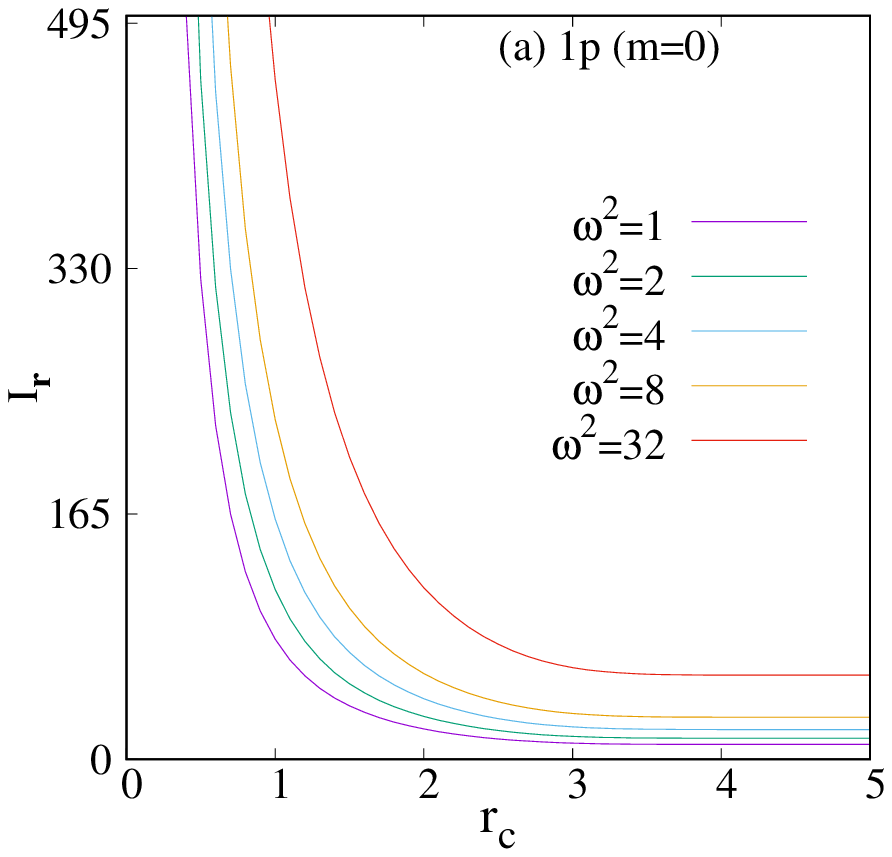}
\end{minipage}%
\hspace{0.2in}
\begin{minipage}[c]{0.45\textwidth}\centering
\includegraphics[scale=0.65]{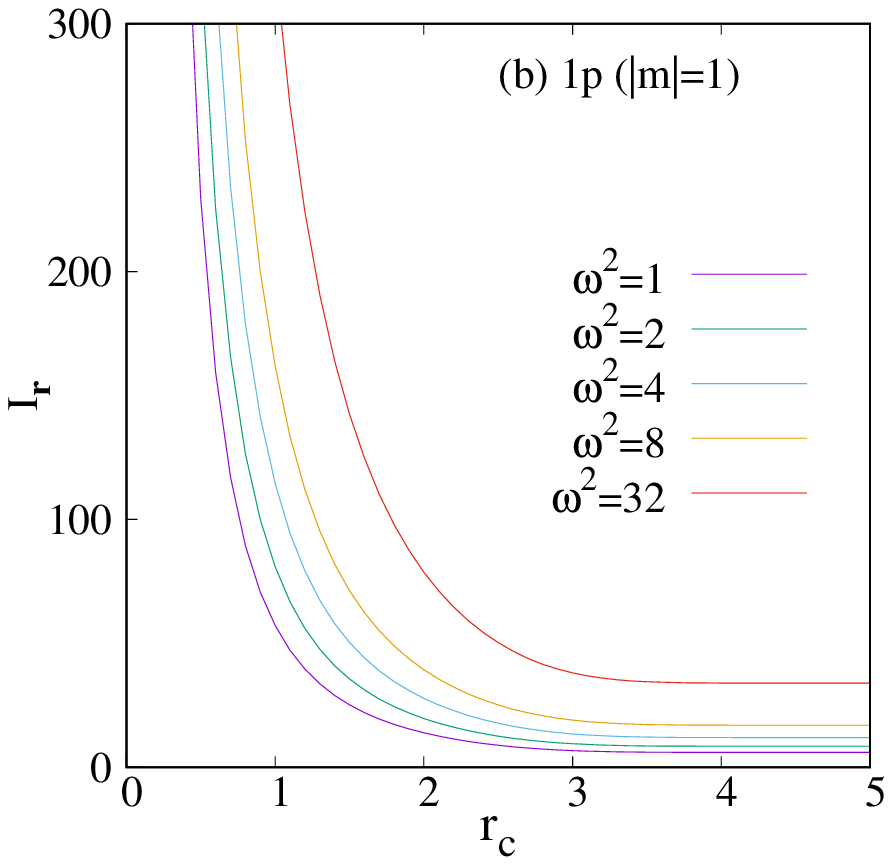}
\end{minipage}%
\caption{Plots of $I_{\rvec}$ and $I_{\pvec}$ in bottom ((a), (b)) and top ((c),(d)) panels in 3DCHO with $r_c$, of 
$1p$ state at five selected $\omega^{2} (1, 2, 4, 8, 32)$. Left, right columns correspond to $|m|=0$, 1 respectively \cite{mukherjee18}.}
\end{figure}

Next we move to $S$ in Table~12, where $S_{\rvec}, S_{\pvec}$ and $S_{t}$ are demonstrated for $1s,2s,1p,2p,1d,2d$ states of 3DCHO
at same set of $r_c$ of the previous table. Akin to 
$R^{\alpha}_{\rvec}$, $S_{\rvec}$ also yield $(-)$ve values for all six states at very low $r_c$, and then continuously 
increase until reaching the limiting 3DQHO value. This trend of $S_{\rvec}$ also interprets the delocalization of 3DCHO with 
progress in $r_c$. Like $R^{\beta}_{\pvec}$, $S_{\pvec}$ also offers a reverse nature (continuous decrease with $r_c$) of 
$S_{\rvec}$ (and $R^{\alpha}_{\rvec}$); from an initial $(+)$ve value, consistently falls off to reach 3DQHO. $S_{t}$'s for both 
$1s,1p$ states decrease to attain 3DQHO result. Interestingly, for $1d$ state it first declines, attains a minimum value, then rises 
to a maximum and ultimately reaches to 3DQHO values. But for $2s,2p,2d$ states it first increases, then reaches a maximum and 
finally merges to borderline 3DQHO values.

\begingroup           
\begin{table}
\caption{$R_{\rvec}^{\alpha},R_{\pvec}^{\beta}$ and $R_{t}^{\alpha,\beta}$ for lowest three $l$ (having $n_r=1,2$) states in 3DCHO 
at eight selected values of $r_c$.  $R_{t}^{\alpha,\beta}$ for all these states obey the lower bound given in Eq.~(57) \cite{mukherjee18a}.}
\centering
\begin{tabular}{>{\scriptsize}l>{\scriptsize}l>{\scriptsize}l>{\scriptsize}l>{\scriptsize}l>{\scriptsize}
l>{\scriptsize}l>{\scriptsize}l<{\scriptsize}}
\hline
$r_c$  &    $R_{\rvec}^{\alpha}$     & $R_{\pvec}^{\beta}$  &  $R_{t}^{\alpha,\beta}$  &  
$r_c$  &    $R_{\rvec}^{\alpha}$     & $R_{\pvec}^{\beta}$  &  $R_{t}^{\alpha,\beta}$  \\
\cline{1-4} \cline{5-8}
\multicolumn{4}{c}{$1s$}    &      \multicolumn{4}{c}{$2s$}    \\
\cline{1-4} \cline{5-8}
0.1      & $-$6.0366917844 & 12.247171  & 6.210480 & 0.1   & $-$6.0653334752 & 14.251561 & 8.186227   \\
0.2      & $-$3.9572613535 & 10.167740  & 6.210479 & 0.2   & $-$3.9858896952 & 12.172113 & 8.186223   \\
0.5      & $-$1.2088403559 & 7.4192722  & 6.210431 & 0.5   & $-$1.2369267194 &  9.422993 & 8.186067   \\
1.0      & 0.8636306014   & 5.3460818  & 6.209712 & 1.0     &  0.8438971150 &  7.339578 & 8.183475   \\
5.0      & 3.6326806767   & 2.5410652  & 6.173745 & 5.0     &  4.5764993107 &  2.616203 & 7.192702   \\
8.0      & 3.6326909163   & 2.5410540  & 6.173744 & 8.0     &  4.5767695172 &  2.614825 & 7.191594   \\
$\infty$ & 3.6326909163   & 2.5410540  & 6.173744 &$\infty$ &  4.5767695172 &  2.614825 & 7.191594   \\
\cline{1-4} \cline{5-8}
\multicolumn{4}{c}{$1p$}    &      \multicolumn{4}{c}{$2p$}    \\
\cline{1-4} \cline{5-8}
0.1    &  $-$6.1671363542 & 12.806502 & 6.639365   & 0.1      & $-$6.29011971 & 14.178463  & 7.888343  \\
0.2    &  $-$4.0876997330 & 10.727065 & 6.639366   & 0.2      & $-$4.21067738 & 12.099025  & 7.888348 \\
0.5    &  $-$1.3390273792 & 7.978413  & 6.639386   & 0.5      & $-$1.46177311 &  9.350298  & 7.888525 \\
1.0    &   0.7373200936   & 5.902381  & 6.639701   & 1.0      & 0.61816051   &  7.273091  & 7.891251 \\
5.0    &   3.8830108378   & 2.834951  & 6.717917   & 5.0      & 4.57588095   &  2.982808  & 7.558689 \\
8.0    &   3.8830566606 & 2.834947  & 6.718004   & 8.0      & 4.57683522    &  2.982008  & 7.558843 \\
$\infty$ & 3.8830566606 & 2.834947  & 6.718004   &$\infty$  & 4.57683522    &  2.982008  & 7.558843 \\
\cline{1-4} \cline{5-8}
\multicolumn{4}{c}{$1d$}    &      \multicolumn{4}{c}{$2d$}    \\
\cline{1-4} \cline{5-8}
0.1    & $-$6.1181191683  &13.228954  & 7.110835 & 0.1   & $-$6.2478841627   & 14.295117 & 8.047233 \\
0.2    & $-$4.0386800101  &11.149516  & 7.110836 & 0.2   & $-$4.1684423307   & 12.215678 & 8.047236 \\
0.5    & $-$1.2899046230  & 8.400780  & 7.110875 & 0.5   & $-$1.4195583633   & 9.466934 & 8.047376 \\
1.0    &  0.7880364400    & 6.323450  & 7.111487 & 1.0     & 0.6600654534    & 7.389472 & 8.049538 \\
5.0    &  4.2284239258    & 3.048859  & 7.277282 & 5.0     & 4.8017869530    & 3.233068 & 8.034855 \\
8.0    &  4.2285908429   & 3.047026  & 7.275616 & 8.0     & 4.8046032500  & 3.230664 & 8.035267 \\
$\infty$& 4.2285908429   & 3.047026  & 7.275616 &$\infty$ & 4.8046032500  & 3.230664 & 8.035267 \\
\hline
\end{tabular}
\end{table}
\endgroup

Figure~14 delineates behavioral trends of $S_{\rvec}$, $S_{\pvec}$, $S_{t}$ with $r_c$, in segments (a)-(c), for same five states of 
Fig.~12 (done for $R$). One notices that, both Figs.~12 and 14 offer qualitatively analogous changes with respect to $r_c$ 
variation; thus signalling comparable trends in $R$ and $S$. For all these states $S_{\rvec}$'s tend to progress with $r_c$ and 
then reach the corresponding $r$-space 3DQHO, while $S_{\pvec}$'s reduce before attaining that.  The last panel portrays that, for 
$1s,1p$ states $S_{t}$ decreases with $r_c$, while for the remaining three states, $S_t$ passes through a minimum. It is worth 
noting that, in the $1p$ state, $R_t$ progresses and $S_t$ regresses after $r_c=2.65$. In both occasions, however, as usual they 
all gain the limiting 3DQHO behavior at a sufficiently large $r_c$. 

In Fig.~15, $S_{\rvec}$ (a), $S_{\pvec}$ (b), $S_{t}$ (c) of $l=0-4$ states are plotted with respect to $n_r$ at 
same four values of $r_c$ in Fig.~13, in panels marked with (A)-(D) from bottom to top. Again, these graphs produce resembling 
shape and propensity as in Fig.~13. Thus in harmony with $R^{\alpha}_{\rvec}$ (for all five $l$ at the lowest $r_c=0.1$), 
$S_{\rvec}$ gets lowered in A(a), while $S_{\pvec}$s and $S_{t}$s accelerate with $n_r$ in A(b) and A(c), respectively. This
supports our previous observation (of $R$ in Fig.~13) that, at very low $r_c$, the effect of confinement is more prominent in 
high-lying states, signifying an intensification of quantum nature in such a scenario. As usual, like $R^{\alpha}_{\rvec}$ here 
also, the first column (a) shows the appearance of maximum in $S_{\rvec}$ plot with gradual progress of $r_c$ from the bottom to 
top. Their positions switch to right as $r_c$ enhances. This observation indicates that, at $r_c \rightarrow \infty$ the system 
behaves like 3DQHO.

At this point we proceed for the last measure ($E$) of this study in Table~13. A cross-section of $E_{\rvec}, E_{\pvec}, E_{t}$ 
for $1s,2s,1p,2p,1d,2d$ states of 3DCHO (same set of $r_c$ as used in previous table) is offered. One notices that, $E_{\rvec}$ 
reduces (as opposed to $R^{\alpha}_{\rvec}, S_{\rvec}$) while $E_{\pvec}$ progresses (in contrast to $R^{\beta}_{\pvec}, 
S_{\pvec}$) with rise in $r_c$. However, the nature of variation of $E_{t}$ with $r_c$ varies from state to state. For 
$1s,1p,1d$ states, it enhances with $r_c$, while $2s,2p, 2d$ states, it goes through a minimum. These changes in $E_{\rvec}, 
E_{\pvec}$ and $E_{t}$ with $r_c$ are demonstrated pictorially in Fig.~16, in left (a), middle (b) and right (c) panels for first five 
node-less states. It is found that, $E_{t}$ for $1s,1p,1d$ states advances with $r_c$, while for $1f,1g$ states it decreases. 
Interestingly, at large $r_c$, both  $E_{\rvec}, E_{\pvec}$ decline with growth in $l$. Finally, Fig.~17, depicts $E_{\rvec}, 
E_{\pvec}, E_{t}$ (in columns (a),(b),(c)) for $l=0-4$ states as functions of $n_r$ at four different $r_c$ (in segments A-D). At 
the lowest $r_c$ considered, $E_{\rvec}$ advances with $n_r$. However, the first column (a) suggests that, a minimum appears in 
$E_{\rvec}$ graphs as $r_c$ is increased. Also the positions of these minima shift towards right with rise in $r_c$. On the 
contrary, for all concerned $r_c$, both $E_{\pvec}$, $E_{t}$ decay with $n_r$.   
     
All the discussed measures of Eq.~(72) have been investigated with change of $r_c$, choosing two different and widely used values 
of $b$ ($1,\frac{2}{3}$). Note that, for $b=1$, $C_{ES}^{(2)}$ transforms to $C_{LMC}$; similarly $C_{IS}^{(1)}$ coincides with 
$C_{IS}$ at $b=\frac{2}{3}$. In order to simplify the discussion, a few words may be devoted to the notation used. A uniform symbol 
$C_{order_{s}, disorder_{s}}^b$ is used; where the two subscripts refer to two order ($E, I$) and disorder ($S, R$) parameters. 
Another subscript $s$ is used to specify the space; \emph{viz.}, $r, p$ or $t$ (total). Two scaling parameters $b=\frac{2}{3}, 1$ 
are identified with superscripts (1), (2). These measures are examined systematically for $1s,~1p,~1d,~2s,~1f,~2p,~1g$ and $2d$ 
states in conjugate spaces. In this part we will discuss the outcome extracted from a recent complexity analysis undertaken by us. 
The detailed results with relevant figures and tables will be published elsewhere separately. 

Now we lay our focus on $C_{E_{\rvec}S_{\rvec}}^{(1)},C_{E_{\pvec}S_{\pvec}}^{(1)}$ and $C_{E_{\rvec}S_{\rvec}}^{(2)},
C_{E_{\pvec}S_{\pvec}}^{(2)}$ respectively. $C_{E_{\rvec}S_{\rvec}}^{(1)}$ decreases and $C_{E_{\pvec}S_{\pvec}}^{(1)}$ increases 
with rise of $r_c$, before reaching the borderline 3DQHO result. As mentioned earlier, the 3DCHO model acts as an intermediate 
between the PISB and 3DQHO \cite{mukherjee18}. That means, an enhancement in $r_c$ should lead the system towards delocalization 
(equilibrium), which is supported by the declining nature of $C_{E_{\rvec}S_{\rvec}}^{(1)}$ with progress in $r_c$. At a 
fixed $l$, $C_{E_{\rvec}S_{\rvec}}^{(1)}$ progresses with $n_r$. It may be noted that in a 3DCHO, number of nodes varies with $n_r$ 
only (in contrast to, say the H atom, where it depends on both $n=n_r$ and $l$). Hence in this case, the increment in number of 
nodes takes the system towards the side of ordering. But no such pattern is observed for 
$C_{E_{\pvec}S_{\pvec}}^{(1)}$. The variation of $C_{E_{\rvec}S_{\rvec}}^{(2)},~C_{E_{\pvec}S_{\pvec}}^{(2)}$ with change of $r_c$ 
reveals that, former advances with growth of $r_c$ indicating that this is more inclined towards order. On the other 
hand, at first $C_{E_{\pvec}S_{\pvec}}^{(2)}$ decreases with $r_c$, attains a minimum and finally merges to 3DQHO. The ordering of 
$C_{E_{\rvec}S_{\rvec}}^{(2)}$ regarding $n_r$ quantum number is akin to $C_{E_{\rvec}S_{\rvec}}^{(1)}$. This observation 
complements our previous conjecture that, an increase in number of nodes takes the system towards disequilibrium. It is noticed 
that, after a certain $r_c$ ($\approx 3$), both $C_{E_{\rvec}S_{\rvec}}^{(2)},C_{E_{\pvec}S_{\pvec}}^{(2)}$ show analogous 
nature (increase to reach their respective limiting value). Here, exploring these two sets of complexity measure, namely,  
$C^{(1)}_{ES}$ and $C^{(2)}_{ES}$, it is evident that, $C_{E_{\rvec}S_{\rvec}}^{(1)}$ and 
$C_{E_{\pvec}S_{\pvec}}^{(1)}$ complement each other better (as former decreases, later increases with $r_c$) and more tends
towards disorder (equilibrium), whereas the other set is inclined towards the order-direction. Thus, $C_{ES}^{(1)}$ with 
$b=\frac{2}{3}$ explains the system better. It is important to point out that, when 3DCHO approaches to 3DQHO, 
$C_{E_{\rvec}S_{\rvec}}^{(1)}$ becomes equal to $C_{E_{\pvec}S_{\pvec}}^{(1)}$.      
                 
\begin{figure}                         
\begin{minipage}[c]{0.32\textwidth}\centering
\includegraphics[scale=0.4]{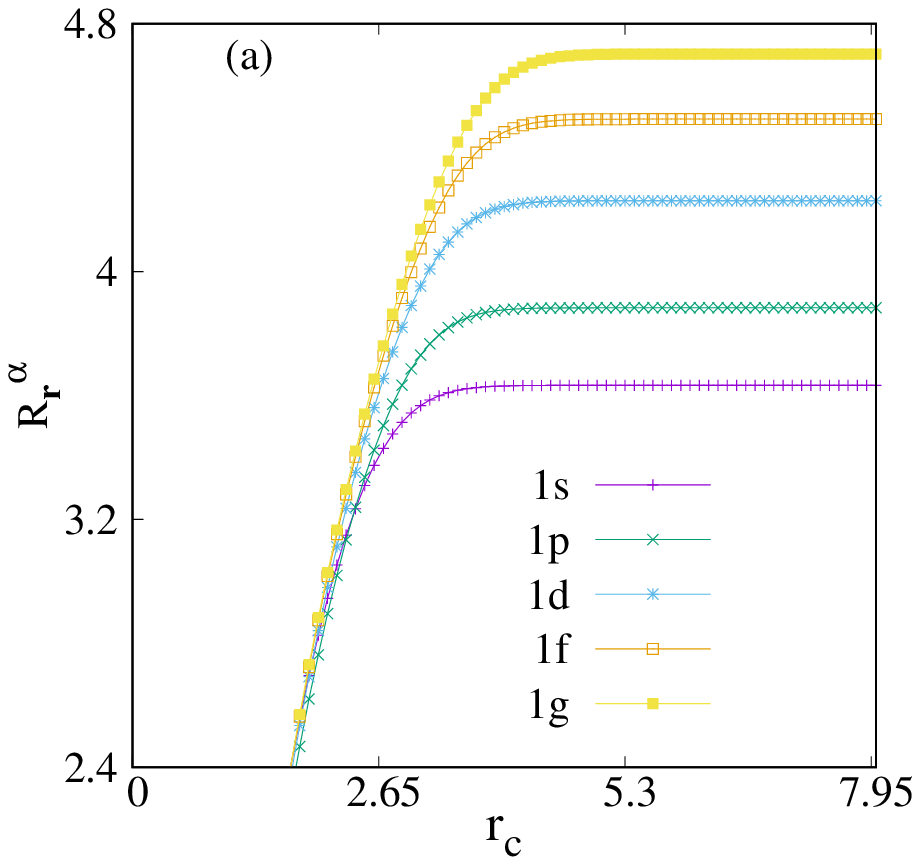}
\end{minipage}%
\hspace{0.02in}
\begin{minipage}[c]{0.32\textwidth}\centering
\includegraphics[scale=0.4]{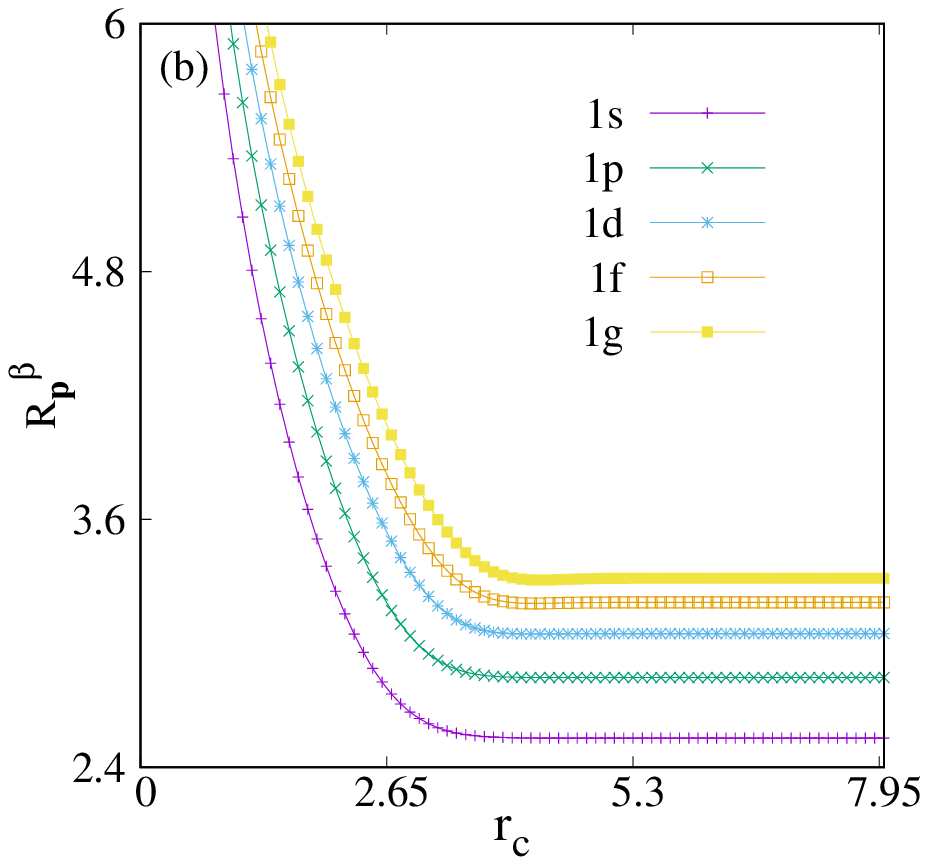}
\end{minipage}%
\hspace{0.02in}
\begin{minipage}[c]{0.32\textwidth}\centering
\includegraphics[scale=0.4]{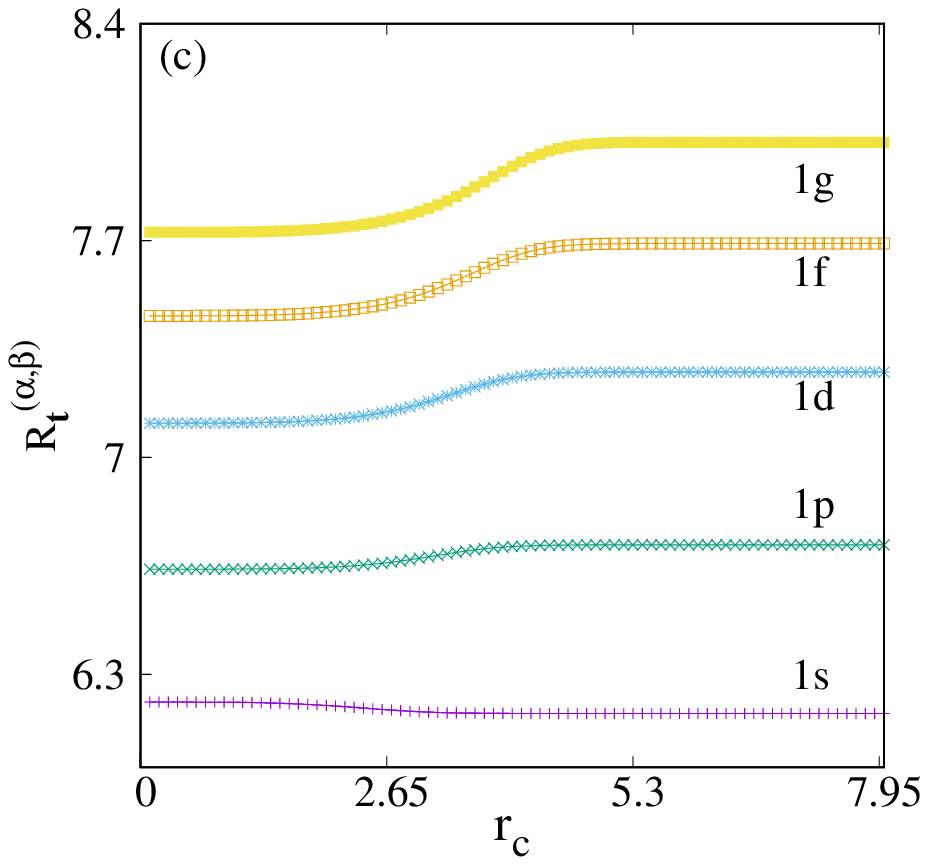}
\end{minipage}%
\caption{Plots of $R_{\rvec}^{\alpha}$, $R_{\pvec}^{\beta}$, $R_{t}^{(\alpha, \beta)}$ against $r_c$ for first five circular 
states of 3DCHO in panels (a), (b), (c) respectively \cite{mukherjee18a}. See text for details.}
\end{figure}  

In the same direction, now we interpret the pattern changes of $C_{E_{\rvec}R_{\rvec}}^{(1)},C_{E_{\pvec}R_{\pvec}}^{(1)}$ with $r_{c}$. 
At first, $C_{E_{\rvec}R_{\rvec}}^{(1)}$ declines with growth of $r_c$, then passes through a minimum and finally converges to respective 
3DQHO result. This minimum gets flatter with rise of both $n_r,~l$. Here also $C_{E_{\rvec}R_{\rvec}}^{(1)}$ shows similar trend
as observed in $C_{E_{\rvec}S_{\rvec}}^{(1)}$, \emph{viz.}, (i) like $C_{E_{\rvec}S_{\rvec}}^{(1)}$, 
$C_{E_{\rvec}R_{\rvec}}^{(1)}$ is more aligned towards disorder (ii) at a particular $l$, they progress with $n_r$. The first point
suggests that, like $C_{E_{\rvec}S_{\rvec}}^{(1)}$, $C_{E_{\rvec}R_{\rvec}}^{(1)}$ also interprets that, with increase in $r_c$ 
the system progresses towards 3DQHO from PISB via 3DCHO. The second point says that, an increase/decrease in number of nodes leads 
towards the direction of order/disorder respectively. Like $C_{E_{\pvec}S_{\pvec}}^{(1)}$, $C_{E_{\pvec}R_{\pvec}}^{(1)}$ increases with 
growth in $r_c$. Moreover, at a definite $l$, $C_{E_{\pvec}R_{\pvec}}^{(1)}$ falls off with $n_r$. At smaller $r_c$ region ($\approx 3$), 
both $C_{E_{\rvec}R_{\rvec}}^{(2)},C_{E_{\pvec}R_{\pvec}}^{(2)}$ vary slowly. At around same $r_c$, the former jumps and latter drops to 
attain the borderline 3DQHO limit. For both $C_{E_{\rvec}R_{\rvec}}^{(2)}$ and $C_{E_{\pvec}R_{\pvec}}^{(2)}$ absolute values of the 
slope of the curve 
enhance as $n_r$ grows (fixed $l$) and decrease with growth of $l$ (fixed $n_r$). The dependence of $C_{E_{\rvec}R_{\rvec}}^{(2)}$ 
on $n_r$ is similar to $C_{E_{\rvec}R_{\rvec}}^{(1)}$. This again validates the dependency of this complexity measure on number of 
nodes of the system. Once again one may conclude that, out of $C^{(1)}_{ER}$ and $C^{(2)}_{ER}$, the former offers a more 
complete knowledge about the characteristics of 3DCHO. Apart from that, $C_{E_{\pvec}R_{\pvec}}^{(1)}$ shows more lucid trend than 
$C_{E_{\pvec}S_{\pvec}}^{(1)}$ with respect to dependence of quantum number $n_r$. Hence, in practice, $C^{(1)}_{ER}$ may possibly 
be considered a better measure of complexity than $C^{(1)}_{ES}$.

\begin{figure}                                            
\begin{minipage}[c]{0.33\textwidth}\centering
\includegraphics[scale=0.4]{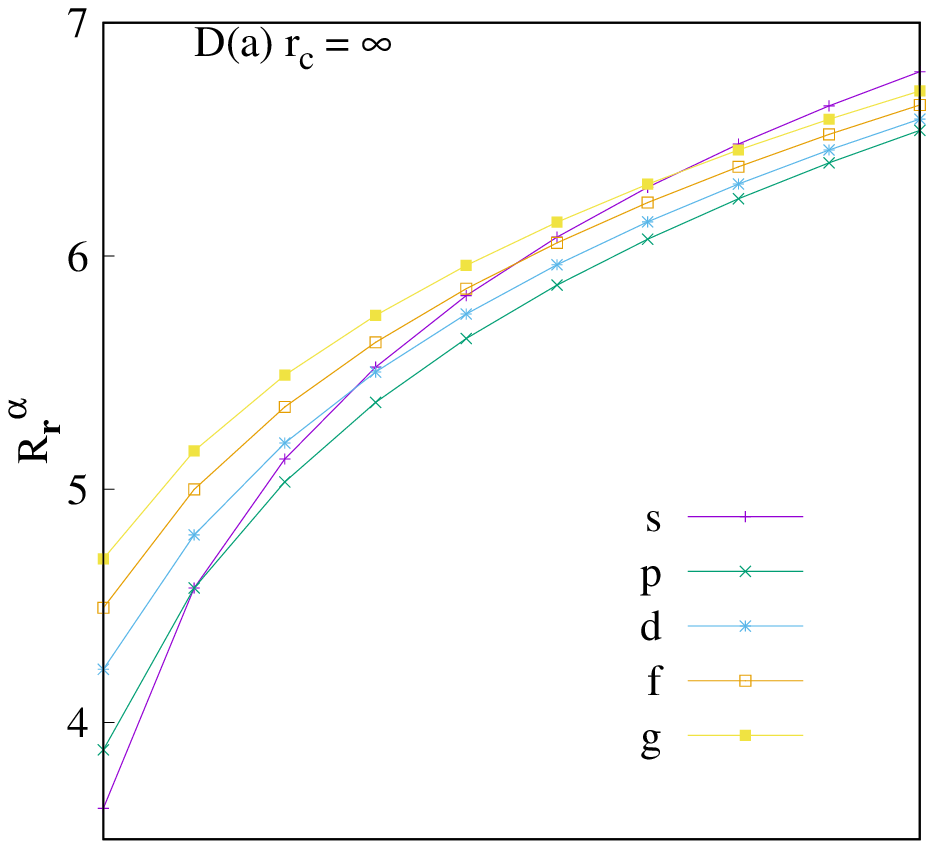}
\end{minipage}%
\begin{minipage}[c]{0.33\textwidth}\centering
\includegraphics[scale=0.4]{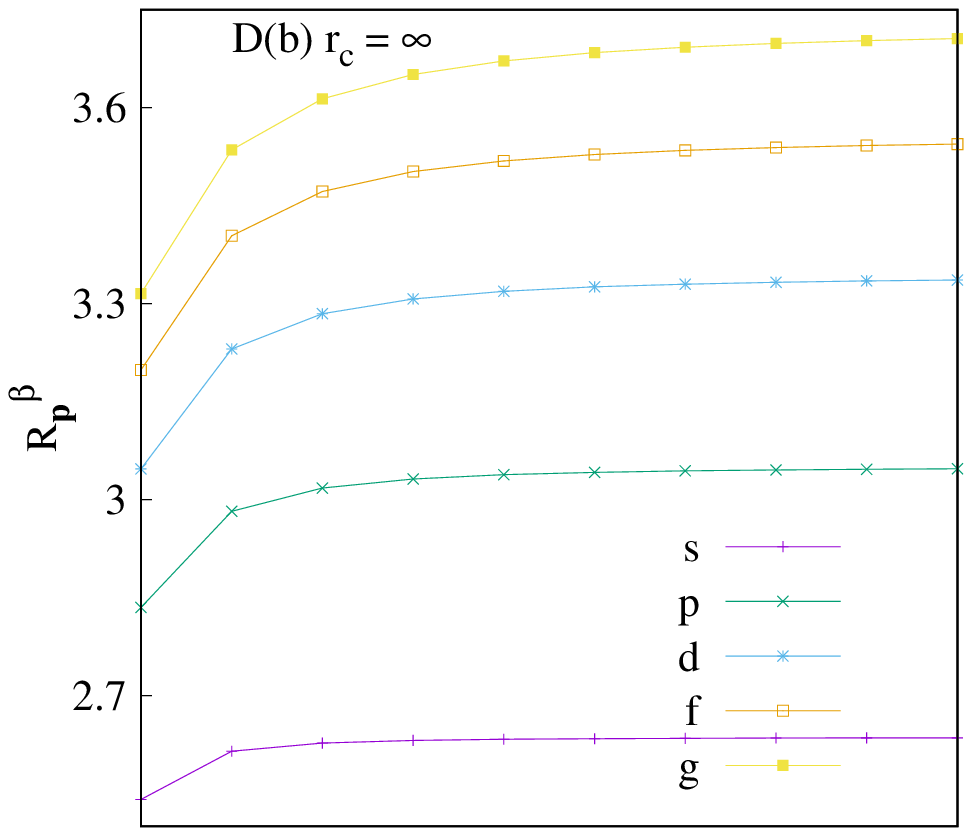}
\end{minipage}%
\begin{minipage}[c]{0.33\textwidth}\centering
\includegraphics[scale=0.4]{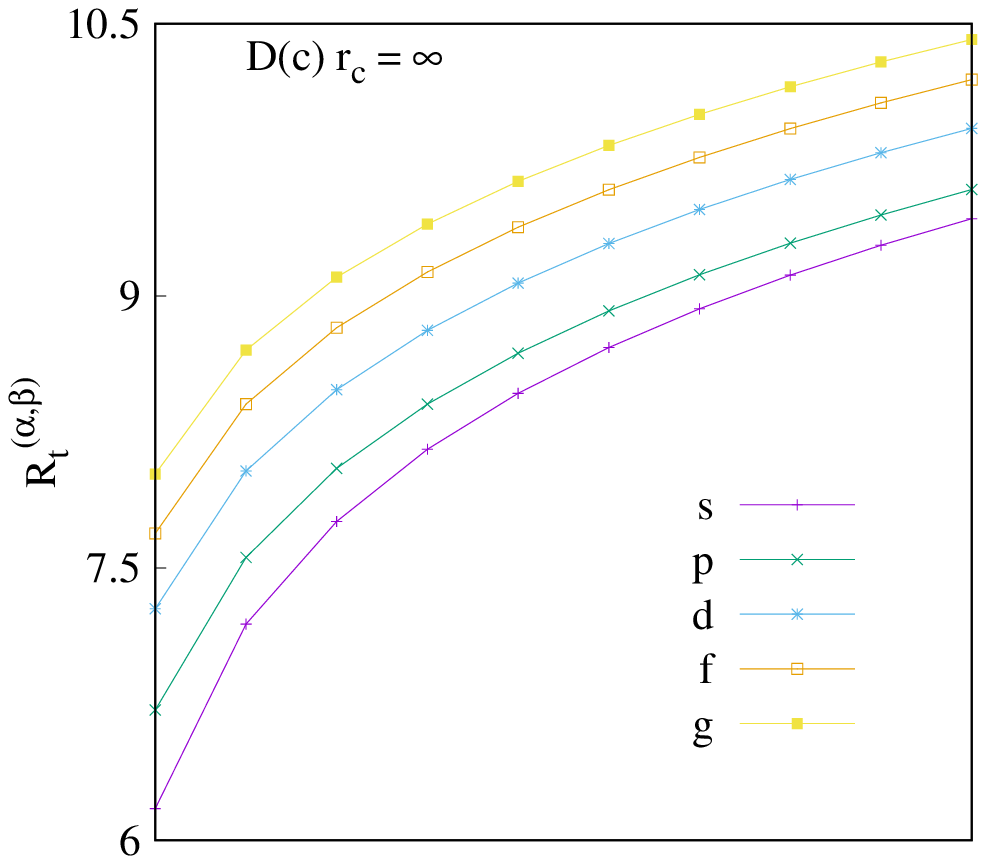}
\end{minipage}%
\\[15pt]
\begin{minipage}[c]{0.33\textwidth}\centering
\includegraphics[scale=0.4]{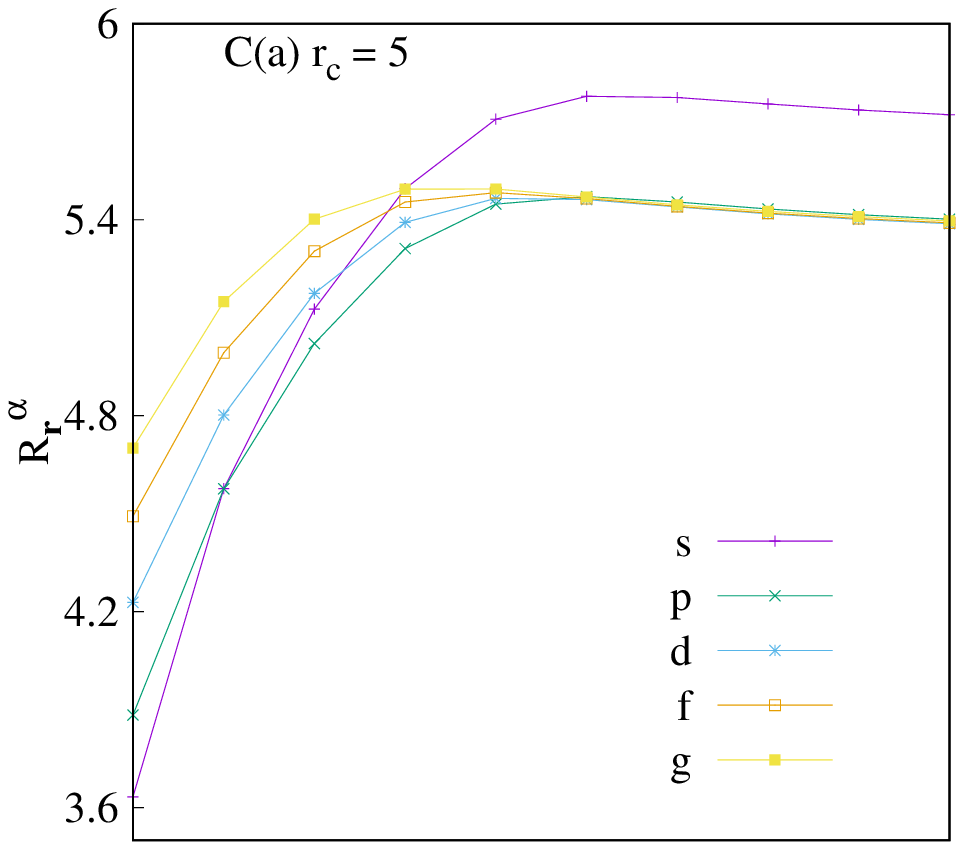}
\end{minipage}
\begin{minipage}[c]{0.33\textwidth}\centering
\includegraphics[scale=0.4]{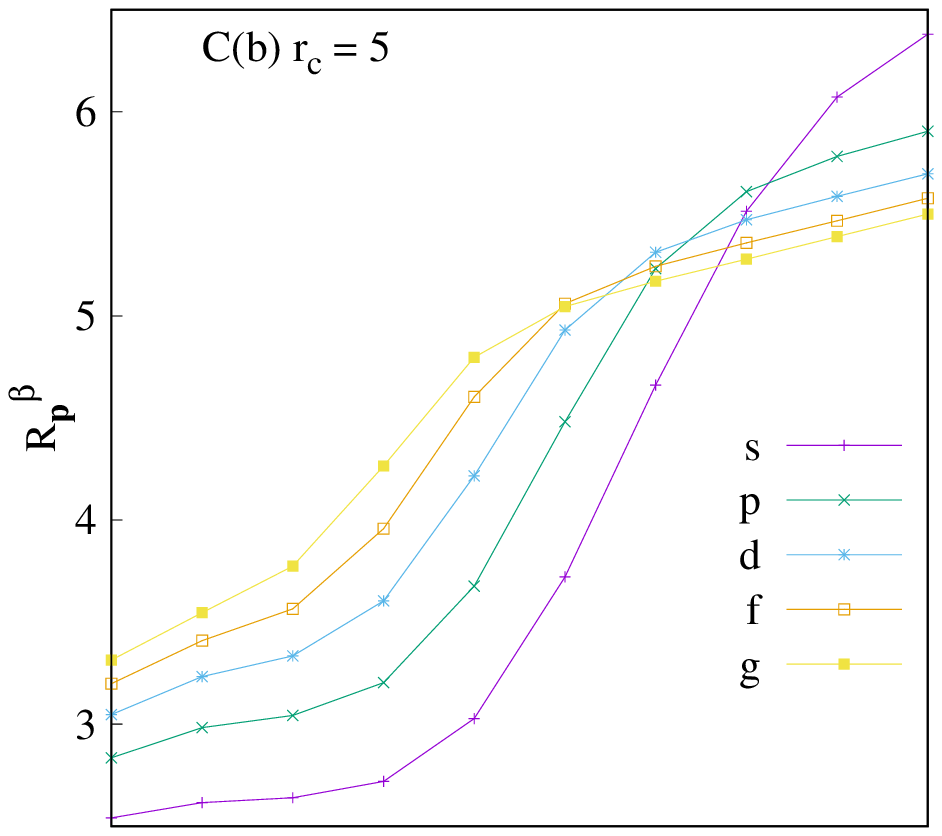}
\end{minipage}%
\begin{minipage}[c]{0.33\textwidth}\centering
\includegraphics[scale=0.4]{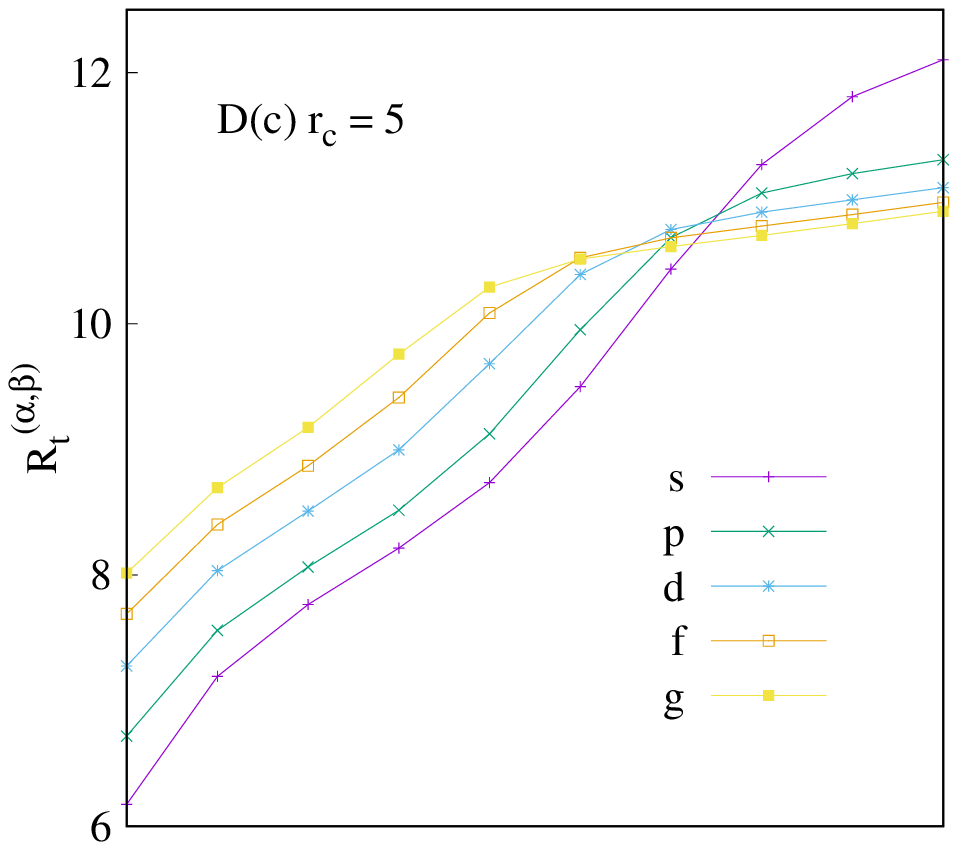}
\end{minipage}%
\\[15pt]
\begin{minipage}[c]{0.33\textwidth}\centering
\includegraphics[scale=0.4]{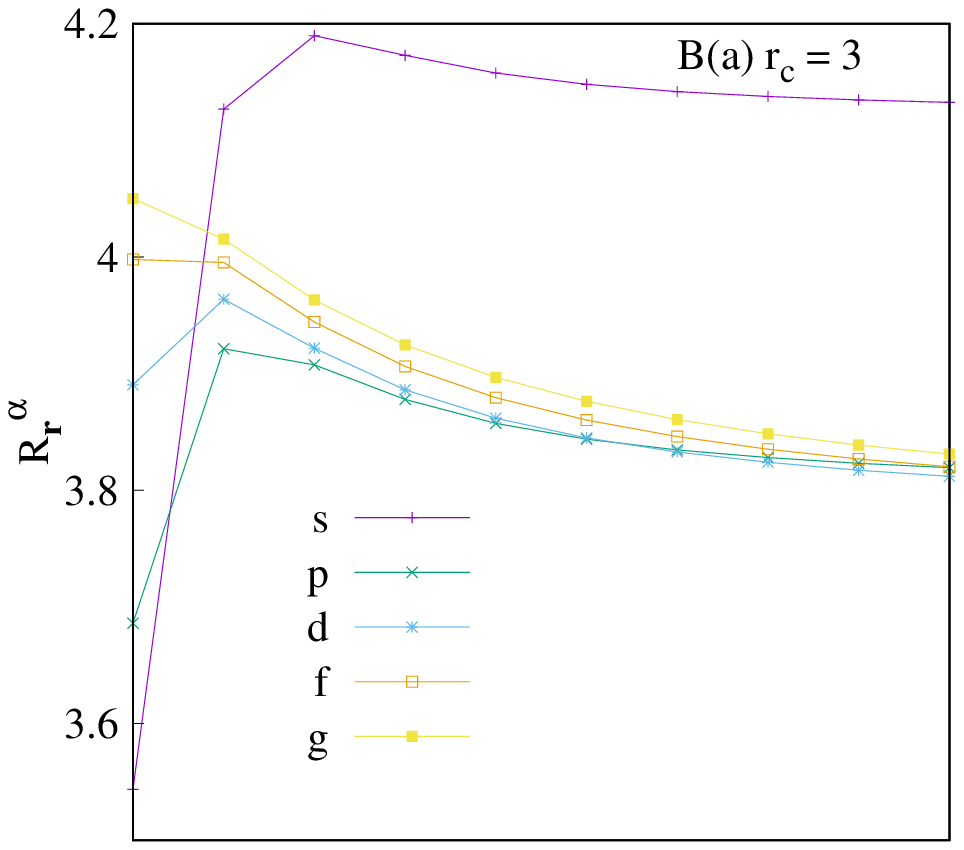}
\end{minipage}%
\begin{minipage}[c]{0.33\textwidth}\centering
\includegraphics[scale=0.4]{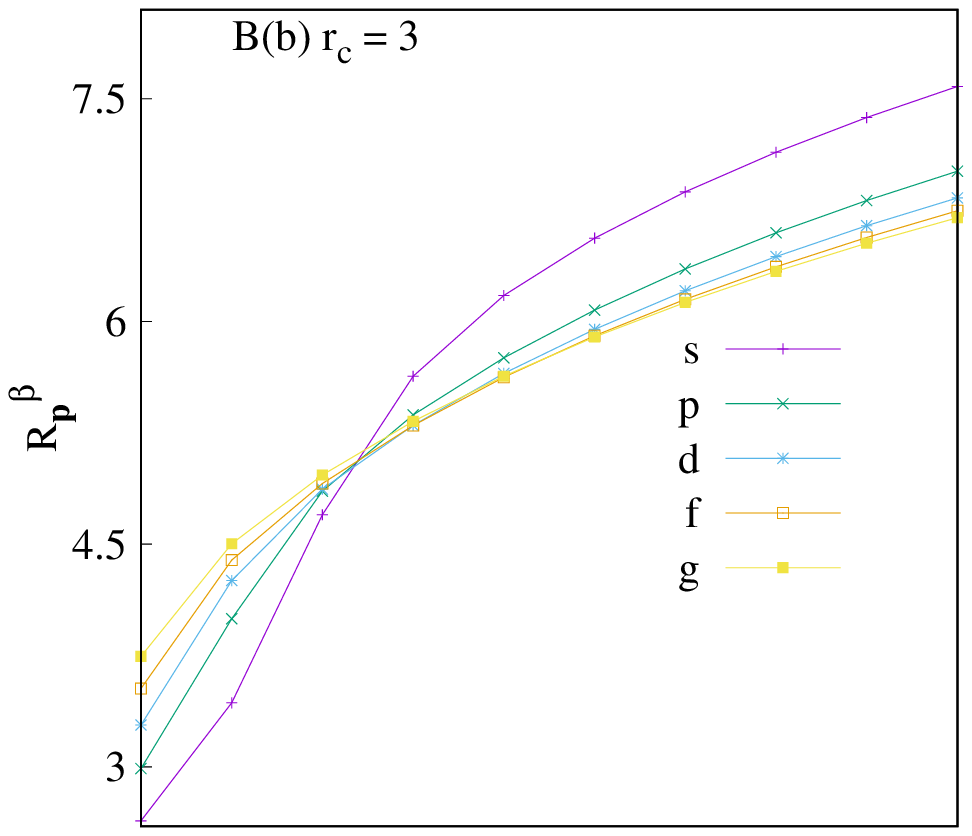}
\end{minipage}%
\begin{minipage}[c]{0.33\textwidth}\centering
\includegraphics[scale=0.4]{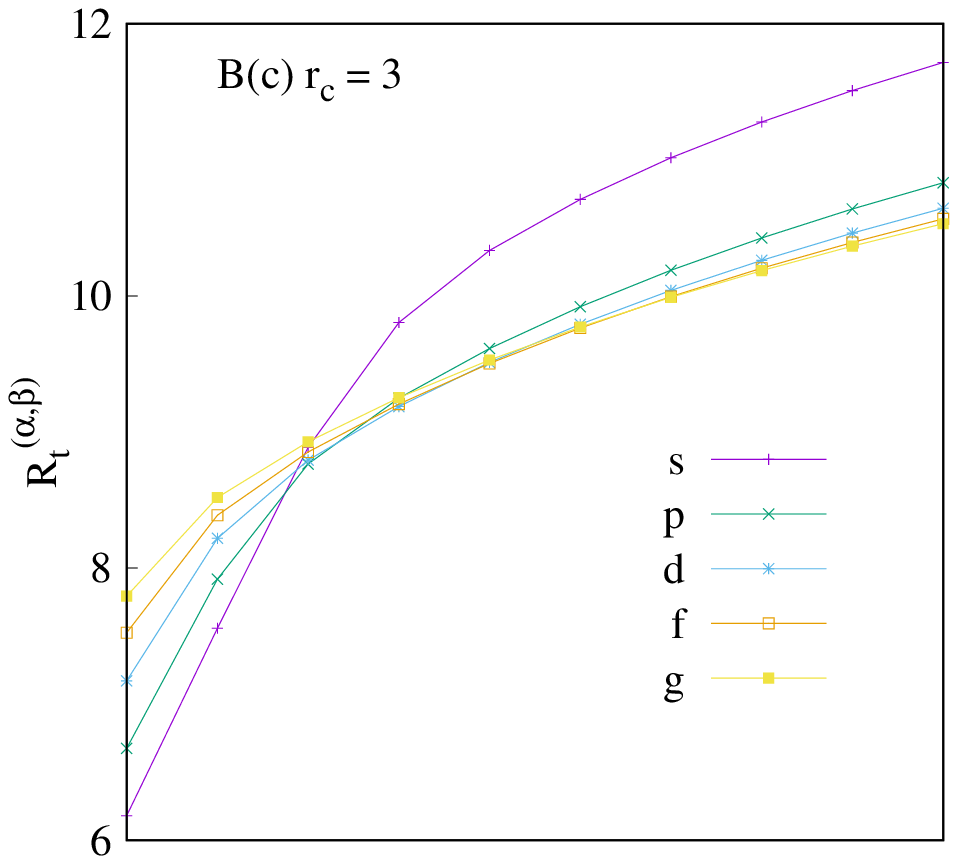}
\end{minipage}%
\\[15pt]
\begin{minipage}[c]{0.33\textwidth}\centering
\includegraphics[scale=0.44]{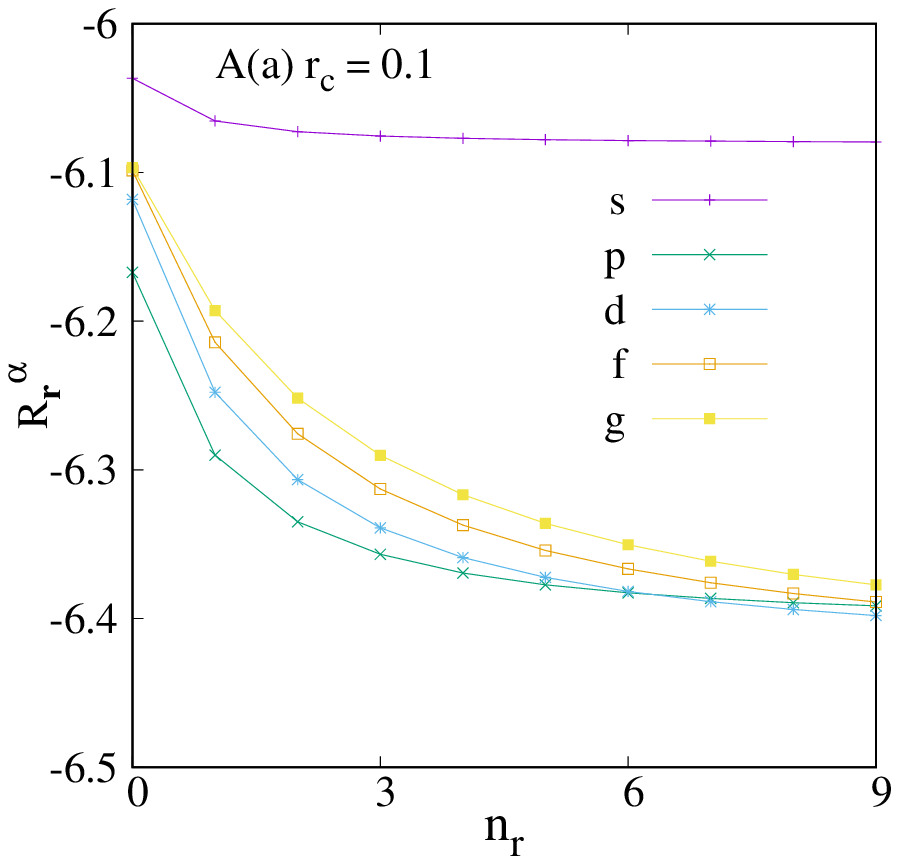}
\end{minipage}%
\begin{minipage}[c]{0.33\textwidth}\centering
\includegraphics[scale=0.44]{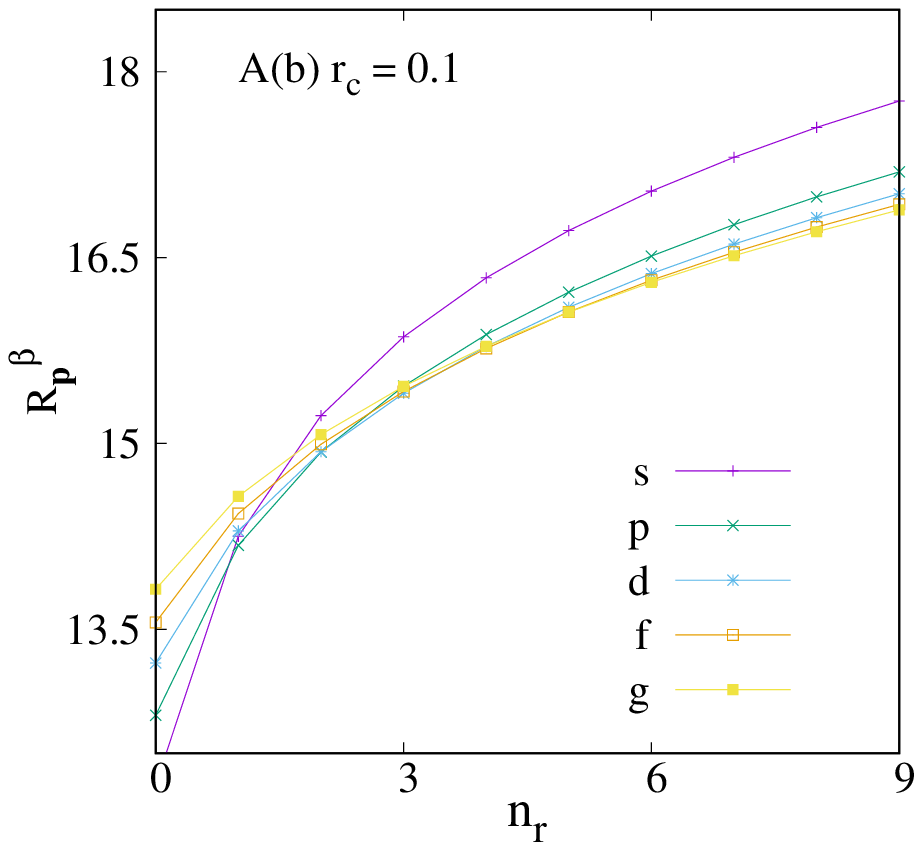}
\end{minipage}%
\begin{minipage}[c]{0.33\textwidth}\centering
\includegraphics[scale=0.44]{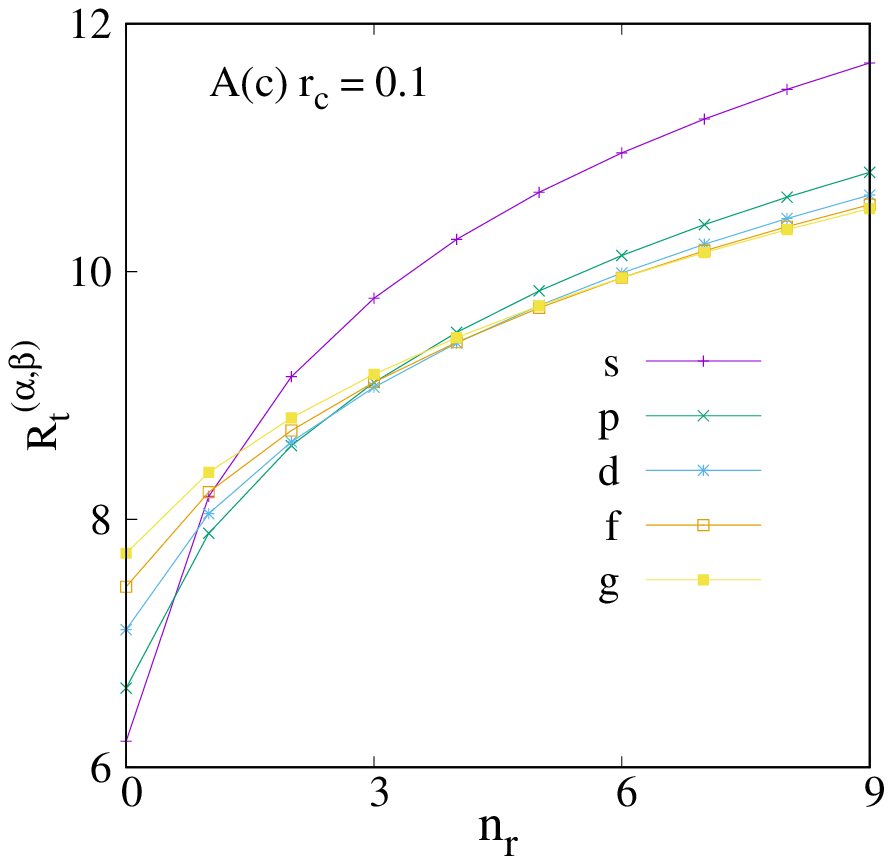}
\end{minipage}%
\caption{Plot of $R^{\alpha}_{\rvec}$ (a), $R^{\beta}_{\pvec}$ (b) and $R_{t}^{(\alpha, \beta)}$ (c) versus $n_{r}$ (at 
$\omega=1$) for $s,p,d,f,g$ states at four particular $r_{c}$'s of 3DCHO, namely, $0.1,3,5,\infty$ in panels (A)-(D). 
$R_{t}^{(\alpha,\beta)}$'s for all these states obey the lower bound given in Eq.~(57) \cite{mukherjee18a}. For more details, consult text.}
\end{figure}

Now we discuss variations of $\{C_{I_{\rvec}S_{\rvec}}^{(1)},C_{I_{\pvec}S_{\pvec}}^{(1)}\}$ and 
$\{C_{I_{\rvec}S_{\rvec}}^{(2)},C_{I_{\pvec}S_{\pvec}}^{(2)}\}$ with changes in $r_{c}$ for all the states mentioned above. Interestingly. 
in this occasion, the nature 
of alteration of $\{C_{I_{\rvec}S_{\rvec}}^{(1)},C_{I_{\pvec}S_{\pvec}}^{(1)}\}$ with $r_c$ varies from state to state. At a fixed 
$n_r$, both $C_{I_{\rvec}S_{\rvec}}^{(1)}, C_{I_{\pvec}S_{\pvec}}^{(1)}$ increase with $l$.
Further, at a fixed $l$, they elevate with growth in $n_r$. This behavior of $C_{I_{\rvec}S_{\rvec}}^{(1)}$ with $n_r$, is again 
connected to the nodal structure in $C_{ES}$ and $C_{ER}$. On the contrary, 
$C_{I_{\rvec}S_{\rvec}}^{(2)}$, $C_{I_{\pvec}S_{\pvec}}^{(2)}$ increase and decrease with growth in $r_c$ and finally approach to 
respective 3DQHO. This obviously conveys that, $C_{I_{\rvec}S_{\rvec}}^{(2)}$ is a descriptor of disequilibrium. At a particular 
$n_r$ both $C_{I_{\rvec}S_{\rvec}}^{(2)}$, $C_{I_{\pvec}S_{\pvec}}^{(2)}$ enhance with 
advancement in $l$. Also, at a certain $l$, they enhance with improvement of $n_r$. However, this pattern is in agreement with 
$C_{ES}$ and $C_{ER}$. In case of 3DCHO, both 
$\{C_{I_{\rvec}S_{\rvec}}^{(1)},~C_{I_{\pvec}S_{\pvec}}^{(1)}\}$ and $\{C_{I_{\rvec}S_{\rvec}}^{(2)},
C_{I_{\pvec}S_{\pvec}}^{(2)}\}$ are unable to explain the delocalization of the system from PISB to 3DQHO through the increase in $r_c$. 
    
\begingroup           
\begin{table}
\caption{ $S_{\rvec}, S_{\pvec}$ and $S_{t}$ for lowest three $l$ (having $n_r=1,2$) states in 3DCHO at eight selected values of 
$r_c$. $S_t$ for all these states obey the lower bound given in Eq.~(53) \cite{mukherjee18a}. }
\centering
\begin{tabular}{>{\scriptsize}l>{\scriptsize}l>{\scriptsize}l>{\scriptsize}l>{\scriptsize}l>{\scriptsize}
l>{\scriptsize}l>{\scriptsize}l<{\scriptsize}}
\hline
$r_c$  &    $S_{\rvec}$     & $S_{\pvec}$  &  $S_{t}$  &  
$r_c$  &    $S_{\rvec}$     & $S_{\pvec}$  &  $S_{t}$  \\
\cline{1-4} \cline{5-8}
\multicolumn{4}{c}{$1s$}    &      \multicolumn{4}{c}{$2s$}    \\
\cline{1-4} \cline{5-8}
0.1      & $-$6.232173222  & 12.8494    & 6.6172       & 0.1       & $-$6.4460987687 & 14.6389      &8.1928     \\
0.2      & $-$4.152747179  & 10.7700    & 6.6172       & 0.2       & $-$4.3666534417 & 12.5595      &8.1928     \\
0.5      & $-$1.404504328  & 8.0214     & 6.6168       & 0.5       & $-$1.6176276192 & 9.8106       &8.1929     \\
1.0      & 0.6652222004    & 5.9458     & 6.6110       & 1.0       &  0.4641636149 & 7.731        &8.195      \\
5.0      & 3.2170947394    & 3.217094 & 6.434189 & 5.0       &  4.150729546 & 4.1510       &8.3017     \\
8.0      & 3.2170948239    & 3.217094 & 6.434189 & 8.0       &  4.150745547 & 4.15074      &8.30148    \\
$\infty$ & 3.2170948239    & 3.217094 & 6.434189 &$\infty$   &  4.150745547 & 4.15074     &8.30148\\
\cline{1-4} \cline{5-8}
\multicolumn{4}{c}{$1p$}    &      \multicolumn{4}{c}{$2p$}    \\
\cline{1-4} \cline{5-8}
0.1    &   $-$6.38738206 & 13.418       & 7.030       & 0.1      & $-$6.651966568   & 14.7283        &  8.0763         \\
0.2    &   $-$4.30794705 & 11.338       & 7.030       & 0.2      & $-$4.572523919   & 12.6489        &  8.0763         \\
0.5    &   $-$1.55934019 & 8.5894       & 7.0300     & 0.5      & $-$1.823606736   & 9.9000         &  8.0763         \\
1.0    &   0.51599338    & 6.5114       & 7.0273      & 1.0      &  0.256528223     & 7.82132        &  8.07784        \\
5.0    &   3.4874566574  & 3.4875       & 6.9750    & 5.0      &  4.1477548396    & 4.1483         &  8.2960         \\
8.0    &   3.4874576660  & 3.487457     & 6.974915 & 8.0      &  4.14786196159   & 4.147861       &  8.295723       \\
$\infty$ & 3.4874576660  & 3.487457     & 6.974915 &$\infty$  &  4.14786196159   & 4.147861  &  8.295723   \\
\cline{1-4} \cline{5-8}
\multicolumn{4}{c}{$1d$}    &      \multicolumn{4}{c}{$2d$}    \\
\cline{1-4} \cline{5-8}
0.1    &  $-$6.3553068427 & 14.0035      & 7.6481       & 0.1     & $-$6.5939939435     & 15.0676        & 8.4736       \\
0.2    &  $-$4.2758683878 & 11.9241      & 7.6481       & 0.2     & $-$4.5145520348     & 12.9881         & 8.4736        \\
0.5    &  $-$1.52712156  & 9.1753       & 7.6481       & 0.5     & $-$1.7656649279     & 10.2393        & 8.4736       \\
1.0    &  0.5503764295    & 7.0964       & 7.6467       & 1.0     &  0.3140078818     & 8.1605         & 8.4745       \\
5.0    &  3.8426303929    & 3.8427   & 7.6853     & 5.0     &  4.3885945973     & 4.3909         & 8.7794       \\
8.0    &  3.8426381378    & 3.842638 & 7.685276   & 8.0     &  4.389113529281   & 4.389113       & 8.778227      \\
$\infty$& 3.8426381378    & 3.842638 & 7.685276   &$\infty$ &  4.389113529281   & 4.389113 & 8.778227 \\
\hline
\end{tabular}
\end{table}
\endgroup

Finally, we mention a few words regarding the behavioral pattern in our last complexity measure, \emph{viz.,} 
$C_{I_{\rvec}R_{\rvec}}^{(1)},C_{I_{\pvec}R_{\pvec}}^{(1)}$, $C_{I_{\rvec}R_{\rvec}}^{(2)},C_{I_{\pvec}R_{\pvec}}^{(2)}$, with 
alterations in $r_{c}$. It is found that, $C_{I_{\rvec}R_{\rvec}}^{(1)}$ progresses slowly with $r_c$ to reach respective 3DQHO values, 
implying that, 
$C_{I_{\rvec}R_{\rvec}}^{(1)}$ is an order parameter. At a suitable $n_r$, this quantity advances with $l$. In a parallel way, at 
constant $l$, $C_{I_{\rvec}R_{\rvec}}^{(1)}$ accumulates with $n_r$. This again means that with increase in number of nodes, 
the system moves towards order. Besides this, it can be concluded that, for node-less (like $1s,1p,1d,1f,1g$) states,  
$C_{I_{\pvec}R_{\pvec}}^{(1)}$ diminishes with progression in $r_c$. But for states having one radial node (such as $2s,2p,2d$), there 
appears a maximum in $C_{I_{\pvec}R_{\pvec}}^{(1)}$, which gets right shifted with increase in $l$. It can be said that, 
$C_{I_{\rvec}R_{\rvec}}^{(2)},C_{I_{\pvec}R_{\pvec}}^{(2)}$, accelerate and decelerate respectively with growth in 
$r_c$. Hence, $C^{(2)}_{IR}$ is an indicator of order. Like $C^{(1)}_{IR}$, both $C_{I_{\rvec}R_{\rvec}}^{(2)}$, 
$~C_{I_{\pvec}R_{\pvec}}^{(2)}$ enhance with emergence of $n_r$ and $l$ quantum numbers. A closer investigation conveys 
that, neither $C^{(1)}_{IR}$ nor $C^{(2)}_{IR}$ can explain the transformation of system from PISB to 3DQHO via 3DCHO. 
It is hoped that, this study would be useful for future references and would inspires further work along this direction. 

\begin{figure}                      
\begin{minipage}[c]{0.32\textwidth}\centering
\includegraphics[scale=0.42]{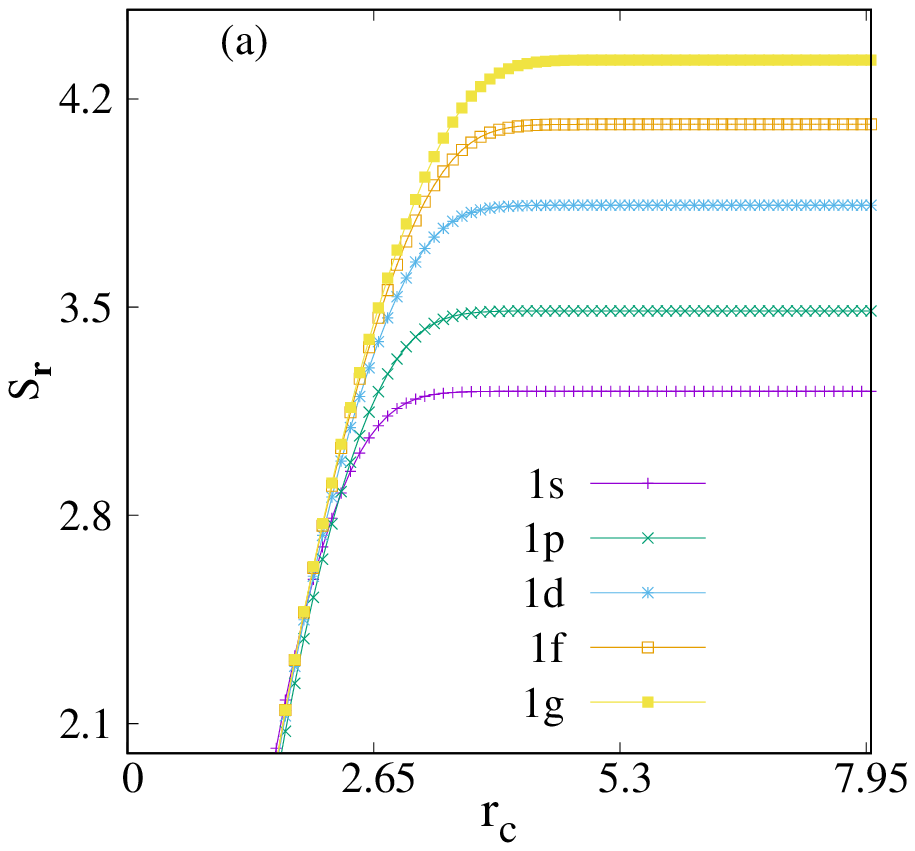}
\end{minipage}%
\hspace{0.02in}
\begin{minipage}[c]{0.32\textwidth}\centering
\includegraphics[scale=0.42]{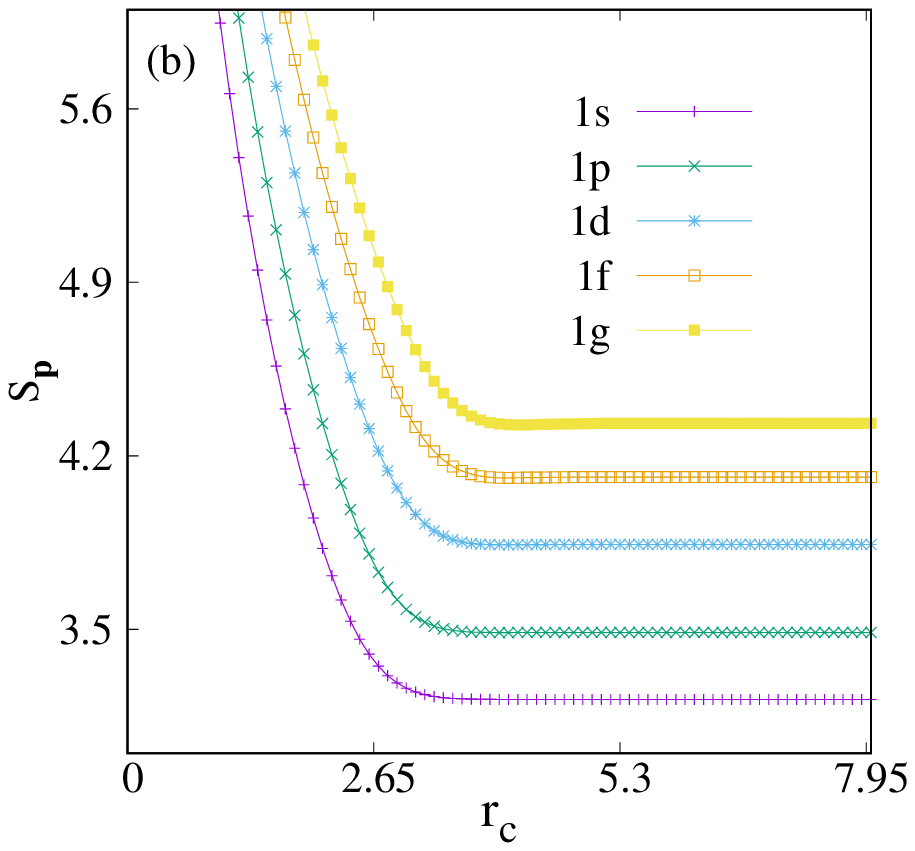}
\end{minipage}%
\hspace{0.02in}
\begin{minipage}[c]{0.32\textwidth}\centering
\includegraphics[scale=0.42]{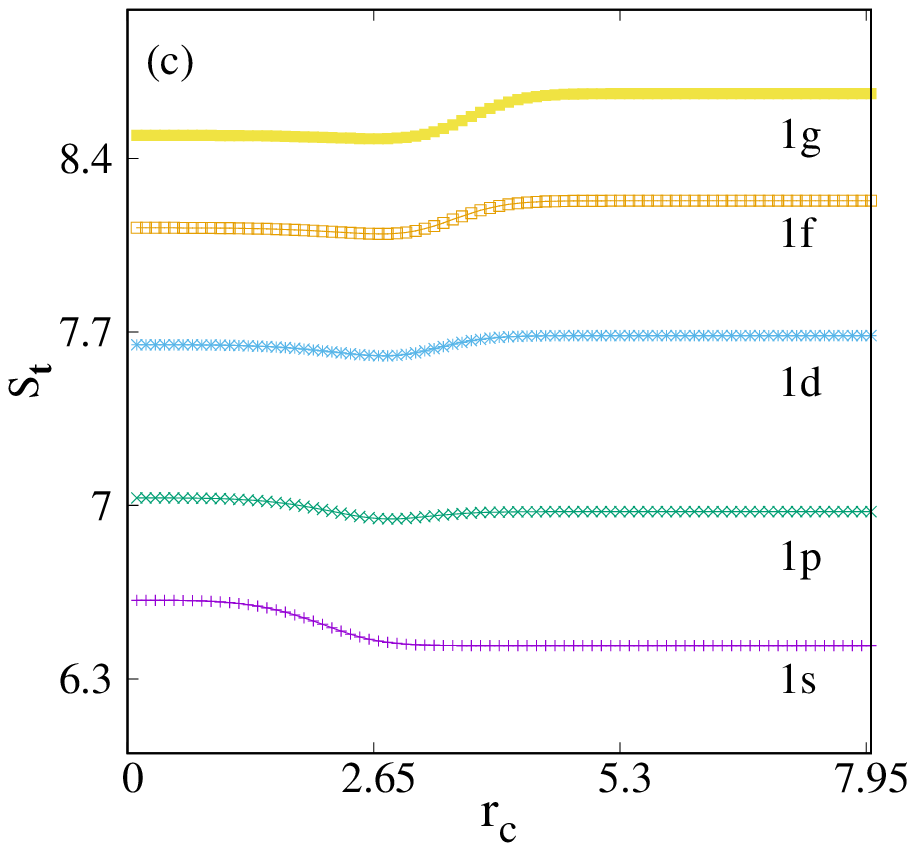}
\end{minipage}%
\caption{Plots of $S_{\rvec}$, $S_{\pvec}$, $S_{t}$ against $r_c$ for first five circular states of 3DCHO in panels (a), (b), (c) 
respectively \cite{mukherjee18a}. See text for details.}
\end{figure}

\subsection{Relative Information}
\subsubsection{1D Quantum harmonic oscillator (1DQHO)}
At first we would like to explore IR in model 1DQHO in $x, p$ spaces. The 
underlying potential is characterized by the expression: $v(x)= \frac{1}{2} \omega^2 x^2$ (mass $m$ is set to unity throughout), 
where $\omega$ signifies angular frequency. The normalized $x$-space wave function is expressed as ($H_n (x)$ refers to Hermite 
polynomial),
\begin{equation}
\psi_{n}(x)=\left(\frac{\omega}{\sqrt{2}\pi}\right)^\frac{1}{4}\frac{1}{\sqrt{2^{n}n!}} \ H_{n}\left(\frac{\sqrt{\omega}}
{2^{\frac{1}{4}}}x\right)e^{-\frac{\omega}{2\sqrt{2}}x^{2}}.
\end{equation}
Choosing $n=0$ as reference, $\frac{\sqrt{\omega}}{2^{\frac{1}{4}}}x=y$, and using definition of IR in Eq.~(73), one gets, 
\begin{equation}
\mathrm{IR}_{x}=\frac{\omega}{\sqrt{2}} \ \frac{1}{2^{n-1}n!}\int_{0}^{\infty} \ H_{n}^{2}(y) \ 
e^{-y^{2}}\left[\frac{\psi^{\prime}_{n}(y)}
{\psi_{n}(y)}-\frac{\psi^{\prime}_{0}(y)}{\psi_{0}(y)}\right]^{2}\mathrm{d}y. \setlength{\arrayrulewidth}{1mm}
\setlength{\tabcolsep}{18pt}
\renewcommand{\arraystretch}{1.5}
\end{equation}
Here the suffix ``x" denotes a position-space quantity. Now use of recurrence relation $H_{n}^{\prime}(y)=2nH_{n-1}(y)$, in 
conjunction with orthonormality condition of Hermite polynomial, 
$\int_{0}^{\infty}e^{-y^{2}}H_{m}(y)H_{k}(y) \mathrm{d}y=2^{m-1}m!\sqrt{\pi}\delta_{mk}$ ($\delta_{mn}=1$ when $m=k$, and $0$ 
otherwise) produces,  
\begin{equation}
\mathrm{IR}_{x} =4n^{2}\frac{\omega}{\sqrt{2}} \ \frac{1}{2^{n-1}n!}\int_{0}^{\infty}\left[H_{n-1}(y)\right]^{2}
e^{-y^{2}}\mathrm{d}y =4\sqrt{2} \ \omega n. 
\end{equation}
Thus Eq.~(82) suggests that, IR$_{x}$ in $n$th state may be obtained from a knowledge of $(n-1)$th-state wave function. Clearly, 
it increases linearly with state index $n$, with a positive slope of $4\sqrt{2}\omega$. This is in agreement with the fact that 
in this system, localization as well as fluctuation enhance with $\omega$. So IR$_{x}$ result simply complements this. 

\begin{figure}                                            
\begin{minipage}[c]{0.33\textwidth}\centering
\includegraphics[scale=0.4]{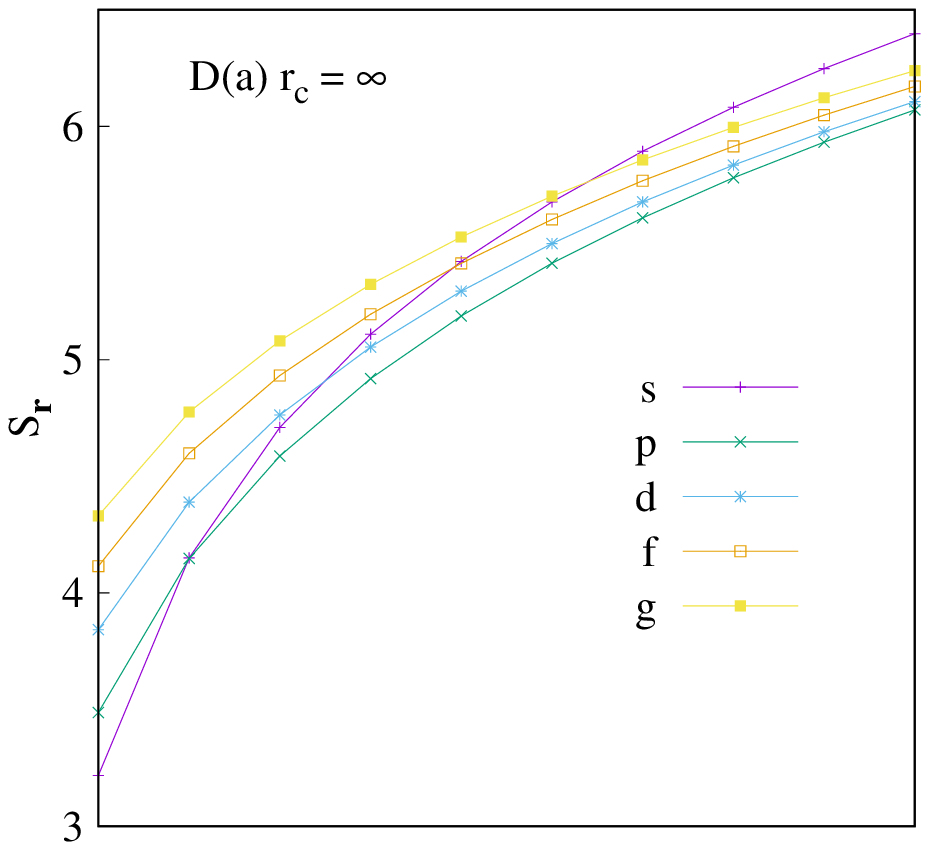}
\end{minipage}%
\begin{minipage}[c]{0.33\textwidth}\centering
\includegraphics[scale=0.4]{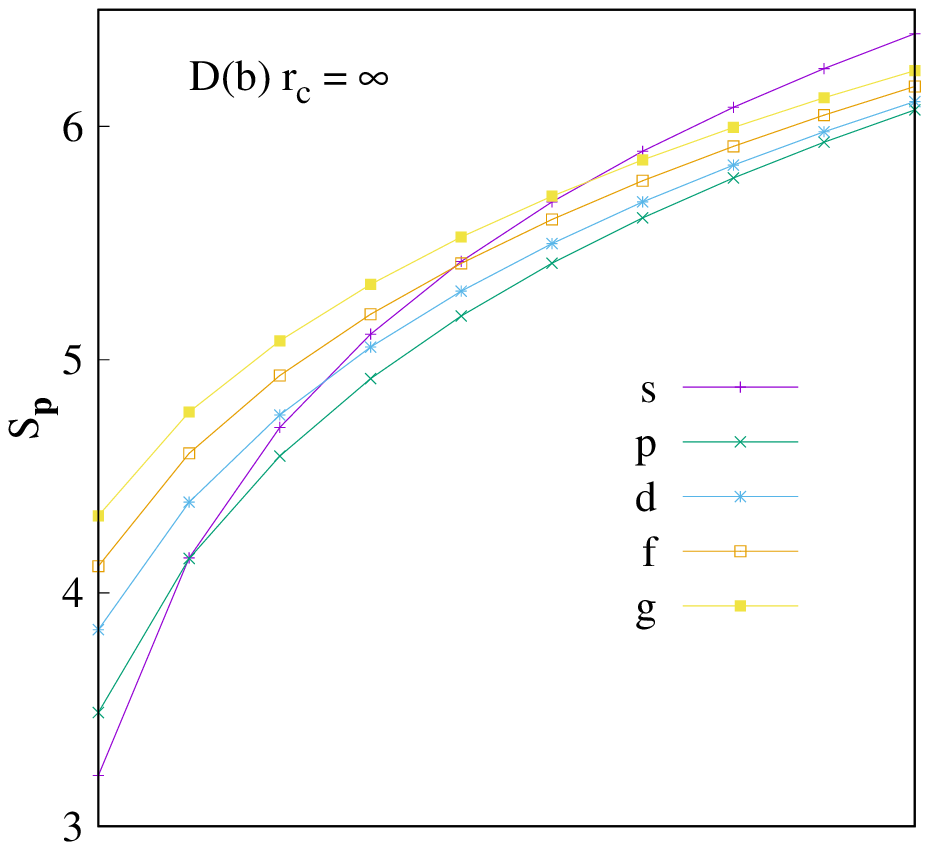}
\end{minipage}%
\begin{minipage}[c]{0.33\textwidth}\centering
\includegraphics[scale=0.4]{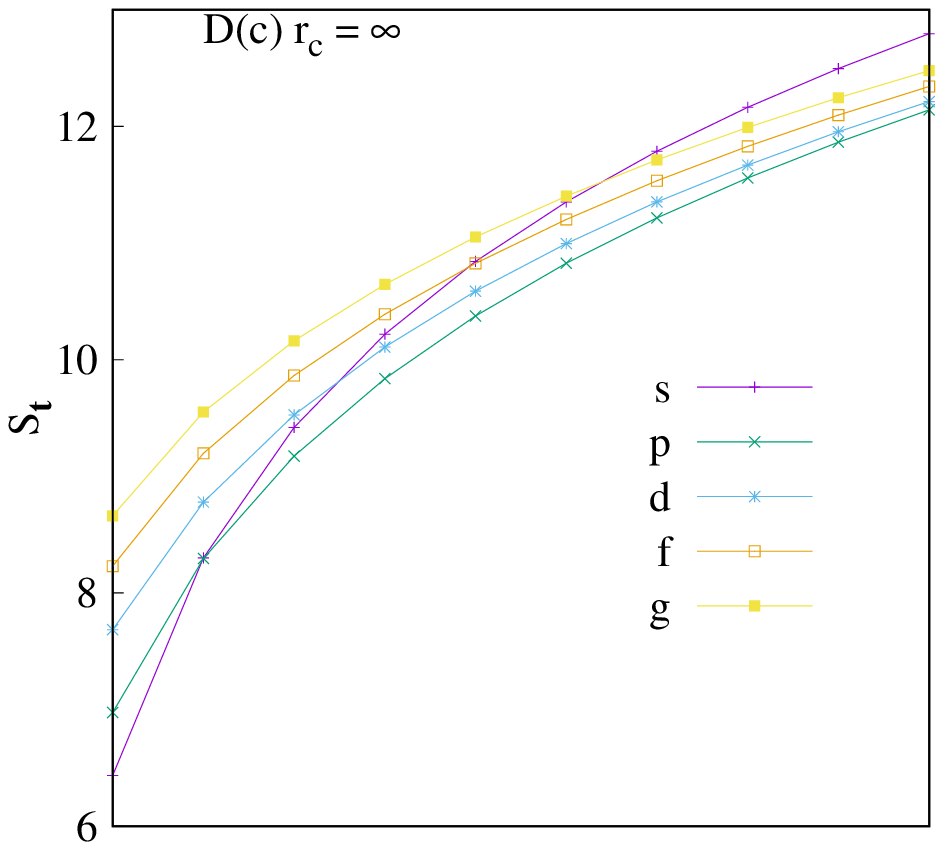}
\end{minipage}%
\\[15pt]
\begin{minipage}[c]{0.33\textwidth}\centering
\includegraphics[scale=0.4]{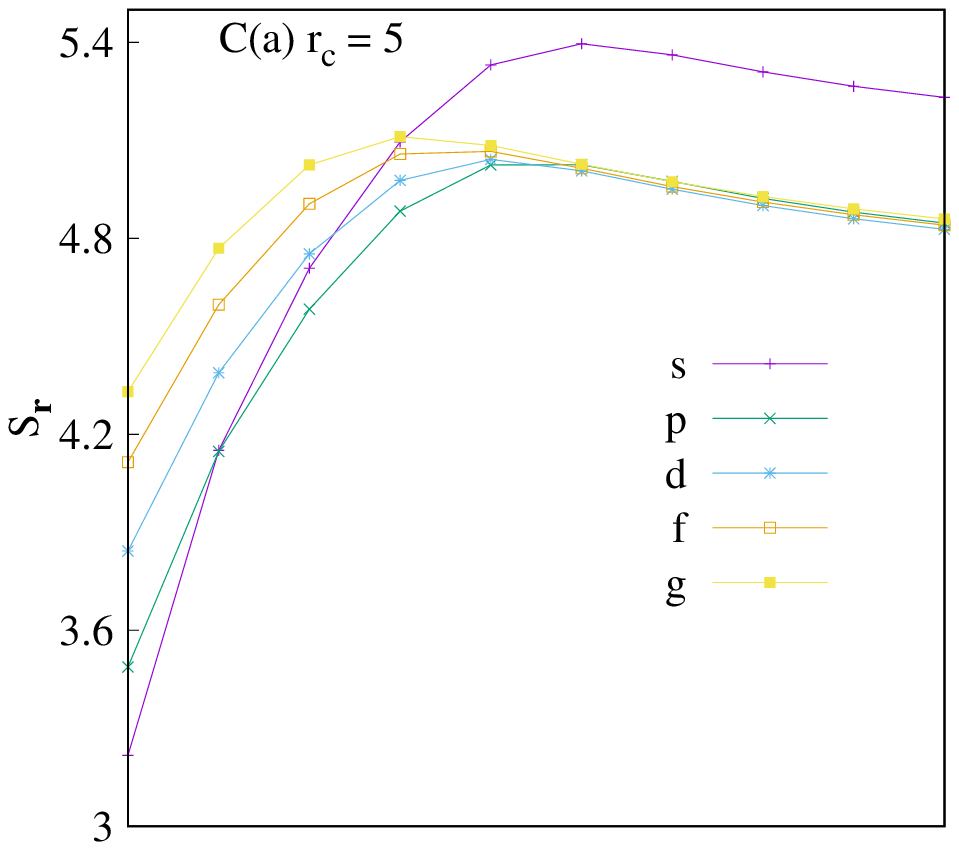}
\end{minipage}
\begin{minipage}[c]{0.33\textwidth}\centering
\includegraphics[scale=0.4]{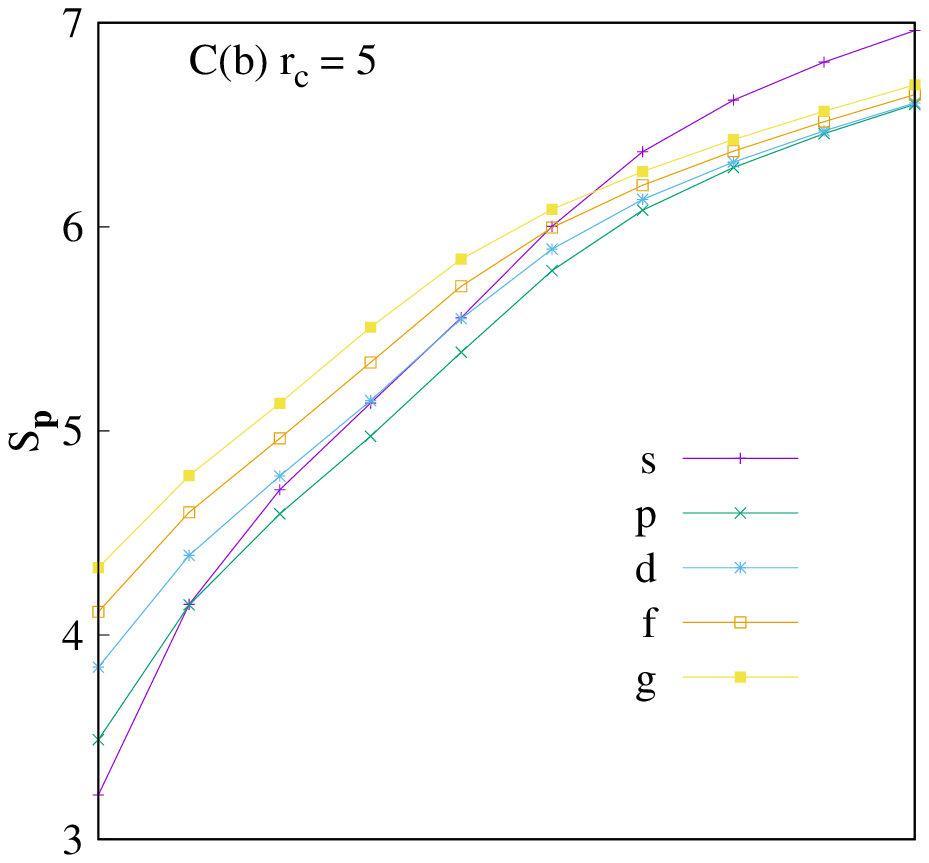}
\end{minipage}%
\begin{minipage}[c]{0.33\textwidth}\centering
\includegraphics[scale=0.4]{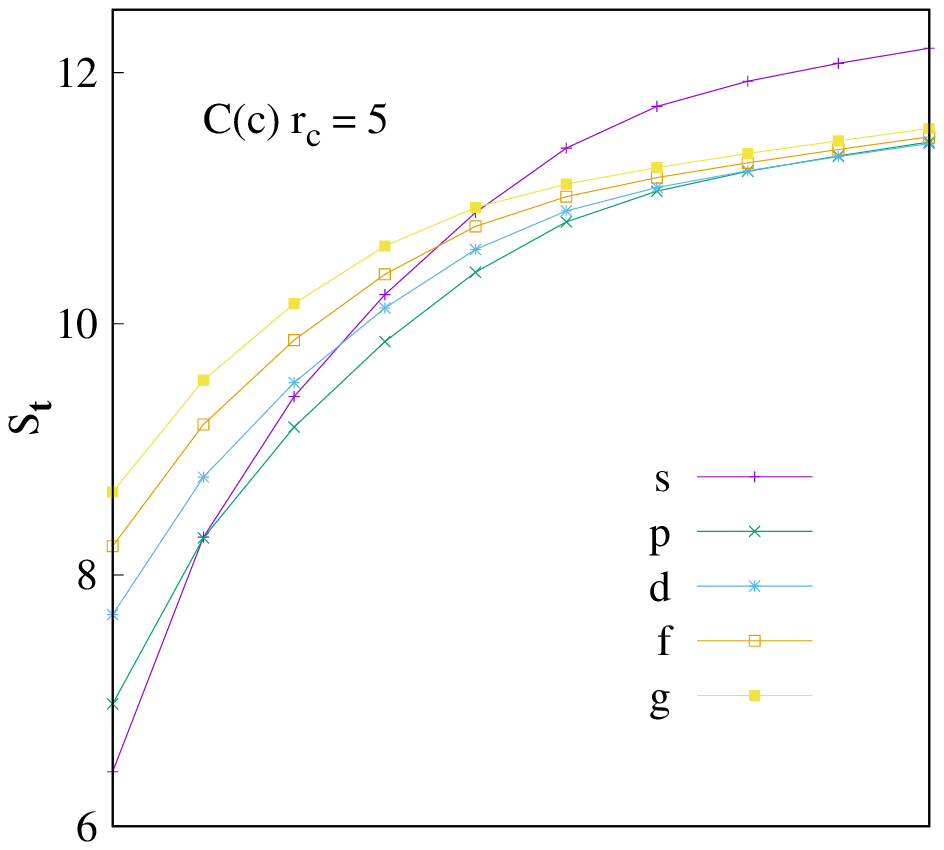}
\end{minipage}%
\\[15pt]
\begin{minipage}[c]{0.33\textwidth}\centering
\includegraphics[scale=0.4]{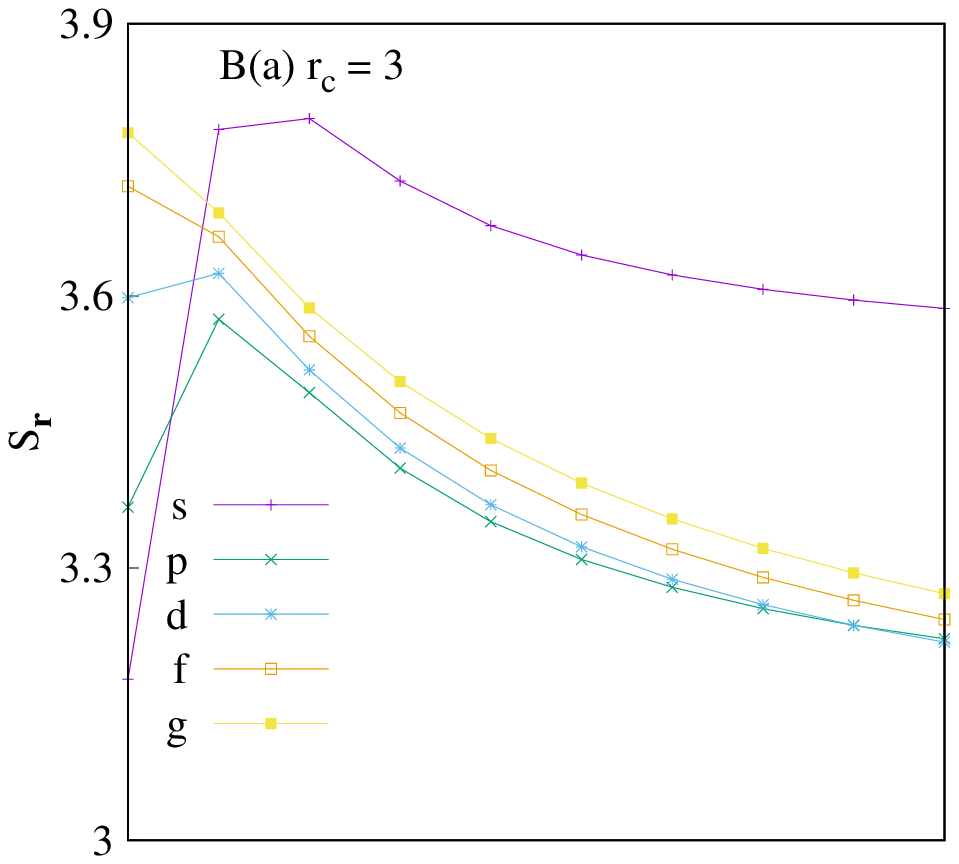}
\end{minipage}
\begin{minipage}[c]{0.33\textwidth}\centering
\includegraphics[scale=0.4]{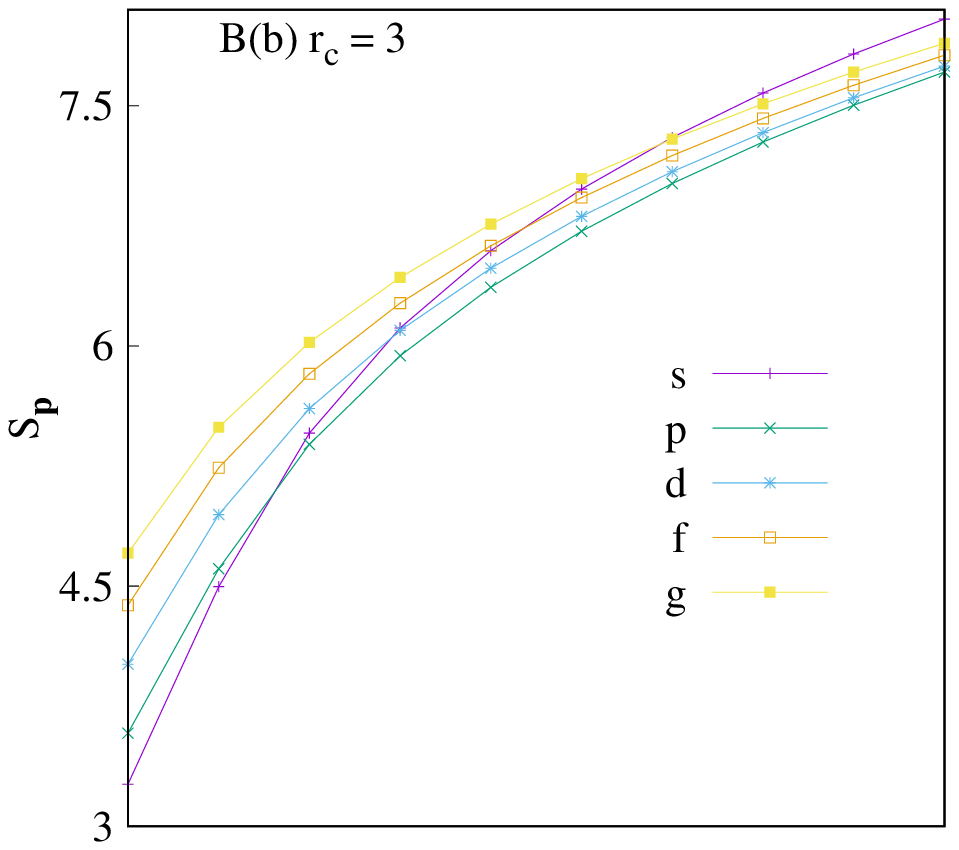}
\end{minipage}%
\begin{minipage}[c]{0.33\textwidth}\centering
\includegraphics[scale=0.4]{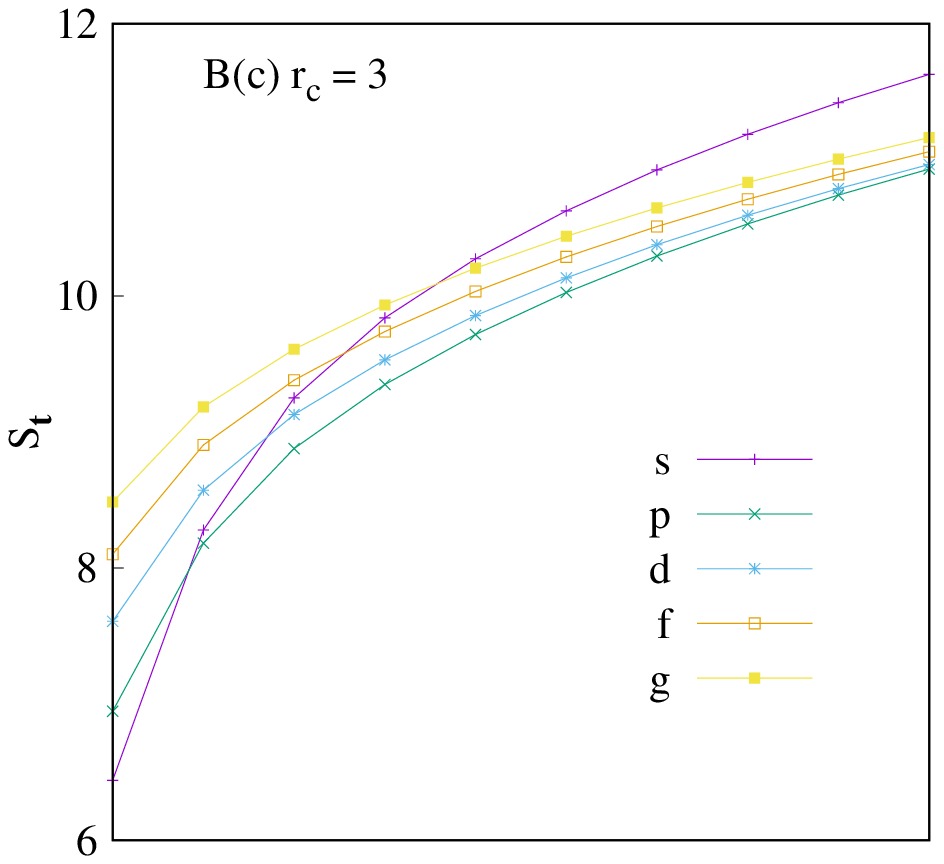}
\end{minipage}%
\\[15pt]
\begin{minipage}[c]{0.33\textwidth}\centering
\includegraphics[scale=0.44]{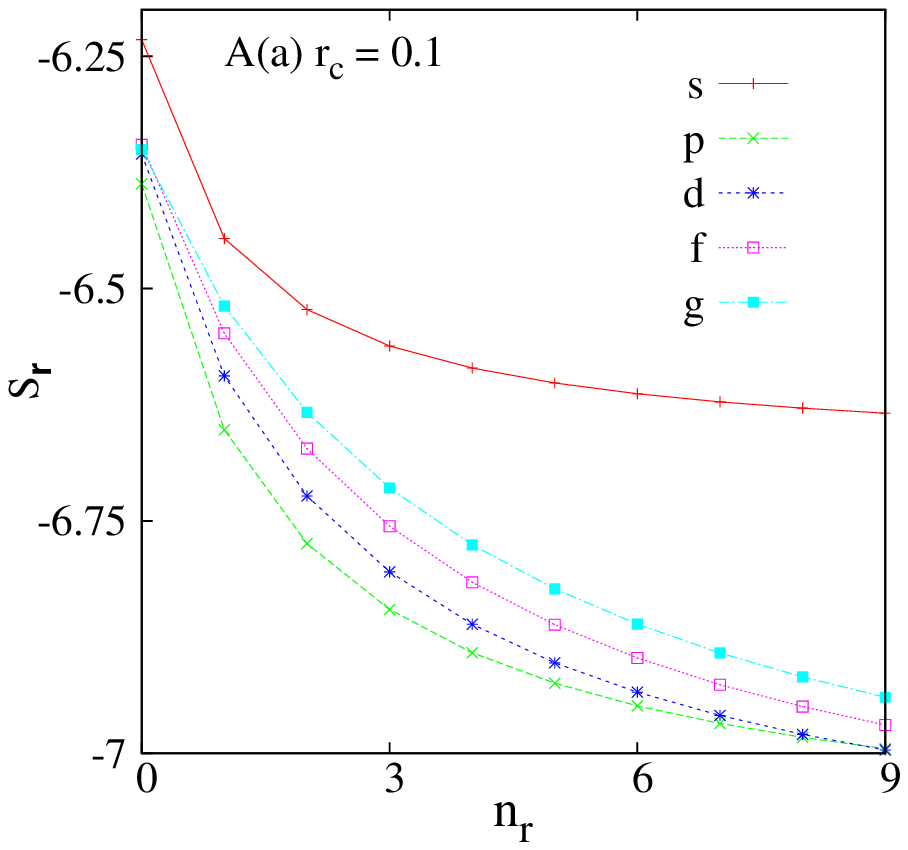}
\end{minipage}%
\begin{minipage}[c]{0.33\textwidth}\centering
\includegraphics[scale=0.44]{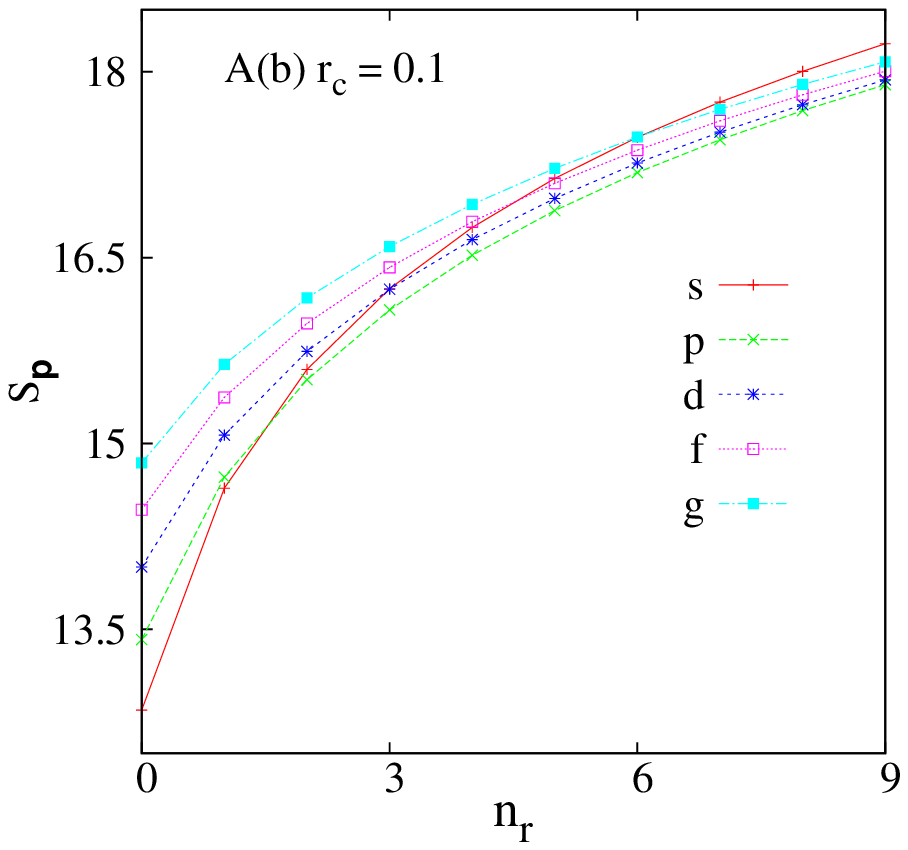}
\end{minipage}%
\begin{minipage}[c]{0.33\textwidth}\centering
\includegraphics[scale=0.44]{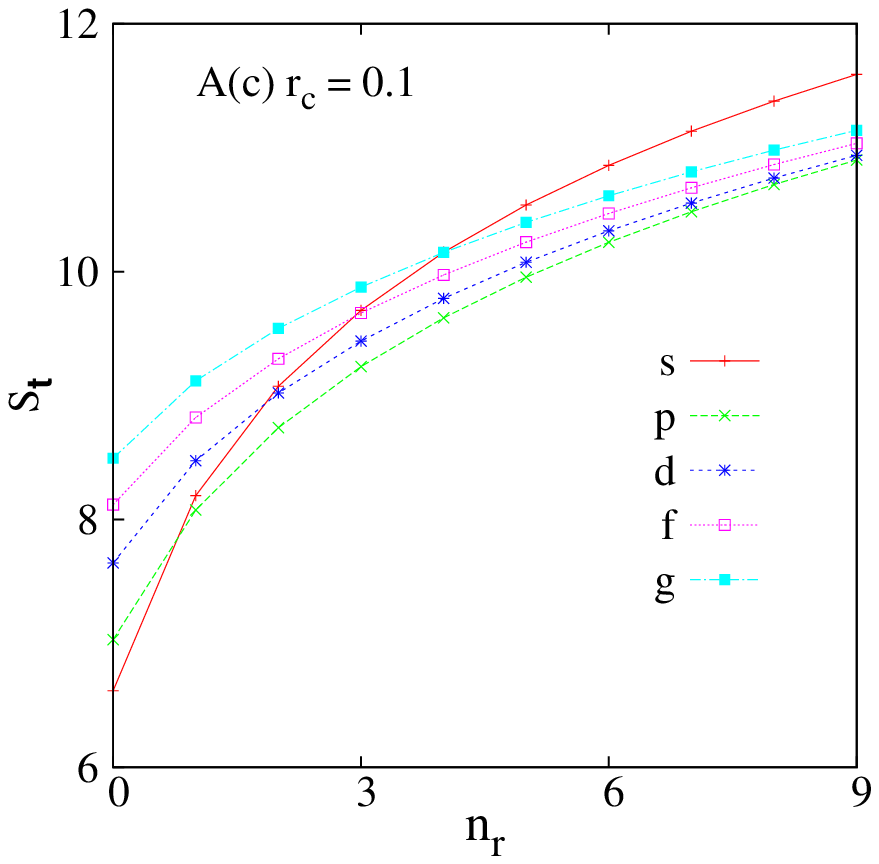}
\end{minipage}%
\caption{Plot of $S_{\rvec}$ (a), $S_{\pvec}$ (b) and $S_{t}$ (c) versus $n_{r}$  (at $\omega=1$) for $s,p,d,f,g$ states at four 
particular $r_{c}$'s of 3DCHO, namely, $0.1,3,5,\infty$ in panels (A)-(D). $S_{t}$'s for all these states obey the lower bound 
given in Eq.~(53) \cite{mukherjee18a}. For more details, consult text.}
\end{figure}

Now we move on to $p$ space, where the normalized wave function is in the form of, 
\begin{equation}
\psi_{n}(p)=\left(\frac{\sqrt{2}}{\omega \pi}\right)^\frac{1}{4}\frac{1}{\sqrt{2^{n}n!}} \ H_{n}\left(\frac{2^{\frac{1}{4}}}
{\sqrt{\omega}}p\right)\  e^{-\frac{1}{\sqrt{2}\omega} \ p^{2}}. 
\end{equation}   
Again, considering $n=0$ state as reference, setting $\frac{2^{\frac{1}{4}}}{\sqrt{\omega}}p=g$, using the recurrence relation 
$H_{n}^{\prime}(g)=2nH_{n-1}(g)$ and invoking orthonormality condition of Hermite polynomial, one can derive
the following expression for IR$_{p}$ after some straightforward algebra ($p$ subscript indicates $p$-space quantity), namely, 
\begin{equation}
\mathrm{IR}_{p}=4n^{2}\sqrt{\frac{2}{\pi}} \ \frac{1}{2^{n}n!}\frac{1}{\omega}\int_{0}^{\infty}[H_{n-1}(g)]^{2}e^{-g^{2}}
\mathrm{d}g =\frac{8\sqrt{2}}{\omega} \ n. 
\end{equation}
Thus, similar to IR$_{x}$, here also IR$_{p}$ of a particular oscillator state can be recovered from the wave function of adjacent 
lower state. Equation~(84) tells that, progress of IR$_{p}$ with $n$ is again linear like its $x$-space counterpart, slope of 
the straight line in this case being $\frac{8\sqrt{2}}{\omega}$. It is inversely proportional to $\omega$ in accordance with the 
fact that, with increase in oscillation, localization as well as fluctuation predominates. At the special value of $\omega=\sqrt{2}$, 
IR$_{x}$, IR$_{p}$ become equal ($8n$). Apart from that, like the total energy difference in a 1DQHO, $\Delta 
\mathrm{E_{n}}(=\mathrm{E_{n+1}}-\mathrm{E_{n}})$, the difference of IR between two successive states also remains constant, i.e.,
$\Delta (\mathrm{IR}_{x})= \mathrm{IR}_{x}(n+1)-\mathrm{IR}_{x}(n)=4\sqrt{2}\omega$, and $\Delta (\mathrm{IR}_{p})=\mathrm{IR}_{p}
(n+1)-\mathrm{IR}_{p}(n)=\frac{8\sqrt{2}}{\omega}$.  

\begingroup           
\begin{table}
\caption{$E_{\rvec}, E_{\pvec}, E_{t}$ for $1s,2s,1p,2p,1d,2d$ states in 3DCHO at eight selected $r_c$ \cite{mukherjee18a}.}
\centering
\begin{tabular}{>{\tiny}l|>{\tiny}l>{\tiny}l>{\tiny}l|>{\tiny}l|>{\tiny}l>{\tiny}l>{\tiny}l<{\tiny}}
\hline
$r_c$  &    $E_{\rvec}$     & $E_{\pvec}$  &  $E_{t}$  &  
$r_c$  &    $E_{\rvec}$     & $E_{\pvec}$  &  $E_{t}$  \\
\cline{1-4} \cline{5-8}
\multicolumn{4}{c}{$1s$}    &      \multicolumn{4}{c}{$2s$}    \\
\cline{1-4} \cline{5-8}
0.1      & 672.0719164	& 0.00000398	& 0.0026791    & 0.1       & 1453.1909702895	  &    0.00000057  &  0.00082825 \\
0.2      & 84.01080088	& 0.00003189	& 0.0026791    & 0.2       & 181.6485572148	  &    0.00000455  &  0.00082825 \\
0.5      & 5.3814002356	& 0.00049808	& 0.00268041 & 0.5       & 11.6246913489	  &    0.00007124  &  0.00082821  \\
1.0      & 0.6818097823	& 0.00396012	& 0.00270005 & 1.0       & 1.4515093698	  &    0.00057014  &  0.00082757  \\
5.0      & 0.0634936361	& 0.06349340	& 0.00403142 & 5.0       & 0.0406758398	  &    0.04064829  &  0.00165340  \\
8.0      & 0.0634936347	& 0.06349363	& 0.00403144 & 8.0       & 0.040675749	  &    0.04067560  &  0.00165430  \\
$\infty$ & 0.0634936347	& 0.06349363	& 0.00403144 &$\infty$   & 0.0406756097	  &    0.04067560  &  0.00165450  \\
\cline{1-4} \cline{5-8}
\multicolumn{4}{c}{$1p$}    &      \multicolumn{4}{c}{$2p$}    \\
\cline{1-4} \cline{5-8}
0.1      & 803.22700816   &	0.00000227	& 0.00182938 & 0.1      & 1454.974575234  & 0.00000058 & 0.00085788  \\
0.2      & 100.40423515   &	0.00001822	& 0.00182940 & 0.2      & 181.871793182   & 0.00000471 & 0.00085788   \\
0.5      & 6.4281046025   &	0.00028463	& 0.00182965 & 0.5      & 11.6397197874	  & 0.00007369 & 0.00085777   \\
1.0      & 0.8078468658   &	0.00226961	& 0.00183350 & 1.0      & 1.454810759	  & 0.00058839 & 0.00085600   \\
5.0      & 0.0476202385   &	0.04761162	& 0.00226727 & 5.0      & 0.0324128865	  & 0.03237583 & 0.00104939   \\
8.0      & 0.047620224    &	0.04762022	& 0.00226768 & 8.0      & 0.032411515	  & 0.03241151 & 0.00105050   \\
$\infty$ & 0.047620224    &	0.04762022	& 0.00226768 &$\infty$  & 0.032411515	  & 0.03241151 & 0.00105050   \\
\cline{1-4} \cline{5-8}
\multicolumn{4}{c}{$1d$}    &      \multicolumn{4}{c}{$2d$}    \\
\cline{1-4} \cline{5-8}
0.1    &  851.25726418	& 0.00000137 & 0.00117301 & 0.1     & 1368.86082394   & 0.00000047	& 0.00064967 \\
0.2    &  106.40757650	& 0.00001102 & 0.00117301 & 0.2     & 171.107627763   & 0.00000379	& 0.00064967  \\
0.5    &  6.8111728555	& 0.00017222 & 0.00117305 & 0.5     & 10.9509523492   & 0.00005931	& 0.00064960  \\
1.0    &  0.8535077417	& 0.00137508 & 0.00117364 & 1.0     & 1.3689869817    & 0.00047372	& 0.00064852  \\
5.0    &  0.0357152613	& 0.03571436 & 0.00127554 & 5.0     & 0.0249812774    & 0.02492595	& 0.00062268  \\
8.0    &  0.0357151695	& 0.03571516 & 0.00127557 & 8.0     & 0.0249755634    & 0.02497556	& 0.00062377  \\
$\infty$& 0.0357151695	& 0.03571516 & 0.00127557 &$\infty$ & 0.0249755634    & 0.02497556	& 0.00062377  \\
\hline
\end{tabular}
\end{table}
 
\subsubsection{Isotropic 3D quantum harmonic oscillator (3DQHO)}
We start from the normalized $r$-space wave function given as, 
\begin{equation}
\psi_{n_{r},l}(r)= \sqrt{\frac{2 \ \omega^{l+\frac{3}{2}} \ n_{r}!}{\Gamma(n_{r}+l+\frac{3}{2})}} \ \ 
r^{l} \ e^{-\frac{\omega r^{2}}{2}} \ L_{n_{r}}^{l+\frac{1}{2}}(\omega r^{2}).
\end{equation} 
In the above, $L_n^{\alpha}(x)$ represents the associated Laguerre polynomial. Now, using Eq.~(85) and substituting
$\omega r^{2}=u$, first part of Eq.~(76) yields,     
\begin{equation}
\mathrm{IR}_{\rvec}=\frac{16 \ \omega \ n_{r}!}{\Gamma(n_{r}+l+\frac{3}{2})} \int_{0}^{\infty} u^{l+\frac{3}{2}} e^{-u} 
\left[L_{n_{r}}^{l+\frac{1}{2}}(u)\right]^{2} \left(\frac{\psi_{n_{r},l}^{\prime}(u)}{\psi_{n_{r},l}(u)}-
\frac{\psi_{n_{r_{1}},l}^{\prime}(u)}{\psi_{n_{r_{1}},l}(u)}\right)^{2} \mathrm{d}u. 
\end{equation} 
Now, using the recurrence relation $\ \frac{d}{du}L_{n_r}^{l+\frac{1}{2}} (u)=-L_{n_r-1}^{l+\frac{3}{2}} (u)$, the 
ratios of wave functions given in the parentheses may be simplified as, 
\begin{equation}
\frac{\psi_{n_{r},l}^{\prime}(u)}{\psi_{n_{r},l}(u)}= 
\frac{l}{2u}-\frac{1}{2}-\frac{L_{n_{r}-1}^{l+\frac{3}{2}} (u) }{L_{n_{r}}^{l+\frac{1}{2}} (u) }, 
\end{equation} 
the right-hand side of which, for a node-less state becomes $\left(\frac{l}{2u}-\frac{1}{2}\right)$, because the ratio of 
polynomials vanishes. Next one may invoke the usual orthonormality relation, 
\begin{equation} 
\int_{0}^{\infty}u^{k} e^{-u} L_{i}^{k}(u) L_{j}^{k}(u) \mathrm{d}u=\frac{(i+k)!}{i!} \ \delta_{ij}. 
\end{equation}
to obtain the final form of IR$_{\rvec}$ as below,
\begin{equation}
\mathrm{IR}_{\rvec}= \frac{16 \ \omega \ n_r!}{\Gamma(n_{r}+l+\frac{3}{2})} \  \int_{0}^{\infty} u^{l+\frac{3}{2}} e^{-u} 
\left[L_{n_{r}-1}^{l+\frac{3}{2}}\right]^{2} \mathrm{d}u =16\omega \ n_{r} = 8\omega (n-l).  
\end{equation} 
Equation~(89) predicts that, IR$_{\rvec}$ in a 3DQHO, like its 1D counterpart, is also a linear function of $n$; however in this
occasion the slope is $8\omega$ and intercept is negative. For a fixed $l$, the slope progresses and intercept decreases with 
$\omega$ respectively. Further, at a certain $\omega$, the 
intercept declines with rise of $l$. In this scenario, the particle gets more and more localized with progress of $\omega$. The 
relative fluctuation with respect to reference state advances with $\omega$. Dependence of IR$_{\rvec}$ on $\omega$ is 
reminiscent to that of $r$-space Fisher information, I$_{\rvec}$ \cite{romera05}---both quantities escalate with $\omega$. 

\begin{figure}                         
\begin{minipage}[c]{0.31\textwidth}\centering
\includegraphics[scale=0.41]{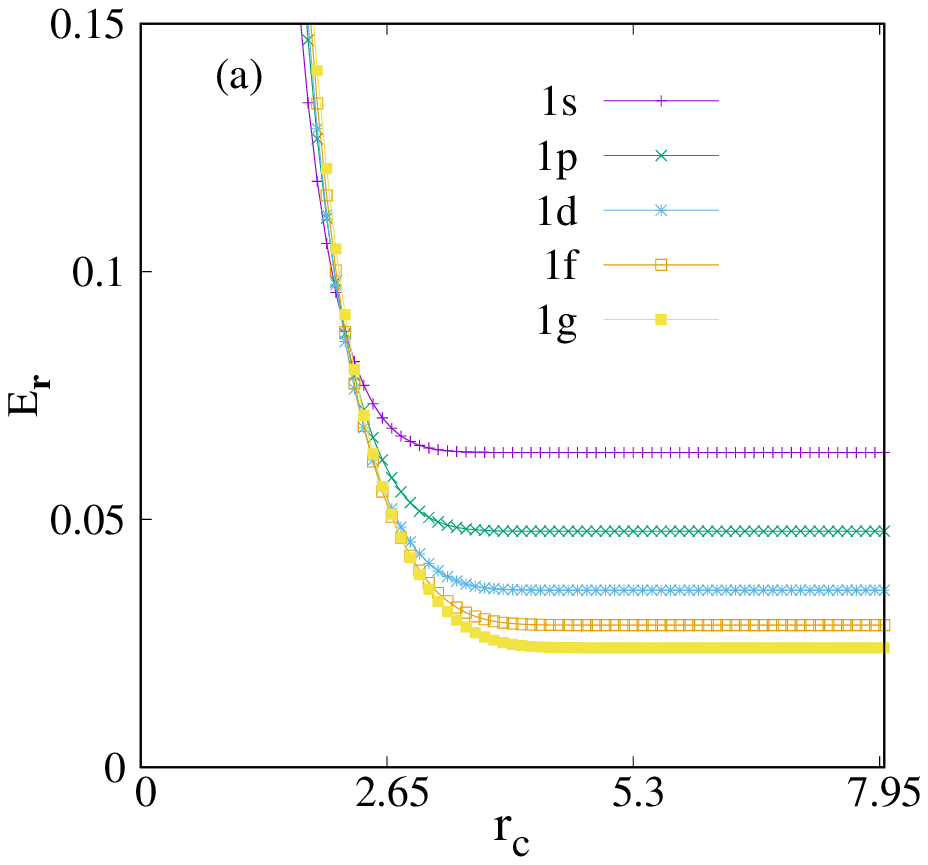}
\end{minipage}%
\hspace{0.02in}
\begin{minipage}[c]{0.31\textwidth}\centering
\includegraphics[scale=0.41]{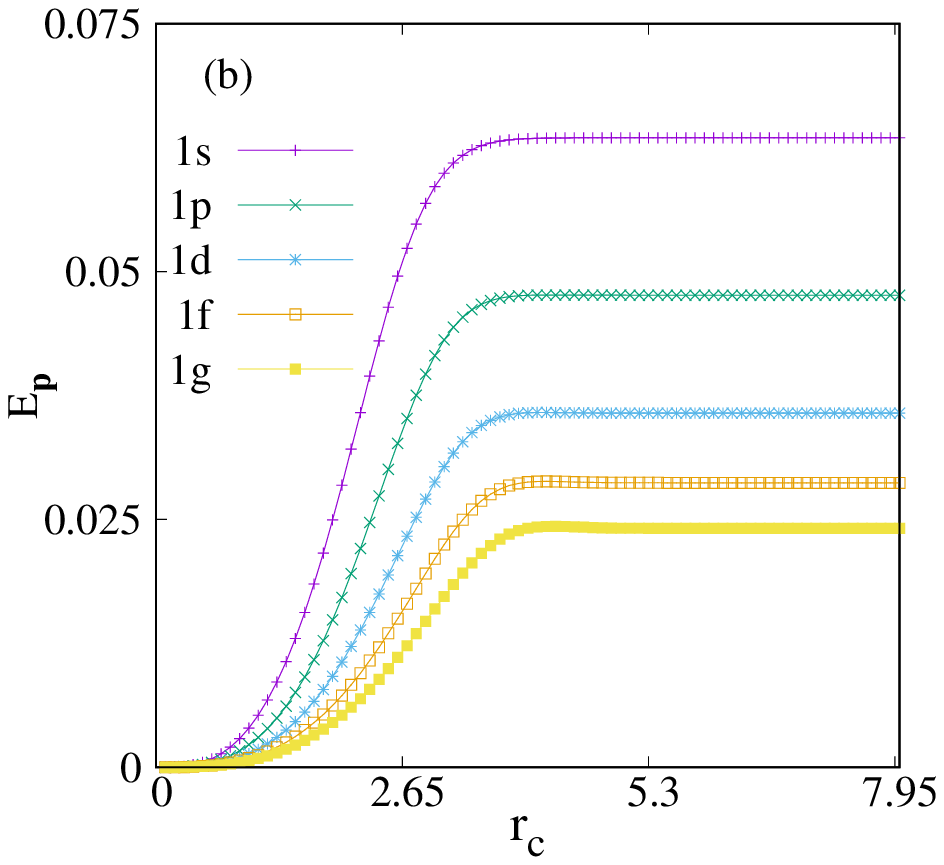}
\end{minipage}%
\hspace{0.02in}
\begin{minipage}[c]{0.31\textwidth}\centering
\includegraphics[scale=0.41]{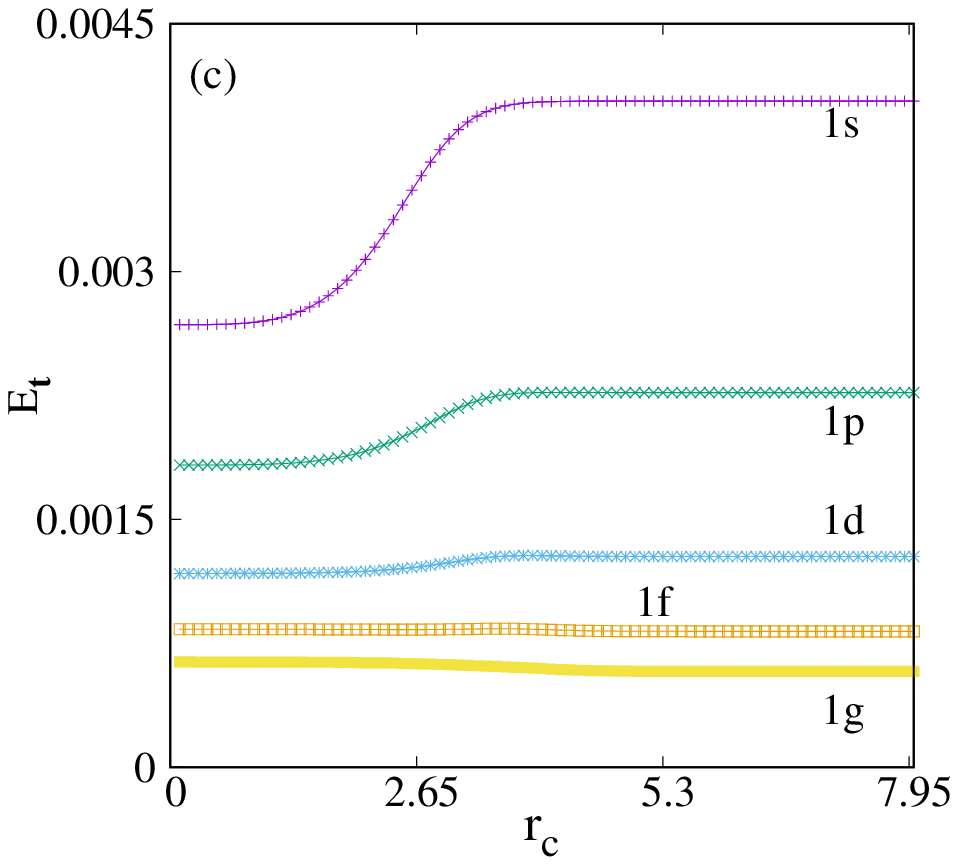}
\end{minipage}%
\caption{Plots of $E_{\rvec}$, $E_{\pvec}$, $E_{t}$ against $r_c$ for first five circular states of 3DCHO in panels (a), (b), (c) 
respectively \cite{mukherjee18a}. See text for details.}
\end{figure}                    

Analogously, in $p$-space the normalized wave function can be expressed as \cite{yanez94},  
\begin{equation}
\psi_{n_{r},l}(p)=\sqrt{\frac{2 \ n_{r}!}{\Gamma(n_{r}+l+\frac{3}{2}) \ \ \omega^{l+\frac{3}{2}}}} \ \ 
p^{l} \ e^{-\frac{p^{2}}{2\omega}} \ L_{n_{r}}^{l+\frac{1}{2}}\left(\frac{ p^{2}}{\omega}\right). 
\end{equation}
Substituting $\frac{p^{2}}{\omega}=\chi$, and going through some simple algebraic steps, one can derive, 
\begin{equation}
\mathrm{IR}_{\pvec}=\frac{16 \ n_{r}!}{\Gamma(n_{r}+l+\frac{3}{2}) \ \omega}\ \int_{0}^{\infty} 
\chi^{l+\frac{3}{2}} e^{-{\chi}} 
\left[L_{n_{r}}^{l+\frac{1}{2}}(\chi)\right]^{2} \left(\frac{\psi_{n_{r},l}^{\prime}(\chi)}{\psi_{n_{r},l}(\chi)}-
\frac{\psi_{n_{r_{1}},l}^{\prime}(\chi)}{\psi_{n_{r_{1}},l}(\chi)}\right)^{2} \mathrm{d}\chi.
\end{equation}
Applying similar arguments as discussed earlier for IR$_{\rvec}$ produces,  
\begin{equation}
\frac{\psi_{n_{r},l}^{\prime}(\chi)}{\psi_{n_{r},l}(\chi)}-\frac{\psi_{n_{r_{1}},l}^{\prime}(\chi)}{\psi_{n_{r_{1}},l}(\chi)}=
-\frac{L_{n_{r}-1}^{l+\frac{3}{2}} (\chi) }{L_{n_{r}}^{l+\frac{1}{2}} (\chi) }. 
\end{equation}
which upon applying in Eq.~(91), leads to the following final expression, namely, 
\begin{equation}
\mathrm{IR}_{\pvec}=16\left[\frac{n_{r}!}{\Gamma(n_{r}+l+\frac{3}{2}) \ \omega}\right]\int_{0}^{\infty} 
{\chi}^{l+\frac{3}{2}}e^{-\chi}\left[L_{n_{r}-1}^{l+\frac{3}{2}}\right]^{2} \mathrm{d} \chi
      =\frac{16}{\omega}n_{r} = \frac{8}{\omega}(n-l).
\end{equation}
Equation~(93) implies that, IR$_{\pvec}$, like IR$_{\rvec}$, also linearly changes with $n$; the slope and intercept are 
$\frac{16}{\omega}$ and $-\frac{16l}{\omega}$ respectively. An increase in $\omega$ facilitates localization and 
fluctuation too. IR$_{\rvec}$ progresses with $\omega$, while IR$_{\pvec}$ decreases, signifying higher fluctuation at 
larger $n_r$. Once again IR$_{\rvec}$, IR$_{\pvec}$ of a given $n_r,l$-state may be calculated from $(n_{r}-1),(l+1)$-state wave 
functions; this holds true in both spaces. We close the discussion by pointing out that, in parallel to 1D case, here also both 
$\Delta (\mathrm{IR}_{\rvec})$, $\Delta (\mathrm{IR}_{\pvec})$, at a given $l$, depend only on $\omega$ and remain unaltered with 
respect to $n$. They are expressed as in the following, 
\begin{equation}
\begin{aligned}
\Delta (\mathrm{IR}_{\rvec}) & = \mathrm{IR}_{\rvec}(n+1,~l)-\mathrm{IR}_{\rvec}(n,~l)=8\omega \\    
\Delta (\mathrm{IR}_{\pvec}) & = \mathrm{IR}_{\pvec}(n+1,~l)-\mathrm{IR}_{\pvec}(n,~l)=\frac{8}{\omega}. 
\end{aligned}
\end{equation}

\begin{figure}                                         
\begin{minipage}[c]{0.32\textwidth}\centering
\includegraphics[scale=0.35]{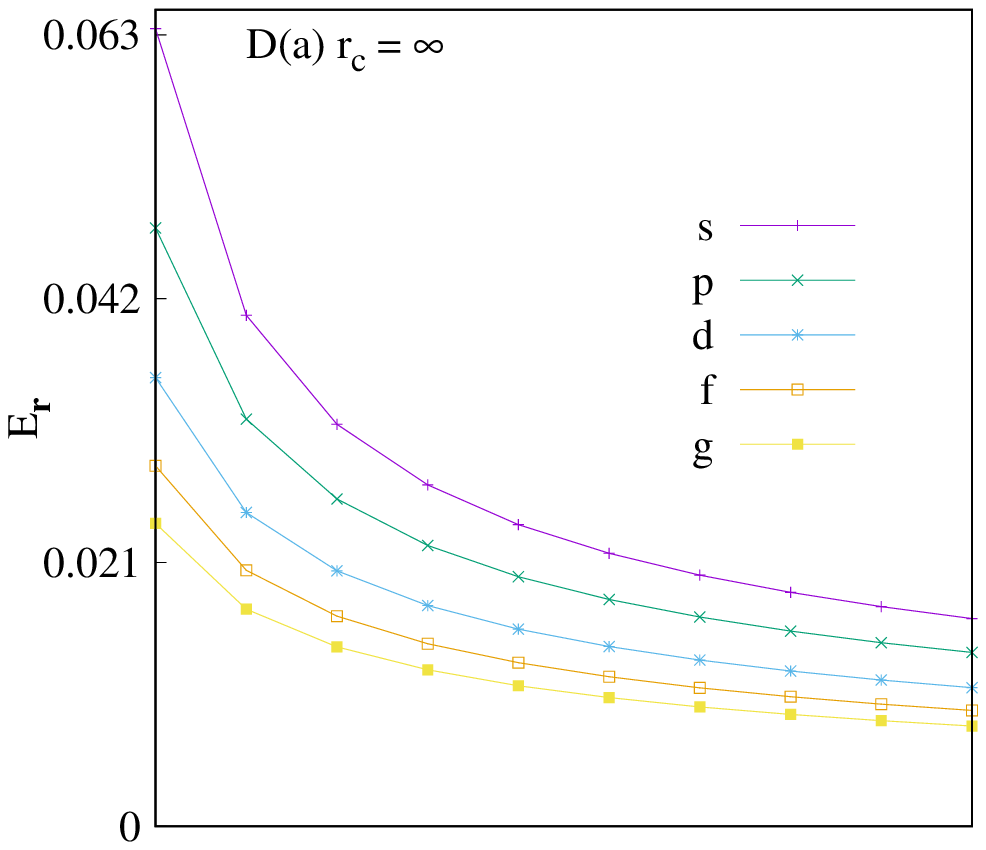}
\end{minipage}%
\begin{minipage}[c]{0.32\textwidth}\centering
\includegraphics[scale=0.35]{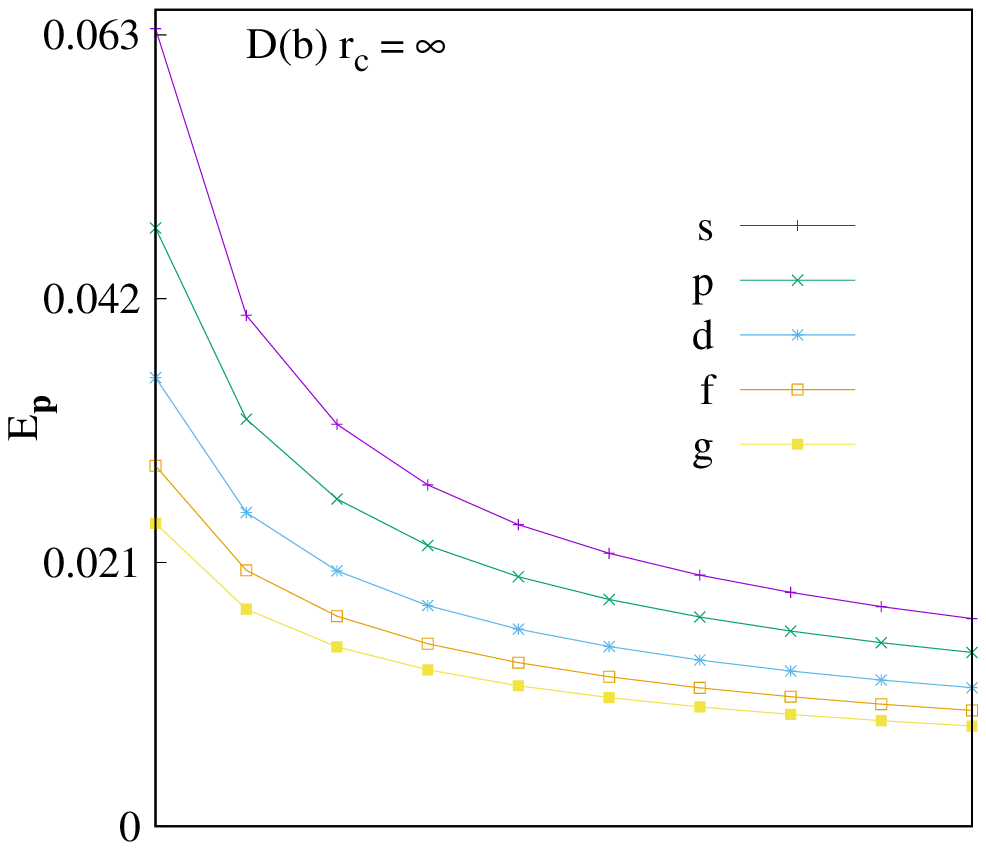}
\end{minipage}%
\begin{minipage}[c]{0.32\textwidth}\centering
\includegraphics[scale=0.35]{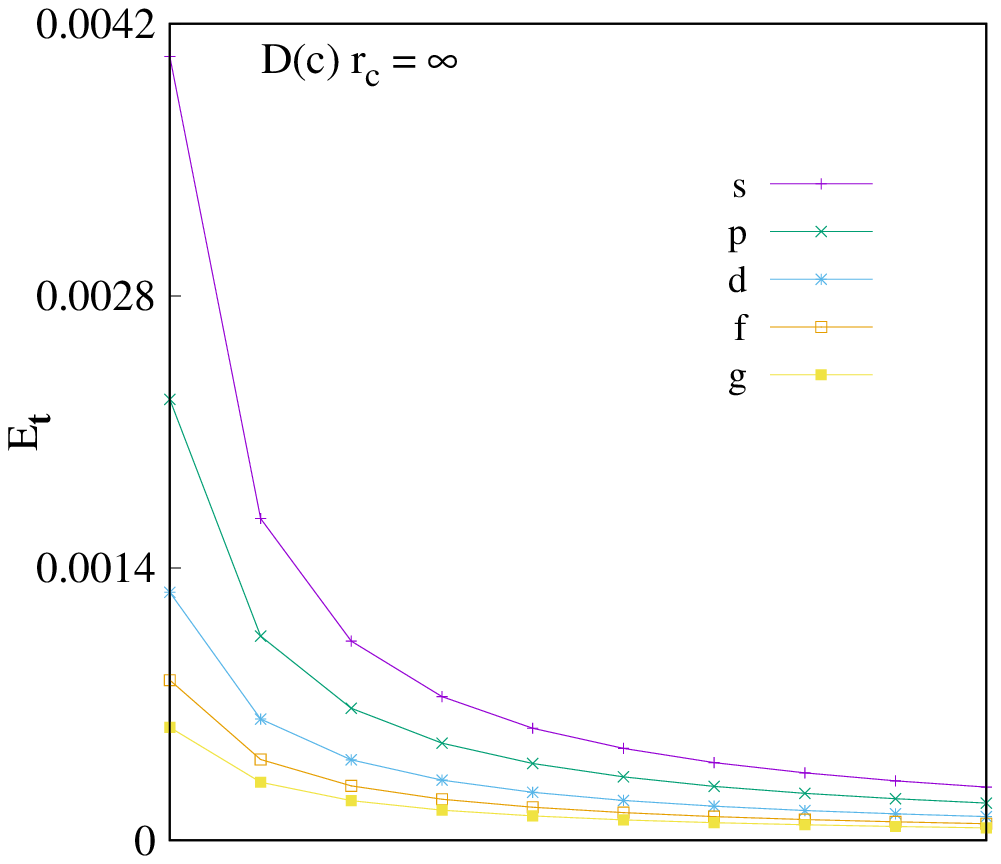}
\end{minipage}%
\\[15pt]
\begin{minipage}[c]{0.32\textwidth}\centering
\includegraphics[scale=0.35]{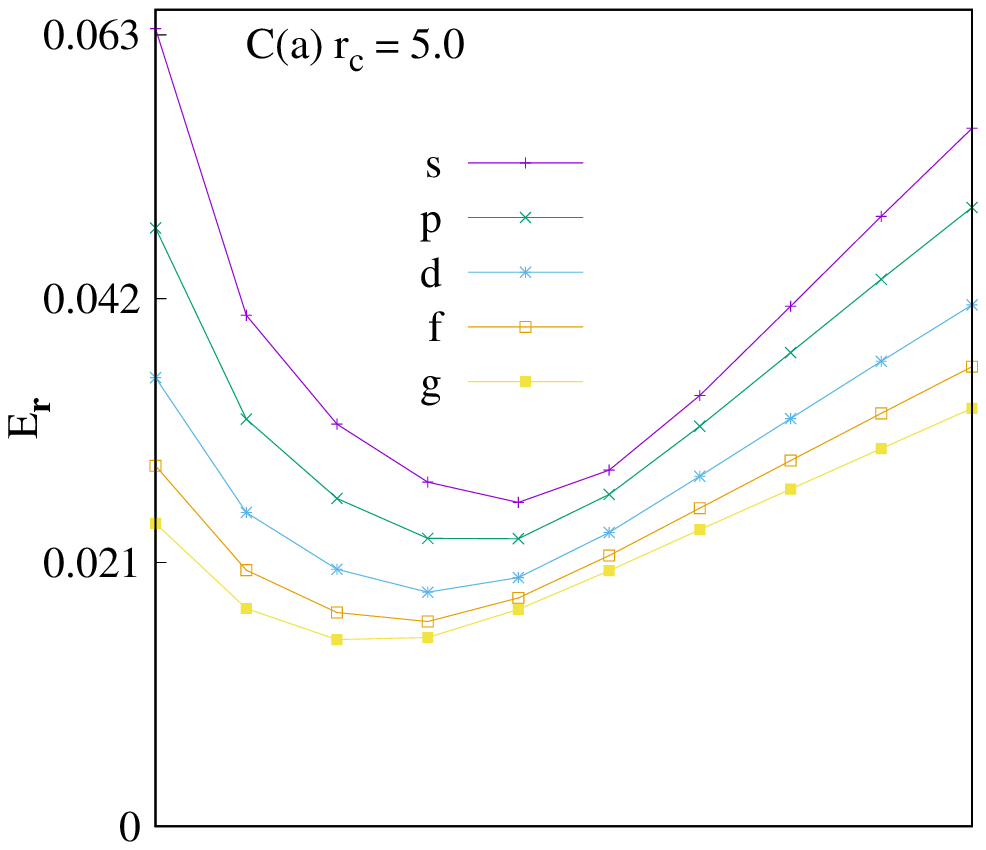}
\end{minipage}
\begin{minipage}[c]{0.32\textwidth}\centering
\includegraphics[scale=0.35]{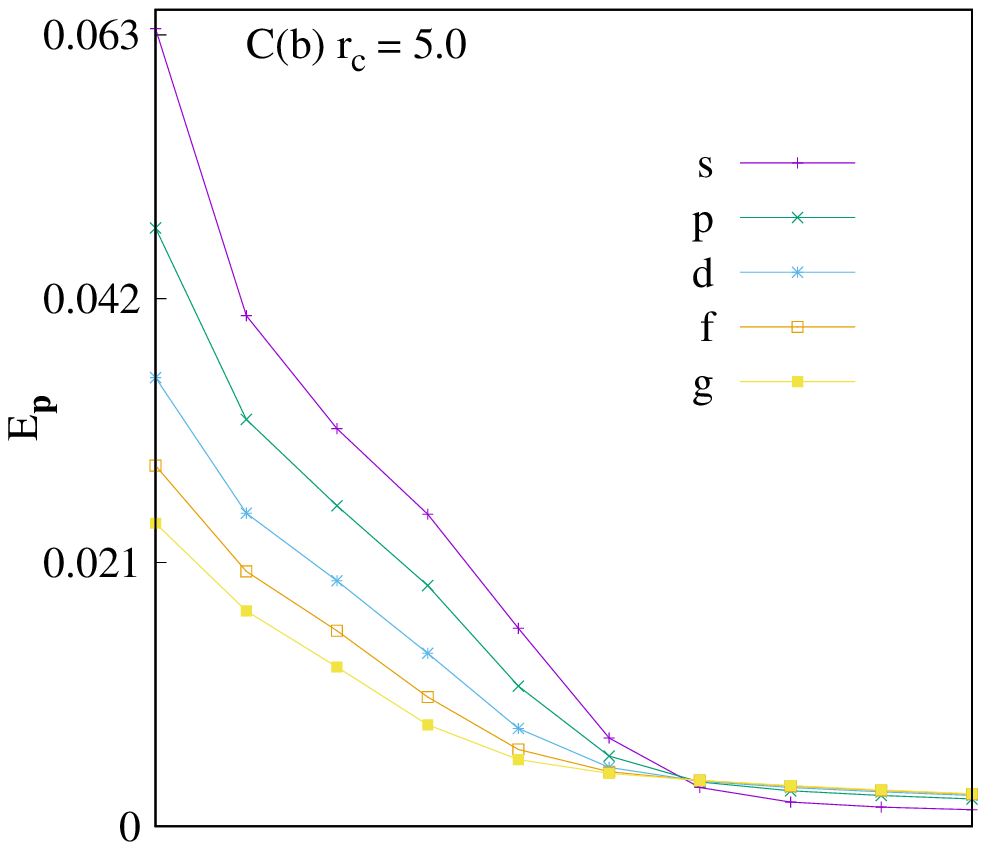}
\end{minipage}%
\begin{minipage}[c]{0.32\textwidth}\centering
\includegraphics[scale=0.35]{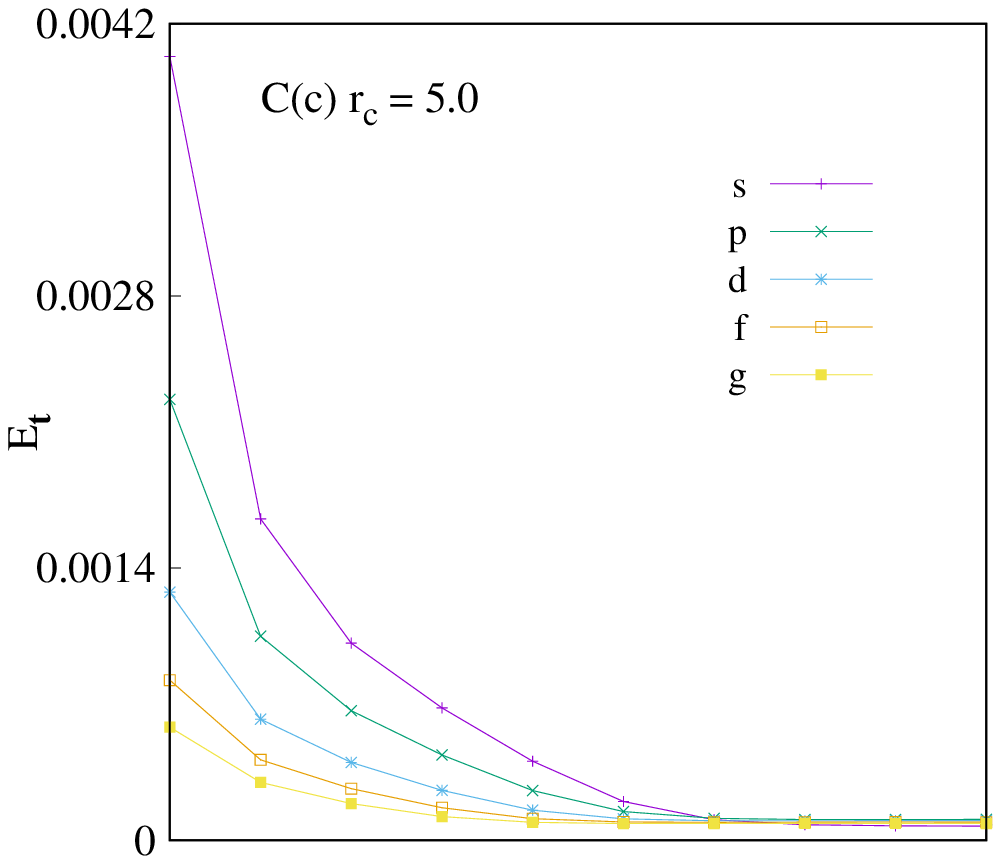}
\end{minipage}%
\\[15pt]
\begin{minipage}[c]{0.32\textwidth}\centering
\includegraphics[scale=0.35]{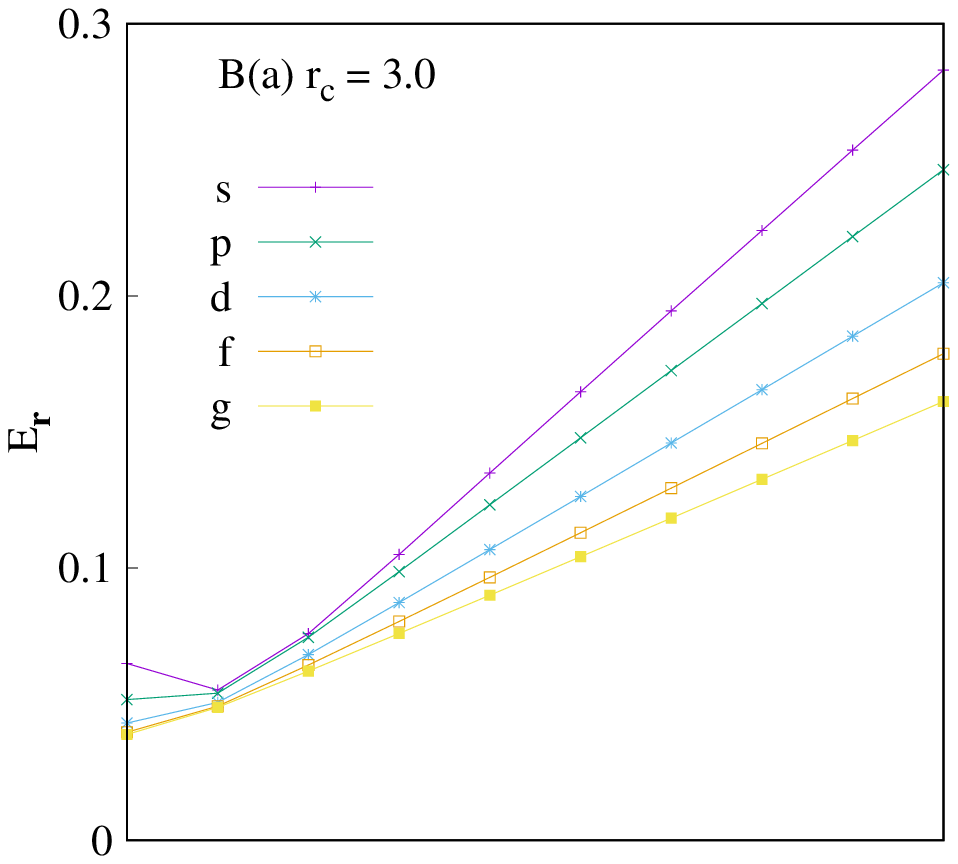}
\end{minipage}%
\begin{minipage}[c]{0.32\textwidth}\centering
\includegraphics[scale=0.35]{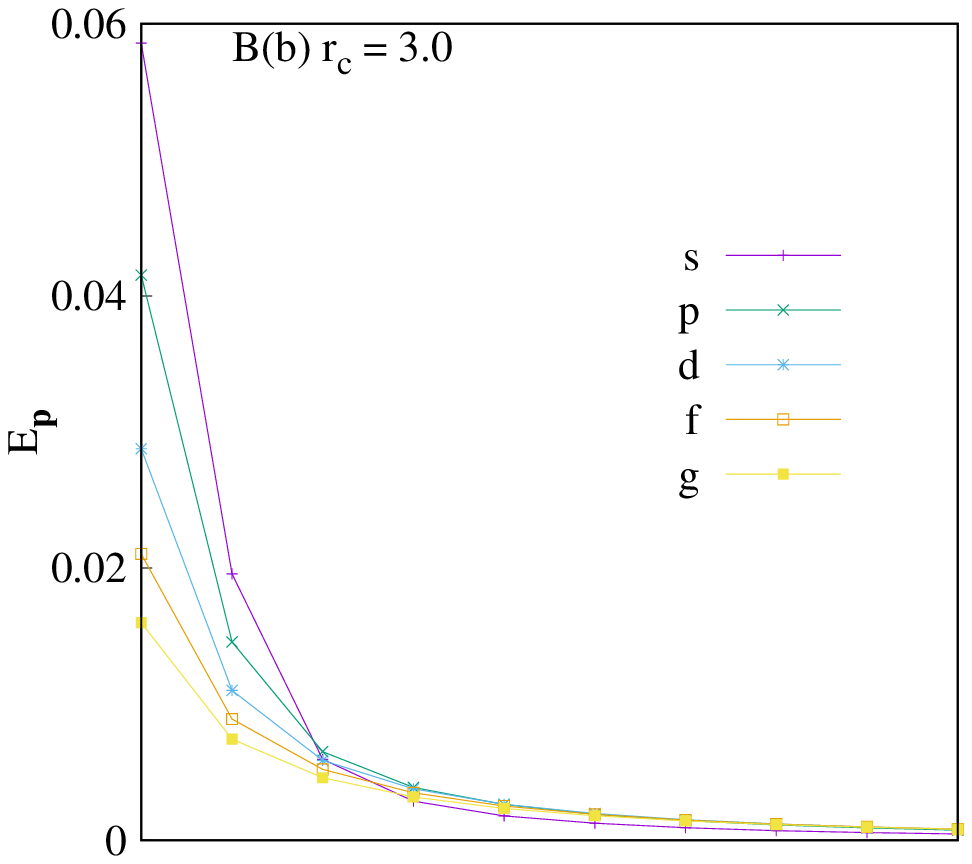}
\end{minipage}%
\begin{minipage}[c]{0.32\textwidth}\centering
\includegraphics[scale=0.35]{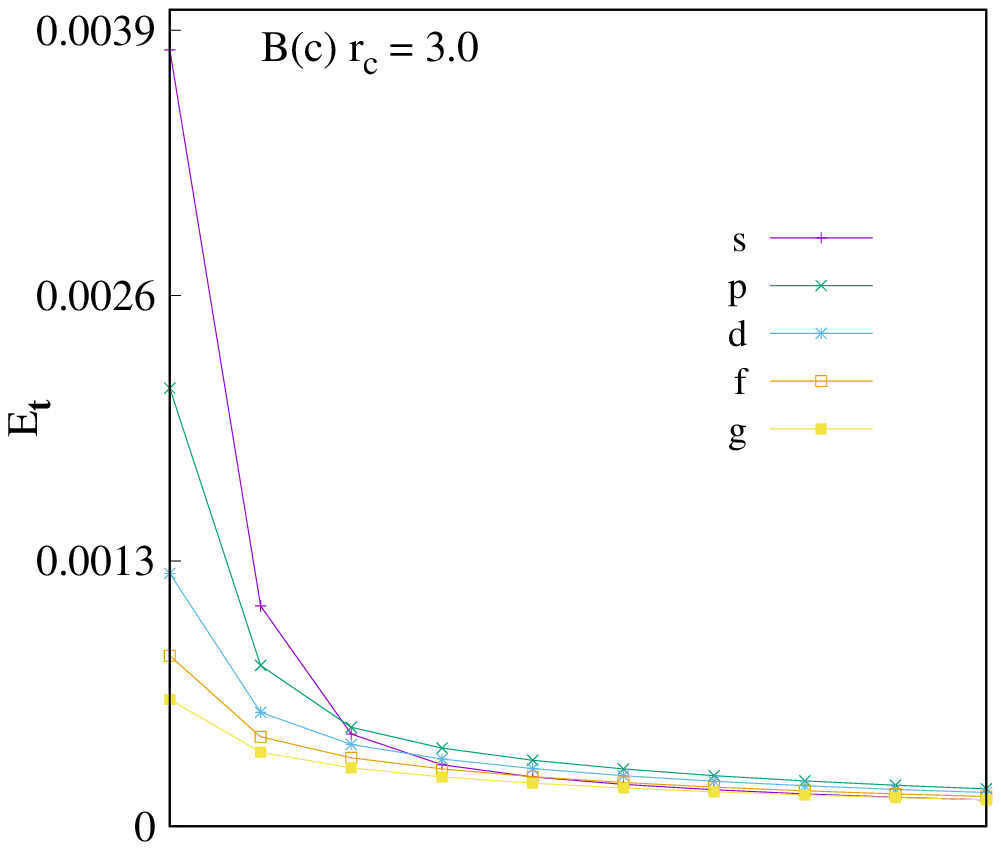}
\end{minipage}%
\\[15pt]
\begin{minipage}[c]{0.32\textwidth}\centering
\includegraphics[scale=0.37]{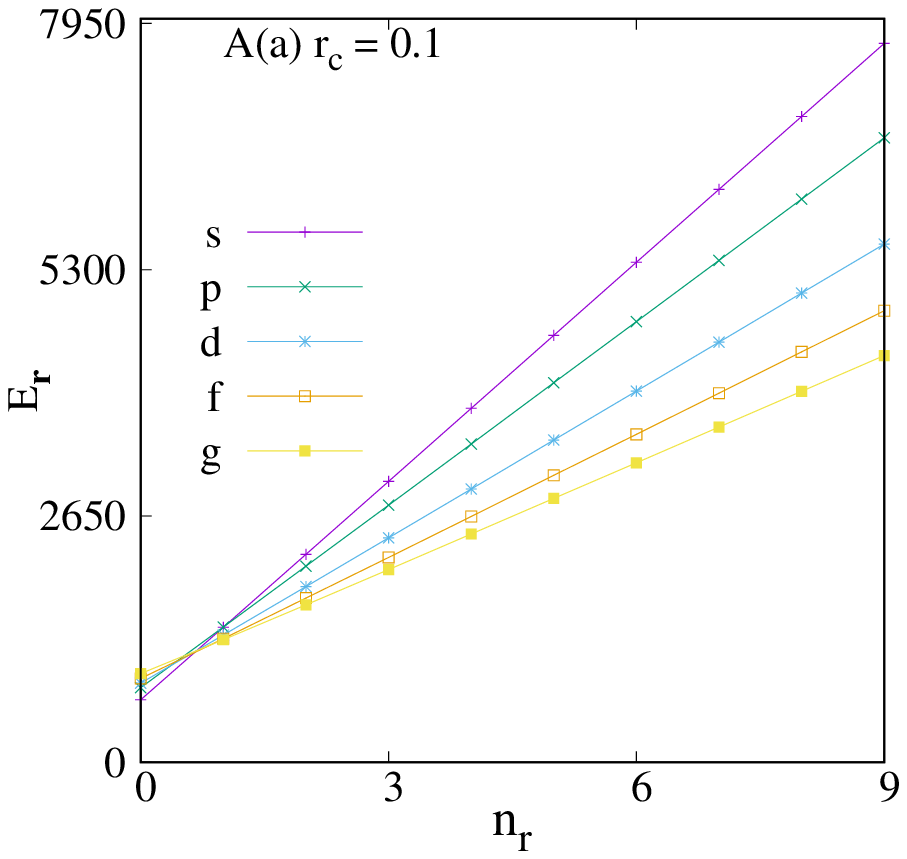}
\end{minipage}%
\begin{minipage}[c]{0.32\textwidth}\centering
\includegraphics[scale=0.37]{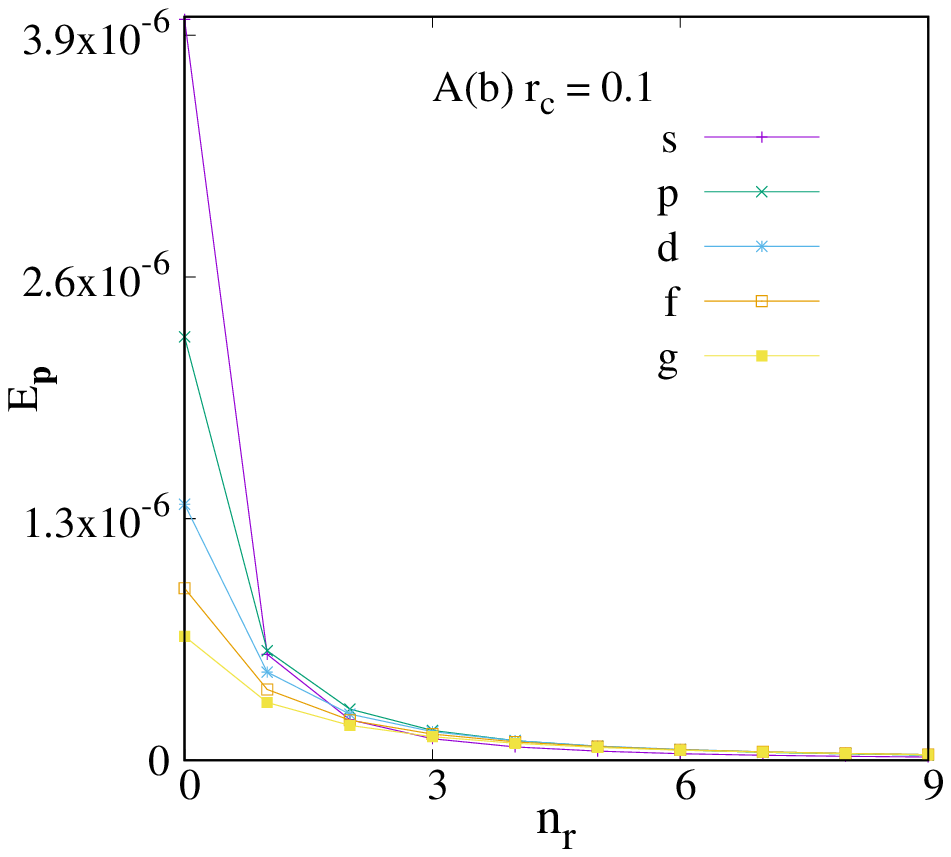}
\end{minipage}%
\begin{minipage}[c]{0.32\textwidth}\centering
\includegraphics[scale=0.37]{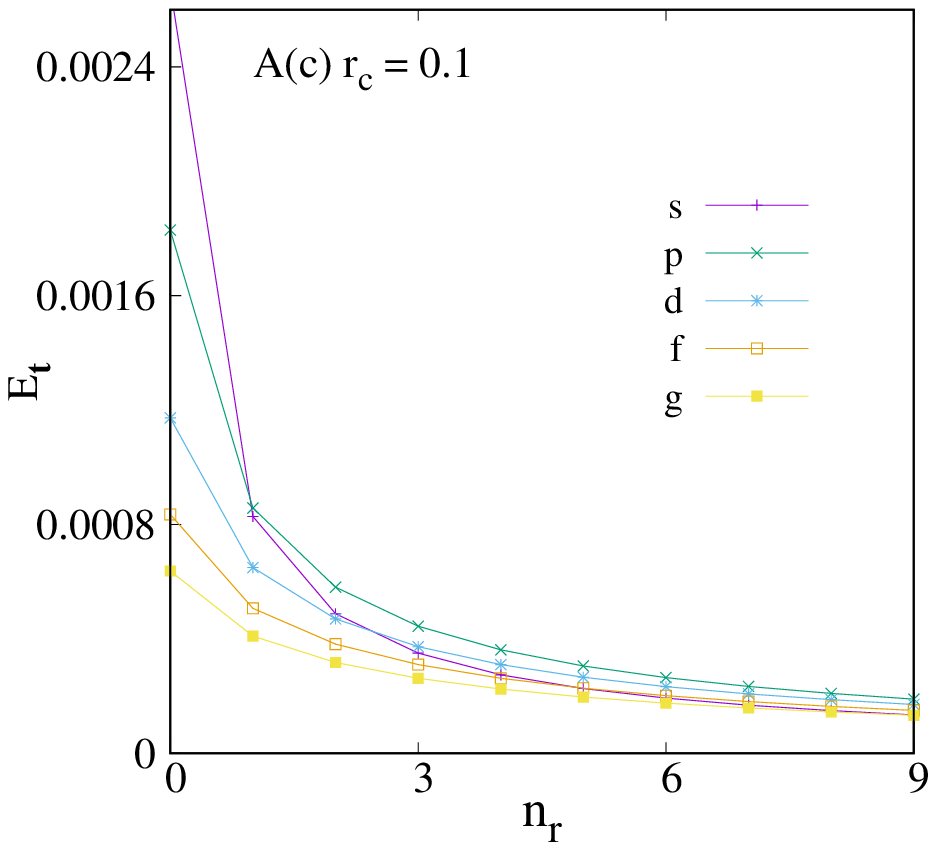}
\end{minipage}%
\caption{Plot of $E_{\rvec}$ (a), $E_{\pvec}$ (b) and $E_{t}$ (c) versus $n_{r}$ (at $\omega=1$) for $s,p,d,f,g$ 
states at four particular $r_{c}$'s of 3DCHO, namely, $0.1,3,5,\infty$ in panels (A)-(D) \cite{mukherjee18a}. For more details, consult text.} 
\end{figure}

\subsection{A Virial-like theorem}
Recently a Virial-like equation has been derived by the authors for confined quantum systems \cite{mukherjee19}. It has the form,
\begin{equation} \label{eq:virial}
\begin{aligned}
\langle \hat{T}^{2} \rangle_{n}-\langle \hat{T} \rangle^{2}_{n} & = \langle \hat{V}^{2} \rangle_{n}-\langle \hat{V} 
\rangle^{2}_{n} \\
(\Delta \hat{T}_{n})^{2} & =   \langle \hat{V} \rangle_{n} \langle \hat{T} \rangle_{n}
-\langle \hat{T}\hat{V} \rangle_{n} = (\Delta \hat{V}_{n})^{2} = \langle \hat{T} \rangle_{n} \langle \hat{V} \rangle_{n}
-\langle \hat{V}\hat{T} \rangle_{n}.
\end{aligned}
\end{equation}
This relation interprets that, the magnitude of error incurred in $\langle \hat{T} \rangle_{n}$ and $\langle \hat{V} \rangle_{n}$
are identical. Now, one can easily conclude that, $\mathcal{E}_{n}$ is a sum of two average quantities but still provides 
exact result. It is due to the cancellation of errors between $\langle \hat{T} \rangle_{n}$ and $\langle \hat{V} \rangle_{n}$.
In a confinement condition, validity of Eq.~(\ref{eq:virial}) can be verified by deriving the expressions of 
$\langle \hat{T} \hat{V} \rangle_{n,\ell}$, $\langle \hat{V} \hat{T} \rangle_{n,\ell}$, $\langle \hat{V}^{2} \rangle_{n,\ell}$ and 
$\langle \hat{V} \rangle_{n,\ell}$ (other integrals remain unchanged). 

\begingroup           
\begin{table}[h]
\caption{$\mathcal{E}_{n}, \left(\Delta V_{n}\right)^{2}, \left(\Delta T_{n}\right)^{2}, \langle T \rangle_{n}\langle V 
\rangle_{n}$ $-$ $\langle TV \rangle_{n}, \langle T \rangle_{n}\langle V \rangle_{n}$ $-$ $\langle VT \rangle_{n}$ values for 
$n=0,1$ states in 1DCHO at four ($0.1, 0.5, 1, \infty$) values of $x_{c}$. See text for detail.}
\centering
\begin{tabular}{>{\scriptsize}l<{\scriptsize}|>{\scriptsize}l|>{\scriptsize}l>{\scriptsize}l>{\scriptsize}l
>{\scriptsize}l<{\scriptsize}}
\hline
$n$ & Property           &  $x_c=0.1$ & $x_c=0.5$ & $x_c=1$ & $x_c=\infty$   \\
\hline
      & $\mathcal{E}_{0}^{\dagger}$                                                   & 123.3707084678  & 4.9511293232  
& 1.2984598320   & 0.4999999999   \\
      & $\left(\Delta V_{0}\right)^{2}$                                               & 0.000000600468  & 0.0003747558  
& 0.0058688193    & 0.1299999999 \\ 
0     & $\left(\Delta T_{0}\right)^{2}$                                               & 0.000000600466  & 0.0003747558  
& 0.0058688193   & 0.1299999999 \\
      & $\langle T \rangle_{0}\langle V \rangle_{0}-\langle TV \rangle_{0}$     & 0.000000600466  & 0.0003747558  
& 0.0058688193    & 0.1299999999 \\
      & $\langle T \rangle_{0}\langle V \rangle_{0}-\langle VT \rangle_{0}$     & 0.000000600466  & 0.0003747558  
& 0.0058688193    & 0.1299999999 \\
\hline
      & $\mathcal{E}_{1}^{\ddag}$                                                     & 493.481633417 & 19.7745341792 
& 5.0755820152  &  1.499999999 \\
      & $\left(\Delta V_{1}\right)^{2}$                                               & 0.00000085445 & 0.00053374630 
& 0.0084865378  &  0.374999999 \\ 
1     & $\left(\Delta T_{1}\right)^{2}$                                               & 0.00000085434 & 0.00053374630 
& 0.0084865378  &  0.374999999 \\
      & $\langle T \rangle_{1}\langle V \rangle_{1}-\langle TV \rangle_{1}$     & 0.00000085434 & 0.00053374630 
& 0.0084865378  &  0.374999999 \\
      & $\langle T \rangle_{1}\langle V \rangle_{1}-\langle VT \rangle_{1}$     & 0.00000085434 & 0.00053374630 
& 0.0084865378  &  0.374999999 \\             
\hline
\end{tabular}
\begin{tabbing} 
$^{\dagger}$Literature results \cite{roy15} of $\mathcal{E}_{0}$ for $x_{c}=0.1,0.5,1,\infty$ are: 
 123.37070846785, \\ 
4.9511293232541, 1.2984598320321, 0.5 respectively. \\
$^{\ddag}$Literature results \cite{roy15} of $\mathcal{E}_{1}$ for $x_{c}=0.1,0.5,1,\infty$ are: 
 493.48163341761, \\ 
19.774534179208, 5.0755820152268, 1.5 respectively.
\end{tabbing}
\end{table}  

\subsubsection{Symmetrically confined harmonic oscillator (SCHO)}
In a 1DCHO case, the desired expectation values will take following form:
\begin{equation}
\langle \hat{T} \hat{V} \rangle_{n} = \langle \hat{T} v(x) \rangle_{n} + \langle \hat{T}v_{c} \rangle_{n}= \langle \hat{T} 
v(x) \rangle_{n}.
\end{equation}  
One can utilize the property of Reimann integral to simplify, 
\begin{equation}
\begin{aligned}
\langle \hat{T}v_{c} \rangle_{n} & =  \int_{-\infty}^{-x_c} \psi^{*}_{n}(x)\hat{T} v_{c} \psi_{n}(x) \mathrm{d}x + \\
&\int_{-x_c}^{x_c} \psi^{*}_{n}(x)\hat{T} v_{c}\psi_{n}(x)
\mathrm{d}x + \int_{x_c}^{\infty} \psi^{*}_{n}(x)\hat{T} v_{c} \psi_{n}(x) \mathrm{d}x =0 \\
\end{aligned}
\end{equation}
The first and third integrals reduce to \emph{zero} because $\psi_{n}(x)=0$ when $x \ge |x_{c}|$, whereas the second integral 
leads to \emph{zero} as, $v_{c}=0$ inside the box. Similarly, 
\begin{equation}
\langle \hat{V} \hat{T} \rangle_{n} = \langle v(x)\hat{T} \rangle_{n} + \langle v_{c} \hat{T} \rangle_{n}= \langle v(x) 
\hat{T} \rangle_{n} 
\end{equation}
\begin{equation}
\langle \hat{V}^{2} \rangle_{n} = \langle (v(x)^{2} \rangle_{n} +\langle v(x) v_{c} \rangle_{n} + \langle v_{c} v(x) 
\rangle_{n} + \langle v_{c}^{2} \rangle_{n} = \langle (v(x))^{2} \rangle_{n} 
\end{equation}
\begin{equation}
\langle \hat{V} \rangle_{n} = \langle v(x) \rangle_{n} + \langle v_{c} \rangle_{n} = \langle v(x) \rangle_{n}.   
\end{equation}
Thus, for a 1DCHO, with the help of above equations, Eq.~(95) may be recast as, 
\begin{equation} \label{eq:1d}
\left(\Delta \hat{T}_{n}\right)^{2} = \left(\Delta \hat{V}_{n}\right)^{2}  = \langle \hat{T} \rangle_{n} \langle v(x) \rangle_{n}
-\langle v(x)\hat{T} \rangle_{n} = \langle \hat{T} \rangle_{n} \langle v(x) \rangle_{n}
-\langle \hat{T} v(x) \rangle_{n}.
\end{equation}
Thus it is clear from Eq.~(\ref{eq:1d}) that, $v_c$ does not contribute to the desired expectation values. Hence the only difference 
between free and enclosed system is that, in latter case, the boundary has been reduced to a finite region from infinity. 
Representative numerical values of $\mathcal{E}_{n}$, $(\Delta \hat{T}_{n})^{2}$, $(\Delta \hat{V}_{n})^{2}$, 
$\langle T \rangle_{n}\langle V \rangle_{n}-\langle TV \rangle_{n}$ and $\langle T \rangle_{n}\langle V \rangle_{n} 
- \langle VT \rangle_{n}$ are produced in Table~14 for $n=0,~1$ states of 1DCHO at four selected $x_c$ values, namely 
$0.1,~0.5,~1,~\infty$. In all these four $x_c$, $\mathcal{E}_{0}$ and 
$\mathcal{E}_{1}$ remain in excellent agreement with available literature results as compared in \cite{roy15}, and hence not 
repeated here. However, no direct reference could be found for expectation values to compare. It is easily noticed that, in 
both confined and free (last column) scenario, Eq.~(\ref{eq:virial}) is obeyed, as all expectation values offer equal values, 
which affirms the applicability of our newly designed theorem in case of 1DCHO. Additionally, with increase in $x_c$, both 
$\big(\Delta \hat{T}\big)^{2}, \big(\Delta \hat{V}\big)^{2}$ increase, which presumably occurs as the wave function 
delocalizes with $x_c$. Consequently, the difference between mean square and average values of $\hat{T},~\hat{V}$ tends to rise. 
For more details, please consult \cite{mukherjee19}. 

\begingroup           
\begin{table}
\caption{$\mathcal{E}_{n_{r},\ell}$ ,$(\Delta V_{n_{r},\ell})^{2}$ ,$(\Delta T_{n_{r},\ell})^{2}$ ,$\langle T \rangle_{n_{r},\ell}\langle V 
\rangle_{n_{r}, \ell}-\langle TV \rangle_{n_{r},\ell}$ ,$\langle T \rangle_{n_{r},\ell}\langle V \rangle_{n_{r},\ell}-\langle VT 
\rangle_{n_{r}, \ell}$ for $1s,~1p,~2s$ states in 3DCHO at four specific $r_{c}$'s, such as $0.1, 0.5, 1, \infty$.}
\centering
\begin{tabular}{>{\tiny}l<{\tiny}|>{\tiny}l<{\tiny}|>{\tiny}l>{\tiny}l
>{\tiny}l>{\tiny}l<{\tiny}}
\hline
State & Property           &  $r_c=0.1$ & $r_c=0.5$ & $r_c=1$ & $r_c=\infty$  \\
\hline
      & $\mathcal{E}_{1,0}^{\P}$                                                        & 493.481633459 & 19.774534179 
& 5.0755820153  & 1.4999999    \\
      & $\left(\Delta V_{1,0}\right)^{2}$                                               & 0.00000085434  & 0.0005337463 
& 0.0084865378  & 0.3749999  \\ 
$1s$    & $\left(\Delta T_{1,0}\right)^{2}$                                               & 0.00000085434  & 0.0005337463 
& 0.0084865378  & 0.3749999 \\
      & $\langle T \rangle_{1,0}\langle V \rangle_{1,0}$ $-$ $\langle TV \rangle_{1,0}$ & 0.00000085434  & 0.0005337463 
& 0.0084865378 & 0.3749999 \\
      & $\langle T \rangle_{1,0}\langle V \rangle_{1,0}$ $-$ $\langle VT \rangle_{1,0}$ & 0.00000085434  & 0.0005337463 
& 0.0084865378  & 0.3749999 \\
\hline
      & $\mathcal{E}_{1,1}^{\S}$                                                        & 1009.53830080 & 40.428276496   
& 10.282256939  & 2.4999999  \\
      & $\left(\Delta V_{1,1}\right)^{2}$                                               & 0.0000008424  & 0.0005264224  
& 0.0084064867  & 0.6249999 \\ 
$1p$    & $\left(\Delta T_{1,1}\right)^{2}$                                               & 0.00000084238 & 0.0005264224  
& 0.0084064867  & 0.6249999\\
      & $\langle T \rangle_{1,1}\langle V \rangle_{1,1}-\langle TV \rangle_{1,1}$ & 0.00000084238 & 0.0005264224  
& 0.0084064867  & 0.6249999 \\
      & $\langle T \rangle_{1,1}\langle V \rangle_{1,1}-\langle VT \rangle_{1,1}$ & 0.00000084238 & 0.0005264224  
& 0.0084064867  & 0.6249999 \\
\hline
      & $\mathcal{E}_{2,0}^{\ddag}$                                                     & 1973.92248339 & 78.9969211469 
& 19.899696502  & 3.4999999 \\
      & $\left(\Delta V_{2,0}\right)^{2}$                                               & 0.00000182     & 0.00113739969 
& 0.0181584455  & 1.6249999 \\ 
$2s$    & $\left(\Delta T_{2,0}\right)^{2}$                                               & 0.00000182     & 0.00113739969 
& 0.0181584455  & 1.6249999 \\
      & $\langle T \rangle_{2,0}\langle V \rangle_{2,0}-\langle TV \rangle_{2,0}$ & 0.00000182     & 0.00113739969 
& 0.0181584455  & 1.6249999 \\
      & $\langle T \rangle_{2,0}\langle V \rangle_{2,0}-\langle VT \rangle_{2,0}$ & 0.00000182     & 0.00113739969 
& 0.0181584455  & 1.6249999 \\             
\hline
\end{tabular}
\begin{tabbing} 
$^{\P}$Literature results \cite{roy14} of $\mathcal{E}_{1,0}$ for $r_{c}=0.1,0.5,1,\infty$ are: 
 493.48163346, \\ 
19.774534180, 5.0755820154, 1.5 respectively. \\
$^{\S}$Literature results \cite{roy14} of $\mathcal{E}_{1,1}$ for $r_{c}=0.1,0.5,1,\infty$ are: 
 1009.5383008, \\
40.428276496, 10.282256939, 2.5 respectively. \\
$^{\ddag}$ Literature results \cite{roy14} of $\mathcal{E}_{2,0}$ for $r_{c}=0.1,0.5,1,\infty$ are: 
1973.922483399, \\ 
78.996921147, 19.899696502, 3.5 respectively.
\end{tabbing}
\end{table}
\endgroup

\subsubsection{3D Confined harmonic oscillator (3DCHO)}
The relevant expectation values for 3DCHO will now take following forms, 
\begin{equation}
\langle \hat{T}\hat{V} \rangle_{n_{r},\ell}= \langle \hat{T} v(r) \rangle_{n_{r},\ell} + \langle \hat{T}v_{c}(r) 
\rangle_{n_{r},\ell}= \langle \hat{T} v(r) \rangle_{n_{r},\ell}. 
\end{equation}
This happens because $\langle \hat{T}v_{c}(r) \rangle_{n_{r},\ell}=0$, which, in turn occurs, since wave function disappears for
$r \geq r_c$. A similar argument ($\langle v_{c}(r) \hat{T}\rangle_{n_{r},\ell}=0$) leads to the following relation, 
\begin{equation}
\begin{aligned}
\langle \hat{V}\hat{T} \rangle_{n_{r},\ell} & = & \langle v(r) \hat{T} \rangle_{n_{r},\ell} + \langle v_{c}(r) \hat{T}  
\rangle_{n_{r},\ell}= \langle  v(r) \hat{T} \rangle_{n_{r},\ell}  
\end{aligned}
\end{equation}  
Then since $\langle v(r)v_{c}(r) \rangle_{n_{r},\ell} = \langle v_{c}(r)v(r) \rangle_{n_{r},\ell} = \langle v_{c}(r)^{2} 
\rangle_{n_{r},\ell} =0$, we can write, 
\begin{equation}
\begin{aligned}
\langle \hat{V}^{2} \rangle_{n_{r}, \ell} &= \langle v(r)^{2} \rangle_{n_{r},\ell} + \langle v(r)v_{c}(r) \rangle_{n_{r},\ell} + 
\langle v_{c}(r)v(r) \rangle_{n_{r}, \ell} + \langle v_{c}(r)^{2} \rangle_{n_{r},\ell}\\
 &= \langle v(r)^{2} \rangle_{n_{r},\ell}. 
\end{aligned}
\end{equation}
And finally, one can derive (since $\langle v_{c}(r) \rangle_{n_{r},\ell}=0$), 
\begin{equation}
\langle \hat{V} \rangle_{n_{r},\ell} = \langle v(r) \rangle_{n_{r},\ell} + \langle v_{c}(r) \rangle_{n_{r},\ell} =  
\langle v(r) \rangle_{n_{r},\ell}. 
\end{equation}
Thus, for a 3DCHO, Eq.~(\ref{eq:virial}) can be recast in to,  
\begin{eqnarray}
\langle \hat{T}^{2} \rangle_{n_{r},\ell}-\langle \hat{T} \rangle^{2}_{n_{r},\ell}  =  \langle \hat{V}^{2} 
\rangle_{n_{r},\ell}-\langle \hat{V} \rangle^{2}_{n_{r},\ell}  \nonumber \\
(\Delta \hat{T}_{n_{r},\ell})^{2}  =  (\Delta \hat{V}_{n_{r},\ell})^{2} = & \langle \hat{T} 
\rangle_{n_{r},\ell} \langle v(r) \rangle_{n_{r},\ell}
-\langle v(r)\hat{T} \rangle_{n_{r},\ell} \nonumber \\
 = & \langle \hat{T} \rangle_{n_{r},\ell} \langle v(r) \rangle_{n_{r},\ell}
-\langle \hat{T} v(r) \rangle_{n_{r},\ell}. \nonumber 
\end{eqnarray}
This suggests that, similar to 1DCHO, here also the confining potential has no contribution on desired
expectation values; only the boundary in confined system gets shifted to $r_c$, from $\infty$ of corresponding free 
counterpart. It clearly supports the validity of Eq.~(\ref{eq:virial}) in a 3DCHO. As an illustration, Table~15 imprints numerically 
calculated expectation values, for three low-lying ($1s,1p,2s$) states at four chosen values of confinement 
radius, i.e., $0.1,~0.5,~1,~\infty$. This again establishes the efficiency of Eq.~(\ref{eq:virial}) for such potential in both confined 
and free system, as evident from identical values of these quantities at all $r_c$'s--last column signifying the corresponding 
\emph{unconstrained} system. Accurate energies are quoted from GPS results \cite{roy14}. No literature is available 
for average values considered here. Like the 1D case, here also $(\Delta \hat{T}_{n_r,\ell})^{2},~(\Delta \hat{V}_{n_r,\ell})^{2}$ 
increase with $r_c$. A more detailed discussion may be found in \cite{mukherjee19}.  

\section{Conclusion}
Information-based uncertainty measures like $I, S, R, E$ and several complexities in composite $r,p$ spaces have been employed 
to understand the effect of confinement in 3DCHO. Before that, we have also discussed $I, S, E$ and various complexities for 
SCHO and ACHO systems. Apart from that, relative Fisher information for 1DQHO and 3DQHO are reported here. A detailed study reveals that, 
an SCHO can be considered as an interim model between PIB and 1DQHO. Similarly, a 3DCHO may be treated as a two-mode system; at 
$r_{c} \rightarrow 0$ it behaves as PISB and at $r_{c} \rightarrow \infty$ it reduces to a 3DQHO.

It is imperative to mention that at $d_{m}=0$, an ACHO reduces to an SCHO. An increase in $d_{m}$ leads to localization of the 
system in real space. $S_{x}, S_{p}$ can completely interpret the competing nature of ACHO potential. In contrast, $I_{x}, I_{p}$ 
prove inefficient for such analysis. Further, $E_{x}, E_{p}$ provide important knowledge which guides the reader to a better 
understanding of such systems. 

In 3DCHO, at very low $r_{c}$, $R_{\rvec}^{\alpha}$, $S_{\rvec}$ reduce and $E_{\rvec}$ grows up as $n_{r}$ progresses. This 
behavior clearly suggests that the effect of confinement accelerates with increase of number of nodes in the system. In future, 
it will be interesting to study the effect of penetrable boundary on such systems. Further, study of relative information for 
SCHO, ACHO and 3DCHO systems would be highly desirable.      

\section{Acknowledgement}
The authors express their sincere gratitude to Prof. Nadya S. Columbus, President, NOVA Science Publishers, NY, USA, for the 
kind invitation to present our work in this exciting area. We are extremely thankful to the Editor, Prof. Yilun Shang for giving 
us the opportunity to contribute and 
also extending the submission deadline generously. AKR gratefully acknowledges financial support from SERB, Department of Science 
and Technology (DST), New Delhi (sanction no. EMR/2014/000838). NM thanks DST for financial support through NPDF grant 
(PDF/2016/000014CS). It is a pleasure to thank Ms. Sangita Majumdar for giving a careful reading of the manuscript. 

\bibliographystyle{mychicago}
\bibliography{ref}

\begin{thebibliography}{}

\bibitem[\protect\citeauthoryear{{Michels, A.; De Boer, J.; Bijl,
  A.}}{{Michels, A.; De Boer, J.; Bijl, A.}}{1937}]{michels37}
{Michels, A.; De Boer, J.; Bijl, A.} (1937).
\newblock Remarks concerning molecular interaction and their influence on the
  polarizability.
\newblock {\em Physica\/}~{\em 4}, 981.

\bibitem[\protect\citeauthoryear{{Jask\'{o}lski, W.}}{{Jask\'{o}lski,
  W.}}{1996}]{jaskolski96}
{Jask\'{o}lski, W.} (1996).
\newblock Confined many-electron systems.
\newblock {\em Phys. Rep.\/}~{\em 271}, 1.

\bibitem[\protect\citeauthoryear{{Dolmatov, V. K.; Baltenkov, A. S.; Connerade,
  J.-P; Mason, S. T.}}{{Dolmatov, V. K.; Baltenkov, A. S.; Connerade, J.-P;
  Mason, S. T.}}{2004}]{dolmatov04}
{Dolmatov, V. K.; Baltenkov, A. S.; Connerade, J.-P; Mason, S. T.} (2004).
\newblock Structure and photoionization of confined atoms.
\newblock {\em Radiat. Phys. Chem.\/}~{\em 70}, 417.

\bibitem[\protect\citeauthoryear{{Sabin, J., Br\"{a}ndas, E.; and Cruz,
  S.}}{{Sabin, J., Br\"{a}ndas, E.; and Cruz, S.}}{2009}]{sabin09}
{Sabin, J., Br\"{a}ndas, E.; and Cruz, S.} (Ed.) (2009).
\newblock {\em Adv. Quant. Chem.}, Volume 57 \& 58.
\newblock New York: Academic Press.

\bibitem[\protect\citeauthoryear{{Sen, K. D.}}{{Sen, K. D.}}{2014}]{sen14}
{Sen, K. D.} (Ed.) (2014).
\newblock {\em Electronic {S}tructure of {Q}uantum {C}onfined {A}toms and
  {M}olecules}.
\newblock Switzerland: Springer International Publishing.

\bibitem[\protect\citeauthoryear{{Ley-Koo, E.}}{{Ley-Koo, E.}}{2018}]{leykoo18}
{Ley-Koo, E.} (2018).
\newblock Recent progress in confined atoms and molecules: superintegrability
  and symmetry breakings.
\newblock {\em Revista Mexicana de F\'{i}sica\/}~{\em 64}, 326.

\bibitem[\protect\citeauthoryear{{Dolmatov, V. K.}}{{Dolmatov, V.
  K.}}{2009}]{dolmatov09}
{Dolmatov, V. K.} (2009).
\newblock Photoionization of atoms encaged in spherical fullerenes.
\newblock {\em Adv. Quant. Chem.\/}~{\em 58}, 13.

\bibitem[\protect\citeauthoryear{{Charkin, O. P.; Klimenko, N. M.; Charkin, D.
  O.}}{{Charkin, O. P.; Klimenko, N. M.; Charkin, D. O.}}{2009}]{charkin09}
{Charkin, O. P.; Klimenko, N. M.; Charkin, D. O.} (2009).
\newblock {DFT} study of molecules confined inside fullerene and fullerene-like
  cages.
\newblock {\em Adv. Quant. Chem.\/}~{\em 58}, 69.

\bibitem[\protect\citeauthoryear{{Etindele, A. J.; Maezono, R.;
  Melingui-Melono, R. L.; Motapon, O.}}{{Etindele, A. J.; Maezono, R.;
  Melingui-Melono, R. L.; Motapon, O.}}{2017}]{etindele17}
{Etindele, A. J.; Maezono, R.; Melingui-Melono, R. L.; Motapon, O.} (2017).
\newblock Influence of endohedral confinement of atoms on structural and
  dynamical properties of the ${C}_{60}$ fullerene.
\newblock {\em Chem. Phys. Lett.\/}~{\em 685}, 395.

\bibitem[\protect\citeauthoryear{{Auluck, F. C.}}{{Auluck, F.
  C.}}{1941a}]{auluck41}
{Auluck, F. C.} (1941a).
\newblock {Energy levels of an artificially bounded linear oscillator}.
\newblock {\em Proc. Nat. Inst. Sci. India\/}~{\em 7}, 133.

\bibitem[\protect\citeauthoryear{{Auluck, F. C.}}{{Auluck, F.
  C.}}{1941b}]{auluck42}
{Auluck, F. C.} (1941b).
\newblock {The Artificially Bounded Relativistic Linear Oscillator}.
\newblock {\em Proc. Nat. Inst. Sci. India\/}~{\em 7}, 383.

\bibitem[\protect\citeauthoryear{{Chandrasekhar, S.}}{{Chandrasekhar,
  S.}}{1943}]{chandrasekhar43}
{Chandrasekhar, S.} (1943).
\newblock Dynamical friction. {II}. the rate of escape of stars from clusters
  and the evidence for the operation of dynamical friction.
\newblock {\em Astrophys. J.\/}~{\em 97}, 263.

\bibitem[\protect\citeauthoryear{{Dingle, R. B.}}{{Dingle, R.
  B.}}{1952a}]{dingle52}
{Dingle, R. B.} (1952a).
\newblock Some magnetic properties of metals {III. D}iamagnetic resonance.
\newblock {\em Proc. Royal Soc. London A\/}~{\em 212}, 38.

\bibitem[\protect\citeauthoryear{{Dingle, R. B.}}{{Dingle, R.
  B.}}{1952b}]{dingle52a}
{Dingle, R. B.} (1952b).
\newblock Some magnetic properties of metals {IV. P}roperties of small systems
  of electrons.
\newblock {\em Proc. Royal Soc. London A\/}~{\em 212}, 47.

\bibitem[\protect\citeauthoryear{{Grinberg, M.; Jask\'{o}lski, W.; Koepke, Cz.;
  Planelles, J.; Janowicz, M.}}{{Grinberg, M.; Jask\'{o}lski, W.; Koepke, Cz.;
  Planelles, J.; Janowicz, M.}}{1994}]{grinberg94}
{Grinberg, M.; Jask\'{o}lski, W.; Koepke, Cz.; Planelles, J.; Janowicz, M.}
  (1994).
\newblock Spectroscopic manifestation of a confinement-type lattice
  anharmonicity.
\newblock {\em Phys. Rev. B\/}~{\em 50}, 6504(R).

\bibitem[\protect\citeauthoryear{{Auluck, F. C.; Kothari, D. S.}}{{Auluck, F.
  C.; Kothari, D. S.}}{1945}]{auluck45}
{Auluck, F. C.; Kothari, D. S.} (1945).
\newblock The quantum mechanics of a bounded linear harmonic oscillator.
\newblock {\em Proc. Cambridge Phil. Soc.\/}~{\em 41}, 175.

\bibitem[\protect\citeauthoryear{{Baijal, J. S.; Singh, K. K.}}{{Baijal, J. S.;
  Singh, K. K.}}{1955}]{baijal55}
{Baijal, J. S.; Singh, K. K.} (1955).
\newblock The energy levels and transition probabilities for a bounded linear
  harmonic oscillator.
\newblock {\em Prog. Theor. Phys. (Kyoto)\/}~{\em 14}, 214.

\bibitem[\protect\citeauthoryear{{Hull, T. E.; Julius, R. S.}}{{Hull, T. E.;
  Julius, R. S.}}{1956}]{hull56}
{Hull, T. E.; Julius, R. S.} (1956).
\newblock Enclosed quantum mechanical systems.
\newblock {\em Can. J. Phys.\/}~{\em 34}, 914.

\bibitem[\protect\citeauthoryear{{Dean, P.}}{{Dean, P.}}{1966}]{dean66}
{Dean, P.} (1966).
\newblock The constrained quantum mechanical harmonic oscillator.
\newblock {\em Proc. Cambridge Phil. Soc.\/}~{\em 62}, 277.

\bibitem[\protect\citeauthoryear{{Vawter, R.}}{{Vawter, R.}}{1968}]{vawter68}
{Vawter, R.} (1968).
\newblock Effects of finite boundaries in a one-dimensional harmonic
  oscillator.
\newblock {\em Phys. Rev.\/}~{\em 174}, 749.

\bibitem[\protect\citeauthoryear{{Vawter, R.}}{{Vawter, R.}}{1973}]{vawter73}
{Vawter, R.} (1973).
\newblock Energy eigenvalues of a bounded centrally located harmonic
  oscillator.
\newblock {\em J. Math. Phys.\/}~{\em 14}, 1864.

\bibitem[\protect\citeauthoryear{{Consortini, A.; Frieden, B. R.}}{{Consortini,
  A.; Frieden, B. R.}}{1976}]{consortini76}
{Consortini, A.; Frieden, B. R.} (1976).
\newblock Quantum-mechanical solution for the simple harmonic oscillator in a
  box.
\newblock {\em Nuovo Cimento B\/}~{\em 35}, 153.

\bibitem[\protect\citeauthoryear{{Aguilera-Navarro, V.C.; Ley-Koo, E.;
  Zimerman, A. H.}}{{Aguilera-Navarro, V.C.; Ley-Koo, E.; Zimerman, A.
  H.}}{1980}]{navarro80}
{Aguilera-Navarro, V.C.; Ley-Koo, E.; Zimerman, A. H.} (1980).
\newblock Perturbative, asymptotic and {P}ad\'e-approximant solutions for
  harmonic and inverted oscillators in a box.
\newblock {\em J. Phys. A\/}~{\em 13}, 3585.

\bibitem[\protect\citeauthoryear{{Fern\'andez, F. M.; Castro, E.
  A.}}{{Fern\'andez, F. M.; Castro, E. A.}}{1981a}]{fernandez81}
{Fern\'andez, F. M.; Castro, E. A.} (1981a).
\newblock Hyper-virial analysis of enclosed quantum mechanical systems. {I.
  D}irichlet boundary conditions.
\newblock {\em Int. J. Quant. Chem.\/}~{\em 19}, 521.

\bibitem[\protect\citeauthoryear{{Fern\'andez, F. M.; Castro, E.
  A.}}{{Fern\'andez, F. M.; Castro, E. A.}}{1981b}]{fernandez81a}
{Fern\'andez, F. M.; Castro, E. A.} (1981b).
\newblock Hyper-virial analysis of enclosed quantum mechanical systems. {III.
  U}nsymmetrical boundary conditions.
\newblock {\em Int. J. Quant. Chem.\/}~{\em 20}, 623.

\bibitem[\protect\citeauthoryear{{Arteca, G. A.; Maluendes, S. A.; Fern\'andez,
  F. M.; Castro, E. A.}}{{Arteca, G. A.; Maluendes, S. A.; Fern\'andez, F. M.;
  Castro, E. A.}}{1983}]{arteca83}
{Arteca, G. A.; Maluendes, S. A.; Fern\'andez, F. M.; Castro, E. A.} (1983).
\newblock Discussion of several analytical expressions for the eigenvalues of
  the bounded harmonic oscillator and hydrogen atom.
\newblock {\em Int. J. Quant. Chem.\/}~{\em 24}, 169.

\bibitem[\protect\citeauthoryear{{Ta\c{s}eli, H.}}{{Ta\c{s}eli,
  H.}}{1993}]{taseli93}
{Ta\c{s}eli, H.} (1993).
\newblock Accurate computation of the energy spectrum for potentials with
  multiminima.
\newblock {\em Int. J. Quant. Chem.\/}~{\em 46}, 319.

\bibitem[\protect\citeauthoryear{{Vargas, R.; Garza, J.; Vela, A.;}}{{Vargas,
  R.; Garza, J.; Vela, A.;}}{1996}]{vargas96}
{Vargas, R.; Garza, J.; Vela, A.;} (1996).
\newblock Strongly convergent method to solve one-dimensional quantum problems.
\newblock {\em Phys. Rev. E\/}~{\em 53}, 1954.

\bibitem[\protect\citeauthoryear{{Sinha, A.; Roychoudhury, R.}}{{Sinha, A.;
  Roychoudhury, R.}}{1999}]{sinha99}
{Sinha, A.; Roychoudhury, R.} (1999).
\newblock {WKB} and {MAF} quantization rules for spatially confined quantum
  mechanical systems.
\newblock {\em J. Math. Chem.\/}~{\em 73}, 497.

\bibitem[\protect\citeauthoryear{{Sinha, A.}}{{Sinha, A.}}{2000}]{sinha00}
{Sinha, A.} (2000).
\newblock {SWKB} formalism for confined quantum systems.
\newblock {\em Int. J. Quant. Chem.\/}~{\em 79}, 267.

\bibitem[\protect\citeauthoryear{{Montgomery Jr., H. E.; Campoy, G.; Aquino,
  N.}}{{Montgomery Jr., H. E.; Campoy, G.; Aquino, N.}}{2010}]{montgomery10}
{Montgomery Jr., H. E.; Campoy, G.; Aquino, N.} (2010).
\newblock The confined {N}-dimensional harmonic oscillator revisited.
\newblock {\em Phys. Scr.\/}~{\em 81}, 045010.

\bibitem[\protect\citeauthoryear{{Campoy, G.; Aquino, N.; Granados, V.
  N.}}{{Campoy, G.; Aquino, N.; Granados, V. N.}}{2002}]{campoy02}
{Campoy, G.; Aquino, N.; Granados, V. N.} (2002).
\newblock Energy eigenvalues and {E}instein coefficients for the
  one-dimensional confined harmonic oscillators.
\newblock {\em J. Phys. A\/}~{\em 35}, 4903.

\bibitem[\protect\citeauthoryear{{Amore, P.; Fern\'{a}ndez, F. M.}}{{Amore, P.;
  Fern\'{a}ndez, F. M.}}{2010}]{amore10}
{Amore, P.; Fern\'{a}ndez, F. M.} (2010).
\newblock One-dimensional oscillator in a box.
\newblock {\em Eur. J. Phys.\/}~{\em 31}, 69.

\bibitem[\protect\citeauthoryear{{Roy, A. K.}}{{Roy, A. K.}}{2015}]{roy15}
{Roy, A. K.} (2015).
\newblock Quantum confinement in 1{D} systems through an imaginary-time
  evolution method.
\newblock {\em Mod. Phys. Lett. A\/}~{\em 37}, 1550176.

\bibitem[\protect\citeauthoryear{{Aquino, N.; Cruz, E.}}{{Aquino, N.; Cruz,
  E.}}{2017}]{aquino17}
{Aquino, N.; Cruz, E.} (2017).
\newblock The 1-dimensional confined harmonic oscillator revisited.
\newblock {\em Revista Mexicana de F\'{i}sica\/}~{\em 63}, 580.

\bibitem[\protect\citeauthoryear{{Aquino, N.; Casta\~{n}o, E.; Campoy, G.;
  Granados, V.}}{{Aquino, N.; Casta\~{n}o, E.; Campoy, G.; Granados,
  V.}}{2001}]{aquino01}
{Aquino, N.; Casta\~{n}o, E.; Campoy, G.; Granados, V.} (2001).
\newblock Einstein coefficients and dipole moment for the asymmetrically
  confined harmonic oscillator.
\newblock {\em Eur. J. Phys.\/}~{\em 22}, 645.

\bibitem[\protect\citeauthoryear{{Fern\'andez, F. M.; Castro, E.
  A.}}{{Fern\'andez, F. M.; Castro, E. A.}}{1981}]{fernandez81b}
{Fern\'andez, F. M.; Castro, E. A.} (1981).
\newblock Hyper-virial treatment of multidimensional isotropic bounded
  oscillators.
\newblock {\em Phys. Rev. A\/}~{\em 24}, 2883.

\bibitem[\protect\citeauthoryear{{Aguilera-Navarro, V. C.; Gomes, J. F.;
  Zimerman, A. H.; Ley-Koo, E.}}{{Aguilera-Navarro, V. C.; Gomes, J. F.;
  Zimerman, A. H.; Ley-Koo, E.}}{1983}]{navarro83}
{Aguilera-Navarro, V. C.; Gomes, J. F.; Zimerman, A. H.; Ley-Koo, E.} (1983).
\newblock On the radius of convergence of {R}ayleigh-{Schr\"odinger}
  perturbative solutions for quantum oscillators in circular and spherical
  boxes.
\newblock {\em J. Phys. A\/}~{\em 16}, 2943.

\bibitem[\protect\citeauthoryear{{Marin, J. L.; Cruz, S. A.}}{{Marin, J. L.;
  Cruz, S. A.}}{1991}]{marin91}
{Marin, J. L.; Cruz, S. A.} (1991).
\newblock On the use of direct variational methods to study confined quantum
  systems.
\newblock {\em Am. J. Phys.\/}~{\em 59}, 931.

\bibitem[\protect\citeauthoryear{{Dutt, R.; Mukherjee, A.; Varshni, Y.
  P.}}{{Dutt, R.; Mukherjee, A.; Varshni, Y. P.}}{1995}]{dutt95}
{Dutt, R.; Mukherjee, A.; Varshni, Y. P.} (1995).
\newblock Supersymmetric semiclassical approach to confined quantum problems.
\newblock {\em Phys. Rev. A\/}~{\em 52}, 1750.

\bibitem[\protect\citeauthoryear{{Ta\c{s}eli, H.; Zafer, A.}}{{Ta\c{s}eli, H.;
  Zafer, A.}}{1997a}]{taseli97a}
{Ta\c{s}eli, H.; Zafer, A.} (1997a).
\newblock A {F}ourier-{B}essel expansion for solving radial {Schr\"odinger}
  equation in two dimensions.
\newblock {\em Int. J. Quant. Chem.\/}~{\em 61}, 759.

\bibitem[\protect\citeauthoryear{{Ta\c{s}eli, H.; Zafer, A.}}{{Ta\c{s}eli, H.;
  Zafer, A.}}{1997b}]{taseli97b}
{Ta\c{s}eli, H.; Zafer, A.} (1997b).
\newblock Bessel basis with applications: {N}-dimensional isotropic polynomial
  oscillators.
\newblock {\em Int. J. Quant. Chem.\/}~{\em 63}, 936.

\bibitem[\protect\citeauthoryear{{Aquino, N.}}{{Aquino, N.}}{1997}]{aquino97}
{Aquino, N.} (1997).
\newblock The isotropic bounded oscillators.
\newblock {\em J. Phys. A\/}~{\em 30}, 2403.

\bibitem[\protect\citeauthoryear{{Sinha, A.}}{{Sinha, A.}}{2003}]{sinha03}
{Sinha, A.} (2003).
\newblock Three-dimensional confinement: {WKB} revisited.
\newblock {\em J. Math. Chem.\/}~{\em 34}, 201.

\bibitem[\protect\citeauthoryear{{Filho, E. D.; Ricotta, R. M.}}{{Filho, E. D.;
  Ricotta, R. M.}}{2003}]{filho03}
{Filho, E. D.; Ricotta, R. M.} (2003).
\newblock Supersymmetric variational energies of 3{D} confined potentials.
\newblock {\em Phys. Lett. A\/}~{\em 320}, 95.

\bibitem[\protect\citeauthoryear{{Sen, K. D.; Roy, A. K.}}{{Sen, K. D.; Roy, A.
  K.}}{2006}]{sen06}
{Sen, K. D.; Roy, A. K.} (2006).
\newblock Studies on the 3{D} confined potentials using generalized
  pseudospectral approach.
\newblock {\em Phys. Lett. A\/}~{\em 357}, 112.

\bibitem[\protect\citeauthoryear{Roy}{Roy}{2014}]{roy14mpla}
Roy, A.~K. (2014).
\newblock Confinement in 3{D} polynomial oscillators through a generalized
  pseudospectral method.
\newblock {\em Mod. Phys. Lett. A\/}~{\em 29}, 1450104.

\bibitem[\protect\citeauthoryear{{Stevanovi\'c, Lj.; Sen, K.
  D.}}{{Stevanovi\'c, Lj.; Sen, K. D.}}{2008a}]{stevanovic08}
{Stevanovi\'c, Lj.; Sen, K. D.} (2008a).
\newblock {A study of the confined 2{D} isotropic harmonic oscillator in terms
  of the annihilation and creation operators and the infinitesimal operators of
  the {SU}(2) group}.
\newblock {\em J. Phys. A\/}~{\em 41}, 265203.

\bibitem[\protect\citeauthoryear{{Stevanovi\'c, Lj.; Sen, K.
  D.}}{{Stevanovi\'c, Lj.; Sen, K. D.}}{2008b}]{stevanovic08a}
{Stevanovi\'c, Lj.; Sen, K. D.} (2008b).
\newblock {Eigenspectrum properties of the confined 3{D} harmonic oscillator}.
\newblock {\em J. Phys. A\/}~{\em 41}, 225002.

\bibitem[\protect\citeauthoryear{{Serrano, F. A.; Dong, S.-H.}}{{Serrano, F.
  A.; Dong, S.-H.}}{2013}]{serrano13}
{Serrano, F. A.; Dong, S.-H.} (2013).
\newblock Proper quantization rule approach to three-dimensional quantum dots.
\newblock {\em Int. J. Quant. Chem.\/}~{\em 113}, 2282.

\bibitem[\protect\citeauthoryear{{Al-Jaber, S. M.}}{{Al-Jaber, S.
  M.}}{2008}]{aljaber08}
{Al-Jaber, S. M.} (2008).
\newblock A confined {N}-dimensional harmonic oscillator.
\newblock {\em Int. J. Theor. Phys.\/}~{\em 47}, 1853.

\bibitem[\protect\citeauthoryear{{Montgomery, Jr., H. E.; Aquino, N. A.; Sen,
  K. D.}}{{Montgomery, Jr., H. E.; Aquino, N. A.; Sen, K.
  D.}}{2007}]{montgomery07}
{Montgomery, Jr., H. E.; Aquino, N. A.; Sen, K. D.} (2007).
\newblock Degeneracy of confined {D}-dimensional harmonic oscillator.
\newblock {\em Int. J. Quant. Chem.\/}~{\em 107}, 798.

\bibitem[\protect\citeauthoryear{{Poland, D.}}{{Poland, D.}}{2000}]{poland2000}
{Poland, D.} (2000).
\newblock Maximum-entropy calculation of energy distributions.
\newblock {\em J. Chem. Phys.\/}~{\em 112}, 6554.

\bibitem[\protect\citeauthoryear{{Singer, A.}}{{Singer, A.}}{2004}]{singer04}
{Singer, A.} (2004).
\newblock Maximum entropy formulation of the {K}irkwood superposition
  approximation.
\newblock {\em J. Chem. Phys.\/}~{\em 121}, 3657.

\bibitem[\protect\citeauthoryear{{Antoniazzi, A.; Fanelli, D.; Barr\'e, J.;
  Chavanis, P.-H.; Dauxois, T.; Ruffo, S.}}{{Antoniazzi, A.; Fanelli, D.;
  Barr\'e, J.; Chavanis, P.-H.; Dauxois, T.; Ruffo, S.}}{2007}]{anton07}
{Antoniazzi, A.; Fanelli, D.; Barr\'e, J.; Chavanis, P.-H.; Dauxois, T.; Ruffo,
  S.} (2007).
\newblock Maximum entropy principle explains quasistationary states in systems
  with long-range interactions: {T}he example of the {H}amiltonian mean-field
  model.
\newblock {\em Phys. Rev. E\/}~{\em 75}, 011112.

\bibitem[\protect\citeauthoryear{{Bialyniki-Birula, I.; Mycielski,
  J.}}{{Bialyniki-Birula, I.; Mycielski, J.}}{1975}]{birula75}
{Bialyniki-Birula, I.; Mycielski, J.} (1975).
\newblock Uncertainty relation for information entropy in wave mechanics.
\newblock {\em Commun. Math. Phys.\/}~{\em 44}, 129.

\bibitem[\protect\citeauthoryear{Shannon}{Shannon}{1951}]{shannon51}
Shannon, C.~E. (1951).
\newblock Prediction and entropy of printed english.
\newblock {\em Bell System Technical J.\/}~{\em 30}, 50.

\bibitem[\protect\citeauthoryear{{Bialyniki-Birula, I.}}{{Bialyniki-Birula,
  I.}}{2006}]{birula06}
{Bialyniki-Birula, I.} (2006).
\newblock Formulation of uncertainty relations in terms of the {R}\'enyi
  entropies.
\newblock {\em Phys. Rev. A\/}~{\em 74}, 052101.

\bibitem[\protect\citeauthoryear{{Duan, L.-M.; Giedke. G.; Cirac, J. I.;
  Zoller, P.}}{{Duan, L.-M.; Giedke. G.; Cirac, J. I.; Zoller,
  P.}}{2000}]{duan2000}
{Duan, L.-M.; Giedke. G.; Cirac, J. I.; Zoller, P.} (2000).
\newblock Inseparability criterion for continuous variable systems.
\newblock {\em Phys. Rev. Lett.\/}~{\em 84}, 2722.

\bibitem[\protect\citeauthoryear{Simon}{Simon}{2000}]{simon2000}
Simon, R. (2000).
\newblock Peres-{H}orodecki separability criterion for continuous variable
  systems.
\newblock {\em Phys. Rev. Lett.\/}~{\em 84}, 2726.

\bibitem[\protect\citeauthoryear{Roy}{Roy}{2004a}]{roy04pla}
Roy, A.~K. (2004a).
\newblock Calculation of the spiked harmonic oscillators through a generalized
  pseudospectral method.
\newblock {\em Phys. Lett. A\/}~{\em 321}, 231.

\bibitem[\protect\citeauthoryear{Roy}{Roy}{2004b}]{roy04jpg}
Roy, A.~K. (2004b).
\newblock Calculation of the bound states of power-law and logarithmic
  potentials through a generalized pseudospectral method.
\newblock {\em J. Phys. G\/}~{\em 30}, 269.

\bibitem[\protect\citeauthoryear{Roy}{Roy}{2004c}]{roy04jpbhollowb}
Roy, A.~K. (2004c).
\newblock Studies on the hollow states of atomic lithium using a density
  functional approach.
\newblock {\em J. Phys. B\/}~{\em 37}, 4569.

\bibitem[\protect\citeauthoryear{Roy}{Roy}{2005a}]{roy05pramana}
Roy, A.~K. (2005a).
\newblock The generalized pseudospectral approach to the bound states of
  {H}\'ulthen and {Y}ukawa potential.
\newblock {\em Pramana-J. Phys.\/}~{\em 65}, 1.

\bibitem[\protect\citeauthoryear{Roy}{Roy}{2005b}]{roy05ijqc}
Roy, A.~K. (2005b).
\newblock Studies on some singular potentials in quantum mechanics.
\newblock {\em Int. J. Quant. Chem.\/}~{\em 104}, 861.

\bibitem[\protect\citeauthoryear{Roy}{Roy}{2005c}]{roy05jpbhollowb}
Roy, A.~K. (2005c).
\newblock Density functional studies on the hollow resonances in
  {L}i-isoelectronic sequence ({Z}=4-10).
\newblock {\em J. Phys. B\/}~{\em 38}, 1591.

\bibitem[\protect\citeauthoryear{{Roy, A. K.; Jalbout, A. F.}}{{Roy, A. K.;
  Jalbout, A. F.}}{2007}]{roy07}
{Roy, A. K.; Jalbout, A. F.} (2007).
\newblock Ground and excited states of {L}i$^-$, {B}e$^-$ through a
  density-based approach.
\newblock {\em Chem. Phys. Lett.\/}~{\em 445}, 355.

\bibitem[\protect\citeauthoryear{{Roy, A. K.; Jalbout, A. F.; Proynov, E.
  I.}}{{Roy, A. K.; Jalbout, A. F.; Proynov, E. I.}}{2008a}]{roy08jmc}
{Roy, A. K.; Jalbout, A. F.; Proynov, E. I.} (2008a).
\newblock Accurate calculation of the bound states of hellmann potential.
\newblock {\em J. Math. Chem.\/}~{\em 44}, 260.

\bibitem[\protect\citeauthoryear{{Roy, A. K.; Jalbout, A. F.; Proynov, E.
  I.}}{{Roy, A. K.; Jalbout, A. F.; Proynov, E. I.}}{2008b}]{roy08ijqc}
{Roy, A. K.; Jalbout, A. F.; Proynov, E. I.} (2008b).
\newblock Bound state spectra of the 3{D} rational potential.
\newblock {\em Int. J. Quant. Chem.\/}~{\em 108}, 827.

\bibitem[\protect\citeauthoryear{{Roy, A. K.; Jalbout, A. F.}}{{Roy, A. K.;
  Jalbout, A. F.}}{2008}]{roy08theochem}
{Roy, A. K.; Jalbout, A. F.} (2008).
\newblock Bound states of the generalized spiked harmonic oscillator.
\newblock {\em J. Mol. Struct. (Theochem)\/}~{\em 853}, 27.

\bibitem[\protect\citeauthoryear{Roy}{Roy}{2013a}]{roy13ijqc}
Roy, A.~K. (2013a).
\newblock Studies on some exponential-screened {C}oulomb potentials.
\newblock {\em Int. J. Quant. Chem.\/}~{\em 113}, 1503.

\bibitem[\protect\citeauthoryear{Roy}{Roy}{2013b}]{roy13rinp}
Roy, A.~K. (2013b).
\newblock Accurate ro-vibrational spectroscopy of diatomic molecules in a
  {M}orse oscillator potential.
\newblock {\em Results in Physics\/}~{\em 3}, 103.

\bibitem[\protect\citeauthoryear{Roy}{Roy}{2014a}]{roy14fbsys}
Roy, A.~K. (2014a).
\newblock Studies on the bound-state spectrum of {H}yperbolic potential.
\newblock {\em Few-Body Systems\/}~{\em 55}, 143.

\bibitem[\protect\citeauthoryear{Roy}{Roy}{2014b}]{roy14ijqc}
Roy, A.~K. (2014b).
\newblock Ro-vibrational studies of diatomic molecules in a shifted
  {D}eng-{F}an oscillator potential.
\newblock {\em Int. J. Quant. Chem.\/}~{\em 114}, 383.

\bibitem[\protect\citeauthoryear{Roy}{Roy}{2014c}]{roy14mpla_manning}
Roy, A.~K. (2014c).
\newblock Studies on bound-state spectra of {M}anning-{R}osen potential.
\newblock {\em Mod. Phys. Lett. A\/}~{\em 29}, 1450042.

\bibitem[\protect\citeauthoryear{Roy}{Roy}{2014d}]{roy14jmc}
Roy, A.~K. (2014d).
\newblock Ro-vibrational spectroscopy of molecules represented by {T}ietz-{H}ua
  oscillator potential.
\newblock {\em J. Math. Chem\/}~{\em 52}, 1405.

\bibitem[\protect\citeauthoryear{Roy}{Roy}{2015}]{roy15ijqc}
Roy, A.~K. (2015).
\newblock Spherical confinement of {C}oulombic systems inside an impenetrable
  box: {H} atom and the {H}\'ulthen potential.
\newblock {\em Int. J. Quant. Chem.\/}~{\em 115}, 937.

\bibitem[\protect\citeauthoryear{Roy}{Roy}{2016}]{roy16ijqc}
Roy, A.~K. (2016).
\newblock Critical parameters and spherical confinement of {H} atoms in
  screened {C}oulomb potential.
\newblock {\em Int. J. Quant. Chem.\/}~{\em 116}, 953.

\bibitem[\protect\citeauthoryear{{Abramowitz, M.; Stegun, I.}}{{Abramowitz, M.;
  Stegun, I.}}{1964}]{abramowitz64}
{Abramowitz, M.; Stegun, I.} (1964).
\newblock {\em Handbook of Mathematical Functions}.
\newblock New York: Dover.

\bibitem[\protect\citeauthoryear{{Anderson; J. B.}}{{Anderson; J.
  B.}}{1975}]{anderson75}
{Anderson; J. B.} (1975).
\newblock A randon-walk simulation of the {S}chr\"odinger equation: H$_3^+$.
\newblock {\em J. Chem. Phys.\/}~{\em 63}, 1499.

\bibitem[\protect\citeauthoryear{{Kosloff, R.; Tal-Ezer, H.}}{{Kosloff, R.;
  Tal-Ezer, H.}}{1986}]{kosloff86}
{Kosloff, R.; Tal-Ezer, H.} (1986).
\newblock A direct relaxation method for calculating eigenfunctions and
  eigenvalues of the {S}chr\"odinger equation on a grid.
\newblock {\em Chem. Phys. Lett.\/}~{\em 127}, 223.

\bibitem[\protect\citeauthoryear{{Lehtovaara, L.; Toivanen, J.; Eloranta,
  J.}}{{Lehtovaara, L.; Toivanen, J.; Eloranta, J.}}{2007}]{lehtovaara07}
{Lehtovaara, L.; Toivanen, J.; Eloranta, J.} (2007).
\newblock Solution of time-dependent {S}chr\"odinger equation by the
  imaginary-time propagation method.
\newblock {\em J. Comput. Phys.\/}~{\em 221}, 148.

\bibitem[\protect\citeauthoryear{{Chin, S. A.; Janecek, S.; Krotscheck,
  S.}}{{Chin, S. A.; Janecek, S.; Krotscheck, S.}}{2009}]{chin09}
{Chin, S. A.; Janecek, S.; Krotscheck, S.} (2009).
\newblock Any order imaginary time propagation method for solving the
  {S}chr\"odinger equation.
\newblock {\em Chem. Phys. Lett.\/}~{\em 470}, 342.

\bibitem[\protect\citeauthoryear{{Sudiarta, I. W.; Wallace Geldart, D.
  J.}}{{Sudiarta, I. W.; Wallace Geldart, D. J.}}{2009}]{sudiarta09}
{Sudiarta, I. W.; Wallace Geldart, D. J.} (2009).
\newblock The finite difference time domain method for computing the
  single-particle density matrix.
\newblock {\em J. Phys. A\/}~{\em 42}, 285002.

\bibitem[\protect\citeauthoryear{{Strickland, M.; Yager-Elorriaga,
  D.}}{{Strickland, M.; Yager-Elorriaga, D.}}{2010}]{strickland10}
{Strickland, M.; Yager-Elorriaga, D.} (2010).
\newblock A parallel algorithm for solving the 3d {S}chr\"odinger equation.
\newblock {\em J. Comput. Phys.\/}~{\em 229}, 6015.

\bibitem[\protect\citeauthoryear{{Luukko, P. J. J.; R\"as\"anen, E. }}{{Luukko,
  P. J. J.; R\"as\"anen, E. }}{2013}]{luukko13}
{Luukko, P. J. J.; R\"as\"anen, E. } (2013).
\newblock Imaginary time propagation code for large-scale two-dimensional
  eigenvalue problems in magnetic fields.
\newblock {\em Comput. Phys. Commun.\/}~{\em 184}, 769.

\bibitem[\protect\citeauthoryear{{Roy, A. K.; Dey, B. K.; Deb, B. M.}}{{Roy, A.
  K.; Dey, B. K.; Deb, B. M.}}{1999}]{roy99}
{Roy, A. K.; Dey, B. K.; Deb, B. M.} (1999).
\newblock Direct calculation of ground-state electronic densities and
  properties of noble gas atoms through a single time-dependent hydrodynamical
  equation.
\newblock {\em Chem. Phys. Lett.\/}~{\em 308}, 523.

\bibitem[\protect\citeauthoryear{{Roy, A. K.; Chu, S. I.}}{{Roy, A. K.; Chu, S.
  I.}}{2002}]{roy02}
{Roy, A. K.; Chu, S. I.} (2002).
\newblock Quantum fluid dynamics approach for electronic structure calculation:
  {A}pplication to the study of ground-state properties of rare gas atoms.
\newblock {\em J. Phys. B\/}~{\em 35}, 2075.

\bibitem[\protect\citeauthoryear{{Roy A. K.; Gupta N.; Deb B. M.}}{{Roy A. K.;
  Gupta N.; Deb B. M.}}{2002}]{roy02a}
{Roy A. K.; Gupta N.; Deb B. M.} (2002).
\newblock Time-dependent quantum-mechanical calculation of ground and excited
  states of anharmonic and double-well oscillators.
\newblock {\em Phys. Rev. A\/}~{\em 65}, 012109.

\bibitem[\protect\citeauthoryear{{Gupta N.; Roy A. K.; Deb, B. M.}}{{Gupta N.;
  Roy A. K.; Deb, B. M.}}{2002}]{gupta02}
{Gupta N.; Roy A. K.; Deb, B. M.} (2002).
\newblock One-dimensional multiple-well oscillators: {A} time-dependent quantum
  mechanical approach.
\newblock {\em Pramana J. Phys.\/}~{\em 59}, 575.

\bibitem[\protect\citeauthoryear{{Wadehra, A.; Roy, A. K.; Deb, B. M.
  }}{{Wadehra, A.; Roy, A. K.; Deb, B. M. }}{2003}]{wadehra03}
{Wadehra, A.; Roy, A. K.; Deb, B. M. } (2003).
\newblock Ground and excited states of one-dimensional self-interacting
  nonlinear oscillators through time-dependent quantum mechanics.
\newblock {\em Int. J. Quant. Chem.\/}~{\em 91}, 597.

\bibitem[\protect\citeauthoryear{{Roy, A. K.; Thakkar, A. J.; Deb, B. M.
  }}{{Roy, A. K.; Thakkar, A. J.; Deb, B. M. }}{2005}]{roy05}
{Roy, A. K.; Thakkar, A. J.; Deb, B. M. } (2005).
\newblock Low-lying states of two-dimensional double-well potentials.
\newblock {\em J. Phys. A\/}~{\em 38}, 2189.

\bibitem[\protect\citeauthoryear{Roy}{Roy}{2014}]{roy14jmc_itp}
Roy, A.~K. (2014).
\newblock Ground and excited states of spherically symmetric potentials through
  an imaginary-time evolution method: application to spiked harmonic
  oscillators.
\newblock {\em J. Math. Chem\/}~{\em 52}, 2645.

\bibitem[\protect\citeauthoryear{{Hammond, B. L.; Lester Jr., W. A.; Reynolds,
  P. J.}}{{Hammond, B. L.; Lester Jr., W. A.; Reynolds, P.
  J.}}{1994}]{hammond94}
{Hammond, B. L.; Lester Jr., W. A.; Reynolds, P. J.} (1994).
\newblock {\em Monte Carlo Methods in Ab Initio Quantum Chemistry}.
\newblock Singapore: World Scientific.

\bibitem[\protect\citeauthoryear{{Roy, A. K.; Chu, S. I.}}{{Roy, A. K.; Chu, S.
  I.}}{2002a}]{roy02exc}
{Roy, A. K.; Chu, S. I.} (2002a).
\newblock Density functional calculations on singly- and doubly-excited rydberg
  states of many-electron systems.
\newblock {\em Phys. Rev. A\/}~{\em 65}, 052508.

\bibitem[\protect\citeauthoryear{{Roy, A. K.; Chu, S. I.}}{{Roy, A. K.; Chu, S.
  I.}}{2002b}]{roy02qfd}
{Roy, A. K.; Chu, S. I.} (2002b).
\newblock Quantum fluid dynamics approach for strong field processes:
  {A}application to the study of multiphoton ionization and high-order harmonic
  generation of {H}e and {N}e in intense laser fields.
\newblock {\em Phys. Rev. A\/}~{\em 65}, 043402.

\bibitem[\protect\citeauthoryear{{Mukherjee, N; Roy, A. K.}}{{Mukherjee, N;
  Roy, A. K.}}{2018}]{mukherjee18c}
{Mukherjee, N; Roy, A. K.} (2018).
\newblock Information-entropic measures in free and confined hydrogen atom.
\newblock {\em Int. J. Quant. Chem.\/}~{\em 118}, e25596.

\bibitem[\protect\citeauthoryear{{Cover, T. M.; Thomas, J. A.}}{{Cover, T. M.;
  Thomas, J. A.}}{2006}]{cover06}
{Cover, T. M.; Thomas, J. A.} (2006).
\newblock {\em Elements of Information Theory}.
\newblock New York: John Wiley \& Sons, Ltd.

\bibitem[\protect\citeauthoryear{{Nielsen, M. A.; Chuang, I. L.}}{{Nielsen, M.
  A.; Chuang, I. L.}}{2010}]{nielsen10}
{Nielsen, M. A.; Chuang, I. L.} (2010).
\newblock {\em Quantum Computation and Quantum Information}.
\newblock Cambridge: Cambridge University Press.

\bibitem[\protect\citeauthoryear{{Ram\'{\i}rez, J. C.; Soriano, C.; Esquivel,
  R. O.; Sagar, R. P.; H\^o, M.; Smith, V. H.}}{{Ram\'{\i}rez, J. C.; Soriano,
  C.; Esquivel, R. O.; Sagar, R. P.; H\^o, M.; Smith, V. H.}}{1997}]{ramirez97}
{Ram\'{\i}rez, J. C.; Soriano, C.; Esquivel, R. O.; Sagar, R. P.; H\^o, M.;
  Smith, V. H.} (1997).
\newblock Jaynes information entropy of small molecules: {N}umerical evidence
  of the {C}ollins conjecture.
\newblock {\em Phys. Rev. A\/}~{\em 56}, 4477.

\bibitem[\protect\citeauthoryear{{Delle, L. S.}}{{Delle, L.
  S.}}{2015}]{delle15}
{Delle, L. S.} (2015).
\newblock Shannon entropy and many-electron correlations: {T}heoretical
  concepts, numerical results, and {C}ollins conjecture.
\newblock {\em Int. J. Quant. Chem.\/}~{\em 115}, 1396.

\bibitem[\protect\citeauthoryear{{Ghiringhelli, L. M.; Hamilton, I. P.; Delle,
  L. S}}{{Ghiringhelli, L. M.; Hamilton, I. P.; Delle, L.
  S}}{2010}]{ghiringhelli10}
{Ghiringhelli, L. M.; Hamilton, I. P.; Delle, L. S} (2010).
\newblock Interacting electrons, spin statistics, and information theory.
\newblock {\em J. Chem. Phys.\/}~{\em 132}, 014106.

\bibitem[\protect\citeauthoryear{{Ghiringhelli, L. M.; Delle, L. S.; Mosna, R.
  A.; Hamilton, I. P.}}{{Ghiringhelli, L. M.; Delle, L. S.; Mosna, R. A.;
  Hamilton, I. P.}}{2010}]{ghiringhelli10a}
{Ghiringhelli, L. M.; Delle, L. S.; Mosna, R. A.; Hamilton, I. P.} (2010).
\newblock Information-theoretic approach to kinetic-energy functionals: the
  nearly uniform electron gas.
\newblock {\em J. Math. Chem.\/}~{\em 48}, 78.

\bibitem[\protect\citeauthoryear{{Nagy, \'A.}}{{Nagy, \'A.}}{2014}]{nagy14}
{Nagy, \'A.} (2014).
\newblock Fisher and {S}hannon information in orbital-free density functional
  theory.
\newblock {\em Int. J. Quant. Chem.\/}~{\em 115}, 1392.

\bibitem[\protect\citeauthoryear{{Alipour, M.}}{{Alipour,
  M.}}{2015}]{alipour15}
{Alipour, M.} (2015).
\newblock Making a happy match between orbital-free density functional theory
  and information energy density.
\newblock {\em Chem. Phys. Lett.\/}~{\em 635}, 210.

\bibitem[\protect\citeauthoryear{{Delle, L. S.}}{{Delle, L.
  S.}}{2009a}]{delle09}
{Delle, L. S.} (2009a).
\newblock On the scaling properties of the correlation term of the electron
  kinetic functional and its relation to the {S}hannon measure.
\newblock {\em Euro Phys. Lett.\/}~{\em 86}, 40004.

\bibitem[\protect\citeauthoryear{{Delle, L. S.}}{{Delle, L.
  S.}}{2009b}]{delle09a}
{Delle, L. S.} (2009b).
\newblock Erratum: {O}n the scaling properties of the correlation term of the
  electron kinetic functional and its relation to the {S}hannon measure.
\newblock {\em Euro Phys. Lett.\/}~{\em 88}, 19901.

\bibitem[\protect\citeauthoryear{{ Mohajeri, A.; Alipour, M.}}{{ Mohajeri, A.;
  Alipour, M.}}{2009}]{mohajeri09}
{ Mohajeri, A.; Alipour, M.} (2009).
\newblock Information energy as an electron correlation measure in atomic and
  molecular systems.
\newblock {\em Int. J. Quantum Inf.\/}~{\em 07}, 801.

\bibitem[\protect\citeauthoryear{{Grassi, A.}}{{Grassi, A.}}{2011}]{grassi11}
{Grassi, A.} (2011).
\newblock A relationship between atomic correlation energy of neutral atoms and
  generalized entropy.
\newblock {\em Int. J. Quant. Chem.\/}~{\em 111}, 2390.

\bibitem[\protect\citeauthoryear{{Flores-Gallegos, N.}}{{Flores-Gallegos,
  N.}}{2016}]{gallegos16}
{Flores-Gallegos, N.} (2016).
\newblock Informational energy as a measure of electron correlation.
\newblock {\em Chem. Phys. Lett.\/}~{\em 666}, 62.

\bibitem[\protect\citeauthoryear{{Alipour, M.; Badooei, Z.}}{{Alipour, M.;
  Badooei, Z.}}{2018}]{alipour18}
{Alipour, M.; Badooei, Z.} (2018).
\newblock Toward electron correlation and electronic properties from the
  perspective of information functional theory.
\newblock {\em J. Phys. Chem. A\/}~{\em 122}, 6424.

\bibitem[\protect\citeauthoryear{{Barghathi, H.; Herdman, C.; Maestro, A.
  D.}}{{Barghathi, H.; Herdman, C.; Maestro, A. D.}}{2018}]{barghathi18}
{Barghathi, H.; Herdman, C.; Maestro, A. D.} (2018).
\newblock R\'enyi generalization of the accessible entanglement entropy.
\newblock {\em Phys. Rev. Lett.\/}~{\em 121}, 150501.

\bibitem[\protect\citeauthoryear{{Noorizadeh, S.; Shakerzadeh,
  E.}}{{Noorizadeh, S.; Shakerzadeh, E.}}{2010}]{noorizadeh10}
{Noorizadeh, S.; Shakerzadeh, E.} (2010).
\newblock {Shannon entropy as a new measure of aromaticity, {S}hannon
  aromaticity}.
\newblock {\em Phys. Chem. Chem. Phys.\/}~{\em 12}, 4742.

\bibitem[\protect\citeauthoryear{{R\'enyi, A.}}{{R\'enyi, A.}}{1961}]{renyi61}
{R\'enyi, A.} (1961).
\newblock On measures of entropy and information.
\newblock {\em Proc. Fourth Berkeley Symp. on Math. Statist. and Prob.\/}, 547.

\bibitem[\protect\citeauthoryear{{R\'enyi, A.}}{{R\'enyi, A.}}{1970}]{renyi70}
{R\'enyi, A.} (1970).
\newblock {\em Probability Theory}.
\newblock Amsterdam: North-Holland Pub. Company.

\bibitem[\protect\citeauthoryear{{Varga, I.; Pipek, J.}}{{Varga, I.; Pipek,
  J.}}{2003}]{varga03}
{Varga, I.; Pipek, J.} (2003).
\newblock R\'enyi entropies characterizing the shape and the extension of the
  phase space representation of quantum wave functions in disordered systems.
\newblock {\em Phys. Rev. E\/}~{\em 68}, 026202.

\bibitem[\protect\citeauthoryear{{Renner, R.; Gisin, N.; Kraus. B.}}{{Renner,
  R.; Gisin, N.; Kraus. B.}}{2005}]{renner05}
{Renner, R.; Gisin, N.; Kraus. B.} (2005).
\newblock Information-theoretic security proof for quantum-key-distribution
  protocols.
\newblock {\em Phys. Rev. A\/}~{\em 72}, 012332.

\bibitem[\protect\citeauthoryear{{L\'evay, P.; Nagy, S.; Pipek, J.}}{{L\'evay,
  P.; Nagy, S.; Pipek, J.}}{2005}]{levay05}
{L\'evay, P.; Nagy, S.; Pipek, J.} (2005).
\newblock Elementary formula for entanglement entropies of fermionic systems.
\newblock {\em Phys. Rev. A\/}~{\em 72}, 022302.

\bibitem[\protect\citeauthoryear{{Verstraete, F.; Cirac, J. I.}}{{Verstraete,
  F.; Cirac, J. I.}}{2006}]{verstraete06}
{Verstraete, F.; Cirac, J. I.} (2006).
\newblock Matrix product states represent ground states faithfully.
\newblock {\em Phys. Rev. B\/}~{\em 73}, 094423.

\bibitem[\protect\citeauthoryear{{Bialas, A.; Czyz, W.; Zalewski, K.}}{{Bialas,
  A.; Czyz, W.; Zalewski, K.}}{2006}]{bialas06}
{Bialas, A.; Czyz, W.; Zalewski, K.} (2006).
\newblock Moments of the {W}igner function and {R}\'enyi entropies at
  freeze-out.
\newblock {\em Phys. Rev. C\/}~{\em 73}, 034912.

\bibitem[\protect\citeauthoryear{{Salcedo, L. L.}}{{Salcedo, L.
  L.}}{2009}]{salcedo09}
{Salcedo, L. L.} (2009).
\newblock Phase-space localization of antisymmetric functions.
\newblock {\em J. Math. Phys.\/}~{\em 50}, 012106.

\bibitem[\protect\citeauthoryear{{Nagy, \'A.; Romera. E.}}{{Nagy, \'A.; Romera.
  E.}}{2009}]{nagy09}
{Nagy, \'A.; Romera. E.} (2009).
\newblock Maximum {R}\'enyi entropy principle and the generalized
  {T}homas-{F}ermi model.
\newblock {\em Phys. Lett. A\/}~{\em 373}, 844.

\bibitem[\protect\citeauthoryear{{McMinis, J.; Tubman, N. M.}}{{McMinis, J.;
  Tubman, N. M.}}{2013}]{mcminis13}
{McMinis, J.; Tubman, N. M.} (2013).
\newblock R\'enyi entropy of the interacting fermi liquid.
\newblock {\em Phys. Rev. B\/}~{\em 87}, 081108(R).

\bibitem[\protect\citeauthoryear{{Liu, S.-B.; Rong, C.-Y.; Wu, Z.-M.; Lu,
  T.}}{{Liu, S.-B.; Rong, C.-Y.; Wu, Z.-M.; Lu, T.}}{2015}]{liu15}
{Liu, S.-B.; Rong, C.-Y.; Wu, Z.-M.; Lu, T.} (2015).
\newblock Moments of the {W}igner function and {R}\'enyi entropies at
  freeze-out.
\newblock {\em Acta. Phys. Chim. Sin.\/}~{\em 31}, 2057.

\bibitem[\protect\citeauthoryear{{Kim, I.}}{{Kim, I.}}{2018}]{kim18}
{Kim, I.} (2018).
\newblock { R\'enyi-$\alpha$ entropies of quantum states in closed form:
  {G}aussian states and a class of non-{G}aussian states}.
\newblock {\em Phys. Rev. E\/}~{\em 97}, 062141.

\bibitem[\protect\citeauthoryear{Frieden}{Frieden}{2004}]{frieden04}
Frieden, B.~R. (2004).
\newblock {\em Science from {F}isher Information: {A} Unification}.
\newblock Cambridge: Cambridge University Press.

\bibitem[\protect\citeauthoryear{{Sen, K. D.}}{{Sen, K. D.}}{2011}]{sen11}
{Sen, K. D.} (Ed.) (2011).
\newblock {\em Statistical {C}omplexity: {A}pplications in {E}lectronic
  {S}tructure}.
\newblock Netherland: Springer International Publishing.

\bibitem[\protect\citeauthoryear{{Romera, E.; S\'anchez-Moreno, P.; Dehesa, J.
  S.}}{{Romera, E.; S\'anchez-Moreno, P.; Dehesa, J. S.}}{2005}]{romera05}
{Romera, E.; S\'anchez-Moreno, P.; Dehesa, J. S.} (2005).
\newblock The {F}isher information of single-particle systems with a central
  potential.
\newblock {\em Chem. Phys. Lett.\/}~{\em 414}, 468.

\bibitem[\protect\citeauthoryear{{Mukherjee, N.; Majumdar, S.; Roy, A.
  K.}}{{Mukherjee, N.; Majumdar, S.; Roy, A. K.}}{2018}]{mukherjee18d}
{Mukherjee, N.; Majumdar, S.; Roy, A. K.} (2018).
\newblock Fisher information in confined hydrogen-like ions.
\newblock {\em Chem. Phys. Lett.\/}~{\em 691}, 449.

\bibitem[\protect\citeauthoryear{{Mukherjee, N.; Roy, A. K.}}{{Mukherjee, N.;
  Roy, A. K.}}{2018}]{mukherjee18}
{Mukherjee, N.; Roy, A. K.} (2018).
\newblock Fisher information in confined isotropic harmonic oscillator.
\newblock {\em Int. J. Quant. Chem.\/}~{\em 118}, e25727.

\bibitem[\protect\citeauthoryear{{Patil, S. H.; Sen, K. D.; Watson, N. A.;
  Montgomery Jr., H. E.}}{{Patil, S. H.; Sen, K. D.; Watson, N. A.; Montgomery
  Jr., H. E.}}{2007}]{patil07}
{Patil, S. H.; Sen, K. D.; Watson, N. A.; Montgomery Jr., H. E.} (2007).
\newblock Characteristic features of net information measures for constrained
  {C}oulomb potentials.
\newblock {\em J. Phys. B\/}~{\em 40}, 2147.

\bibitem[\protect\citeauthoryear{{Ghosal, A.; Mukherjee, N.; Roy, A.
  K.}}{{Ghosal, A.; Mukherjee, N.; Roy, A. K.}}{2016}]{ghosal16}
{Ghosal, A.; Mukherjee, N.; Roy, A. K.} (2016).
\newblock Information-entropic measures of a quantum harmonic oscillator in
  symmetric and asymmetric confinement within an impenetrable box.
\newblock {\em Ann. Phys. (Berlin)\/}~{\em 528}, 796.

\bibitem[\protect\citeauthoryear{{Onicescu, O.}}{{Onicescu,
  O.}}{1966}]{onicescu66}
{Onicescu, O.} (1966).
\newblock Energie informationnelle.
\newblock {\em Comptes Rendus Hebdomadaires des Seances de l’Academie des
  Sciences Serie A\/}~{\em 263}, 841.

\bibitem[\protect\citeauthoryear{{Pardo, L.}}{{Pardo, L.}}{1986}]{pardo86}
{Pardo, L.} (1986).
\newblock {Order-$\alpha$ weighted information energy}.
\newblock {\em Info. Sciences\/}~{\em 40}, 155.

\bibitem[\protect\citeauthoryear{{Bhatia, P.}}{{Bhatia, P.}}{1997}]{bhatia97}
{Bhatia, P.} (1997).
\newblock On measures of information energy.
\newblock {\em Info. Sciences\/}~{\em 97}, 233.

\bibitem[\protect\citeauthoryear{{Noughabi, H. A.; Chahkandi, M.}}{{Noughabi,
  H. A.; Chahkandi, M.}}{2015}]{noughabi15}
{Noughabi, H. A.; Chahkandi, M.} (2015).
\newblock Informational energy and its application in testing normality.
\newblock {\em Ann. Data Sci.\/}~{\em 2}, 391.

\bibitem[\protect\citeauthoryear{{Alcoba, D. R.; Torre, A.; Lain, L.;
  Massaccesi, G. E.; O{\~{n}}a, O. B.; Ayers, P. W.; Van Raemdonck, M.;
  Bultinck, P.; Van Neck, D.}}{{Alcoba, D. R.; Torre, A.; Lain, L.; Massaccesi,
  G. E.; O{\~{n}}a, O. B.; Ayers, P. W.; Van Raemdonck, M.; Bultinck, P.; Van
  Neck, D.}}{2016}]{alcoba16}
{Alcoba, D. R.; Torre, A.; Lain, L.; Massaccesi, G. E.; O{\~{n}}a, O. B.;
  Ayers, P. W.; Van Raemdonck, M.; Bultinck, P.; Van Neck, D.} (2016).
\newblock Performance of {S}hannon-entropy compacted {N}-electron wave
  functions for configuration interaction methods.
\newblock {\em Theor. Chem. Acc.\/}~{\em 135}, 153.

\bibitem[\protect\citeauthoryear{{Zozor, S.; Portesi, M.; Vignat, C.}}{{Zozor,
  S.; Portesi, M.; Vignat, C.}}{2008}]{zozor07}
{Zozor, S.; Portesi, M.; Vignat, C.} (2008).
\newblock Some extensions of the uncertainty principle.
\newblock {\em Physica A\/}~{\em 387}, 4800.

\bibitem[\protect\citeauthoryear{{Shiner, J. S.; Landsberg, P. T.; Davison,
  M.}}{{Shiner, J. S.; Landsberg, P. T.; Davison, M.}}{1999}]{shiner99}
{Shiner, J. S.; Landsberg, P. T.; Davison, M.} (1999).
\newblock Simple measure for complexity.
\newblock {\em Phys. Rev. E\/}~{\em 59}, 1459.

\bibitem[\protect\citeauthoryear{{Catal\'an, R. G.; Garay, J.; L\'opez-Ruiz,
  R.}}{{Catal\'an, R. G.; Garay, J.; L\'opez-Ruiz, R.}}{2002}]{catalan02}
{Catal\'an, R. G.; Garay, J.; L\'opez-Ruiz, R.} (2002).
\newblock Features of the extension of a statistical measure of complexity to
  continuous systems.
\newblock {\em Phys. Rev. E\/}~{\em 66}, 11102.

\bibitem[\protect\citeauthoryear{{S\'anchez, J. R.; L\'opez-Ruiz,
  R.}}{{S\'anchez, J. R.; L\'opez-Ruiz, R.}}{2005}]{sanchez05}
{S\'anchez, J. R.; L\'opez-Ruiz, R.} (2005).
\newblock A method to discern complexity in two-dimensional patterns generated
  by coupled map lattices.
\newblock {\em Physica A\/}~{\em 355}, 633.

\bibitem[\protect\citeauthoryear{{Landsberg, P. T.; Shiner; J. S.}}{{Landsberg,
  P. T.; Shiner; J. S.}}{1998}]{landsberg98}
{Landsberg, P. T.; Shiner; J. S.} (1998).
\newblock Disorder and complexity in an ideal non-equilibrium fermi gas.
\newblock {\em Phys. Lett. A\/}~{\em 245}, 228.

\bibitem[\protect\citeauthoryear{{Romera, E.; Dehesa, J. S.}}{{Romera, E.;
  Dehesa, J. S.}}{2004}]{romera04}
{Romera, E.; Dehesa, J. S.} (2004).
\newblock The {F}isher-{S}hannon information plane, an electron correlation
  tool.
\newblock {\em J. Chem. Phys.\/}~{\em 120}, 8906.

\bibitem[\protect\citeauthoryear{{Sen, K. D.; Antol\'in, J.; Angulo, J.
  C.}}{{Sen, K. D.; Antol\'in, J.; Angulo, J. C.}}{2007}]{sen07}
{Sen, K. D.; Antol\'in, J.; Angulo, J. C.} (2007).
\newblock Fisher-{S}hannon analysis of ionization processes and isoelectronic
  series.
\newblock {\em Phys. Rev. A\/}~{\em 76}, 032502.

\bibitem[\protect\citeauthoryear{{Sen, K. D.; Antol\'in, J.; Angulo, J.
  C.}}{{Sen, K. D.; Antol\'in, J.; Angulo, J. C.}}{2008}]{angulo08}
{Sen, K. D.; Antol\'in, J.; Angulo, J. C.} (2008).
\newblock Fisher-{S}hannon plane and statistical complexity of atoms.
\newblock {\em Phys. Rev. A\/}~{\em 372}, 670.

\bibitem[\protect\citeauthoryear{{Antol\'in, J.; Angulo, J. C.}}{{Antol\'in,
  J.; Angulo, J. C.}}{2009}]{antolin09}
{Antol\'in, J.; Angulo, J. C.} (2009).
\newblock Complexity analysis of ionization processes and isoelectronic series.
\newblock {\em Int. J. Quant. Chem.\/}~{\em 109}, 586.

\bibitem[\protect\citeauthoryear{{Calbet, X.; L\'opez-Ruiz, R.}}{{Calbet, X.;
  L\'opez-Ruiz, R.}}{2001}]{calbet01}
{Calbet, X.; L\'opez-Ruiz, R.} (2001).
\newblock Tendency towards maximum complexity in a nonequilibrium isolated
  system.
\newblock {\em Phys. Rev. E\/}~{\em 63}, 066116.

\bibitem[\protect\citeauthoryear{{Martin, M. T.; Plastino, A.; Rosso, A.
  O.}}{{Martin, M. T.; Plastino, A.; Rosso, A. O.}}{2006}]{martin06}
{Martin, M. T.; Plastino, A.; Rosso, A. O.} (2006).
\newblock Generalized statistical complexity measures: {G}eometrical and
  analytical properties.
\newblock {\em Physica A\/}~{\em 369}, 439.

\bibitem[\protect\citeauthoryear{{Romera E., Nagy, \'A.}}{{Romera E., Nagy,
  \'A.}}{2008}]{romera08}
{Romera E., Nagy, \'A.} (2008).
\newblock R\'enyi information of atoms.
\newblock {\em Phys. Lett. A\/}~{\em 372}, 4918.

\bibitem[\protect\citeauthoryear{{Angulo, J. C.; Antol\'in, J.}}{{Angulo, J.
  C.; Antol\'in, J.}}{2008}]{angulo08a}
{Angulo, J. C.; Antol\'in, J.} (2008).
\newblock Atomic complexity measures in position and momentum spaces.
\newblock {\em J. Chem. Phys.\/}~{\em 128}, 164109.

\bibitem[\protect\citeauthoryear{{Esquivel, R. O.; Angulo, J. C.; Antol\'in,
  J.; Dehesa, J. S.; L\'opez-Rosa, S.; Flores-Gallegos, N.}}{{Esquivel, R. O.;
  Angulo, J. C.; Antol\'in, J.; Dehesa, J. S.; L\'opez-Rosa, S.;
  Flores-Gallegos, N.}}{2010}]{esquivel10}
{Esquivel, R. O.; Angulo, J. C.; Antol\'in, J.; Dehesa, J. S.; L\'opez-Rosa,
  S.; Flores-Gallegos, N.} (2010).
\newblock Analysis of complexity measures and information planes of selected
  molecules in position and momentum spaces.
\newblock {\em Phys. Chem. Chem. Phys.\/}~{\em 12}, 7108.

\bibitem[\protect\citeauthoryear{{Welearegay, M. A.; Balawender, R.; Holas
  A.}}{{Welearegay, M. A.; Balawender, R.; Holas A.}}{2014}]{welearegay14}
{Welearegay, M. A.; Balawender, R.; Holas A.} (2014).
\newblock Information and complexity measures in molecular reactivity studies.
\newblock {\em Phys. Chem. Chem. Phys.\/}~{\em 16}, 14928.

\bibitem[\protect\citeauthoryear{{Esquivel, R. O.; Molina-Esp\'iritu, M.;
  Angulo, J. C.; Antol\'in, J.; Flores-Gallegos, N.; Dehesa, J. S.}}{{Esquivel,
  R. O.; Molina-Esp\'iritu, M.; Angulo, J. C.; Antol\'in, J.; Flores-Gallegos,
  N.; Dehesa, J. S.}}{2011}]{esquivel11}
{Esquivel, R. O.; Molina-Esp\'iritu, M.; Angulo, J. C.; Antol\'in, J.;
  Flores-Gallegos, N.; Dehesa, J. S.} (2011).
\newblock Information-theoretical complexity for the hydrogenic abstraction
  reaction.
\newblock {\em Mol. Phys.\/}~{\em 109}, 2353.

\bibitem[\protect\citeauthoryear{{Molina-Esp\'iritu, M.; Esquivel, R. O.;
  Angulo, J. C.; Antol\'in, J.; Dehesa, J. S.}}{{Molina-Esp\'iritu, M.;
  Esquivel, R. O.; Angulo, J. C.; Antol\'in, J.; Dehesa, J.
  S.}}{2012}]{molina12}
{Molina-Esp\'iritu, M.; Esquivel, R. O.; Angulo, J. C.; Antol\'in, J.; Dehesa,
  J. S.} (2012).
\newblock Information-theoretical complexity for the hydrogenic identity
  $s_{N}2$ exchange reaction.
\newblock {\em J. Math. Chem.\/}~{\em 50}, 1882.

\bibitem[\protect\citeauthoryear{{Esquivel, R. O.; Molina-Esp\'iritu, M.;
  Dehesa, J. S.; Angulo, J. C.; Antol\'in, J.}}{{Esquivel, R. O.;
  Molina-Esp\'iritu, M.; Dehesa, J. S.; Angulo, J. C.; Antol\'in,
  J.}}{2012}]{esquivel12}
{Esquivel, R. O.; Molina-Esp\'iritu, M.; Dehesa, J. S.; Angulo, J. C.;
  Antol\'in, J.} (2012).
\newblock Concurrent phenomena at the transition region of selected elementary
  chemical reactions: {A}n information‐theoretical complexity analysis.
\newblock {\em Int. J. Quant. Chem.\/}~{\em 112}, 3578.

\bibitem[\protect\citeauthoryear{{Kullback, S.; Leibler, R. A.}}{{Kullback, S.;
  Leibler, R. A.}}{1951}]{kullback51}
{Kullback, S.; Leibler, R. A.} (1951).
\newblock On information and sufficiency.
\newblock {\em Ann. Math. Stat.\/}~{\em 22}, 79.

\bibitem[\protect\citeauthoryear{{Kullback, S.}}{{Kullback,
  S.}}{1959}]{kullback78}
{Kullback, S.} (1959).
\newblock {\em Information Theory and Statistics}.
\newblock New York: John Wiley \& Sons.

\bibitem[\protect\citeauthoryear{{Sagar, R. P.; Guevara, N. L.}}{{Sagar, R. P.;
  Guevara, N. L.}}{2008}]{sagar08}
{Sagar, R. P.; Guevara, N. L.} (2008).
\newblock Relative entropy and atomic structure.
\newblock {\em J. Mol. Struct. (Theochem)\/}~{\em 857}, 72.

\bibitem[\protect\citeauthoryear{{Nagy, \'A.; Romera, E.}}{{Nagy, \'A.; Romera,
  E.}}{2009}]{nagy09a}
{Nagy, \'A.; Romera, E.} (2009).
\newblock Relative {R}\'enyi entropy for atoms.
\newblock {\em Int. J. Quant. Chem.\/}~{\em 109}, 2490.

\bibitem[\protect\citeauthoryear{{Villani, C.}}{{Villani,
  C.}}{2000}]{villani00}
{Villani, C.} (2000).
\newblock {\em Topics in Optimal Transportation, in Graduate Studies in
  Mathematics}, Volume~58.
\newblock New York: American Mathematical Society.

\bibitem[\protect\citeauthoryear{{Yamano, T.}}{{Yamano, T.}}{2013}]{yamano13}
{Yamano, T.} (2013).
\newblock Relative {F}isher information for {M}orse potential and isotropic
  quantum oscillators.
\newblock {\em J. Math. Phys.\/}~{\em 54}, 113301.

\bibitem[\protect\citeauthoryear{{S\'anchez-Moreno, P.; Zarzo, A.; Dehesa, J.
  S.}}{{S\'anchez-Moreno, P.; Zarzo, A.; Dehesa, J. S.}}{2012}]{sanchez12}
{S\'anchez-Moreno, P.; Zarzo, A.; Dehesa, J. S.} (2012).
\newblock Jensen divergence based on {F}isher's information.
\newblock {\em J. Phys. A\/}~{\em 45}, 125305.

\bibitem[\protect\citeauthoryear{{Toscani, G.}}{{Toscani,
  G.}}{2017}]{toscani17}
{Toscani, G.} (2017).
\newblock Poincar\'e-type inequalities for stable densities.
\newblock {\em Ric. Mat.\/}~{\em 66}, 15.

\bibitem[\protect\citeauthoryear{{Yamano, T.}}{{Yamano, T.}}{2013}]{yamano13a}
{Yamano, T.} (2013).
\newblock de bruijn-type identity for systems with flux.
\newblock {\em Eur. Phys. J. B\/}~{\em 86}, 363.

\bibitem[\protect\citeauthoryear{{Frieden, B. R.; Plastino, A.; Plastino, A.
  R.; Soffer, B. H.}}{{Frieden, B. R.; Plastino, A.; Plastino, A. R.; Soffer,
  B. H.}}{1999}]{frieden99}
{Frieden, B. R.; Plastino, A.; Plastino, A. R.; Soffer, B. H.} (1999).
\newblock Fisher-based thermodynamics: {I}ts {L}egendre transform and concavity
  properties.
\newblock {\em Phys. Rev. E\/}~{\em 60}, 48.

\bibitem[\protect\citeauthoryear{{Antol\'in, J.; Angulo, J. C.; L\'opez-Rosa,
  S.}}{{Antol\'in, J.; Angulo, J. C.; L\'opez-Rosa, S.}}{2009}]{antolin09a}
{Antol\'in, J.; Angulo, J. C.; L\'opez-Rosa, S.} (2009).
\newblock Fisher and {J}ensen-{S}hannon divergences: {Q}uantitative comparisons
  among distributions. {A}pplication to position and momentum atomic densities.
\newblock {\em J. Chem. Phys.\/}~{\em 130}, 074110.

\bibitem[\protect\citeauthoryear{{Lev\"am\"aki, H.; Nagy, \'A; Vilja, I.;
  Kokko, K.; Vitos, L.}}{{Lev\"am\"aki, H.; Nagy, \'A; Vilja, I.; Kokko, K.;
  Vitos, L.}}{2017}]{levamaki17}
{Lev\"am\"aki, H.; Nagy, \'A; Vilja, I.; Kokko, K.; Vitos, L.} (2017).
\newblock Kullback-{L}eibler and relative {F}isher information as descriptors
  of locality.
\newblock {\em Int. J. Quant. Chem.\/}~{\em 118}, e25557.

\bibitem[\protect\citeauthoryear{{Flego, S. P.; Plastino, A.; Plastino, A.
  R.}}{{Flego, S. P.; Plastino, A.; Plastino, A. R.}}{2011}]{flego11}
{Flego, S. P.; Plastino, A.; Plastino, A. R.} (2011).
\newblock Non-linear relativistic diffusions.
\newblock {\em Physica A\/}~{\em 390}, 2776.

\bibitem[\protect\citeauthoryear{{Frieden, B. R.; Soffer, B. H.}}{{Frieden, B.
  R.; Soffer, B. H.}}{2010}]{frieden10}
{Frieden, B. R.; Soffer, B. H.} (2010).
\newblock Weighted {F}isher informations, their derivation and use.
\newblock {\em Phys. Lett. A\/}~{\em 374}, 3895.

\bibitem[\protect\citeauthoryear{{Flego, S. P.; Plastino, A.; Plastino, A.
  R.}}{{Flego, S. P.; Plastino, A.; Plastino, A. R.}}{2011}]{flego11a}
{Flego, S. P.; Plastino, A.; Plastino, A. R.} (2011).
\newblock Fisher information, the {H}ellmann-{F}eynman theorem, and the
  {J}aynes reciprocity relations.
\newblock {\em Ann. Phys.\/}~{\em 326}, 2533.

\bibitem[\protect\citeauthoryear{{Venkatesan, R. C.; Plastino,
  A.}}{{Venkatesan, R. C.; Plastino, A.}}{2014}]{venkatesan14}
{Venkatesan, R. C.; Plastino, A.} (2014).
\newblock Legendre transform structure and extremal properties of the relative
  {F}isher information.
\newblock {\em Phys. Lett. A\/}~{\em 378}, 1341.

\bibitem[\protect\citeauthoryear{{Venkatesan, R. C.; Plastino,
  A.}}{{Venkatesan, R. C.; Plastino, A.}}{2015}]{venkatesan15}
{Venkatesan, R. C.; Plastino, A.} (2015).
\newblock {Hellmann-{F}eynman connection for the relative {F}isher
  information}.
\newblock {\em Ann. Phys.\/}~{\em 359}, 300.

\bibitem[\protect\citeauthoryear{{Mukherjee, N.; Roy, A. K.}}{{Mukherjee, N.;
  Roy, A. K.}}{2018}]{mukherjee18b}
{Mukherjee, N.; Roy, A. K.} (2018).
\newblock Relative {F}isher information in some central potentials.
\newblock {\em Ann. Phys. (N.Y.)\/}~{\em 398}, 190.

\bibitem[\protect\citeauthoryear{{Yamano, T.}}{{Yamano, T.}}{2018}]{yamano18}
{Yamano, T.} (2018).
\newblock Relative {F}isher information of hydrogen-like atoms.
\newblock {\em Chem. Phys. Lett.\/}~{\em 691}, 196.

\bibitem[\protect\citeauthoryear{{Laguna, H. G.; Sagar, R. P.}}{{Laguna, H. G.;
  Sagar, R. P.}}{2014}]{laguna14}
{Laguna, H. G.; Sagar, R. P.} (2014).
\newblock Quantum uncertainties of the confined harmonic oscillator in
  position, momentum and phase‐space.
\newblock {\em Ann. Phys. (Berlin)\/}~{\em 526}, 555.

\bibitem[\protect\citeauthoryear{{Griffiths, D. J.}}{{Griffiths, D.
  J.}}{2004}]{griffiths04}
{Griffiths, D. J.} (2004).
\newblock {\em Introduction to Quantum Mechanics}.
\newblock New Jersey: Pearson Prentice Hall.

\bibitem[\protect\citeauthoryear{{Mukherjee, N.; Roy, A.; Roy, A.
  K.}}{{Mukherjee, N.; Roy, A.; Roy, A. K.}}{2015}]{mukherjee15}
{Mukherjee, N.; Roy, A.; Roy, A. K.} (2015).
\newblock Information entropy as a measure of tunneling and quantum confinement
  in a symmetric double-well potential.
\newblock {\em Ann. Phys. (Berlin)\/}~{\em 527}, 825.

\bibitem[\protect\citeauthoryear{{Mukherjee, N.; Roy, A. K.}}{{Mukherjee, N.;
  Roy, A. K.}}{2016}]{mukherjee16}
{Mukherjee, N.; Roy, A. K.} (2016).
\newblock Quantum confinement in an asymmetric double-well potential through
  energy analysis and information-entropic measure.
\newblock {\em Ann. Phys. (Berlin)\/}~{\em 528}, 412.

\bibitem[\protect\citeauthoryear{{Cohen-Tannoudji, C.; Diu, B.; Laloe,
  F.;}}{{Cohen-Tannoudji, C.; Diu, B.; Laloe, F.;}}{1978}]{cohen78}
{Cohen-Tannoudji, C.; Diu, B.; Laloe, F.;} (1978).
\newblock {\em Quantum Mechanics}.
\newblock New York: Wiley.

\bibitem[\protect\citeauthoryear{Gueorguiev}{Gueorguiev}{2006}]{Gueorguiev06}
Gueorguiev, V. G.;~Rau, A. R. P. D. J.~P. (2006).
\newblock Confined one-dimensional harmonic oscillator as a two-mode system.
\newblock {\em American Journal of Physics\/}~{\em 74}, 394.

\bibitem[\protect\citeauthoryear{{Mukherjee, N.; Roy, A. K.}}{{Mukherjee, N.;
  Roy, A. K.}}{2018}]{mukherjee18a}
{Mukherjee, N.; Roy, A. K.} (2018).
\newblock Information{$-$}entropic measures in confined isotropic harmonic
  oscillator.
\newblock {\em Adv. Theory Simul.\/}~{\em 1}, 1800090.

\bibitem[\protect\citeauthoryear{{Y\'a\~nez, R. J.; Van Assche, W.; Dehesa, J.
  S.}}{{Y\'a\~nez, R. J.; Van Assche, W.; Dehesa, J. S.}}{1994}]{yanez94}
{Y\'a\~nez, R. J.; Van Assche, W.; Dehesa, J. S.} (1994).
\newblock Position and momentum information entropies of the {D}-dimensional
  harmonic oscillator and hydrogen atom.
\newblock {\em Phys. Rev. A\/}~{\em 50}, 3065.

\bibitem[\protect\citeauthoryear{Mukherjee}{Mukherjee}{2019}]{mukherjee19}
Mukherjee, N.;~Roy, A.~K. (2019).
\newblock Quantum mechanical virial-like theorem for confined quantum systems.
\newblock {\em Phys. Rev. A\/}~{\em 99}, 022123.

\bibitem[\protect\citeauthoryear{{Roy, A. K.}}{{Roy, A. K.}}{2014}]{roy14}
{Roy, A. K.} (2014).
\newblock Confinement in 3{D} polynomial oscillators through a generalized
  pseudospectral method.
\newblock {\em Mod. Phys. Lett. A\/}~{\em 29}, 1450104.

\end{thebibliography}
\label{lastpage-01}
\end{document}